\documentclass[twoside,a4paper,11pt]{book} 
\usepackage{amsmath,amssymb,graphicx,psfrag,fancyhdr,epsfig,oldgerm} 
\usepackage{cite,here,exscale}

\graphicspath{{figures/}}

\newcommand{\phm}{\phantom{-}}
\newcommand{\phone}{\phantom{1}}

\DeclareMathOperator{\Dfp}{\textfrak{D}_{\rm FP}}
\DeclareMathOperator{\Dpar}{\textfrak{D}_{\rm par}}
\DeclareMathOperator{\D}{\textfrak{D}}
\DeclareMathOperator{\Dt}{{\cal D}}
\DeclareMathOperator{\Dtfp}{{\cal D}_{\rm FP}} 
\DeclareMathOperator{\Dtpar}{{\cal D}_{\rm par}} 

\DeclareMathOperator{\DW}{D_W}
\DeclareMathOperator{\R}{R}
\DeclareMathOperator{\Tr}{Tr}
\DeclareMathOperator{\tr}{tr}

\renewcommand{\i}{\rm i}
\newcommand{\la}{\lower.7ex\hbox{$\;\stackrel{\textstyle<}{\sim}\;$}}

\newcommand{\cpages}{\clearpage{\thispagestyle{empty}\cleardoublepage}}

\begin{document}

\allowdisplaybreaks[1]

\frontmatter

\input macros.sty

\pagestyle{empty}
  \vspace{2cm}
  \sffamily\upshape\mdseries
  \noindent
  {\Huge \textbf{Chiral Measurements in Quenched\vspace*{0.4cm}\\Lattice QCD with Fixed Point\vspace*{0.4cm}\\Fermions}}

  \vspace{9.5cm}

  \noindent
  Inauguraldissertation \\
  der Philosophisch-naturwissenschaftlichen Fakult\"at \\
  der Universit\"at Bern

  \vspace{1.5cm}

  \noindent
  vorgelegt von

  \vspace{4mm}

  \noindent  {\large \textbf{Thomas J\"org}}

  \vspace{3mm}

  \noindent
  von Winterthur (ZH)

  \vspace{1.5cm}

  \noindent
  Leiter der Arbeit: \parbox[t]{7cm}{Prof.~Dr.~P.~Hasenfratz \\
    Institut f\"ur theoretische Physik \\
    Universit\"at Bern}
  
  \cleardoublepage


\pagestyle{fancy}

\tableofcontents

\mainmatter

\addcontentsline{toc}{chapter}{Abstract}
\chapter*{Abstract\markboth{Abstract}{}}
\label{cha:abstract}

The low-energy sector of Quantum Chromodynamics (QCD), which is
dominated by a very strong interaction between the quarks and the
gluons, is an excellent example of how an analytical and a numerical
approach can work together and thereby mutually increase their
predictive power.
 
Chiral Perturbation Theory ($\chi$PT) offers a systematic method to
predict analytical dependencies of different physical quantities in the
low-energy region of QCD. It is, however, an effective theory which is
based on the -- generally accepted -- assumption that chiral symmetry
is spontaneously broken in massless QCD and relies on data from other
sources to fix the low-energy constants in its Lagrangian.

Lattice QCD, which is the only non-perturbative regularization of QCD
known at the moment, is heavily based on numerical methods. It offers
the possibility to calculate quantities in QCD from first principles.
But it has shown to be a very delicate problem to incorporate chiral
symmetry on the lattice. A fact which lead to many technical and
fundamental problems. The recently rediscovered Ginsparg-Wilson relation
is equivalent to chiral symmetry on the lattice and with this a long
standing problem in Lattice QCD was solved. Having chiral fermions on
the lattice leads to a much better control in calculations where
chiral symmetry is important and allows to test the assumption of the
spontaneous breakdown of chiral symmetry in QCD in a cleaner way than
with traditional lattice fermions.

Fixed Point (FP) Dirac operators are designed to solve another big
problem in lattice simulations, the discretization errors introduced
by putting a theory on the lattice. Because FP Dirac operators have
been shown to satisfy the Ginsparg-Wilson relation, they combine two
important virtues one can wish a lattice Dirac operator to have.
 
In this work we construct a parametrization of a FP Dirac operator and
apply it in quenched Lattice QCD. The symmetry requirements for a
lattice Dirac operator are discussed and an efficient way to
make a practical construction of general lattice Dirac operators is
provided. We use such a general lattice Dirac operator to
approximately solve the Renormalization Group equation that defines
the FP Dirac operator in an iterative procedure. We discuss the
properties of this parametrization and show that its breaking of
chiral symmetry is much reduced compared to the most frequently used lattice
Dirac operator, the Wilson Dirac operator. Furthermore, we discuss the
overlap construction with the parametrized FP Dirac operator. With
this construction the remaining chiral symmetry breaking of the
parametrization can be reduced systematically to a desired level. We
show that the locality of the overlap construction with our Dirac
operator is improved with respect to the common overlap construction
with the Wilson Dirac operator. Using the overlap construction with
the parametrized FP Dirac operator we perform several test calculations,
where a precise chiral formulation of the lattice Dirac operator is
needed or desirable. Using the Atiyah-Singer index theorem, which is a consequence
of chiral symmetry, we calculate the quenched topological
susceptibility. The results are consistent with other recent
determinations, but they are not accurate enough to make a controlled
continuum extrapolation. We also confirm recent results about the
local chirality of near-zero modes of the Dirac operator. Finally, we
determine the low-energy constant $\Sigma$ of quenched Chiral
Perturbation Theory, which in full QCD with $3$ light flavors would
correspond to $-\langle\bar\psi\psi \rangle$, the order parameter of
spontaneous symmetry breaking.  


\cpages 
\chapter{Introduction}
\label{cha:intro}

Over the last 30 years Quantum Chromodynamics (QCD) has been very
successful in explaining many phenomena of the strong interactions.
Using standard perturbation theory, which was developed in the
framework of Quantum Electrodynamics (QED), processes involving high
energies can be studied in a systematic manner. However, at energies
below $1\, \gev$ the strong coupling constant $\alpha_s$ gets of
${\cal O}(1)$ such that standard perturbative methods fail -- a fact
which has many consequences.  Not only that the interesting physics of
particles like protons and neutrons, out of which nuclear matter in
our surroundings consists, is governed by the non-perturbative properties of
QCD, but also that confinement of quarks, a profound property of QCD,
makes that in experiments always bound states of quarks and gluons --
the so called hadrons -- are observed. This implies that in order to
compare the perturbative calculations with experiments one typically
needs information about the low-energy behaviour of QCD, e.g.~in form
of hadronic matrix elements.
 
Certain non-perturbative problems of QCD can be treated analytically
using an effective theory for the Goldstone bosons degrees of freedom. This approach assumes
that chiral symmetry is spontaneously broken \cite{Weinberg:1979kz,
Gasser:1982ap,Gasser:1983ky, Gasser:1984yg,Gasser:1985gg}.
This spontaneous breaking of chiral symmetry leads to the appearance
of massless excitations -- the Goldstone bosons -- if the mass of the
quarks is set to $0$. In nature the masses of the lightest quarks are
not $0$, but a few $\mev$ for the up and down quark and roughly $120\,
\mev$ for the strange quark. Hence, chiral symmetry is broken also
explicitly by the quark mass terms. It however shows that these masses
are still small enough compared to the intrinsic scale of QCD,
$\Lambda_{\rm QCD}$, that the effect of the spontaneous breakdown of
chiral symmetry dominates the low-energy structure.  Chiral
Perturbation Theory ($\chi$PT), which is a systematic expansion in the quark
masses and momenta of the light quarks, has been very successful in
predicting many of the low-energy properties of QCD, such as
e.g.~quark mass ratios, scattering lengths and, in general, properties
of the Goldstone bosons ($\pi$,K,$\bar{\rm K}$,$\eta$).  However,
being an effective theory the predictions of $\chi$PT rely on accurate
determinations of low-energy constants, which have to be extracted from
various experimental data.  For certain of these constants this shows
to be rather difficult, if it is possible at all. Furthermore,
$\chi$PT works well only within a limited range of energies, which
moreover may vary from process to process under consideration.
 
The lattice approach to QCD is a radically different way\footnote{QCD
  sum rules are another technique used to get non-perturbative results
  in QCD. As we do not use any sum rule results in this work, we refer
  the reader to the review \cite{Radyushkin:1998du}.} to have access
to the low-energy structure of QCD, as it gives a completely
non-perturbative definition of QCD.  The idea, which has been proposed
by Wilson in 1974 \cite{Wilson:1974sk}, is to discretize the QCD
action by replacing the continuum space-time by a discrete
four-dimensional lattice with a finite lattice spacing $a$. It is the
only definition of QCD beyond the perturbative level known at the
moment and it provides, at least in principle, a tool to study QCD
from first principles. Calculations in Lattice QCD involve the
numerical evaluation of the path integral defining the discretized QCD
action by stochastic sampling (Monte Carlo) techniques.  The way from
the point where one writes down the discretized QCD action to the
point where one can extract accurate physical data which is relevant
for QCD phenomenology has, however, shown to be more tedious and
longer than expected in the early 1980's. In order to understand this
we focus on two fundamental problems in lattice QCD: Chiral symmetry
and discretization errors.

In defining a disretized version of the gluonic as well as the
fermionic part of the Lattice QCD action one has, in fact, an infinite
choice of discretizations. The discretizations which are the most
popular are the Wilson gauge action and the Wilson Dirac operator.
These actions define the simplest way how the theory can be
discretized with the right continuum limit $a \to 0$ and essentially
without destroying the subtle structure of the anomalies of QCD. The
Wilson Dirac operator, however, in order to avoid the problem of
species doubling --- an effect we will discuss below --- introduces a
term that breaks chiral symmetry explicitly. This so-called Wilson
term is an irrelevant operator and therefore chiral symmetry is
restored in the continuum limit, but the costs are high. Due to the
breaking of chiral symmetry the Wilson operator introduces errors of
${\cal O}(a)$ into the lattice discretization and, furthermore, the
quark mass gets additively renormalized, operators in different chiral
representations get mixed and additional renormalization factors have
to be calculated. All these problems are essentially of technical
nature, but they make the extraction of physical data at least very
tedious, thereby hiding the underlying chiral structure of QCD. For
chiral gauge theories like the electroweak sector of the Standard
Model, however, the explicit breaking of chiral symmetry has so far
precluded all attempts to achieve a realistic lattice formulation.
Another approach to put fermions on the lattice, which is used very
often in simulations, are the so-called staggered or Kogut-Susskind
fermions \cite{Kogut:1975ag}.  They partially solve the problem of the
species doubling by reducing the number of doublers from 16 to 4 and
thereby also solve the problem of chiral symmetry breaking to a large
extent by leaving a chiral ${\rm U}(1)$ subgroup invariant.  However,
they introduce a new headache: they break flavor symmetry, which makes
the construction of hadron operators with the correct quantum numbers
rather difficult.  Even though staggered fermions have only ${\cal
  O}(a^2)$ discretization errors, these errors can be large
\cite{Hasenfratz:2001bz}, making it necessary to go far to the
continuum limit in order to get physical results that are not
contaminated too heavily by the discretization\footnote{There are
  several promising attempts to improve the cut-off effects of
  staggered fermions and to reduce the breaking of flavor symmetry
  (see references in \cite{Toussaint:2001zc}).}.

In order to reduce lattice artifacts Symanzik has proposed the idea
that one can add irrelevant higher order terms to the initial action
such that the leading cut-off effects are cancelled
\cite{Symanzik:1983dc,Symanzik:1983gh}.  This idea is implemented in
the L\"uscher-Weisz action \cite{Luscher:1985zq} and the
Sheikholeslami-Wohlert (clover) Dirac operator
\cite{Sheikholeslami:1985ij}. Similarly, other improvement programmes
were pursued, most prominently the non-perturbative clover
improvement, which eliminates all the ${\cal O}(a)$ effects
\cite{Luscher:1996sc}. This programme has shown to be very successful
in the determination of hadron masses, which show a very nice scaling
behaviour, whereas for other quantities the ${\cal O}(a^2)$ cut-off
effects can still be large.

This brings us to the main topic of this work, the fixed point
approach to QCD. Wilson's Renormalization Group (RG) approach offers a
radical way to treat the problem of lattice artifacts
\cite{Wilson:1971ag,Wilson:1971bg,Wilson:1974jj}.  Using so-called
quantum perfect actions one gets entirely rid of the lattice
artifacts. Such actions are unfortunately very difficult to
approximate and to really solve the defining RG equations is out of
reach.  Hasenfratz and Niedermayer have shown \cite{Hasenfratz:1994sp}
that for asymptotically free field theories -- like QCD -- it is
possible to define a so-called classically perfect or fixed point (FP)
action that reproduces the properties of the corresponding classical
theory without discretization errors. In fact, this corresponds to an
on-shell Symanzik improvement at tree-level to all order in $a$.  For
asymptotically free theories, where the continuum limit corresponds to
the limit where the coupling goes to $0$, such a FP action is expected
to be a close approximation to the quantum perfect action even at
non-zero coupling.

The FP Dirac operator $\Dfp$ does not only have reduced cut-off
effects, but as pointed out by Hasenfratz in 1997, it satisfies the
Ginsparg-Wilson (GW) relation \cite{Ginsparg:1982bj}
\begin{align}
  \label{eq:GW} 
  \{ \D,\gamma_5 \} &= \D {\gamma_5} 2 R \D \, ,\\
  \intertext{or equivalently}
  \label{eq:GW_propagator} 
  \, \{ \D^{-1},\gamma_5 \} &=  \gamma_5 2 R \, ,
\end{align}
where $\D$ is the lattice Dirac operator and $R$ is a local, hermitian
operator which commutes with $\gamma_5$.  The GW relation is
\emph{equivalent} to have chiral symmetry on the lattice, as L\"uscher
has shown in 1998 \cite{Luscher:1998pq}. Hence, FP actions offer a
solution to two of the fundamental problems -- chiral symmetry and
cut-off effects -- which have been making the extraction of physical
results from lattice simulations in QCD difficult over the last 20
years. Let us mention at this point that even though this work is
concerned mainly with the chiral aspects of the FP approach, i.e.~its
application to very light quarks, the FP QCD action has also the
potential to be used in lattice simulations with heavy quarks --
especially the charm quark -- because these simulations, which are
performed at momentum scales near the cut-off, are particularly
subject to distortions by the discretized action. The application of
FP actions to this topic, however, remains unexplored also in this
work.

There are other (approximate) solutions to the GW relation which are
essentially focussed on the aspect of chiral symmetry, such as e.g.~the
approach of Gattringer, Hip and Lang to approximate a solution of
the GW relation by a systematic expansion in operators built out of
paths up to a certain length
\cite{Gattringer:2000ja,Gattringer:2000js,Gattringer:2000qu}.  The
most prominent examples are, however, the 5 dimensional domain-wall
(DW) fermions \cite{Kaplan:1992bt,Shamir:1993zy,Furman:1995ky} and the
related overlap fermions
\cite{Frolov:1993ck,Narayanan:1993wx,Narayanan:1994sk}, which are both
based on a proposition made by Kaplan in 1992.  For more informations
on these approaches we refer to the recent reviews in
\cite{Blum:1998ud,Vranas:2000tz,Hernandez:2001yd} for the DW fermions
and to \cite{Hernandez:2001yd} for the overlap fermions.
Both constructions, in contrast to the FP Dirac operator, are defined
in such a way that it is more or less straight forward to implement
them in simulations, even though they are numerically very expensive.
In fact, the costs of simulations with (approximately) chiral fermions
in quenched QCD are roughly the same as full QCD simulations with
Wilson fermions, i.e.~that it is ${\cal O}(10) - {\cal O}(100)$ times
more expensive than quenched simulations with Wilson fermions.

Most of the simulations with (approximately) chiral fermions and also
the largest ones have been performed with DW fermions. Long standing
problems, where chiral symmetry plays an essential r\^{o}le, have been
tackled.  Examples are the weak interaction matrix elements, such as
those relevant for kaon physics, i.e.~the $B$-parameter $B_K$, the
$\Delta I = 1/2$ rule and the parameter of direct CP violation
$\epsilon^\prime/\epsilon$
\cite{AliKhan:2001wr,Noaki:2001un,Blum:2001xb}.  The results of these
very intricate calculations, however, are clearly a deception, as the
sign of $\epsilon^\prime/\epsilon$ measured on the lattice is opposite
to the experimental value and for the magnitude of relevant parameter
of the $\Delta I = 1/2$ rule the two large lattice simulations (RBC,
CP--PACS) differ roughly by a factor of $2$ from each other. This
raises the question where these problems come from. The answer is not
definitively clear, but it has shown that it is rather difficult to
have good control over the remaining breaking of chiral symmetry in
the DW fermion approach, even though the linear extent of the
fifth dimension $N_s$ can be used to extrapolate to the limit
$N_s\to\infty$, where the DW fermions get exactly chiral. But there is
evidence that the convergence is rather slow \cite{Vranas:2000tz}.
There are propositions how this problems can be solved \cite{Hernandez:2000iw,Edwards:2000qv} and also the
overlap fermion approach offers very good possibilities to control the
remaining breaking of chiral symmetry. Future simulations will show
whether the problems of the DW approach are really related to the
remaining symmetry breaking or whether other possible explanations as
e.g.~discussed in \cite{Noaki:2001un, Hasenfratz:2001bz} are more
relevant.

The definition of chiral fermions on the lattice has yielded the hope
that finally chiral gauge theories can be formulated on the lattice
and, indeed, the last years have seen a lot of progress in this area
\cite{Hasenfratz:1998jp,Neuberger:1998fp,Neuberger:1998wv,Hasenfratz:1998ri,
  Luscher:1998pq,Kikukawa:1998pd,Chiu:1998xf,Reisz:1999cm,Reisz:1999ck,
  Fujikawa:1998if,Suzuki:1998yz,Adams:1998eg,Adams:2000yi,Narayanan:1998uu,Niedermayer:1998bi,
  Luscher:1998kn,Fujiwara:1999fi,Kikukawa:2000kd,Luscher:1998du,Suzuki:1999qw,
  Luscher:1999un,Luscher:2000zd,Suzuki:2000ii,Igarashi:2000zi,Suzuki:2000ku}.
There remain, however, several open questions, like the relative weight factors between 
the topological sectors, the fate of CP symmetry and a related question 
about the definition of Majorana fermions \cite{Hasenfratz:2001bz,Fujikawa:2002is,Fujikawa:2002vj}.
Early hopes that a GW like approach might work for
supersymmetry also did not realize either.

Before we embark upon a more detailed discussion of chiral symmetry in
QCD and chiral fermions on the lattice --- in particular fixed point
fermions --- let us give an outline of this work.

In Chapter 2 we discuss the structure of a general lattice Dirac
operator that satisfies all the basic symmetry conditions, which are
gauge symmetry, hermiticity condition, charge conjugation, hypercubic
rotations and reflections. We give examples of terms that can
occur in such a general operator, which are specific for our
parametrization of the FP Dirac operator, and show how one can use
them in a practical application.

The details of the parametrization of the FP Dirac operator are
discussed in Chapter 3. The parametrization turns out to be quite
difficult and a lot of effort has been put into this part of this work,
because the results in the simulations clearly depend on the quality
of the parametrization.

In Chapter 4 we give a summary of the properties of the parametrized
FP Dirac operator. We mainly focus on the chiral properties, however,
also scaling properties are discussed to a certain extent. The main
source of data on scaling properties is the hadron spectroscopy study
which is presented in the PhD thesis of Simon Hauswirth
\cite{Hauswirth_diss:2002}.

Chapter 5 explains Neuberger's overlap formula and its application to the
overlap construction with the parametrized FP Dirac operator.
Furthermore, properties of this specific overlap operator are
discussed. In particular, we show that our overlap construction is
more local than the usual construction with the Wilson Dirac operator.

In Chapter 6 we present results for the quenched topological
susceptibility and the local chirality of near-zero modes.

The determination of the low-energy constant $\Sigma$ of quenched
Chiral Perturbation Theory, which in full QCD with 3 light flavors is
equal to the order parameter of spontaneous chiral symmetry breaking
$\langle \bar{\psi} \psi \rangle$, is given in Chapter 7.

Conclusions and prospects are given in Chapter 8.

Finally, this thesis covers only part of the work that has been done
in collaboration with Simon Hauswirth, Kieran Holland, Peter
Hasenfratz and Ferenc Niedermayer in an ongoing project, where the
parametrized FP Dirac operator is tested and applied to various
calculations in quenched QCD. Parts of it have already been published
in
\cite{Hasenfratz:2000qb,Hasenfratz:2000xz,Hasenfratz:2001hr,Hasenfratz:2001qp}
and much more information can be found in \cite{Hauswirth_diss:2002}.

\section{Chiral Symmetry in QCD}
\label{sec:QCD}

In this section we give a short overview of various aspects and
consequences of the symmetries of QCD. We will, however, not discuss
interesting properties of QCD, such as asymptotic freedom, confinement
and various other topics and refer the reader to one of the many books
about QCD e.g.~\cite{Muta:1998vi,Yndurain:1999ui}. We will keep the
discussion in the framework of QCD, because the discussion of the same
topics in quenched QCD is heavily loaded by various technicalities
that rather hide the structure of chiral symmetry and its
consequences. We will, however, mention the differences between the
quenched approximation and real QCD there where it is needed and we
will, in particular, point to the differences in those chapters where
we measure quantities in quenched QCD.


Symmetries are a very important concept in physics since their
presence always simplifies analysis and in certain cases allows one to
obtain exact or semi-exact results. In QCD the presence of light
quarks are the reason for an approximate symmetry and this allows one
to extract a lot of consequences concerning the dynamics of the
theory.

In Euclidean space, the Lagrangian of QCD with the gluon field strength tensor
${F}_{\mu\nu}$, the quark fields $\psi_f$, the Dirac operator ${\cal
  D}\!\!\!\!/$ and the gauge coupling $g$ reads
\begin{equation}
  \label{LQCD}
  {\cal L}_{QCD} = \frac{1}{4 g^2} {F}^{a}_{\mu\nu} {F}^{a}_{\mu\nu}
  \ +\ \sum_{f=1}^6 \bar\psi_f ({\cal D}\!\!\!\!/ + m_f) \psi_f \ , 
\end{equation} 
where $m_f$ are the quark masses. The up, down, and strange quarks are relatively light, with
masses $m_u \approx 3 \ {\rm MeV},\ m_d \approx 6\ {\rm MeV}$, and
$m_s \approx 120\ {\rm MeV}$\footnote{The light quark masses are
  current quark masses in the $\overline{\rm MS}$ scheme at $\mu = 2
  \, \gev$.}. There is a clear gap between $m_{u,d}$, which are
especially small, and the masses of the charm, bottom and top quark
with $m_c \approx 1.25 \,\gev$, $m_b \approx4.2 \,\gev$ and
$m_t=174.3\pm5.1\, \gev$ \cite{Groom:2000in}.  It makes sense to
consider the chiral limit when $N_f = 2$ or $N_f = 3$ quarks become
massless and the other quark masses are sent to infinity.  In this
limit, the Lagrangian (\ref{LQCD}) is invariant under the
transformations
\begin{align}
  \label{symvect}
  \delta_a \psi &= i \epsilon T_a \psi \,, &  \delta_a \bar \psi &= -i \epsilon\bar\psi T_a \,, \\
\intertext{and}
  \label{symax}
  \delta_a \psi &= i \epsilon \gamma_5 T_a \psi \,, &\delta_a \bar \psi &= i \epsilon \bar\psi T_a \gamma_5 \, ,
\end{align}
where $T_a$ ($a = 0, 1, \ldots, N_f^2 - 1$) are the (hermitian) generators of the
flavor SU$(N_f)$ group. The symmetry in eq.~\eqref{symvect} is the
vector symmetry and eq.~\eqref{symax} represents the axial symmetry.
While the vector symmetry is still present even if the quarks are
given a mass (of the same magnitude for all flavors), the axial
symmetry holds only in the massless theory. The corresponding Noether
currents are
\begin{equation}
  j^\mu_a \ = \ \bar \psi T_a \gamma^\mu \psi, \ \ \ \ j^{\mu 5}_a \ 
  = \ \bar \psi T_a \gamma^\mu \gamma_5 \psi\,.
  \label{curvecax}
\end{equation}
They are conserved upon applying the classical equations of motion.

\subsection{Singlet Axial Anomaly}

Let us discuss first the singlet axial symmetry (with $T_a = 1$).  An
important fact is that this symmetry exists only in the classical
case. The full quantum path integral is not invariant under the
transformations
\begin{equation} \delta \psi = i \epsilon\gamma_5 \psi,\ 
  \ \ \ \ \delta \bar\psi = i \epsilon \bar\psi \gamma_5\ ,
  \label{U1chir}
\end{equation} 
where the flavor index is now omitted.  This explicit symmetry
breaking due to quantum effects can be presented as an operator
identity involving an anomalous divergence,
\begin{equation}
  \label{chiranom}
  \partial_\mu j^{\mu 5} \ =\ \frac{N_f} {32 \pi^2}
  F^{a}_{\mu\nu} \tilde{F}^{a}_{\mu\nu}\ ,
\end{equation} 
where $ \tilde F^{a}_{\mu\nu} = \epsilon_{\mu\nu\alpha\beta}
F^{a}_{\alpha\beta}$.  There are many ways to derive and understand this
relation.  Historically, this so-called Adler-Bell-Jackiw (ABJ)
anomaly was first derived by purely diagrammatic methods and showed up
in the anomalous triangle graph \cite{Adler:1969er}.  Using the index
theorem of Atiyah and Singer \cite{Atiyah:1971rm}, which states
\begin{equation}
  \label{eq:AS_index}
  n_{\rm L} - n_{\rm R} \ =\ Q \ , 
\end{equation}
where $n_{\rm L,R}$ is the number of the left--handed (right--handed)
zero modes and $Q$ is the topological charge of the gauge field
configuration,
\begin{equation}
  \label{eq:topological_charge_gauge}
  Q \ =\ \frac {1}{32\pi^2} \int d^4x \ F^{a}_{\mu\nu}
  \tilde{F}^{a}_{\mu\nu} \,,   
\end{equation}
Fujikawa showed that in the case of the singlet axial symmetry the
anomaly can also be understood in connection with the Jacobian of the
change of the measure in the path integral under a global chiral
transformation \cite{Fujikawa:1979ay}.

The axial anomaly is by far not only of theoretical interest because it
implies that QCD has no singlet axial symmetry and therefore no
associated Goldstone boson can be found in the particle spectrum. This
is the explanation for the large mass of the $\eta^\prime$ particle,
which is 958 MeV, whereas the mass of the $\pi^0$ is 135 MeV. The
absence of a massless flavor singlet particle also provides a linear
relation between the topological susceptibility and the quark mass.
In contrast to this the quenched topological
susceptibility does not vanish in the chiral limit and is equal to the topological susceptibility
of pure SU(3) gauge theory.  Finally, through
the coupling to the electroweak sector of the standard model the axial
anomaly is responsible for the decay $\pi^0\to2\gamma$.

\subsection{Spontaneous Breaking of Non-singlet Chiral Symmetry}

Consider now the whole set of symmetries from eqs.~\eqref{symvect} and
\eqref{symax}.  It is convenient to introduce
\begin{equation}
  \label{psiLR}
  \psi_{\rm L,R} = \ \frac{1}{2} (1 \mp \gamma_5) \psi, \qquad \bar\psi_{\rm L,R}
  = \ \frac 12 \bar\psi (1 \pm \gamma_5)\ 
\end{equation}
and rewrite eqs.~\eqref{symvect} and \eqref{symax} as
\begin{equation}
  \label{chirtrans}
  \psi_{\rm L} \ \to \ V_{\rm L} \psi_{\rm L}, \ \ \ \ \psi_{\rm R} \ \to \ V_{\rm R} \psi_{\rm R}\ ,
\end{equation}
where $V_{\rm L}$ and $V_{\rm R}$ are two different U$(N_f)$ matrices.
The singlet axial transformations with $V_{\rm L} = V_{\rm R}^* =
e^{i\phi}$ are anomalous as in the theory with a single quark flavor.
Therefore, the true fermionic symmetry group of massless QCD is
\begin{equation}
  \label{SULSUR}
  {\cal G} \ = \ {\rm SU}_{\rm L}(N_f) \times {\rm SU}_{\rm R}(N_f) \times {\rm
    U}_{\rm V}(1)\,.  
\end{equation}
There is ample experimental evidence that the symmetry in
eq.~\eqref{SULSUR} is actually spontaneously broken, which means that
the vacuum state is not invariant under the action of the group ${\cal
  G}$. The symmetry ${\cal G}$ is, however, not broken completely. The
vacuum is still invariant under transformations with $V_{\rm L} =
V_{\rm R}$, generated by the vector current.

Thus, the pattern of breaking is
\begin{equation}
  \label{patbreak}
  {\rm SU}_{\rm L}(N_f) \times {\rm SU}_{\rm R}(N_f) \ \to {\rm SU}_{\rm V}(N_f)\,.  
\end{equation}
The vacuum expectation values
\begin{equation}
  \label{condfg}
  \Sigma^{\alpha \beta} = \langle \bar \psi_{\rm R}^\alpha \psi_{\rm L}^\beta \rangle 
\end{equation}
are the order parameters of the spontaneously broken axial symmetry.
The matrix $\Sigma^{\alpha \beta}$ is referred to as the \emph{quark
  condensate} matrix.

The non-breaking of the vector symmetry implies that the matrix order
parameter from eq.~\eqref{condfg} can be cast in the form
\begin{equation}
  \label{conddel}
  \Sigma^{\alpha \beta}\ = \ -\frac{1}{2}\, \Sigma \delta^{\alpha \beta} 
\end{equation}
by group transformations from eq.~\eqref{chirtrans}.  This means that
the general condensate matrix $\Sigma^{\alpha\beta}$ is a unitary
SU$(N_f)$ matrix multiplied by the real constant $\Sigma$.

By Goldstone's theorem, spontaneous breaking of a global continuous
symmetry leads to the appearance of purely massless Goldstone bosons.
Their number coincides with the number of broken generators, which is
$N_f^2 - 1$ in our case.  As it is the axial symmetry which is broken,
the Goldstone particles are pseudoscalars. They are the pions for $N_f
= 2$ or the octet $(\pi, K, \bar K, \eta)$ for $N_f = 3$.  It is a
fundamental and important fact that spontaneous breaking of continuous
symmetries not only creates massless Goldstone particles, but also
fixes the interactions of the latter at low energies: a fact which is
also called soft pion theorem.

The Goldstone particles are massless whereas all other states in the
physical spectrum have nonzero mass. Therefore, we have two distinct
energy scales and one can write down an effective Lagrangian depending
only on slow Goldstone fields with the fast degrees of freedom
corresponding to all other particles being integrated out. The
corresponding effective theory is called chiral perturbation theory.

The Lagrangian of real QCD from eq.~\eqref{LQCD} is not invariant
under the axial symmetry transformations just because quarks have
nonzero masses. However, the symmetry in eq.~\eqref{SULSUR} is still
very much relevant to QCD because some of the quarks happen to be very
light.  For $N_f = 2$, spontaneous breaking of an exact ${\rm SU}_{\rm
  L}(2) \times {\rm SU}_{\rm R}(2)$ symmetry would lead to the
existence of 3 strictly massless pions. As the symmetry is not quite
exact, the pions have a small mass. However, their mass $m_\pi$ goes
to zero in the chiral limit $m_{u,d} \to 0$. This fact is encoded in
the Gell-Mann-Oakes-Renner (GMOR) relation which can be derived from
the chiral Ward identities
\begin{equation}
  \label{GMOR}
  f_\pi^2 m_\pi^2 = 2 (m_u + m_d) \Sigma + {\cal O}(m_q^2)\,.  
\end{equation}
The constant $f_\pi$ appears also in the matrix element $$\langle 0 |
A_\mu^+ |\pi \rangle_p = i f_\pi p_\mu^\pi$$
of the axial-vector
current $A_\mu^+ = \bar d \gamma_\mu \gamma_5 u$ and determines the
charged pion decay rate.  Experimentally, $f_\pi \approx 131 $ MeV.

In contrast to $f_\pi$ it is not possible to give an accurate number
for $\Sigma$ from experimental data and therefore data from lattice
calculations can help to improve the estimates for the scalar
condensate. Closely related is the problem of the overall scale of the
quark masses; $\chi$PT can provide accurate predictions for the quark
mass ratios, but the exact scale can not be determined within this
framework.

This shows that the predictive power of $\chi$PT relies on
measurements which fix its low-energy constants. In certain cases
lattice calculations are the only viable source of data. This brings
us to our next topic, which is the lattice Dirac operator and the
problems to realize chiral symmetry on the lattice. In this short
introduction we do not cover the basic ideas of the lattice approach
to QCD and refer the reader to the books \cite{Montvay:1994cy,
  rothe92:_lattic_gauge_theor,creutz83:_quark_gluon_lattic} or the
more recent review \cite{Gupta:1997nd}.

\section{Chiral Symmetry on the Lattice}
\label{sec:lattice_chiral_symmetry}

This section deals with the general properties of lattice fermions.
After recalling the fermion doubling problem and the Nielsen-Ninomiya
No-Go theorem we shall discuss the recent progress made in formulating
chiral symmetry on the lattice.

\subsection{Fermion Doubling and the Nielsen-Ninomiya No-Go Theorem}

Suppose we want to describe massless free fermions on the lattice. The
fermionic part of the QCD lattice action can be written as
\begin{equation}
  {\cal A}_f = \sum_{x,y}\psibar(x)\,D(x-y)\,\psi(y),
\end{equation}
where~$D$ denotes the lattice Dirac operator. In particular, one would
like to formulate the theory such that~$D$ satisfies the following
conditions
\begin{itemize}
  \itemsep 3pt
\item[(a)] $D(x-y)$ is local, i.e.~the absolute values of its couplings $\rho(x-y)$ are bounded by an
  exponential function $\exp(-\nu |x-y|)$ with $\nu > 0$ (see also Section \ref{subsec:GW}).
\item[(b)] $D(p)=i\gamma_\mu p_\mu +O(ap^2)$
\item[(c)] $D(p)$ is invertible for $p\not=0$
\item[(d)] $\gamma_5\,D+D\,\gamma_5=0$
\end{itemize}
Locality is required in order to ensure renormalizability and
universality of the continuum limit; it ensures that a consistent
field theory is obtained. Furthermore, condition~(c) ensures that no
additional poles occur at non-zero momentum. If this is not satisfied,
as is the case for the so called ``naive'' discretization of the Dirac
operator, additional poles corresponding to spurious fermion states
can appear: this is the famous fermion doubling problem.  Finally,
condition~(d) implies that ${\cal A}_f$ is chirally invariant.

The main conclusion of the Nielsen-Ninomiya No-Go theorem
\cite{Nielsen:1981hk} is that conditions~(a)--(d) cannot be satisfied
simultaneously. Since one is not willing to give up locality and
condition~(b), this implies that one is usually confronted with the
choice of tolerating either doubler states or explicit chiral symmetry
breaking. This is manifest in the two most widely used lattice fermion
formulations: staggered fermions leave a chiral U(1) subgroup
invariant, but only partially reduce the number of doubler species
\cite{Kogut:1975ag}.  Wilson fermions, on the other hand, remove the
doublers entirely at the expense of breaking chiral symmetry
explicitly. This is easily seen from the expression for the free
Wilson Dirac operator
\begin{equation}
  D_W = \frac{1}{2}
  \gamma_\mu\left(\nabla_\mu+\nabla_\mu^*\right) 
  -\frac{1}{2}\nabla_\mu^*\nabla_\mu.
\end{equation}
Using the definitions for the forward and backward lattice
derivatives, $\nabla_\mu$ and $\nabla_\mu^*$, one easily proves
conditions~(a)--(c), while it is obvious that~(d) is not satisfied.

But even though the Nielsen-Ninomiya No-Go theorem is correct, there is
a way to have chiral symmetry on the lattice; the solution is to relax
the condition (d) in a particular way, which we discuss in the
following.

\subsection{The Ginsparg-Wilson Relation and some Consequences}
\label{subsec:GW}

As we already mentioned, formulations of chiral fermions on the
lattice have been found that result from completely different
constructions. However, they are all connected by the fact that they
satisfy the GW relation in eq.~\eqref{eq:GW}.

\begin{figure}[tbp]
  \begin{center}
    \includegraphics[width=5.5cm]{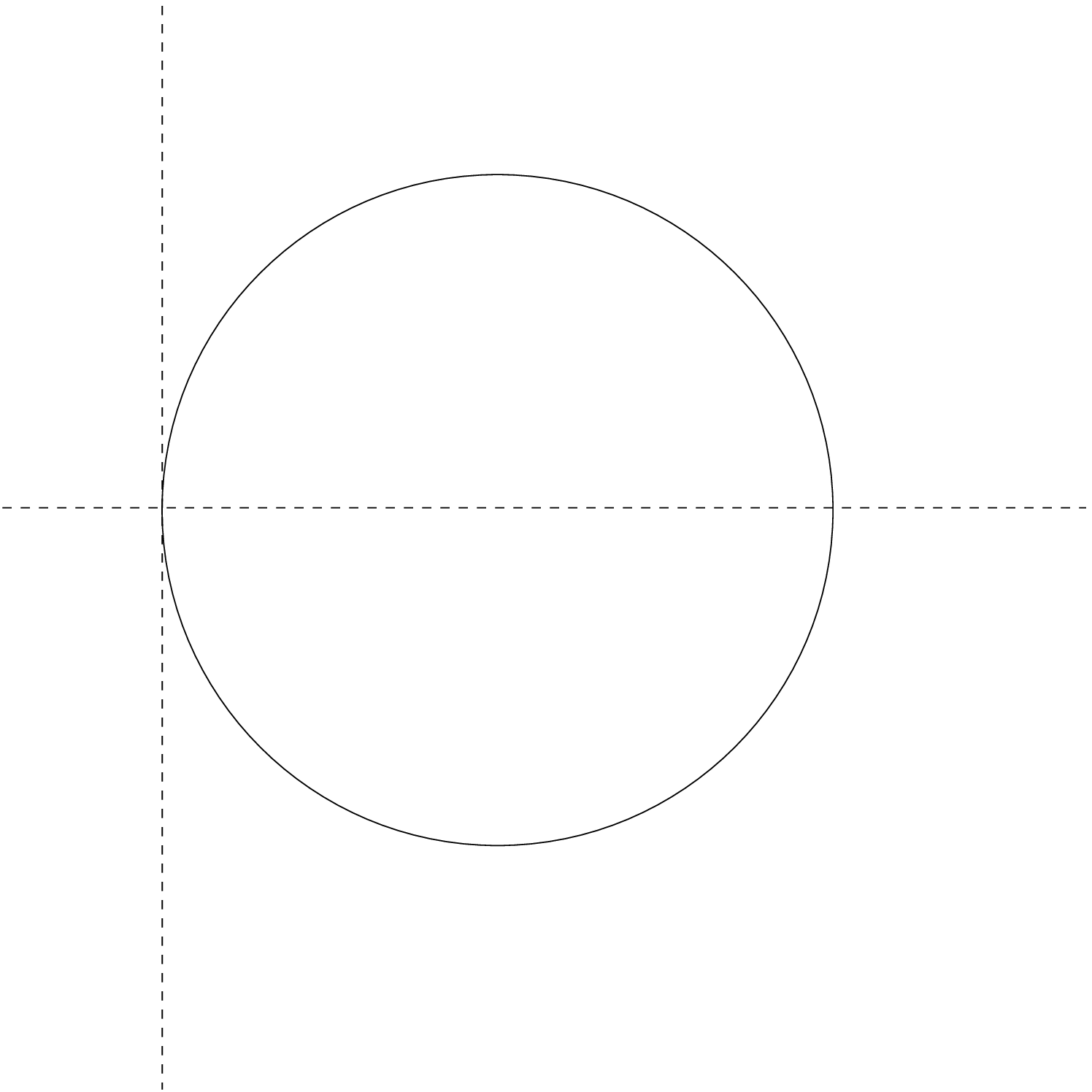} \hspace{0.2cm}
    \includegraphics[width=5.5cm]{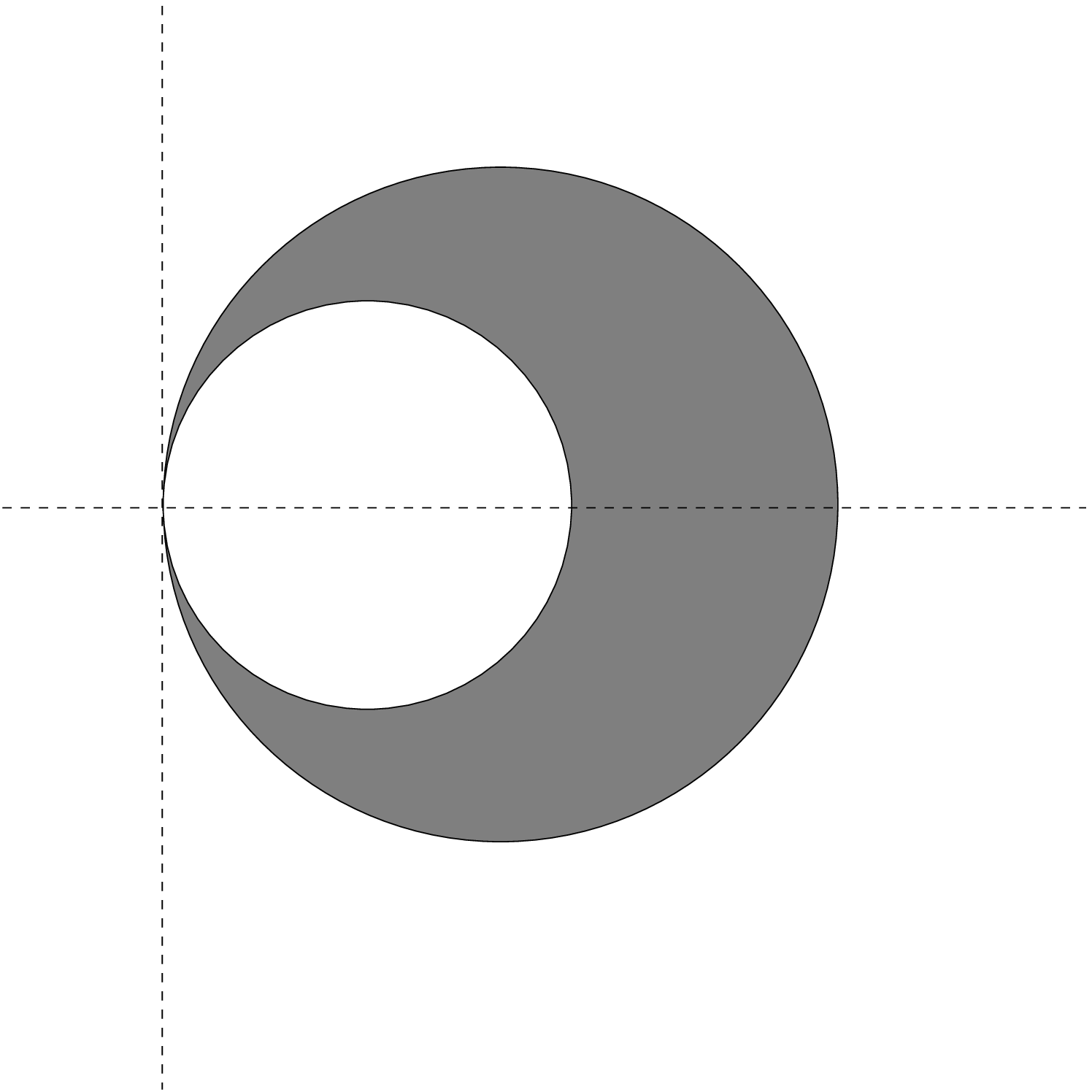}
    \caption{The eigenvalue spectrum of GW Dirac operators in the complex plane. In the 
      case $R \propto 1$ the spectrum lies exactly on a circle (left),
       whereas in the general case the spectrum is bounded by two
      circles and lies in the shaded area (right).}
    \label{fig:eigenvalue_spectra}
  \end{center}
\end{figure}

Before we start the discussion about the GW relation, let us
introduce some notations.  We denote a Dirac operator that satisfies
the GW relation, as it stands in eq.~\eqref{eq:GW}, i.e.~with a general
operator $R$, by $\D$ and, in particular, $\Dfp$ for the FP Dirac
operator.  In the special case when $R=1/2$ we use $\Dt$ to denote the
Dirac operator and the GW relation reads
\begin{align}
  \label{eq:GW_simple}
  \{\Dt, \gamma_5\}  &= \Dt \gamma_5 \Dt \,, \\
  \intertext{or equivalently}
  \label{eq:GW_simple_prop}
  \{\Dt^{-1}, \gamma_5\}  &= \gamma_5 \,. 
\end{align}
From this equation one sees that the Dirac operator $\Dt$ is normal,
i.e.~it commutes with its hermitian conjugate and therefore the
eigenstates of $\Dt$ are simultaneously eigenstates of $\Dt^\dagger$.
The GW relation in combination with the $\gamma_5$-hermiticity of the
Dirac operator implies furthermore that $\Dt$ and $\gamma_5$ commute
in the subspace of the real eigenmodes of $\Dt$. Hence, real
eigenmodes $\psi$ of $\Dt$, in particular zero modes, have definite
chirality, i.e.~$\gamma_5 \psi = \pm \psi$. Another consequence is
that the eigenvalue spectrum of $\Dt$ lies on a circle in the complex
plane with radius $1$ and the origin at $(1,0)$. Note however that
this is not the case for a general $R$. In this case the eigenvalue
spectrum is bounded by 2 circles, one with radius $\lambda_{\rm
  min}^{-1}/2$, the other with radius $\lambda_{\rm max}^{-1}/2$, where
$\lambda_{\rm min}$ and $\lambda_{\rm max}$ are the smallest and the
largest eigenvalue of $R$, respectively\footnote{The operator $R$ is local and
  hermitian and its eigenvalues are real and bounded by a
  constant $\lambda > 0$.}. In all cases the circles touch the
imaginary axis at the origin, as shown in Figure
\ref{fig:eigenvalue_spectra}.

The Dirac operator $\D$ can always be rescaled as follows
\begin{equation}
  \label{eq:D_DT}
  (2 R)^{1/2} \D (2 R)^{1/2} = \Dt \,,
\end{equation}
which is often very convenient to derive relations in a simpler way
and therefore we will use $\Dt$ in our following discussion; even
though all the results -- with the obvious modifications -- are also
valid for $\D$. The rescaled FP solution is denoted by $\Dtfp$.

In order to see that the GW relation is indeed equivalent to chiral
symmetry with a non-zero lattice spacing, we perform an infinitesimal
change of variables $\psi \to \psi + i\epsilon\delta\psi$ and
$\overline{\psi} \to \overline{\psi} + i\epsilon\delta\overline{\psi}$
with a global flavor singlet transformation \cite{Luscher:1998pq}
\begin{eqnarray}
  \delta\psi &=& \gamma_5 \left( 1-\frac{1}{2}\Dt\right)\psi \label{35} \\
  \delta\overline{\psi} &=& 
  \overline{\psi}\left( 1-\frac{1}{2}\Dt\right)\gamma_5  \nonumber \,.
\end{eqnarray}
From the GW relation eq.~\eqref{eq:GW_simple} it follows that the action ${\cal
  A}_f = \overline{\psi} \Dt \psi$ is invariant under this
transformation, i.e.
\begin{equation}
  \delta(\overline{\psi} \Dt \psi ) = 0 \,.
\end{equation}
The corresponding flavor non-singlet transformation is given by
\begin{eqnarray}
  \delta_a \psi &=& T_a \gamma_5 \left( 1-\frac{1}{2} \Dt \right)\psi \label{36} \, , \\
  \delta_a \overline{\psi} &=& 
  \overline{\psi}\left( 1-\frac{1}{2}\Dt \right)\gamma_5 T_a  \nonumber \, .
\end{eqnarray}

Analogous to Fujikawa's observation discussed in Section \ref{sec:QCD}
the fermionic integration measure is {\em not} invariant under the
singlet transformation in eq.~\eqref{35}
\begin{eqnarray}
  \delta[d\overline{\psi}d\psi] &=& {\rm Tr}(\gamma_5 \Dt)
  [d\overline{\psi}d\psi] \label{37} \\
  &=& 2N_{\rm f}(n_{\rm L} - n_{\rm R})[d\overline{\psi}d\psi] \nonumber\, ,
\end{eqnarray}
where the trace Tr denotes the trace over all indices, i.e.~color, flavor,
Dirac and spatial indices.  To derive eq.~\eqref{37} we make use of the
lattice version of the Atiyah-Singer index theorem
\cite{Hasenfratz:1998ri}.  Hence, the measure breaks the flavor
singlet chiral symmetry in a topologically non-trivial gauge field
where the ${\rm index}(\Dt)= n_{\rm L} - n_{\rm R} \ne 0$. Before we
discuss the index theorem in more detail, let us mention that the
non-singlet symmetry is not anomalous, since eq.~(\ref{37}) for this
case contains ${\rm Tr}(\gamma_5 \Dt T)\propto {\rm tr}(T)=0$.

Having a Dirac operator satisfying the GW relation one can easily show
that the following identity, the lattice index theorem, holds
\begin{equation}
  \label{eq:index_theorem}
  {\rm index}(\Dt) = \frac{1}{2} {\rm Tr}(\gamma_5 \Dt)\, .
\end{equation}
Furthermore, one can derive
\begin{equation}
  \label{eq:index_theorem_continuum}
  \frac{1}{2}{\rm tr}(\gamma_5 \Dt)\,  \to \frac{1}{32\pi^2} F^{a}_{\mu\nu} \tilde F^{a}_{\mu\nu}
\end{equation}
in the continuum limit \cite{Fujikawa:1998if,Adams:1998eg}. This shows
that the index of the lattice Dirac operator is indeed connected to
the topological charge or winding number of the gauge fields (see also
eqs.~\eqref{eq:AS_index} and \eqref{eq:topological_charge_gauge}).
Moreover, the index theorem provides a new definition for the
topological charge density on the lattice, namely
\begin{equation}
  \label{eq:top_charge_density}
  q(x) = \frac{1}{2} {\rm tr}(\gamma_5 \Dt(x,x)) \, ,
\end{equation}
where the trace is over Dirac and color indices. 

It has been shown that GW Dirac operators can not be ultra-local 
\cite{Horvath:1998cm,Bietenholz:1999dg,Horvath:1999bk,Horvath:2000az}, 
i.e.~their couplings $\rho(x-y)$ can \emph{not} vanish for all $|x-y| > r$ for
any finite $r$. 
This fact is more a technical than a fundamental problem in simulations, because the
couplings of the Dirac operator -- at least for any acceptable solution of
the GW relation, such as the FP Dirac operator -- fall off
exponentially.  Such Dirac operators are still local in a physical
sense, as the localization range of the couplings measured in physical
units shrinks to 0 in the continuum limit. However, the fact that no
ultra-local solutions exist already gives a hint that it is not easy
to find an approximation to a GW fermion that can be used in numerical
simulations.  But let us discuss now one of the solutions of the GW
relation, the FP Dirac operator and the FP approach to QCD, more in
detail.


\section{The Renormalization Group and Fixed Point Actions}

Field Theories are defined over a large range of scales. However, not
all the scales are equally important for the underlying physics.
Whereas the low energy excitations carry the information on the long
range properties of the theory, the degrees of freedom associated with
very high (and unphysical) scales do influence the physical
predictions only indirectly through a complex cascade process. This
makes the connection between the local form of the interaction
described by the Lagrangian and the final physical predictions obscure
and, moreover, introduces technical difficulties in treating the large
number of degrees of freedom. That is the reason why one attempts to
separate the low energy scales from the scales associated with very
high energies by integrating them out in the path integral.  The
method accomplishing this, taking into account the effect on the
remaining variables exactly, is called the Renormalization Group
transformation (RGT)
\cite{Wilson:1974jj,Wilson:1975mb,Wilson:1975am,Wilson:1993dy,
  Ma:1973zu,Ma:1976mt,Kadanoff:1976wy,Niemeyer:1973aa}.


The repeated use of the RGT on an initial theory gives a sequence of
theories. These theories can be described by a flow trajectory in the
space of couplings.  Fixed points (FP) of this transformation are
theories that reproduce itself under the RGT. Since the correlation
length $\xi$ of the theory scales by the scaling factor of the RGT,
its value has to be 0 or $\infty$ at the fixed point. Yang-Mills
theory has a non-trivial fixed point (the Gaussian FP) with
correlation length $\xi=\infty$ whose exact location in coupling space
$c_1$, $c_2,\ldots$ depends on the RGT used. There is one so-called
relevant coupling whose strength increases as one is starting near this
FP and performing RGTs. In QCD the quark masses are additional
relevant couplings of the RGT. The flow along these relevant scaling
fields whose end-point is the FP is called the renormalized trajectory
(RT). Simulations performed using an action which is on the exact RT would
reproduce continuum physics without any discretization errors. Let us finally
mention that a detailed review on FP actions can be found in \cite{Hasenfratz:1998bb}.

\subsection{Fixed Point Action of QCD}
\label{sec:QCDFP}

We consider the QCD Lagrangian which consists of a SU(3) non-abelian
gauge theory\footnote{The discussion is kept in the physically relevant case where
the number of colors $N_c$ is 3. The discussion might, however, easily be translated to 
an arbitrary number of colors $N_c$.} and a fermionic part describing the quarks in interaction
with the gauge fields in four dimensional Euclidean space defined on a
periodic lattice. The partition function is defined through
\begin{equation}
  \label{eq:partition_function}
 {\cal Z}(\beta) = \int DU D\bar{\psi} D\psi \exp[-\beta {\cal A}_g(U) - {\cal A}_f(U,\bar{\psi},\psi)] ,
\end{equation}
where $DU$ is the invariant group measure, $D\bar{\psi} D\psi$ the
(anti)fermion integration measure and $\beta$ the gauge coupling.  The
lattice actions ${\cal A}_g(U)$ and ${\cal A}_f(U,\bar{\psi},\psi)$
are some lattice regularizations of the corresponding continuum gauge
and fermion action, respectively. We can perform a real space
renormalization group transformation (RGT),
\begin{multline}
  \label{eq:rgt}
  \exp[-(\beta' {\cal A}_g'(V) + {\cal A}_f'(V,\bar{\chi},\chi))] = \\
  \int DU D\bar{\psi} D\psi \exp[-\{\beta ({\cal A}_g(U) + T_g(U,V)) +
  ({\cal A}_f(U,\bar{\psi},\psi) +
  T_f(U,\bar{\chi},\chi,\bar{\psi},\psi))\}],
\end{multline}
where $V$ is the blocked link variable, $\chi$ and $\bar\chi$ the
blocked fermion fields. Finally, $T_g(U,V)$ and
$T_f(U,\bar{\chi},\chi,\bar{\psi},\psi)$ are the blocking kernels
defining the RGT,
\begin{equation}
  \label{eq:blocking_kernel_gauge}
  T_g(U,V) = - \frac{\kappa_g}{N} \sum_{n_B,\mu} \left(\text{Re}\, \text{tr}(V_\mu(n_B)
  Q^\dagger_\mu(n_B)) - {\cal N}_\mu^\beta \right) 
\end{equation}
and
\begin{multline}
  \label{eq:blocking_kernel_fermion}
  T_f(U,\bar{\chi},\chi,\bar{\psi},\psi) = \kappa_f
  \sum_{n_B}\big(\bar{\chi}_{n_B}
  - \sum_{n} \bar{\psi}_n \Omega^\dagger(U)_{n,n_B}\big) \\
  \times \big(\chi_{n_B} - \sum_{n} \Omega(U)_{n_B,n} \psi_n\big) .
\end{multline} 
In eq.~\eqref{eq:blocking_kernel_gauge} $Q_\mu(n_B)$ is a $3 \times 3$
matrix representing some mean of products of link variables
$U_\mu(n)$, connecting the sites $2 n_B$ and $2 (n_B +\hat \mu)$ on
the fine lattice and ${\cal N}_\mu^\beta$ is a normalization constant
ensuring the invariance of the partition function. By optimizing the
averaging function in $Q_\mu$ and the parameter $\kappa_g$, it is
possible to obtain an action on the coarse lattice, which has a short
interaction range. Such an optimization has been done and we refer to
\cite{Blatter:1996ti} for the explicit form of the RGT block
transformation.

The fermionic blocking kernel in
eq.~\eqref{eq:blocking_kernel_fermion} is defined by the averaging
function $\Omega(U)_{n_B,n}$, which connects the fermion fields on the
fine lattice $\psi$ to the fermion fields on the coarse lattice
$\chi$.  It is a generic gauge invariant function with hypercubic
symmetry that on trivial gauge configurations satisfies the
normalization condition $\sum_n \Omega(2 n_B - n) = 2^{\frac{d-1}{2}}$
for $d$ space-time dimensions.

On the critical surface at $\beta \rightarrow \infty$
eq.~\eqref{eq:rgt} is dominated by the gauge part and it becomes a
saddle point problem representing an implicit equation for the FP
action, ${\cal A}_g^{\text{FP}}$,
\begin{equation}
  \label{eq:FP_equation}
  {\cal A}_g^{\text{FP}}(V) = \min_{\{U\}} \left\{{\cal A}_g^{\text{FP}}(U)
    + T(U,V)\right\}.
\end{equation}
The normalization constant in the blocking kernel, ${\cal
  N}_\mu^\beta$, becomes in the limit $\beta \rightarrow \infty$
\begin{equation}
  \label{eq:transformation_kernel_norm}
  {\cal N}_\mu^\infty = \max_{W \epsilon \text{SU}(3) }
  \left\{\text{Re} \, \text{tr}(W Q\mu^{\dagger})\right\}.
\end{equation}
Note that the FP eq.~\eqref{eq:FP_equation} defines the FP gauge
action as well as the field on the fine configuration $U[V]$
implicitly through a minimization condition.

As the gauge part completely dominates the path integral for large values of the gauge coupling $\beta$, 
the fermionic part does not have any influence on the solution of the FP
gauge action. This is, however, completely different for the fermionic
FP action as the gauge fields on the fine as well on the coarse
lattice show up in the fermionic FP equation \cite{Kunszt_diss:97}
\begin{equation}
  \label{eq:RG_propagator_intro}
  \textfrak{D}_{\rm FP}^{-1}(V) = \frac{1}{\kappa_f} +
  \Omega(U[V]) \textfrak{D}_{\rm FP}^{-1}(U[V]) \Omega^{\dagger}(U[V])
\end{equation}
or equivalently
\begin{equation}
  \label{eq:RG_Dirac_intro}
  \Dfp(V) = \kappa_f - \kappa_f^2
  \Omega(U[V])[ \Dfp(U[V]) + \kappa_f \Omega^{\dagger}(U[V])\Omega(U[V])]^{-1} \Omega^{\dagger}(U[V]) \,.
\end{equation}
Combining eq.~\eqref{eq:RG_propagator_intro} with the GW relation from eq.~\eqref{eq:GW_propagator}, we obtain
the following FP equation for the operator $R$ appearing in the GW relation\footnote{Notice that this
definition of $R$ is very specific for the FP approach.} 
\begin{equation}
  \label{eq:FP_R}
  R_{\rm FP}(V) = \frac{1}{\kappa_f} +
  \Omega(U[V]) R_{\rm FP}(U[V]) \Omega^{\dagger}(U[V]) \,.
\end{equation}
The FP eqs.~\eqref{eq:FP_equation}, \eqref{eq:RG_propagator_intro},
\eqref{eq:RG_Dirac_intro} and \eqref{eq:FP_R}, respectively, can be studied analytically
up to quadratic order in the vector potentials
\cite{Blatter:1996ti,Kunszt_diss:97}.  However, for solving the FP
equations on coarse configurations with large fluctuations -- as they
are needed for numerical simulations in QCD -- one has to resort to
numerical methods, and a sufficiently rich parametrization for the
description of the solution is required. The numerical solution of the
eq.~\eqref{eq:FP_equation} defining the FP gauge action is described
in \cite{Wenger_diss:2000, Rufenacht:2001qy}, whereas a substantial
part of this work is devoted to the numerical solution of the
fermionic FP eqs.~\eqref{eq:RG_propagator_intro} and
\eqref{eq:RG_Dirac_intro}\footnote{Even though
  eqs.~\eqref{eq:RG_propagator_intro} and \eqref{eq:RG_Dirac_intro}
  are equivalent their influence in the parametrization shows to be
  rather different, as we will discuss in Chapter
  \ref{cha:parametrization}}.  But before we give an account of the
parametrization problem, we first discuss the properties of a general
Dirac operator, because we need a Dirac operator with a rich structure
to be able to capture some of the very important properties of the FP
Dirac operator.

\cpages
\chapter{General Lattice Dirac Operator}
\label{cha:general_dirac}

In this chapter we discuss the steps to construct Dirac operators
which have arbitrary fermion offsets, gauge paths, a general structure
in Dirac space and satisfy the basic symmetries on the lattice, which
are gauge symmetry, hermiticity condition, charge conjugation,
hypercubic rotations and reflections. We give an extensive set of
examples and provide an efficient factorization of the operators
occurring in the construction of a Dirac operator. Although the
discussion about the lattice Dirac operators is kept very general, all
the examples are specific to the construction of the parametrized FP
Dirac operator $\Dpar$, which we discuss in Chapter
\ref{cha:parametrization}.

\section{Introduction}
\label{cha:general_dirac:sec:intro}

The construction of a general lattice Dirac operator that
satisfies all the basic symmetries (gauge symmetry, $\gamma_5$
hermiticity, charge conjugation, hypercubic rotations and reflections)
is a fundamental kinematic problem. We discuss this problem in $d=4$ in a
very general way. Notice that a similar discussion with 
different notation can be found in \cite{Gattringer:2000ja,Gattringer:2000qu}. 
For any fermion offset and for any of the 16 elements of the Clifford algebra
we describe the steps to find combinations of gauge paths which
satisfy all the basic symmetries. We construct explicitly paths for
all the elements of the Clifford algebra in the offsets of the
hypercube. We show also how to factorize the sum of paths,
i.e.~writing it as a product of sums, in such a way that the
computational problem is manageable even if the number of paths is
large. 

Most of the contents of this chapter are published in \cite{Hasenfratz:2000xz}
as well as in {\tt hep-lat/0003013}, where an easy-to-use Maple code {\tt
  dirac.maple} is provided. This code can be used, if one wants to add
additional terms to a parametrization of a Dirac operator.

Let us add a few remarks at this point:
\begin{enumerate}
\item There are many ways to fix the parameters of the chosen Ansatz
  for the Dirac operator.  As opposed to production runs this problem
  should be treated only once. It is useful to invest effort here,
  since the choice will influence strongly the quality of the results
  in simulations.
\item There are compelling reasons to use the elements of the Clifford
  algebra beyond $1$ and $\gamma^\mu$ in the Dirac operator. For an
  operator $\D$ satisfying the GW relation in eq.~(\ref{eq:GW}), 
  $\tr (\gamma_5 R \D)$ is the topological charge density
  \cite{Hasenfratz:1998ri, Luscher:1998pq}, i.e.~the $\gamma_5$ part of
  $\D$ is obviously important.  Similarly, a $\sigma^{\mu \nu}$ term
  is already required by the leading Symanzik condition.
\item If the parametrization is close to $\Dfp$, the GW relation is
  approximately satisfied, which means the operator $A^\dagger A$
  under the square root in Neuberger's overlap construction is close
  to $1$.  An expansion converges very fast in this case, as we will
  show in Chapter \ref{cha:overlap}.
  In the present situation
  it is, however, unclear whether this speed-up of the convergence
  compensates for additional expenses of the improved operator
  \cite{Hernandez:2001yd,Bietenholz:2001nu,
    Eshof:2002ms}. But, there are many reasons to use
    improved Dirac operators and performance issues in the
    overlap construction are most likely not more important 
    than good scaling properties of the resulting operator. 
\item The basic numerical operation in production runs is $\Dpar
  \eta$, where $\Dpar$ is the parameterized Dirac matrix and $\eta$ is
  a vector.  The matrix elements of $\Dpar$ should be precalculated
  before the iteration starts. Using all the Clifford algebra elements
  and arbitrary gauge paths, the computational cost per offset of the
  operation $\Dpar \eta$ is a factor of $\sim 4$ higher than that of
  the Wilson action.  Using all the points of the hypercube (81
  offsets), the cost per iteration is increased by a factor of
  $4\times 81/8 \sim 40$ relative to the Wilson Dirac operator.  Note
  however, that this number may vary quite a bit depending on the
  underlying computer architecture, as one of the main issues in
  present lattice simulations is rather fast memory access and fast
  communication than fast floating point units \cite{Gottlieb:2001sy}.
\end{enumerate}

\section{Symmetries of the Dirac Operator}

We define the basis of the Clifford algebra as $\Gamma=1,\,\gamma_\mu,
\,i\sigma_{\mu \nu},\,\gamma_5,\, \gamma_\mu \gamma_5$.  We use the
notation S,V,T,P and A for the scalar, vector, tensor, pseudoscalar
and axial-vector elements of the Clifford algebra, respectively.
Notice that the tensor (T) and axial-vector (A) basis elements of the
Clifford algebra are anti-hermitian as one can see from the explicit
representation given in Appendix \ref{cha:clifford}.  It will later be
useful to list the basis elements of the Clifford algebra by a single
index, as $\Gamma_i$, $i=1,\ldots,16$.  We choose the ordering
\begin{equation}
  \label{cord}
  \Gamma_i = 1,\gamma_1,\gamma_2,\gamma_3,\gamma_4, i\sigma_{12},i\sigma_{13},
  i\sigma_{14},i\sigma_{23},i\sigma_{24},i\sigma_{34},\gamma_5,
  \gamma_1\gamma_5,\gamma_2\gamma_5,\gamma_3\gamma_5,\gamma_4\gamma_5.
\end{equation}

Under charge conjugation, the Dirac matrices $\gamma_\mu$ transform as
\begin{equation} 
  \label{cdef}
  {\cal C} \gamma_\mu^T {\cal C}^{-1} = - \gamma_\mu \,.
\end{equation}

Having chosen the T and A basis elements of the Clifford algebra to be
anti-hermitian, all the basis elements have the property that
\begin{equation}
  \label{pleasant}
  {\cal C}\gamma_5 \Gamma^* \gamma_5 {\cal C}^{-1}= \Gamma \,.
\end{equation}

We define a sign $\epsilon_{\Gamma}$ by the relation
\begin{equation}
  \label{epsi}
  \gamma_5 \Gamma^\dagger \gamma_5 =\epsilon_\Gamma \Gamma,
\end{equation}
which gives $\epsilon_{\rm S}=\epsilon_{\rm P}=\epsilon_{\rm A}=1$ and
$\epsilon_{\rm V}=\epsilon_{\rm T}=-1$.

Under reflection of the coordinate axis $\eta$, the basis elements of
the Clifford algebra are transformed as $\Gamma'={\cal P}_\eta \Gamma
{\cal P}_\eta^{-1}$, where ${\cal P}_\eta = \gamma_{\eta}\gamma_5$.
(When $\Gamma$ is written in terms of Dirac matrices $\gamma_\mu$ this
amounts simply to reversing the sign of $\gamma_\eta$.)  The group of
reflections has $2^4=16$ elements.  Under a permutation $(1,2,3,4)\to
(p_1,p_2,p_3,p_4)$ of the coordinate axes the basis elements of the
Clifford algebra transform by replacing $\mu\to p_\mu$ and accordingly
$\gamma_5 \to \epsilon_{p_1 p_2 p_3 p_4} \gamma_5$.

The matrix elements of the Dirac operator $D$ we denote as
\begin{equation}
  D(n,n';U)_{\alpha \alpha'}^{aa'}
\end{equation}
where $n$, $a$ and $\alpha$ refer to the coordinate, color and Dirac
indices, respectively.  From now on, we suppress the color and Dirac
indices.  The gauge configuration $U$ appearing in $D$ is not
necessarily the original one entering the gauge action -- it could be
a smeared configuration $V$. Using the Dirac operator $D(V)$ should be interpreted
as defining a new Dirac operator $D^\prime$ on the thermally  generated configuration $U$:
$D^\prime(U)=D(V)$, where $D^\prime$ has a large number of additional paths relative to $D$.
 Provided the smearing has appropriate
symmetry properties\footnote{All conventional smearing schemes have
  appropriate symmetry properties, in particular also the two smearing
  schemes used in the parametrization of the FP Dirac operator.  A
  discussion of these smearing schemes can be found in Section
  \ref{sec:smearing}.}  all the following constructions in this
chapter remain valid, without any extra modifications.

The Dirac operator should satisfy the following symmetry requirements:

\subsubsection*{Gauge symmetry}

Under the gauge transformation $U_\mu(n) \rightarrow U_\mu(n)^g=
g(n)\,U_\mu(n)\,g(n+\hat{\mu})^\dagger$, where $g(n) \in {\rm SU}(N)$
we have
\begin{equation}
  D(n,n';U) \rightarrow  \,D(n,n';U^g) =
  g(n)\, D(n,n';U)  \,g(n')^\dagger .
  \label{gau}
\end{equation}

\subsubsection*{Translation invariance} 

Translation symmetry requires that $D(n,n+r)$ depends on $n$ only
through the $n$-dependence of the gauge fields.  There is no explicit
$n$-dependence beyond that. In particular, the coefficients in front
of the different paths which enter $D$ do not depend on $n$.

\subsubsection*{Hermiticity}
\begin{equation}
  D(n,n';U) = \gamma_5 \, D(n',n,U)^\dagger \,\gamma_5,
  \label{s1}
\end{equation}
where $\dagger$ is hermitian conjugation in color and Dirac space.

\subsubsection*{Charge conjugation}

\begin{equation}
  D(n,n';U) = {\cal C} \, D(n',n;U^*)^T \,{\cal C}^{-1} \,,
  \label{s2}
\end{equation}
where $T$ is the transpose operation in color and Dirac space.

It will be useful to combine eqs.~\eqref{s1} and \eqref{s2} to obtain
\begin{equation}
  \label{ss2}
  D(n,n';U) = {\cal C} \gamma_5 \, D(n,n';U^*)^* \,\gamma_5 {\cal C}^{-1} \,.
\end{equation}

\subsubsection*{Reflection of the coordinate axis $\eta$} 

\begin{equation}
  D(n,n';U) = {\cal P}^{-1}_\eta \, D(\tilde{n},\tilde{n}';U^{{\cal P}_\eta})
  \,{\cal P}_\eta \,,
  \label{s3}
\end{equation}
where ${\cal P}_\eta = \gamma_{\eta}\gamma_5$ and
$\tilde{n}_\nu=n_\nu$ if $\nu \neq \eta$, while
$\tilde{n}_\eta=-n_\eta$.  The reflected gauge field $U^{{\cal
    P}_\eta}$ is defined as
\begin{eqnarray}
  \label{12}
  U^{{\cal P}_\eta}_\eta(m) & = & U_\eta(\tilde{m}-\hat{\eta})^\dagger \,, \\
  U^{{\cal P}_\eta}_\nu(m) & = & 
  U_\nu(\tilde{m})\,,\qquad  \nu \neq \eta .\nonumber
\end{eqnarray}

\subsubsection*{Permutation of the coordinate axes}

These are defined in a straightforward way, by permuting the Lorentz
indices appearing in $D$.  Note that rotations by $90^\circ$ on a
hypercubic lattice can be replaced by reflections and permutations of
the coordinate axes.

\section{Constructing a Matrix $D$ which satisfies the Symmetries}

To describe a general Dirac operator in compact notations, it is
convenient to introduce the operator $\hat{U}_\mu$ of the parallel
transport for direction $\mu$
\begin{equation}
  \label{Uhat}
  \left(\hat{U}_\mu\right)_{nn'}=U_\mu(n)\delta_{n+\hat{\mu},n'} \,,
\end{equation}
and analogously for the opposite direction
\begin{equation}
  \label{Uhatd}
  \left(\hat{U}_{-\mu}\right)_{nn'}=U_\mu(n-\hat{\mu})^\dagger
  \delta_{n-\hat{\mu},n'} \,.
\end{equation}
Obviously
$\left(\hat{U}_\mu\right)^\dagger=\hat{U}_{-\mu}$\footnote{Note that
  in terms of these operators the forward and backward covariant
  derivatives are: $\partial_\mu=\hat{U}_\mu -1$,
  $\partial_\mu^*=1-\hat{U}_\mu^\dagger$, and $\partial_\mu^*
  \partial_\mu=\hat{U}_\mu + \hat{U}_\mu^\dagger -2$.}.  The Wilson
Dirac operator at bare mass equal to zero reads:
\begin{equation}
  \label{DWdef}
  D_{\rm W}=\frac{1}{2} \sum_\mu \gamma_\mu 
  \left( \hat{U}_\mu - \hat{U}_\mu^\dagger \right)+ 
  \frac{1}{2}r \sum_\mu 
  \left( 2 - \hat{U}_\mu - \hat{U}_\mu^\dagger \right) \,.
\end{equation}
It is also useful to introduce the operator $\hat{U}(l)$ of the
parallel transport along some path $l=[l_1,l_2,\ldots,l_k]$ where
$l_i=\pm 1,\ldots,\pm 4$, by
\begin{equation}
  \label{Ul}
  \hat{U}(l)=\hat{U}_{l_1}\hat{U}_{l_2}\ldots\hat{U}_{l_k} \,.
\end{equation}
In terms of gauge links this is
\begin{equation}
  \label{eq:Ul1}
  \left( \hat{U}(l)\right)_{nn'}=
  \left( U_{l_1}(n)U_{l_2}(n+\hat{l}_1)\ldots \right)\delta_{n+r_l,n'}\,,
\end{equation}
where $r_l=\hat{l}_1+\ldots+\hat{l}_k$ is the offset corresponding to
the path $l$.  Note that $\hat{U}(l)^\dagger=\hat{U}(\bar{l})$ where
$\bar{l}=[-l_k,\ldots,-l_1]$ is the inverse path.  In particular, one
has
\begin{equation}
  \label{Ustaple}
  \left( \hat{U}([2,1,-2])\right)_{nn'}=
  U_2(n)U_1(n+\hat{2})U_2(n+\hat{1})^\dagger \delta_{n+\hat{1},n'}
\end{equation}
for the corresponding staple.

As another example, the Sheikholeslami-Wohlert (or clover) term
\cite{Sheikholeslami:1985ij} introduced to cancel the ${\cal O}(a)$
artifacts is given up to a constant prefactor by
\begin{multline}
  \label{clover}
  i\sigma_{\mu\nu} \left(
    \hat{U}([\mu,\nu,-\mu,-\nu])+\hat{U}([\nu,-\mu,-\nu,\mu])+ \right. \\
  \left. \hat{U}([-\mu,-\nu,\mu,\nu])+\hat{U}([-\nu,\mu,\nu,-\mu])
    -{\rm h.c.} \right) \,.
\end{multline}

We consider a general form of the Dirac operator
\begin{equation}
  \label{Dgen}
  D=\sum_i \Gamma_i \sum_l c^{(i)}_l \hat{U}(l) \,.
\end{equation}
The Dirac indices are carried by $\Gamma_i=1,\gamma_\mu,\ldots$, the
coordinate and color indices by the operators $\hat{U}(l)$.  The Dirac
operator is determined by the set of paths $l$ the sum runs over, and
the coefficients $c^{(i)}_l$.

In the case of $D_{\rm W}$ (see eq.~\eqref{DWdef}), for $\Gamma=1$ one has
$l=[1],[-1],\ldots,[-4]$ and $l=[]$ (the empty path $l=[]$
  corresponds to $\hat{U}([])=1$), while for $\Gamma=\gamma_\mu$:
$l=[\mu]$ and $[-\mu]$.  In the clover term, eq.~(\ref{clover}) for
$\Gamma=i\sigma_{\mu\nu}$: $l=[\mu,\nu,-\mu,-\nu],\ldots$ (altogether
$6\times 4=24$ plaquette products).  As these well known examples
indicate, the coefficients $c_l^{(i)}$ for related paths differ only
in relative signs, which are fixed by symmetry requirements.

Our aim is to give for all $\Gamma$'s and offsets $r$ on the hypercube
a set of paths and to determine the relative sign for paths related to
each other by symmetry transformations.  We give the general rules for
arbitrary offsets and paths as well.

Eqs.~(\ref{pleasant},\ref{ss2}) imply that the coefficients
$c_l^{(i)}$ in eq.~(\ref{Dgen}) are real.  Further, from hermiticity
in this language it follows that the path $l$ and the opposite path
$\bar{l}$ (or equivalently, $\hat{U}(l)$ and $\hat{U}(l)^\dagger$)
should enter in the combination
\begin{equation}
  \label{GUUd}
  \Gamma \left( \hat{U}(l)+\epsilon_\Gamma \hat{U}(l)^\dagger \right) \,,
\end{equation}
where the sign $\epsilon_\Gamma$ is defined by $\gamma_5
\Gamma^\dagger \gamma_5 =\epsilon_\Gamma \Gamma$, eq.~(\ref{epsi}).

The symmetry transformations formulated in terms of matrix elements in
the previous section can be translated to the formalism used here.
The reflections and permutations act on operators $\hat{U}(l)$ in a
straightforward way. Under a reflection of the axis $\eta$ one has
$\hat{U}(l) \to \hat{U}(l')$ where $l'_i=-l_i$ if $|l_i|=\eta$ and
unchanged otherwise.  Under a permutation $(p_1,p_2,p_3,p_4)$ a
component with $l_i=\pm\mu$ is replaced by $l_i=\pm p_\mu$, as
expected.  The number of combined symmetry transformations is
$16\times 24=384$.  We denote the action of a transformation
$\alpha=1,\ldots,384$ by $\Gamma \to \Gamma^{(\alpha)}$,
$\hat{U}(l)\to \hat{U}(l^{(\alpha)})$.  Acting on the expression in
eq.~(\ref{GUUd}) by all 384 elements of the symmetry group and adding
the resulting operators together, the sum will satisfy the required
symmetry conditions for a Dirac operator.

Let us introduce the notation
\begin{equation}
  \label{sym}
  \hat{d}(\Gamma,l)=\frac{1}{\cal N} \sum_\alpha \Gamma^{(\alpha)}
  \left( \hat{U}(l^{(\alpha)}) +
    \epsilon_\Gamma \hat{U}(l^{(\alpha)})^\dagger \right) \,.
\end{equation}
A general Dirac operator will be a linear combination of such terms, unless 
one chooses the coefficients to be gauge invariant functions of the gauge fields,
which can take different values for different offsets generated by the reference path.
We will be more specific about this possibility to extend the construction of a
general Dirac operator using color singlet factors.
The normalization factor ${\cal N}$ will be defined below.  The total
number of terms in eq.~(\ref{sym}) is $2\times 384=768$.  Typically,
however, the number of different terms which survive after the
summation is much smaller.  It can happen that for a choice of
starting $\Gamma$ and $l$ the sum in eq.~(\ref{sym}) is zero. In this
case the given path does not contribute to the Dirac structure
$\Gamma$.

To fix the convention for the overall sign we single out a definite
term in the sum of eq.~(\ref{sym}) and take its sign to be $+1$.
Denote by $\Gamma_0$, $l_0$ the corresponding quantities of this
reference term, and by $r_0=r(l_0)$ the offset of $l_0$.  This term is
specified by narrowing down the set $\{ \Gamma^{(\alpha)},l^{(\alpha)}
\}$ to a single member as follows:
\begin{itemize}
\item[a)] Given an offset $r=r(l)=(r_1,r_2,r_3,r_4)$ the reflections
  and permutations create all offsets $(\pm r_{p_1},\ \pm r_{p_2},\ 
  \pm r_{p_3},\ \pm r_{p_4})$ where $(p_1,p_2,p_3,p_4)$ is an
  arbitrary permutation.  We choose for the reference offset $r_0$ the
  one from the set $\{ r(l^{(\alpha)}) \}$ which satisfies the
  relations $r_{01} \ge r_{02} \ge r_{03} \ge r_{04} \ge 0$.
\item[b)] If several $\Gamma$ matrices are generated to {\em this}
  offset then choose as $\Gamma_0$ the one which comes first in the
  natural order, eq.~(\ref{cord}).
\item[c)] Consider all the paths $\{ l^{(\alpha)} \}$ having offset
  $r_0$ and associated with $\Gamma_0$, i.e.~$r(l^{(\alpha)})=r_0$ and
  $\Gamma^{(\alpha)}=\Gamma_0$.  To single out one path $l_0$ from
  this set, we associate to a path $l=[l_1,l_2,\ldots,l_k]$ a decimal
  code $d_1d_2\ldots d_k$ with digits $d_i=l_i$ if $l_i>0$ and
  $d_i=9+l_i$ for $l_i<0$.  The path with the smallest code will be
  the reference path $l_0$.  (In other words we take the first in
  lexical order defined by the ordering $1,2,3,4,-4,-3,-2,-1$.)
\end{itemize}

Of course, one can take $\Gamma_0$, $l_0$ as the starting $\Gamma$ and
$l$, and we shall refer to the expression in eq.~(\ref{sym}) as
$\hat{d}(\Gamma_0,l_0)$ to indicate that it is associated to a class
rather than to a specific $(\Gamma,l)$.

We turn now to the normalization of $\hat{d}(\Gamma_0,l_0)$.  In
general, there will be $K$ {\em different} paths in the set $\{
l^{(\alpha)} \, | \, \Gamma^{(\alpha)}=\Gamma_0,
r(l^{(\alpha)})=r_0\}$, i.e.~corresponding to the same offset $r_0$
and Dirac structure $\Gamma_0$.  The normalization is fixed by
requiring that the coefficient of the reference term $\Gamma_0
\hat{U}(l_0)$ is $+1/K$.

Consider a simple example explicitly. Let $r_0=(1,0,0,0)$,
$\Gamma_0=i\sigma_{12}$ and $l_0=[2,1,-2]$.  The starting term, in
eq.~(\ref{GUUd}) is $i\sigma_{12} (
\hat{U}([2,1,-2])-\hat{U}([2,-1,-2]))$.  Applying all the 16 different
reflections gives
\begin{equation} 
  8 i\sigma_{12}\left( \hat{U}([2,1,-2])-\hat{U}([2,-1,-2]) 
    - \hat{U}([-2,1,2])+\hat{U}([-2,-1,2]) \right) \,.
\end{equation} 
Applying all the permutations on this expression results in:
\begin{multline}
  \label{pelda}
  \hat{d}(\Gamma_0,l_0) = \frac{1}{{\cal N}} \left \{ 16 i\sigma_{12}
    \left(
      \hat{U}([2,1,-2]) - \hat{U}([-2,1,2])  \right)+ \right. \\
  16 i\sigma_{13} \left(
    \hat{U}([3,1,-3]) - \hat{U}([-3,1,3])  \right)+ \\
  \left. 16 i\sigma_{14}\left( \hat{U}([4,1,-4]) - \hat{U}([-4,1,4])
    \right) + \ldots \right\}\,,
\end{multline}
where only the terms with the offset $r=r_0$ are written out
explicitly.  Their total number is 96.  The whole generated set has 8
different offsets giving altogether 768 terms.  Notice the form of the
contribution in eq.~(\ref{pelda}).  There is a common factor (16 in
this case) multiplying all the different operators.  Only the tensor
elements of the Clifford algebra enter, since we started with a tensor
element. Beyond the common factor the path products have a coefficient
$\pm 1$. These features are general.  The number of different paths
with $\Gamma_0=i\sigma_{12}$ and $r_0=(1,0,0,0)$ is K=2: the paths
$[2,1,-2]$ and $[-2,1,2]$).  The normalization factor in this case is
${\cal N}=16\times 2=32$, so that one has
$\hat{d}(\Gamma_0,l_0)=\frac{1}{2} i\sigma_{12} \left(
  \hat{U}([2,1,-2]) - \hat{U}([-2,1,2]) \right) + \ldots$.

A general Dirac operator is constructed as
\begin{equation}
  \label{Dfd}
  D = \sum_{i=\Gamma_0,l_0} \frac{1}{{\cal N}_i} \sum_\alpha 
 f_i(l^{(\alpha)}) \Gamma_i^{(\alpha)} \left( \hat{U}(l_i^{(\alpha)}) +
    \epsilon_\Gamma \hat{U}(l_i^{(\alpha)})^\dagger \right) \,.
\end{equation}
The coefficients $f_i(l^{(\alpha)})$ which are the free, adjustable
parameters of the Dirac operator are real constants or more generally
gauge invariant, real functions of the gauge fields, respecting
locality, and invariance under the symmetry transformations.  The
requirement that such a function is gauge invariant and real, restricts
the choice to the real part of traces of closed loops of link products.  The invariance
under symmetry transformations can be assured by constructing these functions 
along the same lines as a scalar operator which satisfies the symmetry requirements. 
The following example should make clear what is meant by such a color singlet function. We
consider again the coupling generated by the staples and the tensor
Clifford algebra elements i.e.~$r_0=(1,0,0,0)$, $\Gamma_0 = \i
\sigma_{12}$ and $l_0 = [2,1,-2]$. A simple choice for such a function
which satisfies all the requirements would e.g.~be a polynomial of the trace of all
the plaquettes generated by the operator $r_0=(0,0,0,0)$, $\Gamma_0 =
1$ and $l_0 = [1,2,-1,-2]$. But there are other possible choices, such
as 
\begin{gather}
  f(l^{(i)}) = f(x(l^{(i)})) \notag \\
  \intertext{with}
  x(l^{(i)}) = 1 - \frac{1}{N_c} \tr [X(l^{(i)}) X(l^{(i)})^\dagger] \, ,
 \label{eq:fluctuation-definition}
\end{gather}
where $X$ is the gauge operator\footnote{The Dirac algebra structure is given only to 
indicate the transformation properties, i.e. the operator $X$ is 
defined without the trivial Dirac structure.} generated by
$r_0=(1,0,0,0)$, $\Gamma_0 = 1$ and $l_0 = [2,1,-2]$, $N_c$ the number of colors
and $l^{(i)}$, $i=1,\ldots,n_{\rm offsets}$ denotes the different offsets generated by $l_0$.
This choice
gives the possibility that different offsets generated from the same
$\Gamma_0$ and $l_0$ can have different couplings in the sense that
the value of the function  $f(l^{(i)})$ is not necessarily the same for every
offset\footnote{Neuberger's construction does not generate such
  terms when one starts with the Wilson Dirac operator.}.
In the parametrization of $\Dfp$ and $R$ we always
use a linear function 
\begin{equation}
  \label{eq:linear_fluctuation_function}
  f(x(l^{(i)})) = c_0 + c_1 x(l^{(i)})
\end{equation}
for the couplings.

\section{Tables for Offsets on the Hypercube}

Choosing offsets and paths to be included in the Dirac operator is a
matter of intuition. It is also influenced by considerations on CPU
time and memory requirements.  In
Tables~\ref{tab:refpath0}-\ref{tab:refpath4} we give the reference
paths $l_0$ for offsets on the hypercube and general Dirac structure
that are used in the current implementation of the parametrized FP
Dirac operator $\Dpar$.  The first 3 columns give $\Gamma_0$, $l_0$
and the number of paths $K$ as defined in the previous section.  The
4th column gives those Clifford basis elements which are generated in
eq.~(\ref{sym}) to the offset $r_0$. In Table
\ref{tab:gauge_functions} we give the gauge invariant, real functions
of the gauge fields which are used as fluctuation polynomials in the
current implementation of $\Dpar$.

\begin{table}[p]
  \begin{center}
    \begin{tabular}{|c|c|c|c|c|}
      \hline
      $\Gamma_0  $    & ref. path $l_0$             &  $K$ & $\Gamma$'s
      generated & $c_X$\\
      \hline\hline
      $1$           & $[]$                    &   1 & 1 &     $1$, $0$ \\
      & $[1,2,-1,-2]$           &  48 &   &     $1$, $0$ \\
      \hline
      $\gamma_1$    & $[1,2,-1,-2]$            & 24 & $\gamma^1,\dots,\gamma^4$ & 0\\
      \hline
      $i\sigma_{12}$ & $[1,2,-1,-2]$     &   8 & $i\sigma_{12},\dots,i\sigma_{34}$ &
      $-1$ \\
      \hline
      $\gamma_5$    & $[1,2,-1,-2,3,4,-3,-4]$ & 384 & $\gamma_5$  & $-1/6$ \\
      & $[1,2,3,4,-1,-2,-3,-4]$ & 384 &             & $-1/6$ \\ 
      \hline
      $\gamma_1 \gamma_5$ & $[1,2,-1,-2,3,4,-3,-4]$ & 192  & 
      $\gamma_1\gamma_5,\dots,\gamma_4\gamma_5$ & 0 \\
      & $[2,1,-2,-1,3,4,-3,-4]$ & 192  & & 0 \\ \hline

    \end{tabular}
    \caption{{}Reference paths for different $\Gamma_0$'s for offset (0000).}
    \label{tab:refpath0}
  \end{center}
\end{table}

\begin{table}[p]
  \begin{center}
    \begin{tabular}{|c|c|c|c|c|}
      \hline
      $\Gamma_0$    & ref. path $l_0$             &  $K$ & 
      $\Gamma$'s generated &  $c_X$\\
      \hline\hline
      $1$           & $[1]$                   &   1 & 1   & $8$, $1$\\
      & $[2,1,-2]$              &   6 & & $8$, $1$ \\
      & $[2,3,1,-3,-2]$         &   24 & & $8$, $1$ \\
      \hline
      $\gamma_1$    & $[1]$                   &   1 & $\gamma_1$  & $2$ \\
      & $[2,1,-2]$              &   6 & & $2$ \\
      \hline
      $\gamma_2$    & $[1,2,3,-2,-3]$ & 16 & $\gamma_2,\gamma_3,\gamma_4$ & $0$ \\
      \hline
      $i\sigma_{12}$& $[2,1,-2]$ & 2 & $i\sigma_{12},i\sigma_{13},i\sigma_{14}$
      & $4$ \\
      \hline
      $i\sigma_{23}$ & $[1,2,3,-2,-3]$ & 16 &
      $i\sigma_{23},i\sigma_{24},i\sigma_{34}$ & $-4$ \\
      \hline
      $\gamma_5$     & $[2,1,-2,3,4,-3,-4]$ & 96 & $\gamma_5$ & $4/3$ \\
      \hline
      $\gamma_1 \gamma_5$ & $[2,1,-2,3,4,-3,-4]$  &96   & $\gamma_1\gamma_5$ & $0$\\
      \hline
      $\gamma_2 \gamma_5$ & $[1,3,4,-3,-4]$  & 16 & 
      $\gamma_2\gamma_5, \gamma_3\gamma_5, \gamma_4 \gamma_5$ & $-2$ \\
      \hline
    \end{tabular}
    \caption{{}Reference paths for different $\Gamma_0$'s for offset (1000).}
    \label{tab:refpath1}
  \end{center}
\end{table}

\begin{table}[p]
  \begin{center}
    \begin{tabular}{|c|c|c|c|c|}
      \hline
      $\Gamma_0$    & ref. path $l_0$              &  $K$ & 
      $\Gamma$'s generated &  $c_X$ \\
      \hline\hline
      $1$           & $[1,2]$                   &  2  & 1 & $24$, $6$ \\
      \hline
      $\gamma_1$    & $[1,2]$     & 2 & $\gamma_1, \gamma_2$ & $12$ \\
      \hline
      $\gamma_3$    & $[1,3,2,-3]$  & 8 & $\gamma_3, \gamma_4$& $0$ \\
      \hline
      $i\sigma_{12}$ & $[1,2]$       & 2 & $i\sigma_{12}$ & $-2$ \\
      \hline
      $i\sigma_{13}$ & $[1,3,2,-3]$ & 4& $i\sigma_{13}, i\sigma_{14},
      i\sigma_{23}, i\sigma_{24}$ & $0$ \\
      \hline
      $i\sigma_{34}$ & $[1,2,3,4,-3,-4]$ & 32 & $i\sigma_{34}$ & $-4$ \\
      \hline
      $\gamma_5$    & $[1,2,3,4,-3,-4]$ & 32 & $\gamma_5$ & $-2$ \\
      \hline
      $\gamma_1 \gamma_5$ & $[1,2,3,4,-3,-4]$ & 16 & 
      $\gamma_1\gamma_5, \gamma_2\gamma_5$ & $4$ \\
      \hline
      $\gamma_3 \gamma_5$ & $[1,4,2,-4]$  &8  & 
      $\gamma_3\gamma_5, \gamma_4\gamma_5$ & $-4$ \\
      \hline
    \end{tabular}
    \caption{{}Reference paths for different $\Gamma_0$'s for offset (1100).}
    \label{tab:refpath2}
  \end{center}
\end{table}

\begin{table}[p]
  \begin{center}
    \begin{tabular}{|c|c|c|c|c|}
      \hline
      $\Gamma_0$    & ref. path $l_0$              &  $K$ &
      $\Gamma$'s generated &  $c_X$ \\
      \hline\hline
      $1$           & $ [1,2,3]$                   &  6 & 1 & $32$, $12$\\
      \hline
      $\gamma_1$    & $ [1,2,3]$ &  4 & $\gamma_1, \gamma_2,\gamma_3$ & $24$ \\
      \hline
      $\gamma_4$    & $[1,2,4,3,-4]$       &  24& $\gamma_4$ & $0$ \\
      \hline
      $i\sigma_{12}$ & $[1,2,3]$ & 4 & 
      $i\sigma_{12}, i\sigma_{13}, i\sigma_{23}$ & $-8$ \\
      \hline
      $i\sigma_{14}$ & $[1,4,2,-4,3]$ & 8 & 
      $i\sigma_{14}, i\sigma_{24}, i\sigma_{34}$ & $0$ \\
      \hline
      $\gamma_5$    & $[1,4,2,-4,3]$ & 12& $\gamma_5$ & $-8/3$ \\
      \hline
      $\gamma_1 \gamma_5$ & $[1,4,2,-4,3]$  & 8 & $\gamma_1 \gamma_5, 
      \gamma_2 \gamma_5, \gamma_3 \gamma_5$ & $8$\\
      \hline
      $\gamma_4 \gamma_5$ & $[1,2,3]$  & 6  & $\gamma_4 \gamma_5$ & $-4/3$ \\
      \hline
    \end{tabular}
    \caption{{}Reference paths for different $\Gamma_0$'s for offset (1110).}
    \label{tab:refpath3}
  \end{center}
\end{table}

\begin{table}[p]
  \begin{center}
    \begin{tabular}{|c|c|c|c|c|}
      \hline
      $\Gamma_0  $    & ref. path $l_0$             &  $K$ & $\Gamma$'s
      generated & $c_X$\\
      \hline\hline
      $1$           & $[1,2,3,4]$                    &   24 & 1 & $16$, $8$  \\
      \hline
      $\gamma_1$    & $[1,2,3,4]$ &  12 & $\gamma^1,\dots, \gamma^4$ & $16$ \\
      \hline
      $i\sigma_{12}$ & $[1,2,3,4]$     &  8 & 
      $i\sigma_{12},\dots,i\sigma_{34}$ & $-8$ \\
      \hline
      $\gamma_5$    & $[1,2,3,4]$ & 24 & $\gamma_5$  & $-2/3$ \\
      \hline
      $\gamma_1 \gamma_5$ & $[1,2,3,4]$  &  12 & $\gamma_1 \gamma_5,\dots,
      \gamma_4 \gamma_5$    & $8/3$ \\
      \hline

    \end{tabular}
    \caption{{}Reference paths for different $\Gamma_0$'s for offset (1111).}
    \label{tab:refpath4}
  \end{center}
\end{table}

\begin{table}[p]
  \begin{center}
    \begin{tabular}{|c|c|c|}
      \hline
      offset        & ref. path $l_0$ &   K  \\ \hline \hline
      $(0000)$      & $[1,2,-1,-2]$   &  48  \\ \hline
      $(1000)$      & $[2,1,-2]$      &   6  \\ \hline
      $(1100)$      & $[1,2]$         &   2  \\ \hline
      $(1110)$      & $[1,2,3]$       &   6  \\ \hline
      $(1111)$      & $[1,2,3,4]$     &  24  \\ \hline
    \end{tabular}
    \caption{{}Reference paths for the fluctuation polynomials as defined in
      eqs.~\eqref{Dfd} and \eqref{eq:fluctuation-definition}. 
      Note however, that the
      fluctuation polynomial for the offset $(1000)$ in the current
      implementation of the operator $R$ differs from the one given in
      this table. It is given by the trace of the sum of the
      plaquettes along a particular link $U(l^{(i)})$ 
      i.e.~$x(l^{(i)}) = 1 - 1/N_c {\rm Re}\, \tr [X(l^{(i)}) U(l^{(i)})^\dagger]$, 
      where $N_c$ is the number of
      colors and the operator $X$ is given by the operator for the offset $(1000)$ in this table.}
    \label{tab:gauge_functions}
  \end{center}
\end{table}

It is also of interest how the given Dirac operator behaves for smooth
gauge fields, i.e.~to obtain the leading terms in the formal continuum
limit.  The expression in eq.~(\ref{sym}) generated by given
$\Gamma_0$ and $l_0$ contributes in this limit to one of the
expressions (of type S,V,T,P,A) below
\begin{multline}
  \label{clim}
  \left( \bar{c}_{\rm S} + c_{\rm S} \partial^2 \right) \,; \quad
  c_{\rm V} \gamma_\mu \partial_\mu \,; \quad
  c_{\rm T} \frac{1}{2} \sigma_{\mu\nu} F_{\mu\nu}\,; \\
  c_{\rm P} \frac{1}{4} \gamma_5
  \epsilon_{\mu\nu\rho\sigma}F_{\mu\nu}F_{\rho\sigma}\,; \quad c_{\rm
    A} i \gamma_\mu \gamma_5 \epsilon_{\mu\nu\rho\sigma}\partial_\nu
  F_{\rho\sigma} \,.
\end{multline}
Here $\partial_\mu$ is the covariant derivative in the continuum.  The
coefficients $c_X(\Gamma_0,l_0)$, which determine the continuum
behavior of the Dirac operator, are presented in the last column of
Tables~\ref{tab:refpath0}-\ref{tab:refpath4}.  For $\Gamma_0=1$ the
first and second entries correspond to $\bar{c}_{\rm S}$ and $c_{\rm
  S}$, respectively.  We list their meaning below.

Introduce the notation
\begin{equation}
  \label{eq:constraint_definition}
  C_X=\sum_{\Gamma_0,l_0} f(\Gamma_0,l_0) c_X(\Gamma_0,l_0) \, ,
\end{equation}
where $f(\Gamma_0,l_0)$ is the formal continuum limit of $f_i(l^{(\alpha)})$ defined in eq.~\eqref{Dfd}.
The bare mass $m_0$ is given by
\begin{equation}
  \label{eq:M}
  \bar{C}_{\rm S}=m_0 \,.
\end{equation}
The normalization condition on the $D\sim \gamma_\mu \partial_\mu$
term in $D$ gives
\begin{equation}
  \label{eq:V}
  C_{\rm V}=1 \,.
\end{equation}
The ${\cal O}(a)$ tree level Symanzik condition reads
\begin{equation}
  \label{eq:ST}
  C_{\rm S} +  C_{\rm T}=0 \,.
\end{equation}

The coefficient $C_{\rm P}$ is interesting if the parametrization
attempts to describe (approximately) a GW fermion.  In this case it is
related to the topological charge density $\tr (\gamma_5 \D R)$ 
(see Section \ref{subsec:GW}), 
\begin{equation}
  \label{eq:P}
  \sum_{\Gamma_0,l_0} f(\Gamma_0,l_0) c_{\rm P}(\Gamma_0,l_0) 
  R_{\rm free}(r_0)= \epsilon_5 \frac{1}{32\pi^2} \,.
\end{equation}
Of course, in the expressions above $\Gamma_0$ should be of the
corresponding type (S,T,\ldots).  The operator $R_{\rm free}(r_0)$, whose couplings of
are specified in Appendix \ref{cha:params}, is
given by $R(n,n+r_0;U=1)$ of the GW relation eq.(\ref{eq:GW}) and
$\epsilon_5$ is defined through
\begin{equation}
  \label{eq:epsilon_5}
  \gamma_5 = \epsilon_5 \gamma_1\gamma_2\gamma_3\gamma_4 \, ,
\end{equation}
which means that $\epsilon_5 = -1$ for our choice of the Clifford
algebra elements (see Appendix \ref{cha:clifford}).  The relation \eqref{eq:P} can be derived easily
from the work of Fujikawa in \cite{Fujikawa:1998if}.  In the parametrization of $\Dfp$ 
we make use of the relations \eqref{eq:M}-\eqref{eq:P} to constrain some of the parameters
in the formal continuum limit (see Section \ref{subsec:constraints}).

\section{Factorizing the Paths}
\label{sec:factorizing}

As the Tables~\ref{tab:refpath0}-\ref{tab:refpath4} show the number of
paths, in particularly for the offset $r_0=(0,0,0,0)$, is large. It is
important to calculate the path products efficiently. An obvious
method is to factorize the paths, i.e.~to write the sum of a large
number of paths as a product of sums over shorter paths. We shall try
to factorize in such a way that these shorter paths are mainly
plaquette, or staple products.

Because this factorization is a very technical issue, we defer the
detailed discussion of the factorization of all the paths involved in
the construction of $\Dpar$ and $R$ to Appendix \ref{cha:factorized_contribs}.
We however explain the idea on a simple example.  Consider the
contribution generated by $r_0=(0,0,0,0)$, $\Gamma_0 = \gamma_5$ and
$l_0 = [1,2,-1,-2,3,4,-3,-4]$. It contains together with its hermitian
conjugate 768 different paths of length 8. The sum over all these
terms can, however, be written as
\begin{gather}
  \label{eq:factor_g5_contact}
  \frac{1}{384} \gamma_5 \left. \sum_{\mu\nu\rho\sigma} \right.'
  \frac{1}{4} \epsilon_{\mu\nu\rho\sigma}
  P^{\rm (as)}_{\mu\nu}P^{\rm (as)}_{\rho\sigma} \, , \\
  \intertext{with}
  P^{\rm (as)}_{\mu\nu} = P^{(--)}_{\mu,\nu}-{\rm h.c.} \notag \\
  P^{(--)}_{\mu,\nu} =
  P_{\mu,\nu}-P_{\mu,-\nu}-P_{-\mu,\nu}+P_{-\mu,-\nu} \notag\, ,
\end{gather}
where $P_{\mu,\nu}$ is the plaquette in the $\mu\nu$-direction, which
is a path of length 4.  Hence, the whole contribution can be written
as the sum of 24 different products of plaquette
combinations. The reduction in
computational expense in comparison to the calculation of all the 768
length 8 paths is obvious. The observation that the plaquette combinations $P^{\rm (as)}_{\mu\nu}$,
as well as other combinations of plaquettes and staples,
appear also in other path factorizations makes it favourable to precalculate
them on the whole lattice, before one starts to build up the Dirac operator, thereby
making the concept of factorizing the paths even more attractive.

Using the factorization of paths the building of the matrix $\Dpar$
with all its $41$ different terms and linear fluctuation polynomials
in Tables \eqref{tab:refpath0}-\eqref{tab:gauge_functions} is as
expensive as ${\cal O}(20)$ matrix vector multiplications with $\Dpar$ on PC's and
Alpha workstations. On supercomputers, like the Hitachi SR8000 in
Munich, the cost of building the operator is no longer as
favourable, since such machines can optimize the matrix vector
multiplication much more efficiently and therefore the cost of the
building of $\Dpar$ is as expensive as ${\cal O}(500)$ matrix vector
multiplications. Such a cost becomes a noticeable fraction ($10-20\%$)
of the calculation of a full quark propagator. This shows
that the factorization of the paths is crucial in order to have a
Dirac operator whose build-up time is in a range where it is
computationally still reasonable.


\cpages
\chapter{Parametrization of the FP Dirac Operator}
\label{cha:parametrization}

The construction of a FP Dirac operator $\Dfp$ is a non-trivial task,
since it is defined through a highly non-linear Renormalization Group
equation.  While the solution can be calculated analytically in the
free case \cite{Kunszt_diss:97}, this is clearly not possible for the
interacting case.  Therefore, one has to find a way to approximate the
solution as well as possible, always keeping in mind that the
resulting Dirac operator shouldn't be too expensive for practical use.
In the following, we describe the details of our construction of an
approximation to $\Dfp$. We choose to approximate $\Dfp$ by a general
parametrization of a hypercubic Dirac operator as described in Chapter
\ref{cha:general_dirac}. It is clear that such a hypercubic
parametrization, which we denote by $\Dpar$, is quite a drastic
reduction compared to $\Dfp$, which has an infinite number of
couplings, like any other Dirac operator satisfying the GW relation
\cite{Horvath:1998cm,Bietenholz:1999dg,Horvath:1999bk,Horvath:2000az}. 
Results from e.g.~the O(3)
$\sigma$-model, however, support the hope that a compact parametrization
which encodes the main features of the full FP operator can be found
\cite{Blatter:1996ik}.  The use of a parametrization that contains
the full Clifford algebra is a very natural choice for the
parametrization of $\Dfp$, since all the Clifford elements are
generated through the RGT in eq.~\eqref{eq:RG_Dirac_intro}; a fact
which was already noticed in \cite{Bietenholz:1996cy,Kunszt_diss:97}.
Furthermore, important properties of Dirac operators satisfying the GW
relation, like that the topological charge density is given by ${\rm
  tr} (\gamma_5 R \D)$, imply that the full Clifford algebra is
necessary to achieve a parametrization that reproduces these
properties accurately.

The ${\rm SU}(3)$ gauge configurations we use for the parametrization
and later on in all the production runs are generated exclusively with
a parametrization of the FP gauge action. This gauge action has
scaling properties which are much improved compared to the Wilson
gauge action. A detailed analysis of the properties of this FP gauge
action is given in \cite{Wenger_diss:2000,Niedermayer:2000ts} and for the corresponding
anisotropic FP gauge action see \cite{Rufenacht:2001qy}. Another important part
of the parametrization is the use of a Renormalization Group inspired
smearing technique, which we describe in Section \ref{sec:smearing}.
The smearing of the gauge fields is a very helpful procedure in the
definition of a Dirac operator, because it reduces the unphysical
short-range fluctuations inherent in the gauge fields quite
significantly. This makes that the problems with additive mass
renormalization and chiral symmetry breaking, which are common to most
of the traditional fermion formulations, are already reduced
substantially, without even changing the fermion action 
\cite{DeGrand:1998pr}. The effect of this RG smearing is illustrated
in Figure \ref{fig:RG_smearing}, where the eigenvalue spectrum of the
Wilson Dirac operator for a pair of unsmeared and smeared
configurations is shown. The smearing clearly reduces the additive mass
renormalization and the fluctuations in the small (real) eigenmodes, which
are the reason for the so-called exceptional configurations. On such
an exceptional configuration the quark propagator can not be
calculated, because some of the real eigenmodes get so close to
zero -- notably at a non-zero bare quark mass -- that the inversion
algorithms break down. Finally, we also explain our parametrization of
the operator $R_{\rm FP}$, which is defined through eq.~\eqref{eq:FP_R}.

In the following, we first explain the strategy of the parametrization
and then discuss the ingredients of the parametrization, such as
fitting procedure, the blocking kernel and the smearing in detail. The
fundamental part of any parametrization of a Dirac operator is already
explained in Chapter \ref{cha:general_dirac}, namely the general
structure of a Dirac operator on the lattice and how one can do a
practical construction of such a general Dirac operator. The
parametrizations used in the crucial steps of the parametrization
procedure and the final parametrizations used in the production runs
are specified in Appendix \ref{cha:params}.

\section{The Equations for the FP Dirac Operator}
\label{sec:construction}

In order to construct our approximation to $\Dfp$, we solve the RG
equations that define $\Dfp$ and its inverse as good as possible
within the limitations of the chosen ansatz of the parametrization.
This means that we try to get an approximate solution for the following
two equations, which are actually equivalent as long as $\D_f(U)$ has
no zero mode:
\begin{equation}
  \label{eq:RG_Dirac}
  \D_c(V) = \kappa_f \, 1 
  - \kappa_f^2 \, \Omega(U[V]) [ \D_f(U[V]) + \kappa_f \,
  \Omega^\dagger(U[V])\Omega(U[V]) ]^{-1} \Omega^\dagger(U[V])
\end{equation}
and
\begin{equation}
  \label{eq:RG_propagator}
  \D^{-1}_c(V) = \frac{1}{\kappa_f} +
  \Omega(U[V]) \D^{-1}_f(U[V]) \Omega^{\dagger}(U[V]) \, .
\end{equation}
The labels $c$ and $f$ stand for the coarse and fine lattice and
indicate that the Dirac operators on both sides of the equation are
not the same operators as long one is not in the fixed point, where
the relation $\D_c = \D_f = \Dfp$ holds. The gauge field on
the fine lattice $U[V]$ is determined from the gauge field $V$ on the
coarse lattice by a minimization procedure.  The exact relation of the
U[V] and V fields is encoded in the FP equation for the gauge fields eq.~\eqref{eq:FP_equation}.
The fermionic blocking kernel $\Omega(U)$ and
the parameter $\kappa_f$ are discussed in detail in Section
\ref{sec:blocking_kernel}.

\section{The Fitting Procedure}
\label{sec:initial}

In principle one would like to have an approximate solution of
eqs.~\eqref{eq:RG_Dirac} and \eqref{eq:RG_propagator} which is valid
for a large range of gauge couplings. This however shows to be a
difficult thing to achieve, because the characteristic fluctuations in
the gauge fields change very much with the gauge coupling, as one can
see in Table \ref{tab:avplaq}, where the expectation value of the
trace of the plaquette is shown. For this reason we adopt an iterative
method for the whole parametrization procedure that starts in the
weak coupling regime, i.e.~at large values of $\beta$ where the
fluctuations of the gauge fields are small and finally ends at $\beta
\approx 3.0$, which corresponds to a lattice spacing of $a \approx
0.15$ fm.  This is amongst the coarsest lattice spacings where typical
lattice simulations with fermions are performed.
\begin{table}[tbph]
  \begin{center}
    \begin{tabular}{|c|c|c|c|} \hline
      $\beta$ & $\langle v\rangle $ &  $\langle w\rangle $ & $\langle u \rangle$ \\ \hline\hline
      100     & 2.92 & 2.989 & 2.998 \\
      10      & 2.33 & 2.90  & 2.987 \\
      5       & 1.68 & 2.76  & 2.97  \\
      3.4     & 1.24 & 2.62  & 2.95  \\
      3.0     & 1.16 & 2.56  & 2.94  \\
      2.7     & 1.05 & 2.49  & 2.94  \\ \hline 
    \end{tabular}
  \end{center}
  \caption{Average plaquette values $\langle v\rangle$ for the unsmeared
    coarse configurations $V$, $\langle w \rangle$ for the RG smeared coarse
    configurations (see Section \ref{sec:smearing}) 
    and $\langle u \rangle$ for the minimizing fine configurations $U[V]$ 
    at different values of the gauge coupling $\beta$.} 
  \label{tab:avplaq}
\end{table}

\begin{figure}[htbp]
  \begin{center}
    \psfrag{Df}[c]{$\D_f = \D_{\rm par}^{(n)}$\rule[-10pt]{0mm}{11pt}}
    \psfrag{Df2}[c]{$\D_f = \D_{\rm par}^{(n+1)}$} 
    \psfrag{Dc}[c]{$\D_c$}
    \psfrag{Dc2}[c]{$\D_c$} 
    \psfrag{P}[c]{\footnotesize \rule[0pt]{10pt}{0pt}Parametrization}
    \psfrag{RGT}[c]{RGT}
    \psfrag{RGT1}[c]{RGT}
    \psfrag{UN}[c]{$U^{(n)}$}
    \psfrag{UN1}[c]{\rule[0pt]{10pt}{0pt}$U^{(n+1)}$}
    \psfrag{VN}[c]{$V^{(n)}$}
    \psfrag{VN1}[c]{\rule[0pt]{4pt}{0pt}$V^{(n+1)}$}
    \includegraphics[width=13cm]{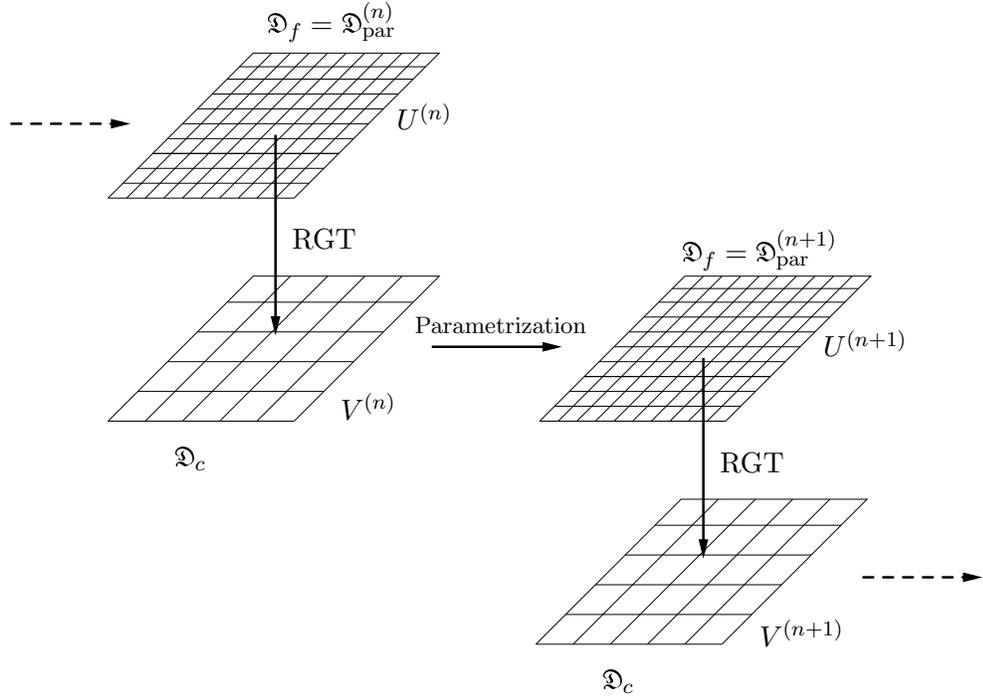}
    \caption{The iterative parametrization procedure used for the parametrization of
      $\Dfp$. Starting with a hypercubic parametrization of the FP
      Dirac operator $\D_f = \D_{\rm par}^{(n)}$ on a set of fine
      configurations $U^{(n)}[V^{(n)}]$, a Renormalization Group
      transformation (RGT) yields $\D_c$ on the corresponding coarse
      configurations $V^{(n)}$.  Subsequently, $\D_c$ is approximated
      using again a hypercubic parametrization which results in
      $\D_{\rm par}^{(n+1)}$. This parametrization is the starting
      point for the next step of the iterative procedure. The crucial
      point is that the fluctuations in the gauge fields are increased
      substantially with each RGT. Hence, $\D_{\rm par}^{(n+1)}$
      describes $\Dfp$ on a larger range of gauge couplings than
      $\D_{\rm par}^{(n)}$. It is also possible to use the resulting
      parametrization $\D_{\rm par}^{(n+1)}$ on the same fine
      configurations $U^{(n+1)}[V^{(n+1)}]= U^{(n)}[V^{(n)}]$ again in
      order to drive $\D_{\rm par}$ closer to the FP.}
    \label{fig:parametrization_sketch}
  \end{center}
\end{figure}

Let us first have a closer look at one step in this iterative
procedure which is also sketched in Figure
\ref{fig:parametrization_sketch}.
Assume that in the $n^{th}$ iteration, we have a Dirac operator $\D_{\rm
  par}^{(n)}$ which is a good approximation to $\Dfp$.  Actually,
$\D_{\rm par}^{(n)}$ has to be a good approximation only on minimized
gauge configurations $U^{(n)}[V^{(n)}]$ which have a certain level of
fluctuations, since we use $\D_{\rm par}^{(n)}$ as $\D_f$ in the RG
equations \eqref{eq:RG_Dirac} and \eqref{eq:RG_propagator},
respectively.  After the RGT, the resulting Dirac operator on the
coarse lattice $\D_c$ is approximated as good as possible within the
chosen ansatz for the parametrization. This Dirac operator $\D_{\rm
  par}^{(n+1)}$ is the parametrization of the next level of the
iterative procedure. The idea behind this step is that the
fluctuations of the gauge fields on the minimized gauge configurations
$U^{(n)}[V^{(n)}]$ are much smaller than the ones on the corresponding
coarse gauge configurations $V^{(n)}$ and thus $\D_{\rm par}^{(n+1)}$
is an approximation to $\Dfp$ on gauge fields with much larger
fluctuations than $\D_{\rm par}^{(n)}$. At this point it should be
noted, that it is much easier to find a good parametrization of $\Dfp$
on configurations with small fluctuations.  Therefore, the step which
leads from $\D_{\rm par}^{(n)}$ to $\D_{\rm par}^{(n+1)}$ is the
crucial step in the whole parametrization procedure and it has to be
done with great care. The next step is now to choose a coarse
configuration $V^{(n+1)}$, such that the fluctuations of the
corresponding minimized configuration $U^{(n+1)}[V^{(n+1)}]$ are
roughly equal to the fluctuations of $V^{(n)}$ and then $\D_{\rm
  par}^{(n+1)}$ can be used as $\D_f$ in the RG equations
\eqref{eq:RG_Dirac} and \eqref{eq:RG_propagator}.  Starting from
minimized configurations which are very far in the continuum limit and
performing these steps several times, this procedure finally yields a
parametrization of $\Dfp$ at intermediate to strong gauge coupling.
This explains roughly the principle of our parametrization procedure.
In the following, we indicate how the intermediate parametrization
steps are done in detail. The Dirac operator used in the first step
and the details of the subsequent steps of the iterative procedure
will be discussed later in this chapter.

\subsection{The Details of the Fit}
\label{subsec:details_of_the_fit}
\begin{figure}[tbph]
  \begin{center}
    \includegraphics[width=8cm]{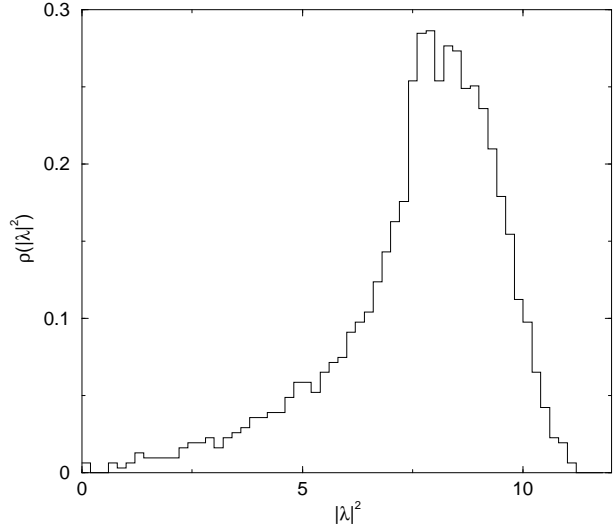}
    \caption{Eigenvalue density $\rho(|\lambda|^2)$ of
      $\D_{\rm par}$ on a $4^4$ lattice at $\beta = 3.0$.}
    \label{fig:density}
  \end{center}
\end{figure}

Let us first focus on the solution of eq.~\eqref{eq:RG_Dirac}, because
in this equation $\D_c$ and not its inverse enters and this is much
simpler to use for parametrization purposes than
eq.~\eqref{eq:RG_propagator}.  But for computational reasons we can
not afford to invert the full operator in eq.~\eqref{eq:RG_Dirac} for
the lattice sizes we use in our parametrization, which are $10^4$ for
the fine i.e.~$5^4$ for the coarse lattice\footnote{For weak gauge
  coupling, i.e.~at $\beta=100$ and $\beta=10$ , $6^4$ lattices for
  the fine and $3^4$ lattices for the coarse configurations are used,
  since the use of larger lattices does not lead to a substantial
  improvement in the parametrization.} and therefore we use this
operator equation acting on a set of normalized vectors. These vectors
are chosen from the following two sets:
\begin{itemize}
\item Random vectors\footnote{Instead of random vectors one may also
    take local vectors, i.e.~vectors with one single non-zero
    entry. It shows that their effect on the parametrization is in
    fact the same as the one of the random vectors. The local vectors
    are, however, less efficient in carrying information about the bulk
    behaviour, which is the reason that we prefer the use of random
    vectors.} with entries from a uniform random distribution in the
  interval $[-1,1]$ for the real and imaginary part. These vectors
  give a large weight to the high-lying modes of the blocked operator,
  because the density of the eigenmodes of the Dirac operator is
  heavily peaked towards the upper end of the spectrum, as illustrated
  in Figure \ref{fig:density}.
\item Low-lying eigenmodes of the operator ${\cal O}\dot{=}\Omega(U)
  \D^\dagger_f(U) \D_f(U) \Omega^\dagger(U)$ determined with the Ritz
  functional method \cite{Bunk:1994, Kalkreuter:1996mm}.  In clear
  contrast to the random vectors these vectors give a large weight to
  the low-lying mode contributions of the blocked operator.  We choose
  the eigenmodes of ${\cal O}$, because this operator is defined on the
  coarse lattice and therefore the determination of its eigenmodes is
  computationally not very expensive. This would not be the case, if
  we chose to determine the low-lying eigenmodes of $\D_f(U)$ and then
  block them to the coarse lattice, even though this would be the more natural choice.
\end{itemize}

Our ansatz for $\Dpar$ contains only linear couplings, i.e.~it can be
written in the following way
\begin{equation}
  \label{eq:linear_operators}
  \Dpar(V) = \sum_{i=1, n_{\rm op}} c_i O_i(V) \, ,
\end{equation}
where the $c_i$ are the couplings and $n_{\rm op}$ the number of
terms, which is $82$ in the present parametrization. In comparison to
eqs.~\eqref{Dfd}-\eqref{eq:linear_fluctuation_function} we have
expanded all the linear fluctuation polynomials in
eq.~\eqref{eq:linear_operators} in order to make the linear structure
obvious.

\subsection{Constraints}
\label{subsec:constraints}

In order to have a Dirac operator which respects properties that are
highly desirable for a decent parametrization of $\Dfp$ we add
several constraints for some of the couplings $c_i$ during the fit.
The properties we enforce at various levels of our parametrization
procedure are relations that provide correct normalizations in the
formal continuum limit for the following quantities:
\begin{quote}
\begin{itemize}
\item[${\cal C}_1$:] The bare quark mass is fixed to $0$.
\item[${\cal C}_2$:] The speed of light in the free energy-momentum
  relation is normalized to $1$.
\item[${\cal C}_3$:] ${\cal O}(a)$ tree level improvement is imposed.
\item[${\cal C}_4$:] The topological charge density $\tr (\gamma_5 R
  \D)$ is normalized to $\frac{\epsilon_5}{32 \pi^2}
  \sum_{\mu\nu\rho\sigma} \epsilon_{\mu \nu \rho \sigma}  F^{a}_{\mu
    \nu} F^{a}_{\rho \sigma}$, where in our case $\epsilon_5 = -1$ due to
  the choice of $\gamma_5 = -
  \gamma_1\gamma_2\gamma_3\gamma_4$\footnote{In steps I-IV of the
    parametrization procedure (see Table \ref{tab:Dparametrization})
    the constraint on the topological charge density was imposed with
    the wrong sign, because we initially overlooked that the results
    in \cite{Fujikawa:1998if} are obtained with a different definition
    of $\gamma_5$. There are,
    however, several reasons why we expect this error to be of minor
    consequence.  The constraint is only imposed on the constant terms
    of the fluctuation polynomials, i.e.~that the linear terms can
    correct for the wrongly imposed constraint, and as the cross check of
    the index calculated with the parametrization IV and the corresponding overlap 
    construction (see Chapter \ref{cha:overlap}) shows, this is indeed the case, i.e.~that the wrong sign
    in constraint ${\cal C}_4$ is straightened out by the linear polynomials (see also
    Section \ref{subsec:index_theorem}). Furthermore, the
    pseudoscalar terms are of very small size compared to the scalar,
    vector and tensor terms and therefore have a limited influence on
    the parametrization procedure described in Section
    \ref{subsec:details_of_the_fit}.  Finally, the final as well as
    all the intermediate parametrizations describe a valid Dirac
    operator that satisfies all the symmetry conditions described in
    Chapter \ref{cha:general_dirac}.}.
\item[${\cal C}_5$:] The free field limit is such that it coincides
  with the hypercubic parametrization of the free massless FP Dirac
  operator given in Appendix \ref{cha:params}. 
\end{itemize}
\end{quote}
Note that these constraints, which are encoded by
eqs.~\eqref{eq:constraint_definition}-\eqref{eq:P}, only affect the
constants in the fluctuation polynomials from
eq.~\eqref{eq:linear_fluctuation_function}, since the linear terms as
defined in eq.~\eqref{eq:fluctuation-definition} and Table
\ref{tab:gauge_functions} vanish in the formal continuum limit.

\subsection{Determination of the Couplings}
\label{subsec:minimization}

We fix the coefficients $c_i$ of our parametrization by minimizing the
following $\chi^2$-function
\begin{equation}
  \label{eq:chi2}
  \chi^2 = \frac{1}{n_{\rm vec}}
  \sum_{\genfrac{}{}{0pt}{}{j=1,n_{\rm vec}^{(i)}}{i=1,n_{\rm conf}}} 
  \big|\big| [\D_c(V^{(i)}) - \Dpar(V^{(i)})] v^{(i,j)} \big|\big|^2 + 
  \sum_{i=1}^5 \lambda_i \chi^2_{{\cal C}_i} \, ,
\end{equation}
where the sum runs over $n_{\rm conf}$ different configurations with a
total of $n_{\rm vec} = \sum_{i=1,n_{\rm conf}} n_{\rm vec}^{(i)}$
vectors $v^{(i,j)}$ used in this parametrization step.  The
constraints ${\cal C}_i$ are enforced with weights $\lambda_i>0$, which
can be freely chosen. Typically, we use $\lambda_i = 10^4$ such that
the constraints are enforced to a high level.

\begin{figure}[tbph]
  \begin{center}
    \includegraphics[width=9cm]{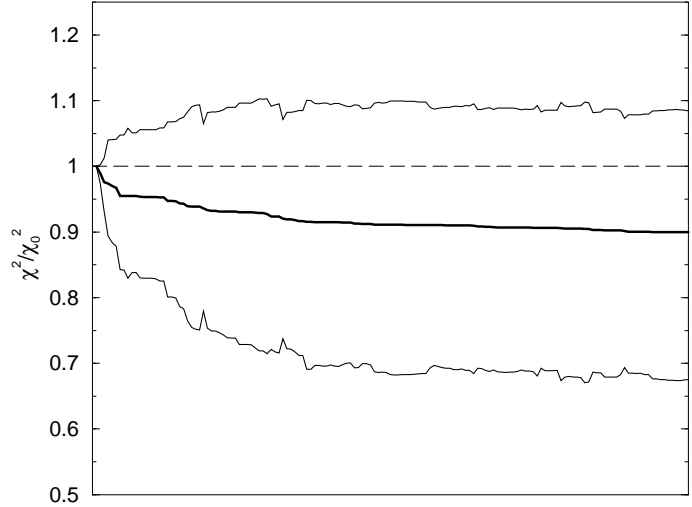}
    \caption{Typical convergence history of the non-linear $\chi^2$-minimization in the
      simulated annealing procedure. Even though the normalized total
      $\chi^2$ (thick line) only decreases by $10\%$, the normalized
      $\chi^2$ for the propagator relation
      eq.~\eqref{eq:RG_propagator} (lower line) decreases by $33\%$,
      while the normalized $\chi^2$ for the relation
      eq.~\eqref{eq:RG_Dirac} (upper line) increases only by $9\%$. Notice
      that the horizontal axis does not exactly 
      correspond to the number of steps in simulated annealing
      procedure. The graph does only show those steps in which the
      total $\chi^2$ decreases.}
    \label{fig:convhist}
  \end{center}
\end{figure}

Due to the linear ansatz in eq.~\eqref{eq:linear_operators} and the fact
that the constraints ${\cal C}_i$ are all linear in the coefficients
$c_i$ this $\chi^2$-minimization amounts simply to a matrix inversion
and therefore poses no serious computational problems. We use the
quasi minimal residual (QMR) matrix inverter to do the actual
calculation \cite{freund_nachtigal:91}. Unfortunately, it shows that
in the region of stronger gauge coupling, i.e.~$\beta < 10$, the
described fitting procedure can no longer cope well with all the
requirements, especially one observes an additive mass renormalization
and, more importantly, a spread in the low-lying eigenmodes that is
too large for the purposes for which we intend to use this Dirac
operator. A solution to this problem can be found within the
fixed point approach itself. Namely, one can give more weight to the
important low-lying modes by including also the FP propagator relation
from eq.~\eqref{eq:RG_propagator} into the fit. We can include this
additional information from the propagator by setting up a different
$\chi^2$-function, which is a combination of both RG
eqs.~\eqref{eq:RG_Dirac} and \eqref{eq:RG_propagator} and the
additional constraints ${\cal C}_i$
\begin{multline}
  \label{eq:chi2_inverse}
  \chi^2 = \frac{1}{n_{\rm vec}} \sum_{\genfrac{}{}{0pt}{}{j=1,n_{\rm
        vec}^{(i)}}{i=1,n_{\rm conf}}}
  \big|\big| [ \D_c(V^{(i)}) - \D_{\rm par}(V^{(i)})] v^{(i,j)} \big|\big|^2 + \\
  \alpha \frac{1}{m_{\rm vec}} \sum_{\genfrac{}{}{0pt}{}{j=1,m_{\rm
        vec}^{(i)}}{i=1,m_{\rm conf}}} \big|\big| [\D^{-1}_c(V^{(i)})-
  \D^{-1}_{\rm par}(V^{(i)})] w^{(i,j)} \big|\big|^2 + \sum_{i=1}^5
  \lambda_i \chi^2_{{\cal C}_i} \, ,
\end{multline}
where $\alpha>0$ is chosen such that the two contributions in the
$\chi^2$-function are approximately equal at the beginning of the
$\chi^2$-minimization. The vectors $ w^{(i,j)}$ are the same type of
random vectors as discussed in Section
\ref{subsec:details_of_the_fit}.  The gauge configurations that are
used for the propagator fit have to be in the $Q=0$ topological
sector, because (approximate) zero modes would dominate the
$\chi^2$-function completely.  As starting point for this
minimization, which is now a highly non-linear problem through the use
of the propagators, we use the coefficients $c_i$ obtained from the
best fit in the corresponding linear $\chi^2$-problem ($\alpha = 0$).
The non-linear optimization is performed with a generic simulated
annealing algorithm \cite{Goffe:1994}. Because our ansatz for $\Dpar$
contains 82 parameters, the parameter space is rather large and each
function evaluation with a new set of parameters involves the
inversion of the Dirac operator on several sources, making this
non-linear optimization computationally very expensive and therefore
an exhaustive search in the parameter space can not be afforded with
our computing resources.  This is however not a big problem, as the
value of the $\chi^2$-function typically decreases quite rapidly in
the first few sweeps through the parameter space and afterwards only
small improvements can be achieved.  It shows that this non-linear
minimization can improve the $\chi^2$ for the propagator relation
eq.~\eqref{eq:RG_propagator} quite substantially, without destroying
the quality of the fit for the relation eq.~\eqref{eq:RG_Dirac}.  An
illustration of these facts is given in Figure \ref{fig:convhist}.
The result of this non-linear minimization is a parametrized Dirac
operator which has only a very small remaining additive mass
renormalization and small fluctuations in the low-lying modes. Furthermore, 
it shows that the FP equations enforce
the GW relation in terms of the propagators \eqref{eq:GW_propagator}
to hold to a good extent even without adding this relation as an
additional constraint during the fit. One even finds that the $\chi^2$
for the propagator relation eq.~\eqref{eq:RG_propagator} and the
$\chi^2$ for the GW relation are highly correlated
\cite{Hauswirth_diss:2002}.

\subsection{The Initial Dirac Operator and the Parametrization Steps}
\label{subsec:parametrization_details}

\begin{figure}[htbp]
  \begin{center}
    \includegraphics[width=10cm]{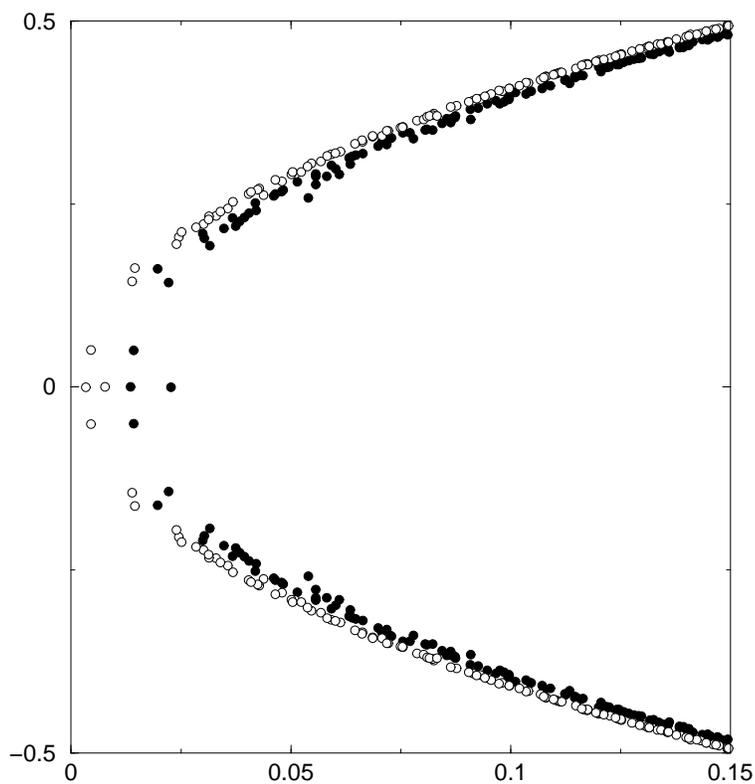}
    \caption{The effect of the overlap reparametrization on minimized 
      $\beta=3.0$ configurations on the low-lying eigenvalues is shown
      (notice the horizontal scale). The initial parametrization
      (filled circles) has a larger additive mass renormalization and
      the fluctuations in the low-lying eigenmodes are apparently
      larger than for the reparametrized operator (open circles).}
    \label{fig:overlap_reparametrization}
  \end{center}
\end{figure}

At the first level of the iterative parametrization procedure shown in
Figure \ref{fig:parametrization_sketch} we use a hypercubic truncation
of the free massless $\Dfp$, which was obtained analytically in
\cite{Kunszt_diss:97} for the overlapping block transformation we use
in the parametrization procedure. This hypercubic parametrization is
explicitly given in Appendix \ref{cha:params}. This operator has
properties that are already much improved compared to the Wilson
Dirac operator, as we show in Chapter \ref{cha:properties}. A more
detailed account on different hypercubic parametrizations of $\Dfp$ in
the free case can be found in \cite{Kunszt_diss:97}.  Starting with
this hypercubic parametrization of the free massless $\Dfp$, the
complete parametrization procedure involves 5 main steps, which are
shown in Table \ref{tab:Dparametrization}.  In step I this hypercubic
parametrization of the free massless $\Dfp$ is used on minimized
$\beta = 100$ configurations. The parametrization procedure yields
$\Dpar$ for thermal configurations at $\beta = 100$. This operator is
inserted again on minimized $\beta = 100$ configurations, until the
$\chi^2$ of the fit does no longer decrease.  In step II the same
iteration is performed on the $\beta = 10$ configurations, starting
with the result of the $\beta = 100$ parametrization. Step III, which
now involves the propagators in the fit, is an iteration on the
$\beta=2.7,\ldots,3.4$ configurations, starting with the result of
$\beta = 10$ parametrization. In step IV a new idea is used, namely a
reparametrization of the Dirac operator on minimized configurations.
This is the only step that deviates from the FP idea, because it
involves an overlap expansion using $\Dpar$. We include this step,
because characteristics of minimized $\beta=3.0$ configurations are
different from thermal configurations and therefore it is difficult to
find a parametrization that has fluctuations of the low-lying
eigenmodes which are small enough for our purposes. Having small
fluctuations on the level of the minimized configurations is crucial,
as the RGT magnifies these fluctuations by roughly a factor 2. More
precisely, we use the same fitting procedure as described in Section
\ref{subsec:details_of_the_fit}, but the vectors used in the fit are
generated by an order $5$ Legendre expansion of the overlap Dirac
operator with $\Dpar$\footnote{The Legendre expansion of the overlap
  Dirac operator with $\Dpar$ is described in detail in Chapter
  \ref{cha:overlap}.}. Since $\Dpar$ is already close to
satisfy the GW relation, the overlap expansion is only a small
correction to $\Dpar$, making the remaining breaking of chiral
symmetry even smaller. The effect of the reparametrization is shown in
Figure \ref{fig:overlap_reparametrization}.  Finally, in step V one
last RGT leads to the final parametrization on $\beta=2.7,\ldots,3.4$
configurations.
\begin{table}[htbp]
  \begin{center}
    \begin{tabular}{|c|c|c|c|c|c|c|c|c|c|c|c|c|}\hline
      step & $\beta$ & size  & $n_{\rm conf}$ & $n_{\rm v}$ & 
      $m_{\rm conf}$ & $m_{\rm v}$ & ${\cal C}_1$ & ${\cal C}_2$ & 
      ${\cal C}_3$ & ${\cal C}_4$ &  ${\cal C}_5$ & $\leftrightarrow$\\ \hline \hline
      I     &    100.0    & $3^4$ & 10  & 5 & - & - & $\times$ &
      $\times$ & $\times$ & $\times$ & $\times$ & $\times$ \\ \hline
      II    &     10.0    & $3^4$ & 10  &10 & - & - & $\times$ & $\times$ 
      & $\times$& $\times$& $\times$& $\times$\\ \hline
      III   &      2.7    & $5^4$ &  2  & 5 & - & - & $\times$ & $\times$ 
      & $\times$& $\times$& $\times$& $\times$\\
            &      2.8    & $5^4$ &  2  & 5 & - & - & & & & & &\\
            &      2.9    & $5^4$ &  2  & 5 & 1 & 3 & & & & & &\\
            &      3.0    & $5^4$ &  2  & 5 & 3 & 3 & & & & & &\\
            &      3.1    & $5^4$ &  2  & 5 & - & - & & & & & &\\
            &      3.2    & $5^4$ &  2  & 5 & - & - & & & & & &\\
            &      3.3    & $5^4$ &  2  & 5 & - & - & & & & & &\\
            &      3.4    & $5^4$ &  2  & 5 & 2 & 3 & & & & & &\\ \hline
      IV   & {\small min.}~3.0    & $6^4$ & 10  & 5& - & - & $\times$ & $\times$ & $\times$ & $\times$ & $\times$ & \\ \hline
      V    &      2.7     & $5^4$ &  2  & 5 & - & - & $\times$ & $\times$ & & & &\\
           &      2.8     & $5^4$ &  2  & 5 & - & - & & & & & &\\
           &      2.9     & $5^4$ &  2  & 5 & 1 & 3 & & & & & &\\
           &      3.0     & $5^4$ &  2  & 5 & 3 & 3 & & & & & &\\
           &      3.1     & $5^4$ &  2  & 5 & - & - & & & & & &\\
           &      3.2     & $5^4$ &  2  & 5 & - & - & & & & & &\\
           &      3.3     & $5^4$ &  2  & 5 & - & - & & & & & &\\
           &      3.4     & $5^4$ &  2  & 5 & 2 & 3 & & & & & &\\ \hline
    \end{tabular} 
    \caption{The $5$ main steps of the parametrization procedure of $\Dpar$.
      The number of configurations for the $n_{\rm v}$ random vectors
      and $n_{\rm v}$ eigenvectors is given by $n_{\rm conf}$, while
      the number of configurations for the $m_{\rm v}$ propagators is
      given by $m_{\rm conf}$. ${\cal C}_1$ - ${\cal C}_5$ indicate
      which of the constraints discussed in Section
      \ref{subsec:constraints} are imposed during the fit. Finally,
      the $\leftrightarrow$ symbol indicates, whether the
      parametrization procedure was iterated on the same
      configurations. As one can see, only steps III and V involve a
      non-linear minimization of the $\chi^2$-function given in
      eq.~\eqref{eq:chi2_inverse}.}
    \label{tab:Dparametrization}
  \end{center}
\end{table}

\section{The Fermionic Blocking Kernel $T_f$}
\label{sec:blocking_kernel}

In order to perform the Renormalization Group transformation in
eqs.~\eqref{eq:RG_Dirac} and \eqref{eq:RG_propagator} numerically, the
averaging function $\Omega(U)$, which defines how the fields $\psi$
from the fine lattice are blocked to the fields $\chi$ on the coarse
lattice, has to be specified.  We choose the overlapping block
transformation discussed in \cite{Kunszt_diss:97}, which has the
property that in the free case all the sites on the fine lattice
contribute equally to the average on the coarse lattice. There are
essentially two ways how one can introduce a block transformation in
the presence of gauge fields. One can fix the gauge and then one can
work with the same definition of the block transformation as in the
free case, i.e.~that one can do a simple averaging of the fields on
the fine lattice.  We find, however, a gauge invariant procedure more
attractive. For the overlapping block transformation gauge invariance
can be achieved by introducing a product of gauge links connecting the
points $n_B$ on the coarse with the points $n$ on the fine lattice,
i.e.~parallel transporting the fine fermion fields before averaging
them. Using the definitions from eq.~\eqref{eq:Ul1} and defining
\begin{align}
  V_{\mu} &= \hat{U}([\mu]) \,,& V_{\mu,\nu} &= \hat{U}([\mu,\nu]) \,, \notag\\
  V_{\mu,\nu,\rho} &= \hat{U}([\mu,\nu,\rho]) \,,&
  V_{\mu,\nu,\rho,\sigma} &= \hat{U}([\mu,\nu,\rho,\sigma]) \,, \notag
\end{align}
as well as the completely symmetric combinations of the shortest paths
to a given offset
\begin{align}
  V^{(sym)}_\mu &= V_{\mu} + V_{-\mu} \notag \\
  V^{(sym)}_{\mu\nu} &=  V_{\mu,\nu} + V_{\mu,-\nu} +  V_{-\mu,\nu} + V_{-\mu,-\nu} \notag\\
  V^{(sym)}_{\mu\nu\rho} &=  V_{\mu,\nu,\rho} + V_{\mu,\nu,-\rho} +  V_{\mu,-\nu,\rho} + V_{\mu,-\nu,-\rho} \notag\\
  &+ V_{-\mu,\nu,\rho} + V_{-\mu,\nu,-\rho} +  V_{-\mu,-\nu,\rho} + V_{-\mu,-\nu,-\rho} \notag\\
  V^{(sym)}_{1234} &= V_{1,2,3,4} + V_{1,2,3,-4} + V_{1,2,-3,4} + V_{1,2,-3,-4} \notag \\
  & + V_{1,-2,3,4} + V_{1,-2,3,-4} + V_{1,-2,-3,4} + V_{1,-2,-3,-4} \notag \\
  & + V_{-1,2,3,4} + V_{-1,2,3,-4} + V_{-1,2,-3,4} + V_{-1,2,-3,-4} \notag \\
  & + V_{-1,-2,3,4} + V_{-1,-2,3,-4} + V_{-1,-2,-3,4} +
  V_{-1,-2,-3,-4} \notag
  \label{eq:defintions_for_block2}
\end{align}
the overlapping block transformation in the presence of gauge fields
can be defined as follows
\begin{equation}
  \Omega(U) = b_f \omega(U) \notag
\end{equation}
with
\begin{multline}
  \label{eq:block_average_fermion}
  \omega(U)_{n_B,n} = \notag\\
  \Big[c_0 1 + c_1 \sum_\mu V^{(sym)}_\mu + c_2 \sum_{\mu<\nu}
  V^{(sym)}_{\mu\nu} + c_3 \sum_{\mu<\nu<\rho} V^{(sym)}_{\mu\nu\rho}
  + c_4 V^{(sym)}_{1234} \Big]_{n_B,n} \, ,
\end{multline}
where the coefficients $c_i$ are given by
\begin{equation}
  \label{eq:block_coefficients}
  c_i = 2^{-d-i}, 
\end{equation}
with the spacetime dimension $d=4$.  The parameter $b_f$ is determined
by the engineering dimension of the field $\psi$ and is therefore
fixed to $b_f = 2^{(d-1)/2}$ \cite{Kunszt_diss:97}.  With all these
definitions the fermionic blocking kernel now reads
\begin{multline}
  T_f(U,\bar{\chi},\chi,\bar{\psi},\psi) = \kappa_f
  \sum_{n_B}\big(\bar{\chi}_{n_B}
  - \sum_{n} \bar{\psi}_n \Omega^\dagger(U)_{n,n_B}\big) \\
  \times \big(\chi_{n_B} - \sum_{n} \Omega(U)_{n_B,n} \psi_n\big) \, .
\end{multline}
This leaves $\kappa_f$ as the only free parameter of the block
transformation.  It can be used to optimize the range of the blocked
Dirac operator $\D_c$. Using a parametrization of $\Dfp$ that has couplings
on the hypercube only, this optimization of the interaction range
becomes a very crucial point in finding a good approximation. In order
to find out for which value of $\kappa_f$ the couplings of blocked
Dirac operator have the fastest fall off, however, is computationally a
very expensive thing to do in the interacting case.  Therefore, we
choose $\kappa_f$ such that the couplings of the corresponding $\Dfp$
in the free case have the fastest fall off, which is at $\kappa_f
\approx 3.45$. Finally, we do not use any smearing in the definition
of the blocking kernel even though, in principle, everything in this
section is valid for smeared configurations.

\section{The Smearing}
\label{sec:smearing}

During the parametrization of $\Dfp$ we use two different types of
smearing schemes for the gauge fields, namely a smearing that is
similar to the APE smearing \cite{Albanese:1987ds} and a smearing
scheme that is closely related to the RG mapping discussed in
\cite{DeGrand:1998ss}. The APE smearing is used practically throughout
the whole parametrization procedure, only in step V when we make the
final parametrization of $\Dfp$ on configurations with gauge couplings
$\beta \sim 2.7,\ldots,3.4$ we use the more sophisticated RG inspired
smearing. Also in the parametrization of $R_{\rm FP}$ the RG inspired smearing
is used only in the last step of the parametrization, whereas in the
other steps the APE smearing is used.

\subsubsection*{Modified APE Smearing}
\label{subsubsec:APE}

The modified APE smearing\footnote{In the original APE smearing the
  third term in eq.~\eqref{eq:APE_smearing} is not present.} is
defined through the following transformation of the gauge links
$U_{\mu}(n)$ into $W_{\mu}(n)$
\begin{equation}
  \label{eq:APE_smearing}
  W_{\mu}(n) = {\cal P}_{\rm SU(3)} \left \{ U_{\mu}(n) + c_1 Q_{\mu}(n) + 
    c_2 Q_{\mu}(n) U^\dagger_{\mu}(n+\hat{\mu}) Q_{\mu}(n)  \right \} \, ,
\end{equation}
where ${\cal P}_{\rm SU(3)}$ denotes the projection to ${\rm SU}(3)$ and
where $Q_{\mu}(n)$ is the average of all the staples minus the
corresponding link
\begin{multline}
  \label{eq:Q_definition}
  Q_{\mu}(n) = \frac{1}{6} \sum_{\lambda \ne \mu}
  [U_\lambda(n)  U_\mu(n+\hat{\lambda}) U^\dagger_\lambda(n+\hat{\mu}) +\\
  U^\dagger_\lambda(n-\hat{\lambda}) U_\mu(n-\hat{\lambda})
  U_\lambda(n-\hat{\lambda}+\hat{\mu})] - U_{\mu}(n) \, .
\end{multline}
This transformation of the original gauge links $U_{\mu}(n)$ into
$W_{\mu}(n)$ has the property that it
reproduces the trivial gauge configuration for any choice of the
parameters $c_1$ and $c_2$.  The resulting gauge configuration $W$ can
be used as starting point for further smearing steps.

In the parametrization of $\Dfp$ we use a $2$-level APE smearing. We do not use more smearing
steps in order not to endanger the locality of the smearing. In
steps I and II we use $c_1=0.23$ and $c_2=-0.25$, whereas in steps III
and IV we use $c_1=-0.10$ and $c_2=-0.70$.  Using these values for the
coefficients the fluctuations measured by the expectation value of the
plaquette trace are reduced substantially.

\subsubsection*{RG Smearing}
\label{subsubsec:RG_smearing}

\begin{figure}[tbhp]
  \begin{center}
    \psfrag{M}[c]{\footnotesize  \rule[0pt]{10pt}{0pt}Minimization} \psfrag{S}[c]{RG
      Smearing} \psfrag{R}[c]{\footnotesize  \rule[0pt]{10pt}{0pt}Blocking}
    \includegraphics[width=\textwidth]{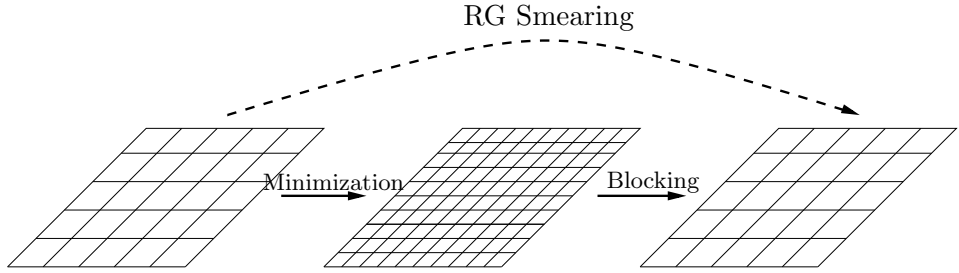}
    \caption{Schematic picture of the idea of the RG smearing procedure. The RG cycle, which
      consists of a minimization step from the configuration $V$ to
      the configuration $U[V]$ and a subsequent blocking step from the
      configuration $U[V]$ to the configuration $\tilde{V}[U[V]]$, is
      approximated by the RG smearing, which is a local transformation
      of the coarse field $V$.}
    \label{fig:smearing}
  \end{center}
\end{figure}

Performing a minimization and subsequently a blocking step one defines
a mapping of a coarse configuration $V$ onto another coarse
configuration $\tilde{V}[U[V]]$; a procedure which is called RG
cycle in \cite{DeGrand:1998de}. The result of the RG cycle
is a configuration that has much smaller UV fluctuations than the original configuration.
It however still encodes
the same long-range properties as the initial configuration due to the
properties of the RG. The basic idea of the RG smearing is to combine
the effect of a minimization and blocking step into one local
transformation of the coarse gauge links as shown in Figure
\ref{fig:smearing}.

In order to approximate such a RG cycle we use a pure gauge operator,
which contains in addition to the simple staples also diagonal
staples.  These are built by first going in the planar or spatial
diagonal directions orthogonal to the direction of the link $\mu$, followed by a step
in direction $\hat{\mu}$ and
finally returning along the corresponding diagonal to $n + \hat{\mu}$.
To be be specific, let us first build the matrices
\begin{align}
  \label{eq:diagonal_links}
  W^{(0)}(n,n+\hat{\nu}) &= 1 \\
  W^{(1)}(n,n+\hat{\nu}) &= U_{\nu}(n) \\
  W^{(2)}(n,n+\hat{\nu}+\hat{\rho}) &= {\cal P}_{\rm SU(3)} \Big\{
  \frac{1}{2}
  [ U_{\nu}(n) U_{\rho}(n+\hat{\nu}) +  U_{\rho}(n) U_{\nu}(n+\hat{\rho})] \Big\}\\
  \label{eq:diagonal_links_end}
  W^{(3)}(n,n+\hat{\nu}+\hat{\rho}+\hat{\lambda}) &= {\cal P}_{\rm
    SU(3)} \Big\{ \frac{1}{6} [ U_{\nu}(n) U_{\rho}(n+\hat{\nu})
  U_{\lambda}(n+\hat{\rho}+ \hat{\nu}) + {\rm perms.}] \Big\} \, .
\end{align}
Here $\nu, \rho$ and $\lambda$ go over all directions (positive and
negative) different from $\mu$ and from each other. In
eqs.~\eqref{eq:diagonal_links}-\eqref{eq:diagonal_links_end} the sum is taken over all shortest
paths leading to the endpoint $n^\prime$ of the corresponding
diagonal. Out of the $W^{(a)}(n,n^\prime)$ we construct the
generalized staples $V^{(a)}(n,n+\hat{\mu})$ given by
\begin{align}
  \label{eq:fuzzy_links_V}
  V^{(1)}(n,n+\hat{\mu}) &= \frac{1}{6} \sum_{n^\prime}
  W^{(1)}(n,n^\prime)
  U_{\mu}(n^\prime) W^{(1)}(n^\prime+\hat{\mu}, n+\hat{\mu}) -  U_{\mu}(n)\\
  V^{(2)}(n,n+\hat{\mu}) &= \frac{1}{12} \sum_{n^\prime}
  W^{(2)}(n,n^\prime)
  U_{\mu}(n^\prime) W^{(2)}(n^\prime+\hat{\mu}, n+\hat{\mu}) -  U_{\mu}(n)\\
  \label{eq:fuzzy_links_V_end}
  V^{(3)}(n,n+\hat{\mu}) &= \frac{1}{8} \sum_{n^\prime}
  W^{(3)}(n,n^\prime) U_{\mu}(n^\prime) W^{(3)}(n^\prime+\hat{\mu},
  n+\hat{\mu}) - U_{\mu}(n) \, .
\end{align}
With these definitions we can now define the fuzzy link by
\begin{multline}
  \label{eq:fuzzy_link}
  W_\mu(n) = {\cal P}_{\rm SU(3)} \Big\{ U_{\mu}(n) + \sum_{m=1}^3 c_m  V^{(m)}(n,n+\hat{\mu}) \\
  + \sum_{m=1}^3 d_m V^{(m)}(n,n+\hat{\mu})
  U^\dagger_{\mu}(n+\hat{\mu}) V^{(m)}(n,n+\hat{\mu}) \Big\} \, .
\end{multline}
The definition of the $V^{(a)}(n,n+\hat{\mu})$ in 
eqs.~\eqref{eq:fuzzy_links_V}-\eqref{eq:fuzzy_links_V_end} ensures that for a trivial gauge
configuration $V^{(a)}(n,n+\hat{\mu})=0$ and therefore $W_\mu(n)$
reduces to $U_\mu(n)$.

The calculation of $W_\mu(n)$ out of $U_{\mu}(n)$ defines one level of
smearing.  Using the resulting configuration as the starting point for
another smearing step this procedure can be repeated as often as
wanted. We denote the result of an $k$-level smearing by
$W^{(k)}_\mu(n)$.  In our parametrization we choose the level of
smearings to be 2, because this is clearly superior to one single
smearing, but nearly as good as 3 levels of smearing as shown in Table
\ref{tab:chi_order}.
\begin{table}[tbph]
  \begin{center}
    \begin{tabular}{|c|c|}\hline
      $k$   & $\chi^2$ \\ \hline \hline
      1     & 0.0133   \\ 
      2     & 0.0052   \\ 
      3     & 0.0050   \\ \hline
    \end{tabular}
    \caption{The dependence of the minimal $\chi^2$, as defined in eq.~\eqref{eq:chi_square_RG_smear},
      on the number of smearing levels $k$.}
    \label{tab:chi_order}
  \end{center}
\end{table}

In order to parametrize the effect of a RG cycle we make a full
non-linear optimization of the parameters $c_m$ and $d_m$. We even
allow the value of these parameters to be different at each level of
smearing. The $\chi^2$-function which has to be minimized is given by
\begin{equation}
  \label{eq:chi_square_RG_smear}
  \chi^2 = \frac{1}{N}\sum^{n_{\rm conf}}_{i=1} \sum^{n_{\rm vol}}_{n=1} \sum_{\mu=1}^{4}
  ||W^{(k)}_{i,\mu}(n) - \tilde{V}_{i,\mu}(n)||^2 \, ,
\end{equation}
where $||M||^2 = \sum_{ij} |M_{ij}|^2$ is the matrix norm and the
normalization is set to $N = 4 N_c^2 n_{\rm conf} n_{\rm vol}$
yielding the square deviation per matrix element.  The non-linear
minimization is done with a simplex algorithm\cite{Nelder:1965,
  Parkinson:1972}. In the fit $10$ $\beta = 3.0$ and $5$ $\beta = 3.4$
configurations are used. The resulting optimal coefficients for the
2-level RG smearing are given in Table \ref{tab:RG_smear_coeffs}.

\begin{figure}[htbp]
  \begin{center}
    \includegraphics[width=8cm]{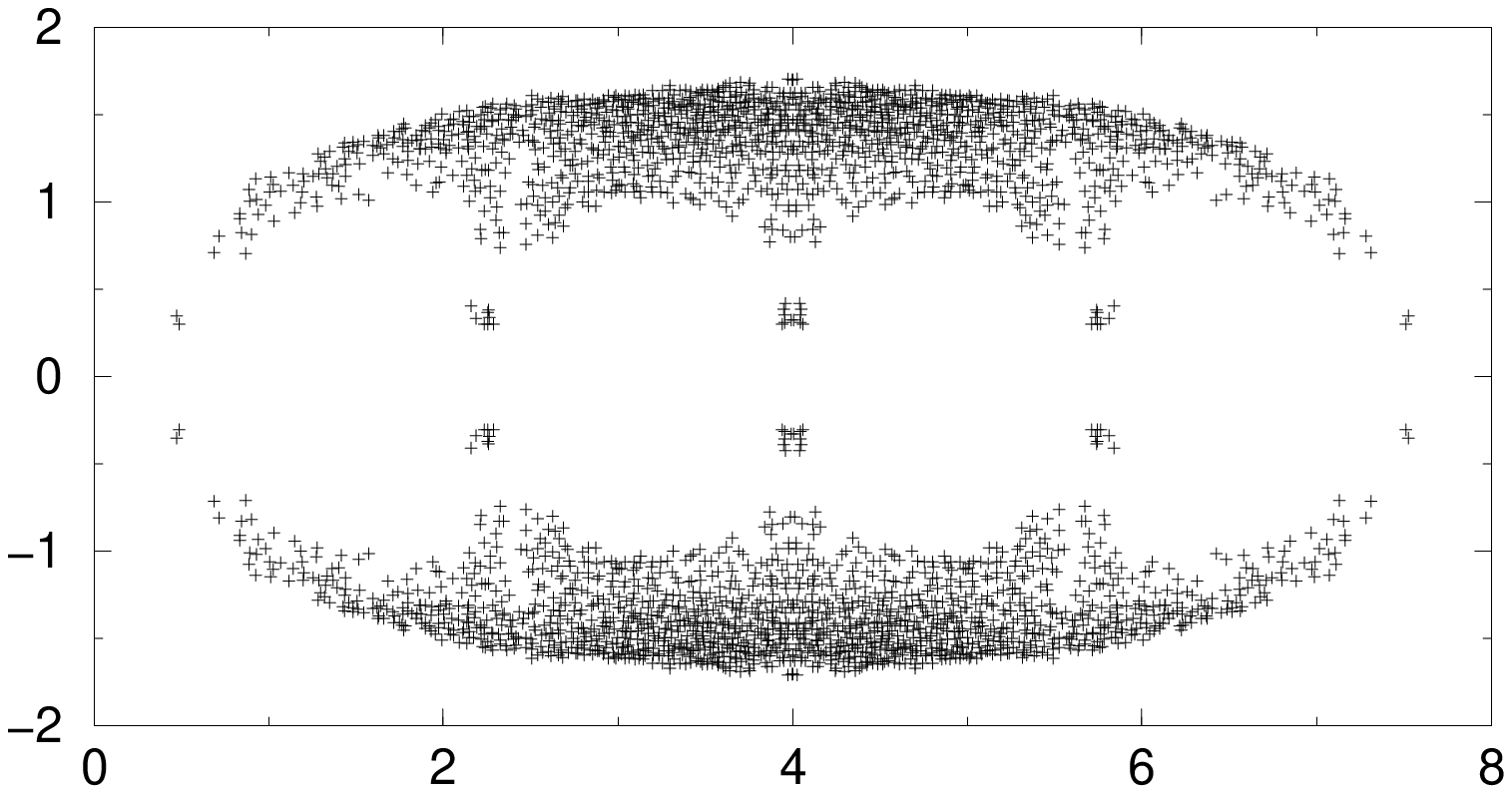}\vspace{0.2cm}
    \includegraphics[width=8cm]{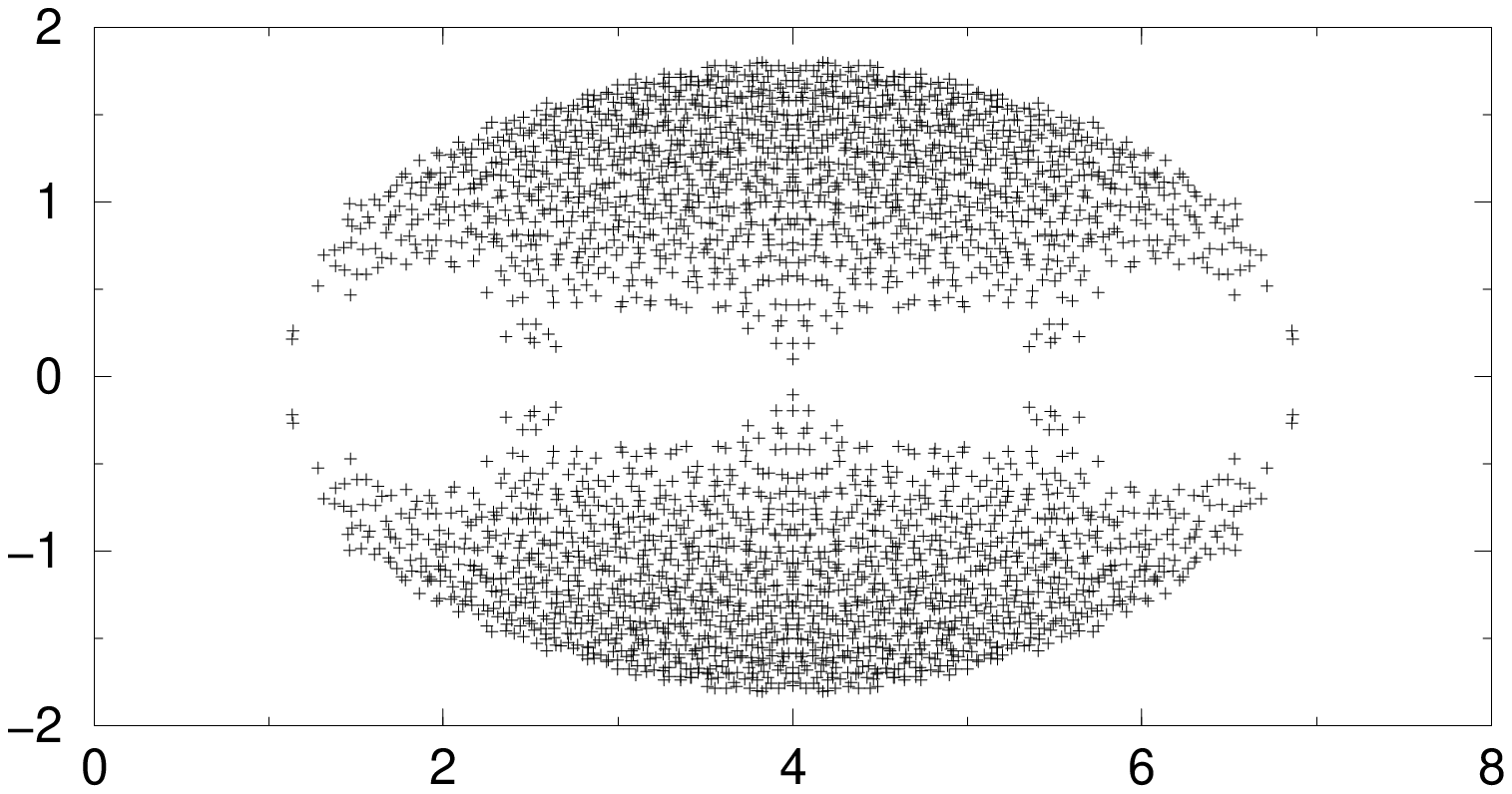}
    \caption{Eigenvalue spectrum of the Wilson Dirac operator on a FP gauge configuration with (top) and
      without RG smearing (bottom).}
    \label{fig:RG_smearing}
  \end{center}
\end{figure}

The effect of the RG smearing on the eigenvalue spectrum of $\DW$ is
illustrated in Figure \ref{fig:RG_smearing}. One can see that the
spectrum calculated on the smeared configurations has clearly a
smaller additive mass renormalization and that the spread of the
low-lying eigenvalues is smaller than on the unsmeared configuration.

\begin{table}[htbp]
  \begin{center}
    \begin{tabular}{|c|c|c|c|c|c|c|}\hline
      k & $c_1$ & $c_2$ & $c_3$ & $d_1$ & $d_2$ & $d_3$  \\ \hline \hline
      1 & 0.109 & 0.169 & 0.098 & 0.002 & 0.018 &-0.014 \\ 
      2 & 0.309 &-0.007 &-0.068 &-0.082 &-0.121 &-0.044 \\ \hline
    \end{tabular}
    \caption{The results for the coefficients $c_i$ and $d_i$ in eq.~\eqref{eq:fuzzy_link} 
      for the 2 different levels of the RG smearing.}
    \label{tab:RG_smear_coeffs}
  \end{center}
\end{table}

\section{The Parametrization of the $R_{\rm FP}$ Operator}
\label{sec:parametrization_R}

A parametrization of the operator $R_{\rm FP}$, which shows up in the GW relation
eq.~\eqref{eq:GW} for $\D=\Dfp$, is needed for many reasons, the simplest being just
to check the accuracy to which $\Dpar$ satisfies this very important
relation. We will also see that a parametrization of $R_{\rm FP}$ is
indispensable, because it enters the definition of the topological
charge density eq.~\eqref{eq:top_charge_density} and is used in the definition of the
overlap operator with $\Dpar$ as we see in Chapter \ref{cha:overlap}; furthermore, it also used in the
definition of the mass term of $\Dpar$ used in the hadron spectroscopy
measurements \cite{Hauswirth_diss:2002}.

Using the RG equation \eqref{eq:RG_propagator} and the GW relation
eq.~\eqref{eq:GW_propagator} the following RG equation can easily
be derived
\begin{equation}
  \label{eq:RGT_R}
  R_{c}(V) = \frac{1}{\kappa_f} +
  \Omega(U[V]) R_{f}(U[V]) \Omega^{\dagger}(U[V]) \, ,
\end{equation}
where the subscripts $c$ and $f$ stand again for coarse and fine
lattice and $\kappa_f$ and $\Omega(U)$ together define the blocking
kernel, which is described in Section \ref{sec:blocking_kernel}.  The
RG equation eq.~\eqref{eq:FP_R} for $\R_{\rm FP}$ has by far a simpler structure than the
corresponding equation for $\Dfp$, because it is a linear equation for
$R_{\rm FP}$. Furthermore, $R_{\rm FP}$ is an operator whose couplings vanish exactly
outside the hypercube for the given block transformation $\Omega(U)$,
i.e.~it is an ultra-local operator.  Note that this is in total
contrast to the RG equation for the FP propagator
\eqref{eq:RG_propagator}, which has exactly the same form of the
operator equation, but the FP propagator itself is an inherently
non-local operator and therefore not suited for parametrization.
That $R_{\rm FP}$ is a hypercubic operator can be seen from the RG equation
\eqref{eq:FP_R} and the fact that $R_{\rm FP}$ is a hypercubic operator in the
free case. A RGT involves the combination $\Omega(U) R_{f}(U)
\Omega^{\dagger}(U)$, which is strictly zero outside the hypercube, if
$R_{f}(U)$ is strictly zero outside the hypercube, i.e.~such a RGT
never takes $R_{c}$ outside the hypercube.

\subsection{The Details of the Fit}
\label{sec:r_fit_details}

The principle of the parametrization procedure for $R_{\rm FP}$ is the same as
for $\Dfp$, i.e.~that one can simply take Figure
\ref{fig:parametrization_sketch} and replace $\D$ by $R$ in order to
have an idea of the basic steps of the parametrization.  However, the
facts that $R_{\rm FP}$ is a hypercubic operator that is trivial in Dirac
space and that the corresponding RG equation \eqref{eq:RGT_R}, through
which it is defined, is of a much simpler structure than the
corresponding RG equations for $\Dfp$ \eqref{eq:RG_Dirac} and
\eqref{eq:RG_propagator}, make that the fitting procedure can be set
up in a simpler way than the one for the Dirac operator. In contrast
to the Dirac operator fit we can use eq.~\eqref{eq:RGT_R} as it
stands, i.e.~that it is computationally feasible to perform the matrix
multiplication $\Omega(U) R_{f}(U) \Omega^{\dagger}(U)$. This allows
to make use of the linear structure of eq.~\eqref{eq:RGT_R} to fit the
5 different types of offsets of the hypercube (denoted by $\alpha$)
independently, i.e.~that we can set up the following
$\chi^2$-functions for the fit
\begin{equation}
  \label{eq:chi2_R}
  \chi^2_{(\alpha)} = \frac{1}{{\cal N}_{(\alpha)}}
  \sum_{i=1,n_{\rm conf}} 
  \big|\big| R^{(\alpha)}_c(V^{(i)}) - R^{(\alpha)}_{\rm par}(V^{(i)}) \big|\big|^2 + 
  \lambda \chi^2_{{\cal C}_{(\alpha)}} \, ,
\end{equation}
where $||M||^2 = \sum_{ij} |M_{ij}|^2$ is the matrix norm. The
normalization is provided by ${\cal N}_{(\alpha)} = n_{\rm conf}
n_{\rm vol} n_{(\alpha)}$, where $n_{(\alpha)}$ is the number of
offsets of the corresponding offset type. The only constraint ${\cal
  C}_{(\alpha)}$ on the fit is that in the formal continuum limit the
coefficients of the free $R_{\rm FP}$ are reproduced. We choose a linear ansatz
for the parametrization of $R_{\rm FP}$, i.e.
\begin{equation}
  \label{eq:linear_ansatz_R}
  R_{\rm par} = \sum_{i=1, n_{\rm op}} c_i O_i(V) \, ,
\end{equation}
where the number of operators $n_{\rm op}$ is $12$ in the present
parametrization in the production run code.  The constraints are again
linear such that the minimization of the $\chi^2$-functions in
eq.~\eqref{eq:chi2_R} amounts simply to matrix inversion. Since this
minimization can be done very fast, it is even possible to optimize
the non-linear parameters $c_i$ of the APE smearing (see Section
\ref{sec:smearing}). This non-linear optimization is performed with
the simplex algorithm \cite{Nelder:1965, Parkinson:1972}.

The $3$ steps we perform in the iterative parametrization procedure,
which are also described in Table \ref{tab:Rparametrization}, are the
following. In step I the free hypercubic $R_{\rm FP}$\footnote{The free hypercubic $R_{\rm par}$ 
actually corresponds to the exact free $R_{\rm FP}$, because the corresponding
RGT equation can be solved exactly.} is used on the minimized
$\beta=5000$ and $\beta=100$ configurations. The fit on the coarse
lattice yields a parametrization that is used on minimized
$\beta=5000$, $\beta=100$ and $\beta=10$ configurations in step II.
This leads to the parametrization that is used in step III on the
minimized $\beta=3$ configurations, where the last blocking step and
fit yields the hypercubic $R_{\rm par}$ used in production runs. In
all steps $3^4$ and $6^4$ configurations are used on the coarse and
fine lattices, respectively. The constraint that fixes the
coefficients in the formal continuum limit to the free $R_{\rm FP}$ is also
kept during the whole parametrization.  In Table
\ref{tab:Rparametrization} the minimal $\chi^2$-values for the
different offsets at the $3$ levels of the parametrization procedure
are given. There one can see that, like for the Dirac operator, the
step which is by far the most difficult is the parametrization on the
$\beta=3$ configurations.

\begin{table}[htbp]
  \begin{center}
    \begin{tabular}{|c|c|c|c|c|c|c|c|}\hline
      step & $\beta$  & $n_{\rm conf}$ & $\chi^2_{(0)}$ & $\chi^2_{(1)}$ & $\chi^2_{(2)}$ & $\chi^2_{(3)}$ & $\chi^2_{(4)}$ \\ \hline \hline
      I     &   5000.0     &   5 & 0.000097 & 0.023826 & 0.003066 & 0.000159 & 0.000006 \\
            &    100.0     &  20 & & & & & \\ \hline 
      II    &   5000.0     &   5 & 0.000338 & 0.091302 & 0.025730 & 0.000809 & 0.000010 \\
            &    100.0     &  20 & & & & & \\ 
            &     10.0     &  20 & & & & & \\ \hline
      III   &      3.0     & 100 & 0.103887 & 1.193480 & 0.251903 & 0.014465 & 0.000786 \\ \hline 
    \end{tabular} 
    \caption{The steps in the parametrization procedure of $R_{\rm FP}$. The values $\chi^2_{(\alpha)}$
      given in units of $10^{-6}$ are the minimum of the
      $\chi^2$-function in eq.~\eqref{eq:chi2_R} for the different
      offsets $(0000),\ldots,(1111)$. In all steps $3^4$ and $6^4$
      configurations are used on the coarse and fine lattices,
      respectively. The constraint that fixes the coefficients in the
      formal continuum limit to the free $R_{\rm FP}$ is kept during the whole
      parametrization.}
    \label{tab:Rparametrization}
  \end{center}
\end{table}

\cpages
\chapter{Properties of the Parametrized FP Dirac Operator}
\label{cha:properties}

In this chapter we try to characterize our parametrizations of $\Dfp$
and $R_{\rm FP}$. From our experience in the fitting procedure, we expect the
parametrization to perform better on gauge configurations with smaller
fluctuations, i.e.~at larger values of the gauge coupling $\beta$.  We
compare properties of $\Dpar$ in different situations.  First of all
we show, how our hypercubic parametrization behaves in the free case,
where also comparisons to exact analytical results are possible. It is
an important test case, because we do not expect a parametrization in
the interacting case to perform better than in the free case.
Furthermore, we discuss properties related to chiral symmetry of the
parametrization used on minimized configurations and the production
run parametrization. In addition, we compare the production run
parametrization to the blocked Dirac operator, defined through the RGT
in eq.~\eqref{eq:RG_Dirac}.  Where it is possible, a comparison to the
Wilson Dirac operator is also given. Finally, we summarize some
results of a larger study of the hadron mass spectrum, which is
presented in more detail in \cite{Hauswirth_diss:2002}. This study
gives first hints for the quality of the parametrized Dirac operator with
respect to scaling.

\section{Free Case}
\label{sec:free}

The properties of the hypercubic approximation of $\Dfp$ in the free
case are certainly of interest, because this approximation is the
starting point for the whole parametrization described in Chapter
\ref{cha:parametrization}.  Moreover, it shows the limitations of our
hypercubic parametrization $\Dpar$, because it is very unlikely that a
parametrization will perform any better in the interacting than in
the free case.  The analytical results used in this section can be
found in \cite{Kunszt_diss:97}.

\subsection{Energy-Momentum Spectrum}
\label{subsec:EM_spectrum}

\begin{figure}[htbp]
  \begin{center}
    \includegraphics[width=9cm]{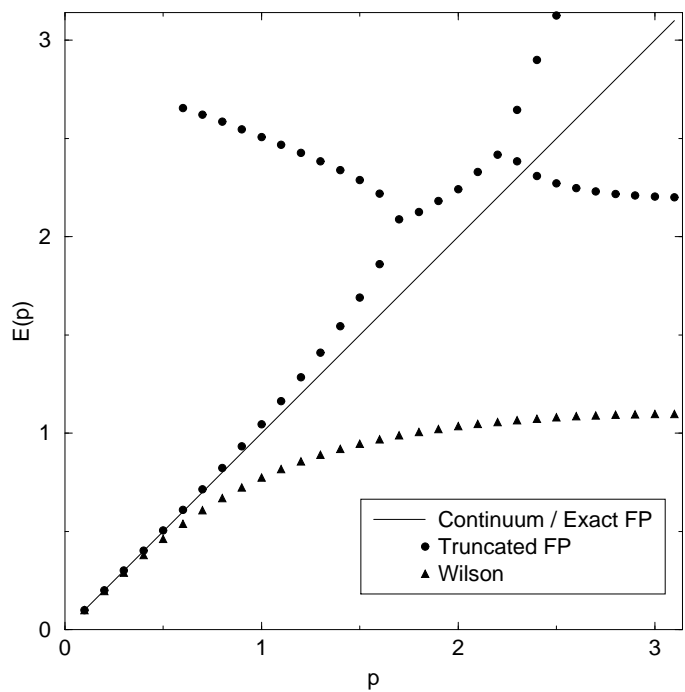}
    \caption{Energy-momentum spectrum of different lattice Dirac operators. The Wilson operator
      clearly deviates earlier from the continuum spectrum, which is
      also the exact FP spectrum, than the hypercubic
      parametrization.}
    \label{fig:EM_spectrum_free}
  \end{center}
\end{figure}

In the free case, one of the few physical quantities that is of
interest is the energy-momentum spectrum. It shows how much the
continuum spectrum is distorted by the discretization of the Dirac
operator. However, the quality how different lattice Dirac operators
approximate the continuum result varies quite a lot. There are simple
discretizations, like the Wilson Dirac operator, which deviate from
the continuum result already at quite small momenta, whereas $\Dfp$
coincides with the continuum result for \emph{all} momenta. The
question is, how well does the hypercubic parametrization perform in
this respect. In Figure \ref{fig:EM_spectrum_free} the energy-momentum
spectrum of $\Dpar$ is compared to the Wilson Dirac operator and to
$\Dfp$.  Even though the spectrum of the truncated FP operator
deviates from the continuum at high momenta, it is clearly closer to
the continuum result than the spectrum of the Wilson Dirac operator
over the whole range of momenta. Note that the high-lying branch
influences the quality of the approximation only very little.

\section{Chiral Properties}
\label{sec:chiral}

How much a Dirac operator deviates from exact chiral symmetry can be
measured by different quantities, such as the residual additive quark
mass renormalization, the occurrence of small $A^\dagger A$
eigenvalues, the breaking of the GW relation on a random vector and
related to this, the non-normality of the Dirac operator.  While
providing a quantitative measure for the breaking of chiral symmetry,
all these quantities do not show the effect of the breaking of the GW
relation as impressively as the eigenvalue spectrum of the Dirac
operator.

We almost exclusively refer to $\Dtpar = (2 R_{\rm
    par})^{1/2}\Dpar (2 R_{\rm par})^{1/2}$ in this section, because it
connects the two operators $\Dpar$ and $R_{\rm par}$ that we have
parametrized and it can be compared more easily to all the
other (nearly) chiral formulations of lattice Dirac operators. We set
up the following notations for the different types of Dirac operators
used in the comparisons below\footnote{The parametrizations for the
  $\Dpar$ and $R_{\rm par}$ operators are given in Appendix
  \ref{cha:params}}:
\begin{itemize}
\item[I] The hypercubic parametrization of the free FP Dirac operator
  together with the exact parametrization of the free $R_{\rm FP}$.
\item[II] The overlap reparametrization used on minimized $\beta=2.7 -
  3.4$ configurations and the corresponding parametrization of $R_{\rm FP}$.
  For the Dirac operator this corresponds to the parametrization
  obtained after step IV in the parametrization procedure and for $R_{\rm FP}$
  it corresponds to the operator obtained after step II in the
  parametrization procedure (see Chapter
  \ref{cha:parametrization}).
\item[III] The blocked Dirac operator that is obtained from the Dirac
  operator II by the RGT given in eqs.~\eqref{eq:RG_Dirac} and
  ~\eqref{eq:RGT_R}.
\item[IV] The parametrizations $\Dpar$ and $R_{\rm par}$ used in all
  the production runs.
\end{itemize}
Furthermore, we denote the Wilson Dirac operator on unsmeared and RG
smeared configurations by $D_W$ and $D_W^S$, respectively.

\subsection{Eigenvalue Spectrum}
\label{subsec:eigenvalue_spectrum}

The eigenvalues of a Dirac operator satisfying the GW relation with
$R=1/2$ lie on a circle in the complex plane with radius $1$ and
center at $(1,0)$. Deviations from the GW relation with $R=1/2$ can
therefore be visualized by the eigenvalue spectrum: the more the
eigenvalues scatter away from the circle, the more the GW relation is
broken. Even though this is not a quantitative statement as the
results we present later on in this chapter, it still represents the
same facts.  All the eigenvalue spectra shown in this section ---
apart from the free spectra, which can be obtained easily in Fourier
space --- have been calculated with the implicitly restarted Arnoldi
method \cite{Sorensen:1992,Lehoucq:1998}.

\begin{figure}[htbp]
  \begin{center}
    \includegraphics[width=9cm]{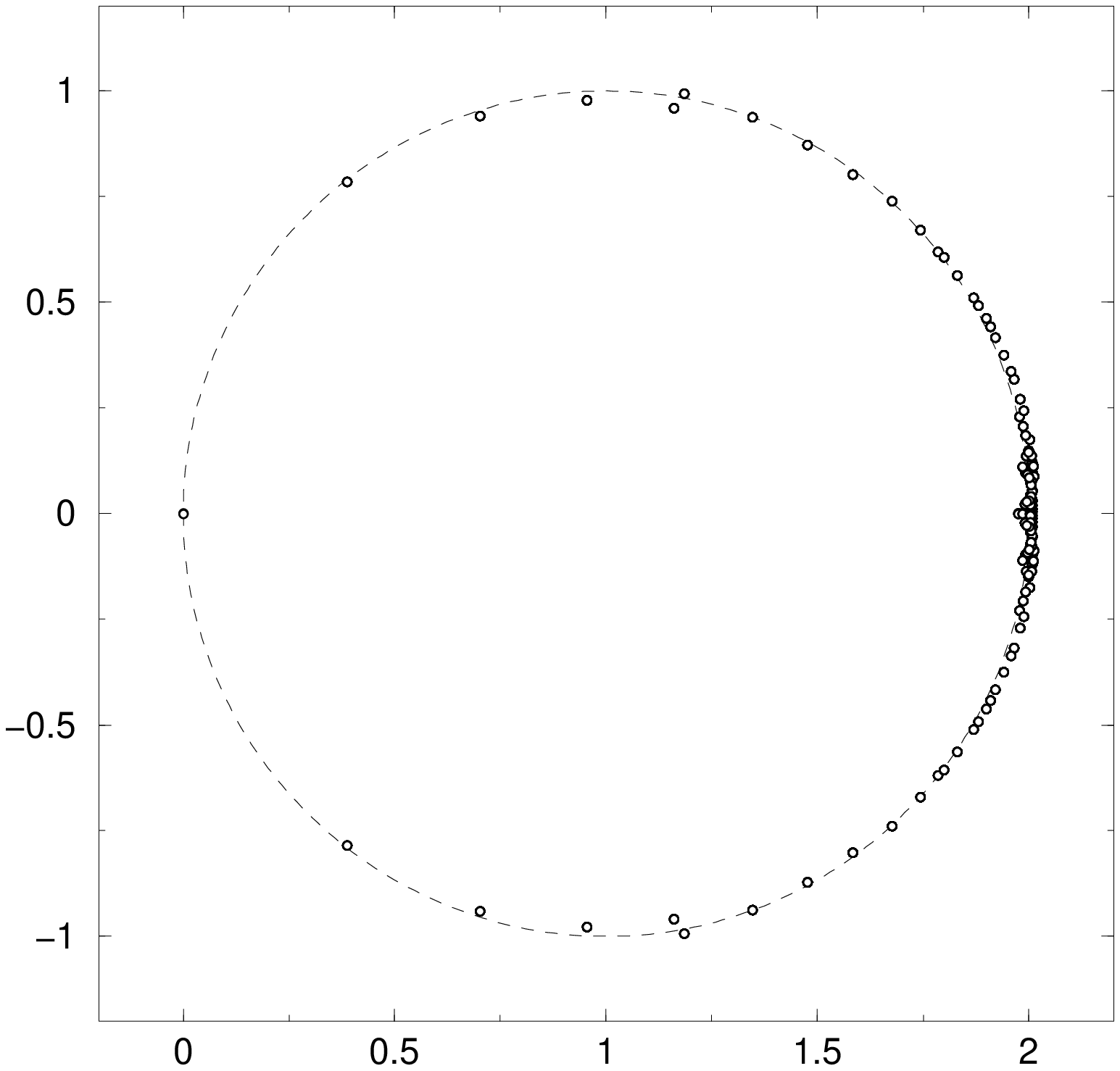}
    \includegraphics[width=9cm]{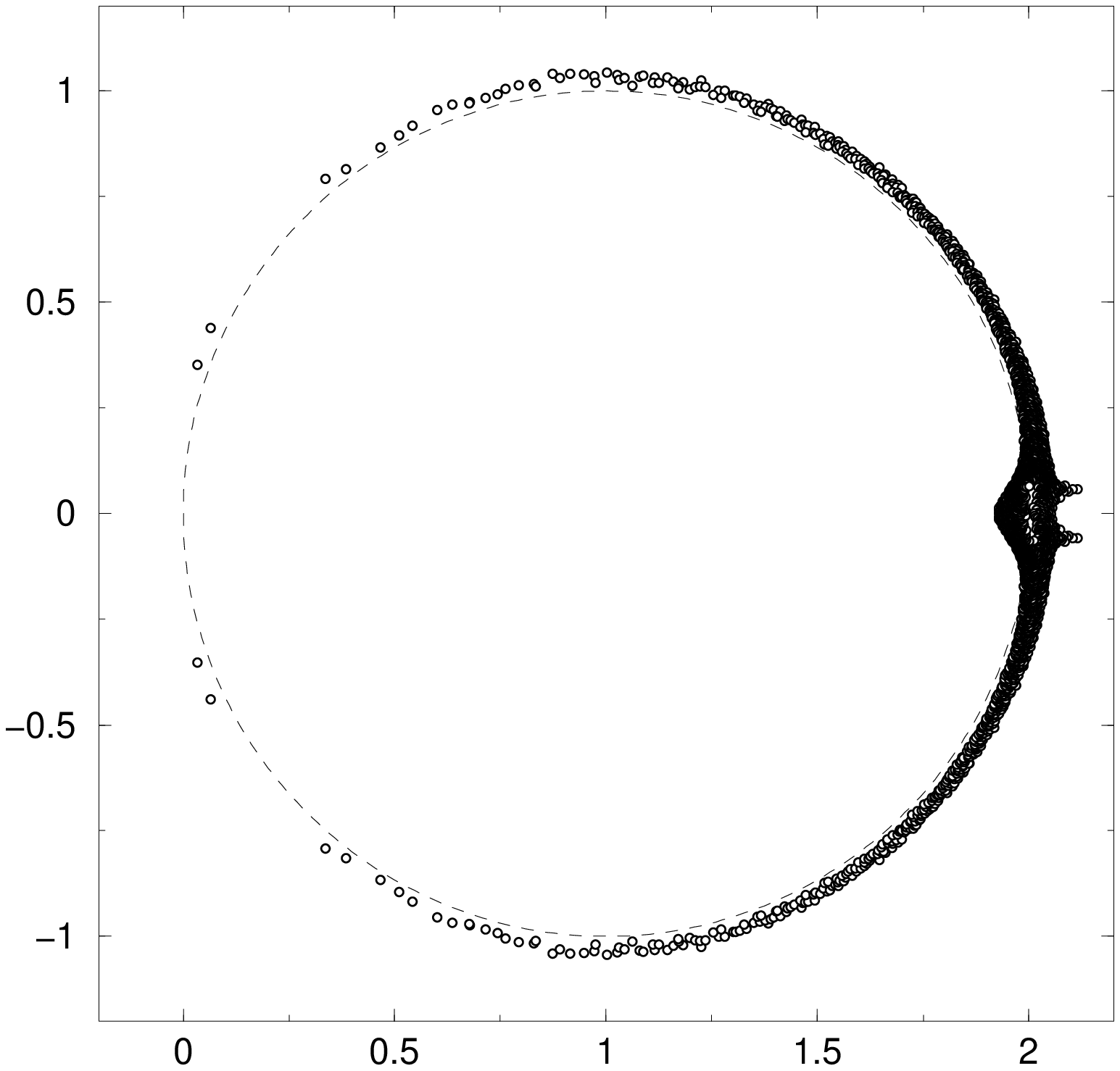}
    \caption{The eigenvalue spectrum of the truncated ${\cal D}_{\rm FP}$ in the free 
      case on a $8^4$ lattice (top) and the eigenvalue spectrum of
      ${\cal D}_{\rm par}$ at $\beta=3.0$ on a $4^4$ lattice (bottom).
      The eigenvalues in the free case are highly degenerate and
      therefore it seems that there are less eigenvalues on the larger
      lattice than on the small lattice, which is however not true.
      The figures show clearly, that the GW relation is broken to a
      lesser extent in the free than in the interacting case; although
      the eigenvalues in the interacting case are still close to the
      circle.}
    \label{fig:free_interacting_circle}
 \end{center}
\end{figure}

In Figure \ref{fig:free_interacting_circle} we compare the
parametrization in the free case with the production run
parametrization at $\beta=3.0$. The eigenvalues in the free case,
which are highly degenerate, are aligned much better and lie closer to
the GW circle than the eigenvalues of the production run
parametrization, in particular in the region of the cut-off, where
most of the eigenvalues are located.  Still, the eigenvalues of the
interacting ${\cal D}_{\rm par}$ lie fairly close to the GW circle,
even in comparison with several other chirally improved Dirac
operators
\cite{Hernandez:2001yd,DeGrand:2001ie,DeGrand:2000tf,Gattringer:2001ia,
  Gattringer:2000qu,Bietenholz:2001nu,Bietenholz:2000cc}.

\begin{figure}[htbp]
  \begin{center}
    \includegraphics[height=7cm]{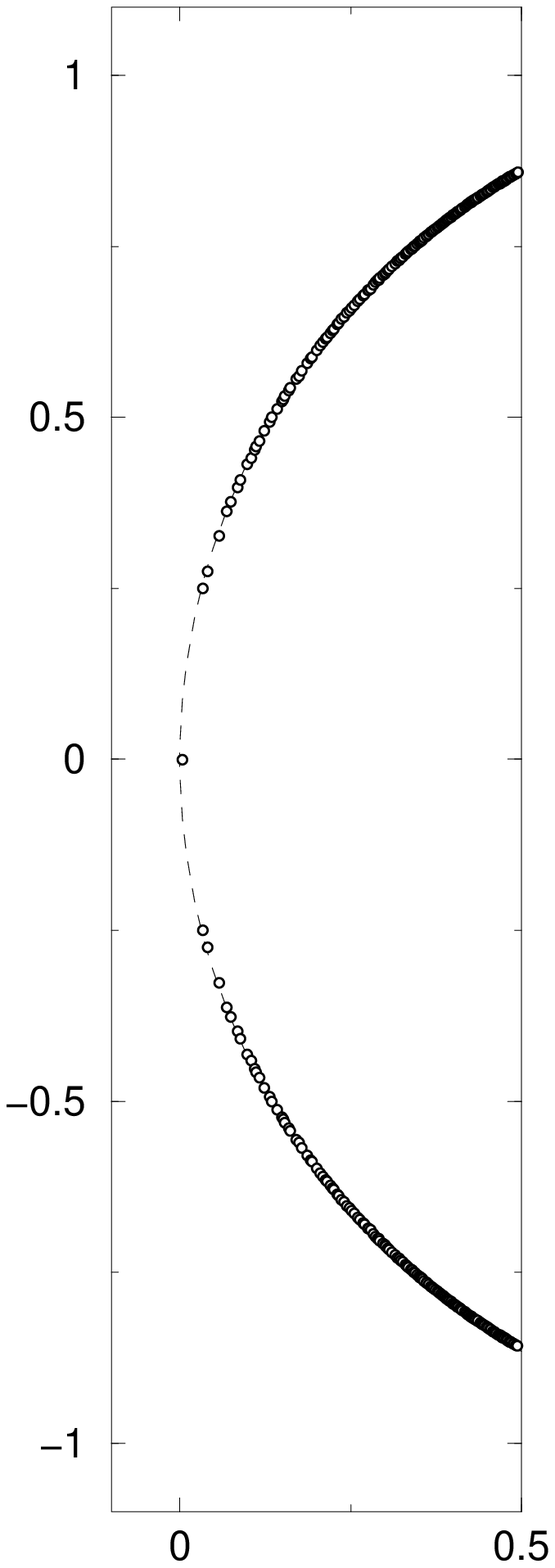}\hspace{0.3cm}
    \includegraphics[height=7cm]{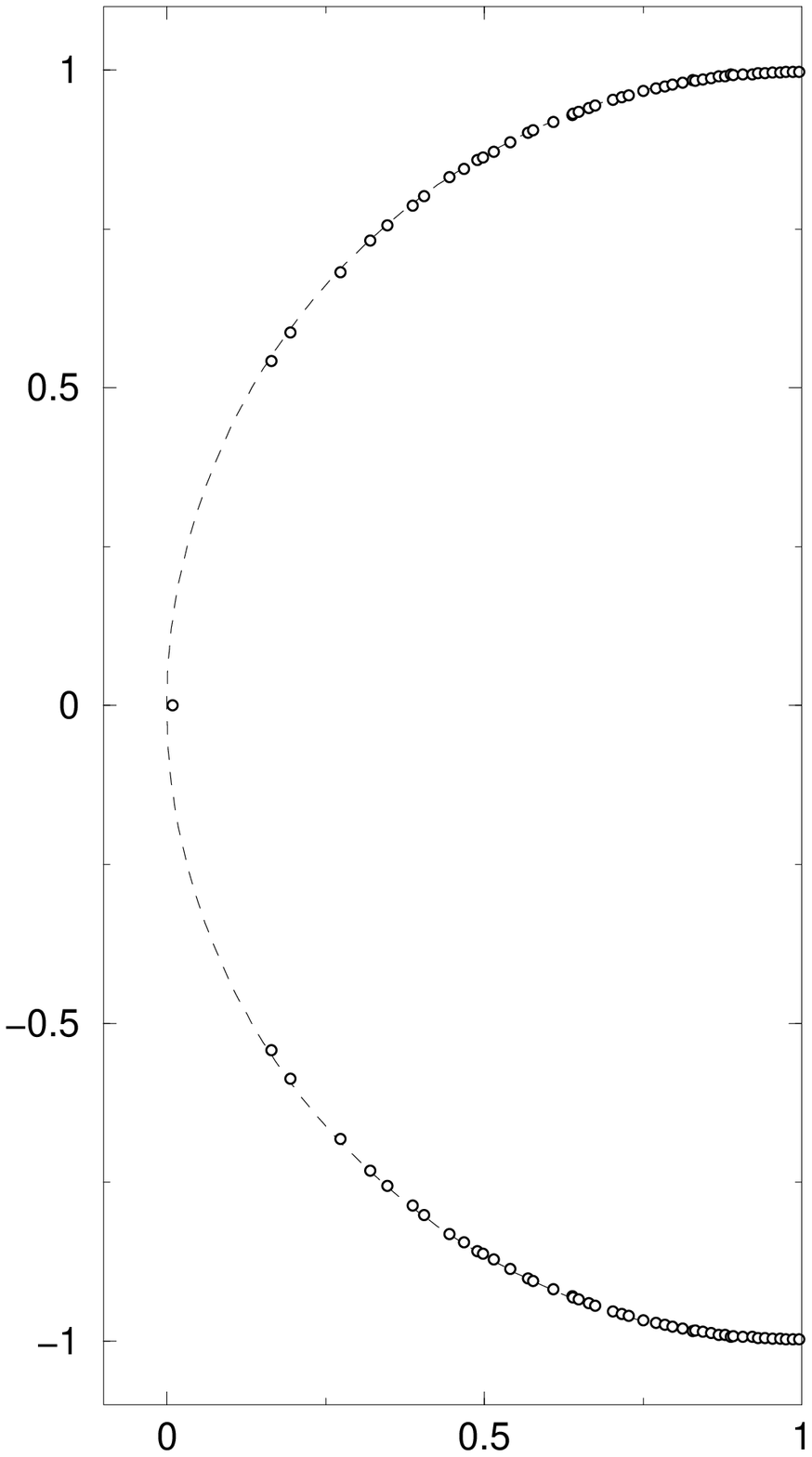}\hspace{0.3cm}
    \includegraphics[height=7cm]{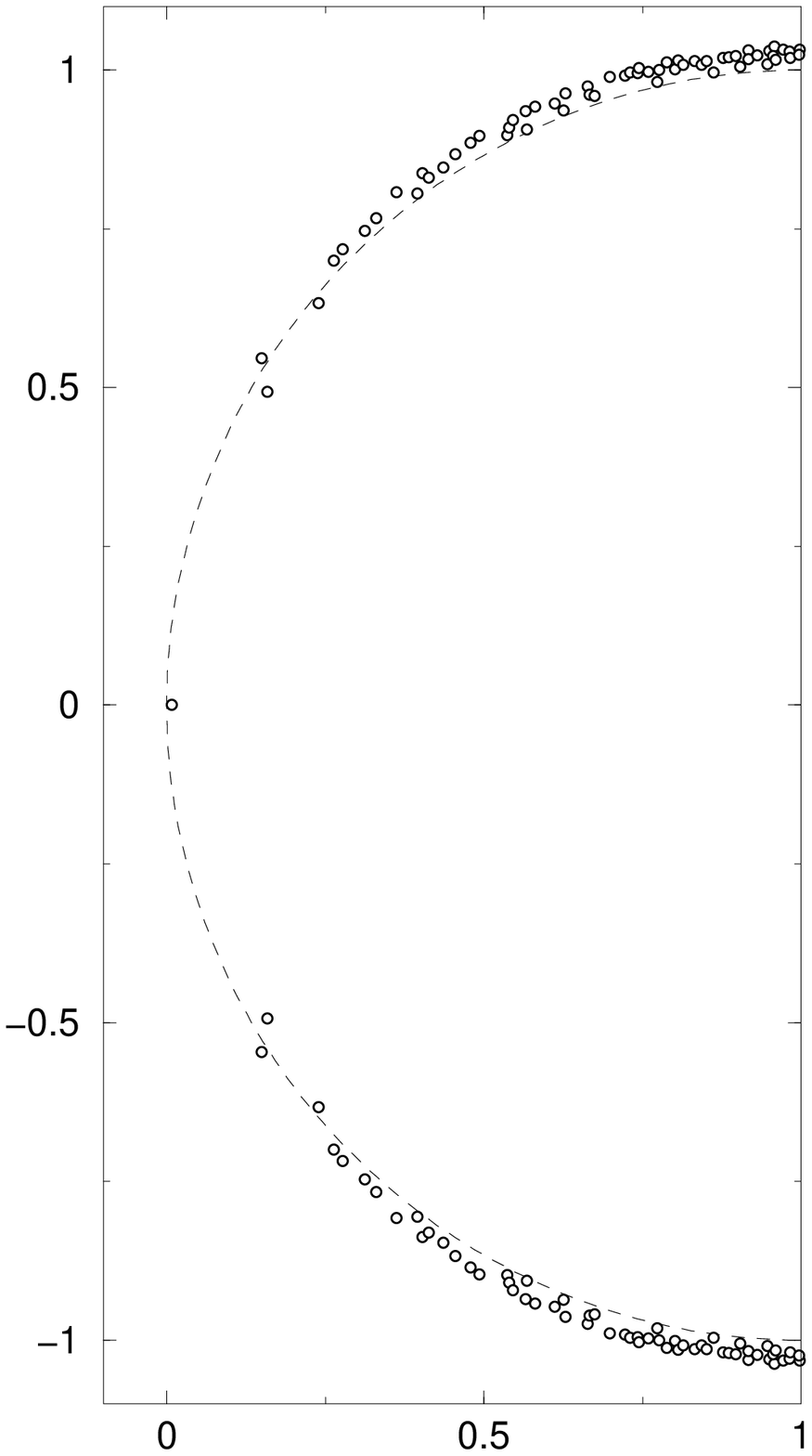}\vspace{0.5cm}
    \includegraphics[height=7cm]{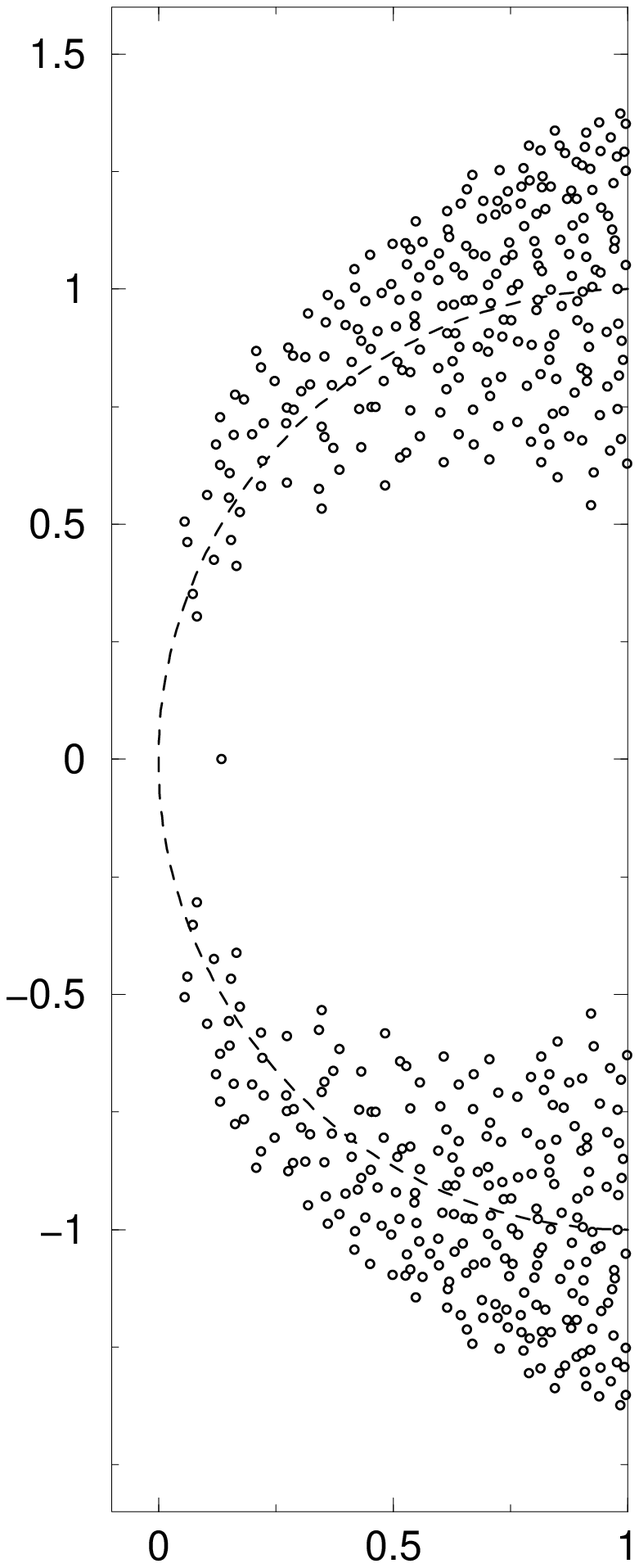}\hspace{0.8cm}
    \includegraphics[height=7cm]{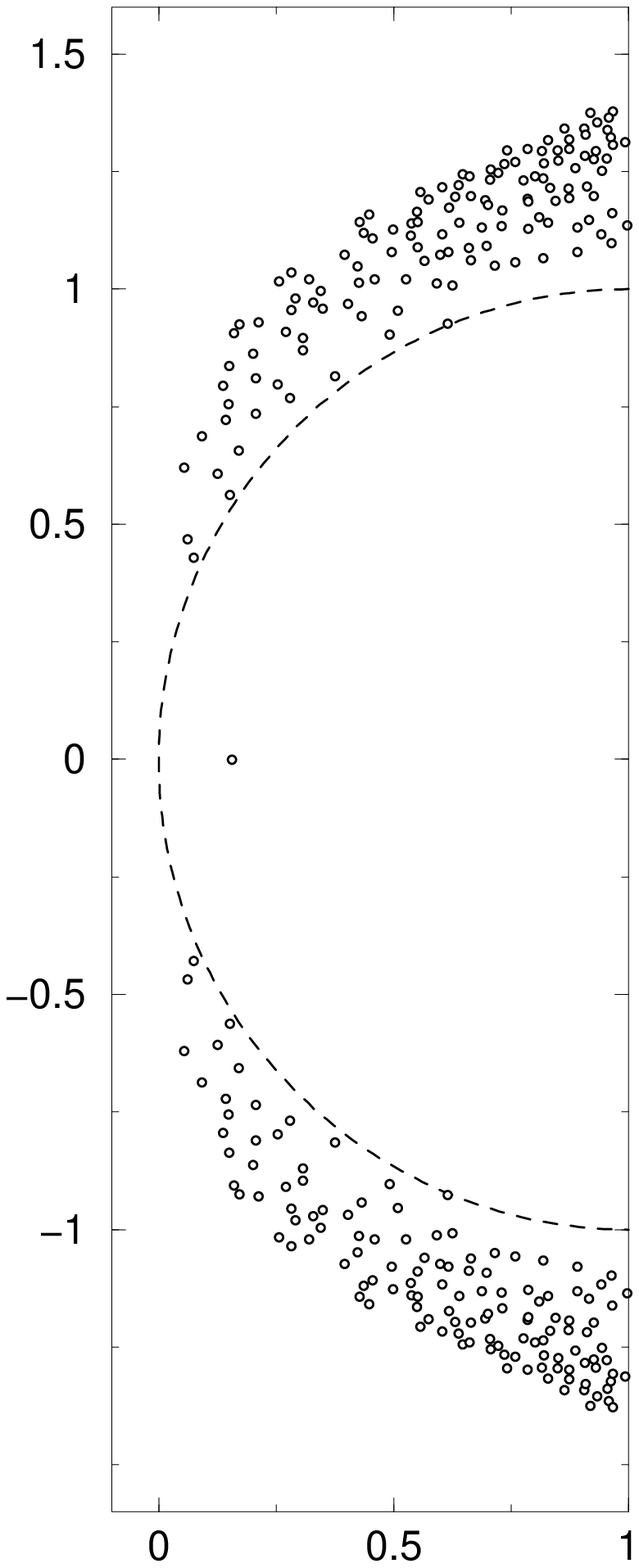}
    \caption{Top row: In these figures the physical branch of the
      eigenvalue spectra on a minimized $10^4$ configuration and the
      corresponding coarse $5^4$ configuration at $\beta = 3.0$ are
      shown.  The eigenvalues on the minimized configuration (left),
      which are calculated with operator II, are clearly aligned to
      the GW circle, which is also the case for the blocked operator
      III (middle). The alignment of the eigenvalues of the final
      parametrization IV (right) is clearly worse than the one of the
      blocked operator III.  However, essential features like the
      small additive mass renormalization are conserved.  Bottom row:
      The eigenvalue spectrum of the Wilson Dirac operator on the same
      $5^4$ configuration at $\beta = 3.0$ is shown without (left) and
      with RG smearing (right). The additive mass renormalization,
      which is roughly 1.2 and 0.6, respectively, is removed for this
      comparison.  The fluctuations of the small (real) modes are
      clearly larger than for $\Dtpar$; however, one has to be aware
      that it is only meaningful to compare the fluctuations close to
      the origin, because the spectrum of a Dirac operator satisfying
      the GW relation with a general R does not necessarily lie on a
      circle (see Section \ref{subsec:GW}).}
    \label{fig:circle_comparison_block_par}
  \end{center}
\end{figure}

In Figure \ref{fig:circle_comparison_block_par} we show the physical
branch, i.e.~the small eigenvalue part, of eigenvalue spectrum of our
parametrization II for a minimized $\beta=3.0$ configuration on a
$10^4$ lattice. The eigenvalues lie very close to the circle.
Performing a RGT, the eigenvalue spectrum of the resulting Dirac
operator III on the corresponding $5^4$ lattice is roughly of the same
quality, as the spectrum on the minimized configuration. This shows
that the Dirac operator, which we approximate in the last step of the
parametrization procedure, has very good chiral properties. In
comparison we show the eigenvalue spectrum of the final
parametrization IV on the same $5^4$ configuration. In this spectrum
the alignment of eigenvalues to the circle is clearly inferior to the
one of the blocked operator, showing the difficulty of parametrizing
$\Dfp$ on rough QCD configurations.  But, one has to keep in mind that
the final parametrization only has a very small additive mass
renormalization and compared to the Wilson Dirac operator, which is
shown in the lower part of this figure, the fluctuations of the small
(real) modes are reduced considerably.

\begin{figure}[htbp]
  \begin{center}
    \includegraphics[width=11cm]{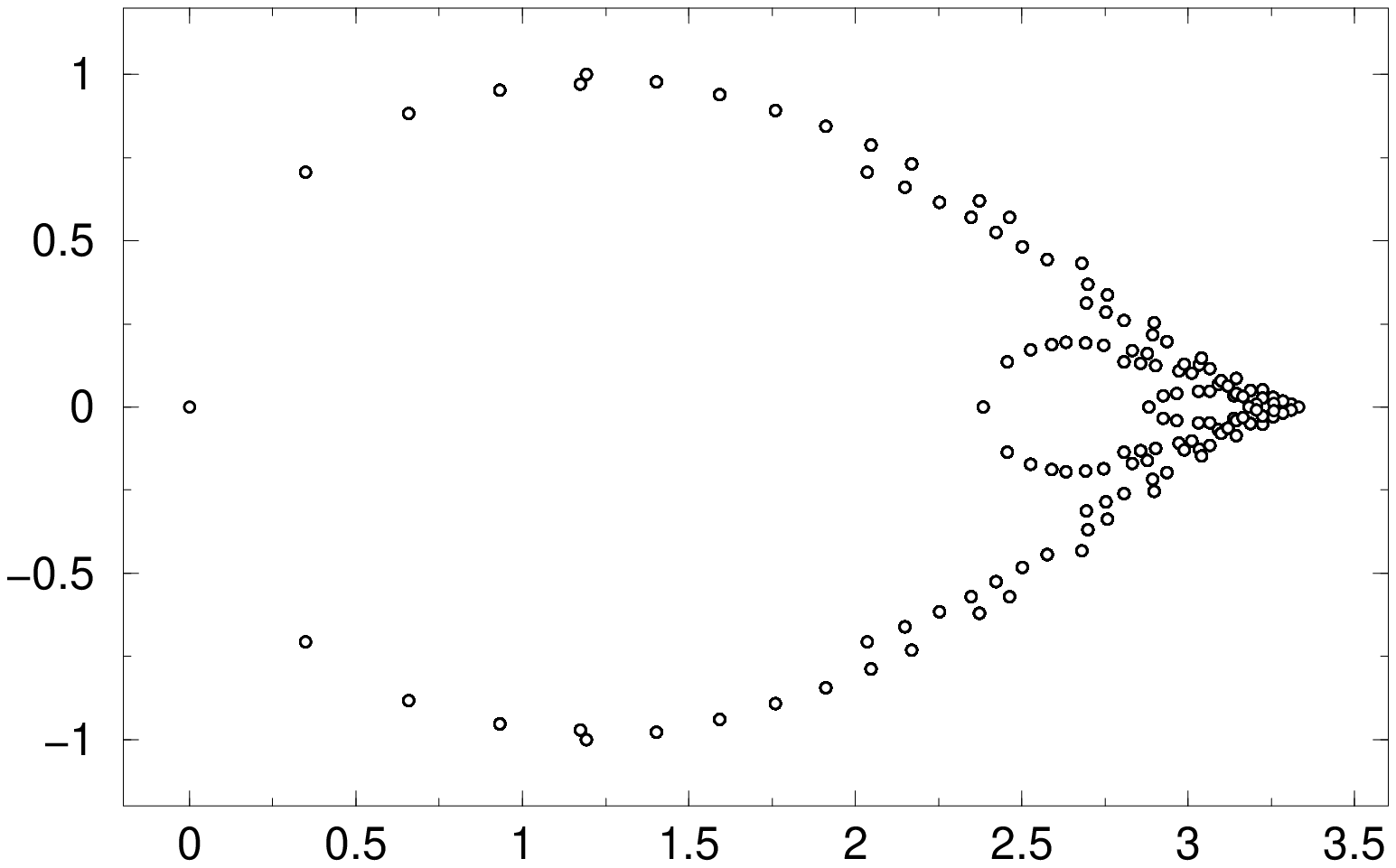}\vspace{1cm}
    \includegraphics[width=11cm]{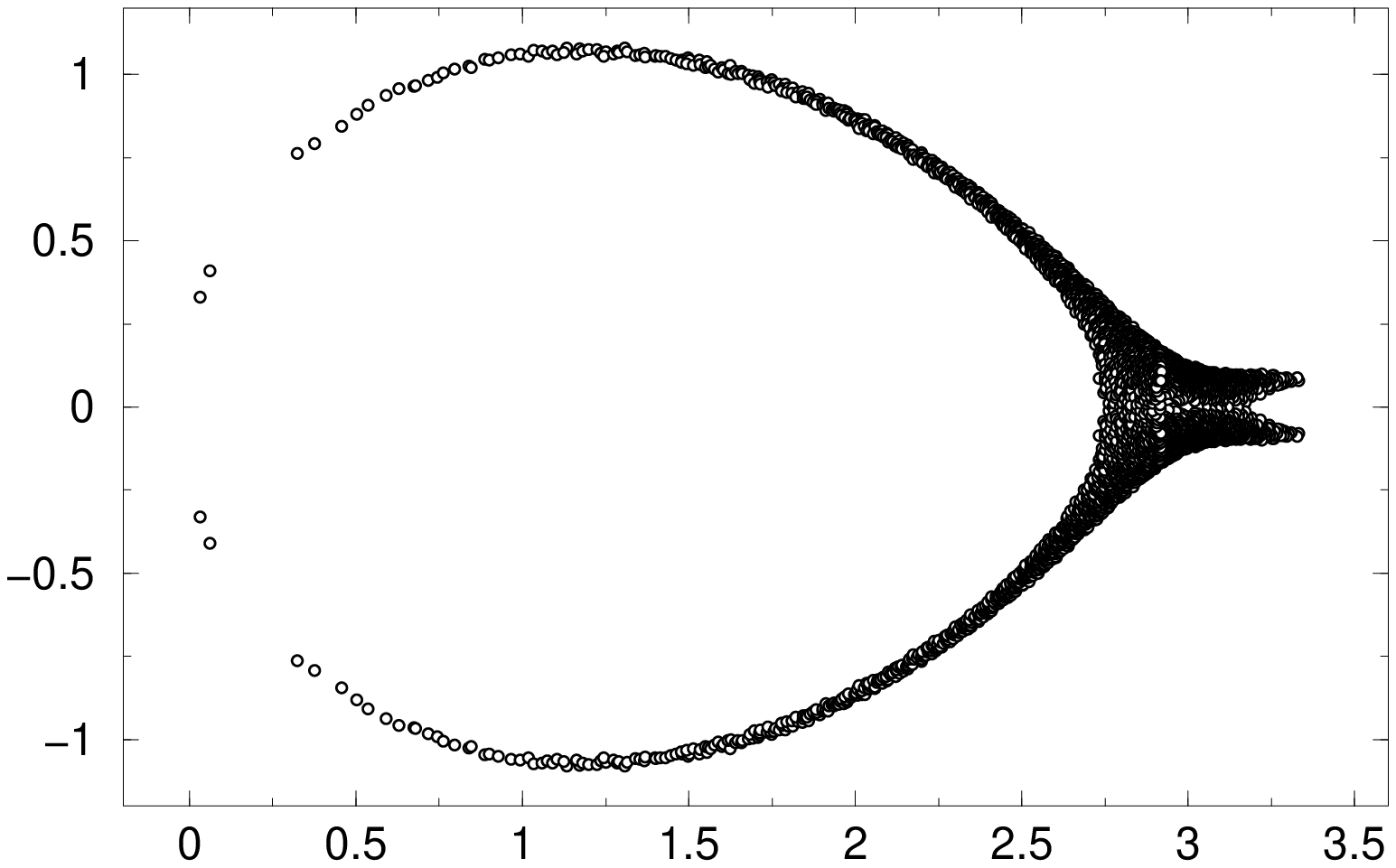}
    \caption{Eigenvalue spectrum of the truncated $\Dfp$ in the free case on a $8^4$ lattice (top) 
      and the eigenvalue spectrum the production run $\Dpar$ at
      $\beta=3.0$ on a $4^4$ lattice (bottom). This figure shows
      clearly that an operator which satisfies (approximately) the GW
      relation with $R$ not proportional to $1$ does not lie on a
      circle (see Section \ref{subsec:GW}).  Notice
      that the eigenvalues in the free case are highly degenerate.}
    \label{fig:free_interacting_Dpar}
 \end{center}
\end{figure}

The eigenvalue spectrum of $\Dpar$ does not lie on a circle (see
Section \ref{subsec:GW}) and therefore can not as
easily be used to get an impression of the amount of the breaking of
chiral symmetry as the eigenvalue spectrum of $\Dtpar$.  But, because
$\Dpar$ is the fundamental operator in the parametrization procedure,
we show its eigenvalue spectrum on $8^4$ lattice in the free case,
where the eigenvalues are highly degenerate, and on a $4^4$ lattice in
the interacting case at $\beta = 3.0$ in Figure
\ref{fig:free_interacting_Dpar}.

\subsection{Breaking of the Ginsparg-Wilson Relation}
\label{subsec:breaking_GW}

The FP Dirac operator has an exact chiral symmetry on the lattice. The
parametrization $\Dpar$, however, satisfies the GW relation no longer
exactly. In order to quantify the degree to which it breaks the GW
relation, we use the following measure, which is also used in the
literature \cite{McNeile:1999ya,Hernandez:2000sb}\footnote{Note that
  in some publications also the square norm $||.||^2$ is used.}
\begin{equation}
  \label{eq:GW_breaking}
  \Delta_{\rm GW} = \big \langle ||({\cal D}_{\rm par} + {\cal D^{\dagger}_{\rm par}} 
    - {\cal D}^{\dagger}_{\rm par} {\cal D}_{\rm par}) v || \big \rangle \, ,
\end{equation}
where $v$ is a normalized random vector. The brackets denote the
expectation value and $||v||$ is the usual vector norm. Table
\ref{tab:gw_breaking} compares $\Delta_{\rm GW}$ of the Dirac
operators I-IV. Let us give a few comments to this table: $\Delta_{\rm
  GW}$ is smaller by one order of magnitude in the free case (operator I) than for
the production run parametrization (IV), whereas the parametrization on the
minimized configurations (II) performs only slightly worse than the free
parametrization on trivial configurations (I). The blocked operator (III) is
better than the parametrization on the minimized configurations (II) and
almost reaches the same quality as the free parametrization (I). The
increase in the breaking of the GW relation with decreasing $\beta$
(i.e.~increasing lattice spacing) is as expected. The fluctuations of
$\Delta_{\rm GW}$ for different vectors and configurations are very
small and finally, $\Delta_{\rm GW}$ does not show any dependence on
the lattice size, as expected.
\begin{table}
  \begin{center}
    \begin{tabular}{|c|c|c|c|c|} \hline
      Type & \rule{0.0mm}{4mm}$\beta$  & $5^4$      & $8^4$      & $10^4$     \\ \hline\hline
      I    &          free       & 0.0293(1)  & 0.0294(1)  & 0.0294(1)  \\ \hline
      II   &          min. 3.4   &    -       &    -       & 0.0525(1)  \\
           &          min. 3.2   &    -       &    -       & 0.0530(1)  \\
           &          min. 3.0   &    -       &    -       & 0.0553(1)  \\
           &          min. 2.7   &    -       &    -       & 0.0604(1)  \\ \hline
      III  &          3.4        & 0.0298(1)  &    -       &    -       \\
           &          3.0        & 0.0308(1)  &    -       &    -       \\
           &          2.7        & 0.0339(3)  &    -       &    -       \\ \hline
      IV   &          3.4        & 0.1782(2)  & 0.1779(2)  & 0.1783(1)  \\
           &          3.2        & 0.1885(2)  & 0.1938(3)  & 0.1947(1)  \\
           &          3.0        & 0.2017(2)  & 0.2176(4)  & 0.2172(1)  \\
           &          2.7        & 0.2819(2)  & 0.2816(3)  & 0.2811(1)  \\ \hline 
    \end{tabular}
  \end{center}
  \caption{The breaking of the GW relation $\Delta_{\rm GW}$, as defined in eq.~\eqref{eq:GW_breaking}, is
    shown for the different types of Dirac operators defined at the beginning of this section.
    The statistical error is given in the brackets.}
  \label{tab:gw_breaking}
\end{table}

\subsection{Breaking of Normality}
\label{subsec:breaking_normality}

A Dirac operator satisfying the GW relation is normal with respect to
$R$, which means
\begin{equation}
  \label{eq:R_normality}
  \D^\dagger R \D = \D R \D^\dagger \quad \text{and} 
  \quad {\cal D} {\cal D}^\dagger = {\cal D}^\dagger {\cal D} \, ,
\end{equation}
respectively.  Hence, ${\cal D}$ and ${\cal D}^\dagger$ are commuting,
which implies that the eigenvectors of ${\cal D}$ are also
eigenvectors of ${\cal D}^\dagger$. A further consequence is that
$[{\cal D},\gamma_5] = 0$ in the subspace of the real eigenmodes and
therefore the real eigenmodes of a $\gamma_5$-hermitian ${\cal D}$
have definite chirality.  Thus, the magnitude of the breaking
of normality indicates to what extent e.g.~the zero modes are chiral.
We measure the non-normality of a Dirac operator $D$ by
\begin{equation}
  \label{eq:normality_breaking}
  \Delta_{\rm n} = \big \langle ||(D D^{\dagger} 
  - D^{\dagger}D ) v || \big \rangle \, ,
\end{equation}
where $v$ is a normalized random vector. Table
\ref{tab:normality_breaking} shows $\Delta_{\rm n}$ for the Dirac
operators II-IV.  The free case is not indicated, since free Dirac
operators are normal. The blocked operator (III) performs slightly worse
than the parametrization on the minimized configurations (II), the
difference is however very small. The fluctuations of $\Delta_{\rm n}$
for different vectors and configurations are again very small.
Finally, $\Delta_{\rm n}$ does not show any dependence on the lattice
size, while the dependence on the lattice spacing is again as
expected, i.e.~the larger the lattice spacing, the larger the breaking
of normality.

\begin{table}
  \begin{center}
    \begin{tabular}{|c|c|c|c|c|} \hline
      Type &  \rule{0.0mm}{4mm}$\beta$ & $5^4$      & $8^4$     & $10^4$     \\ \hline\hline
      II   &           min. 3.4        &    -       &    -      & 0.0196(1)  \\
           &           min. 3.2        &    -       &    -      & 0.0199(1)  \\
           &           min. 3.0        &    -       &    -      & 0.0210(1)  \\
           &           min. 2.7        &    -       &    -      & 0.0235(1)  \\ \hline
      III  &           3.4             & 0.0223(2)  &    -      &    -       \\
           &           3.0             & 0.0251(2)  &    -      &    -       \\
           &           2.7             & 0.0312(5)  &    -      &    -       \\ \hline 
      IV   &           3.4             & 0.0946(2)  & 0.0950(1) & 0.0952(1)  \\          
           &           3.2             & 0.1027(4)  & 0.1043(2) & 0.1050(1)  \\        
           &           3.0             & 0.1163(6)  & 0.1197(2) & 0.1196(1)  \\        
           &           2.7             & 0.1663(10) & 0.1662(3) & 0.1680(2)  \\ \hline 
    \end{tabular}
  \end{center}
  \caption{The breaking of the normality $\Delta_{\rm n}$, as defined in
    eq.~\eqref{eq:normality_breaking}, is shown for the different types of 
    Dirac operators defined at the beginning of this section.}
  \label{tab:normality_breaking}
\end{table}
%

\subsection{Eigenvalues of $A^\dagger A$}
\label{sec:adaggera_ev}

\begin{figure}[htbp]
  \begin{center}
    \includegraphics[width=9cm]{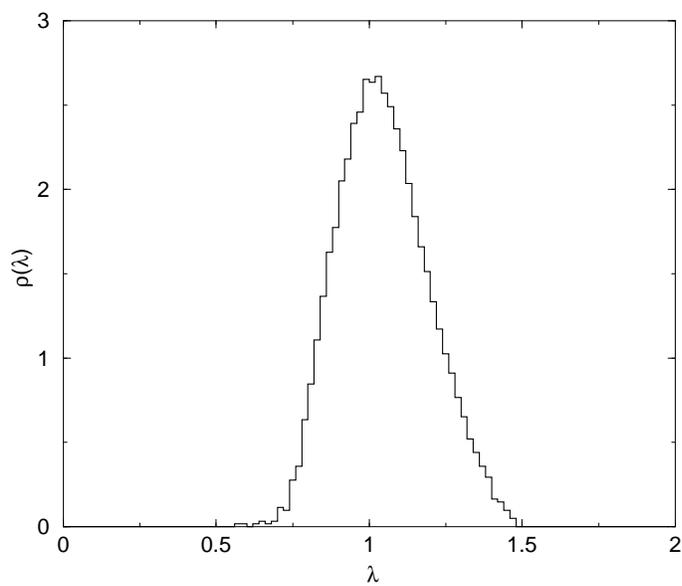}
    \caption{Density $\rho(\lambda)$ of the $A^\dagger A$ eigenvalues 
      of $\Dtpar$ on a $4^4$ configuration at $\beta=3.0$. On larger
      volumes the probability of small $A^\dagger A$ eigenvalues
      increases, which makes that a tail on the left hand side of the
      peak develops. At the upper edge of the spectrum, however, the
      structure of the distribution does not change significantly at
      larger volumes.}
    \label{fig:density_AdaggerA}
  \end{center}
\end{figure}

The GW relation eq.~\eqref{eq:GW_simple} is equivalent to
\begin{equation}
  A^{\dagger} A = 1, \hspace{0.5cm} A = 1 - {\cal D},
\end{equation}
i.e.~that $A$ is a unitary operator. Hence, the breaking of the GW
relation is signaled by the occurrence of $A^\dagger A$ eigenvalues
deviating from 1. In Figure \ref{fig:density_AdaggerA} we show the
density of the $A^\dagger A$ eigenvalues for $\Dtpar$ at $\beta=3.0$
on a $4^4$ lattice. The eigenvalues are distributed between 0.5 and
1.5 with a well-defined peak at 1. For larger physical volumes the
distribution typically will have longer tails, which eventually leads
to the occurrence of very small $A^\dagger A$ eigenvalues.

\begin{figure}[htbp]
  \begin{center}
    \psfrag{FP}{\footnotesize $\Dtpar$}
    \psfrag{DWS}{\footnotesize$D_W^S$} \psfrag{DW}{\footnotesize
      $D_W$} \includegraphics[width=11cm]{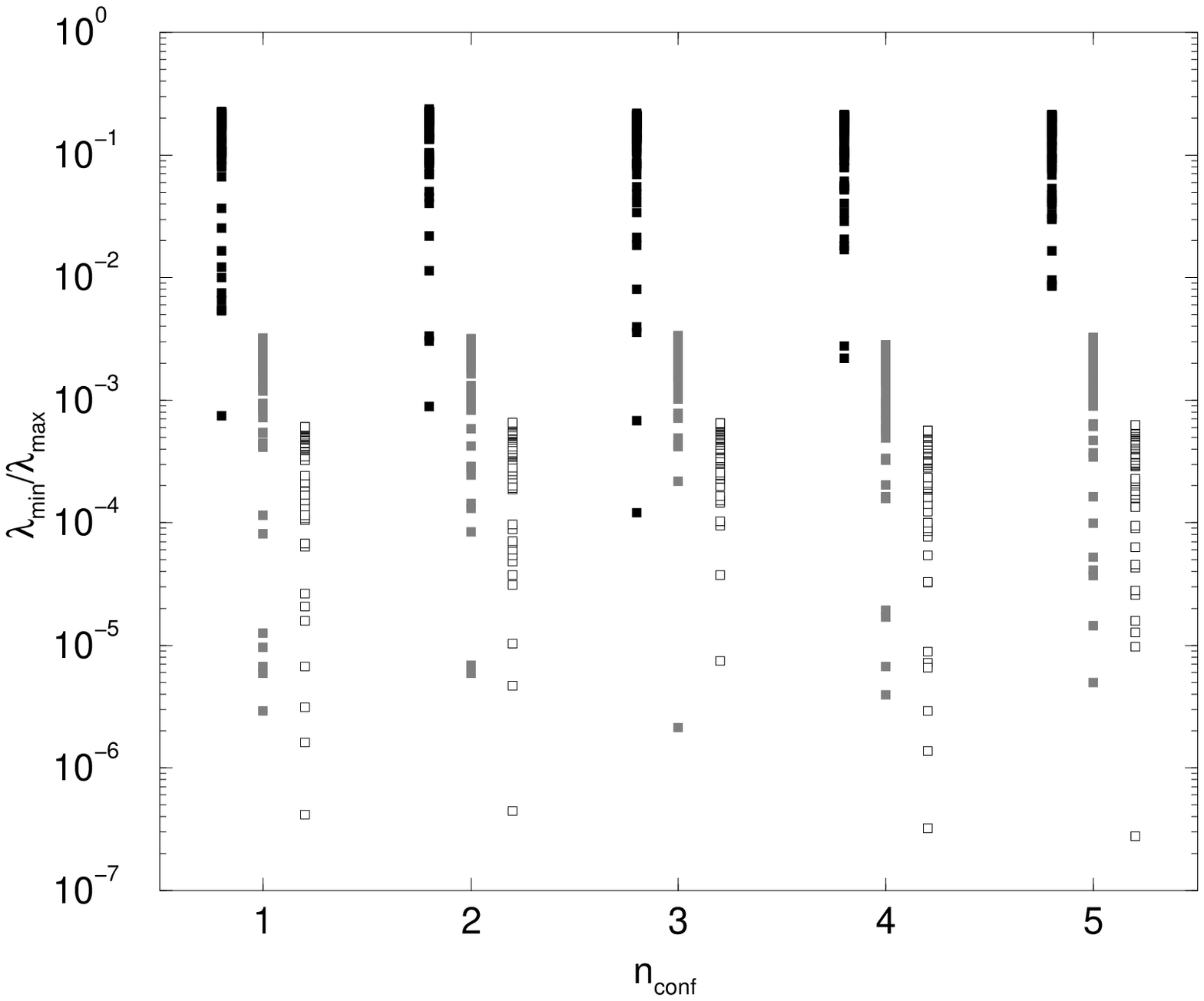}\vspace{.5cm}
    \includegraphics[width=11cm]{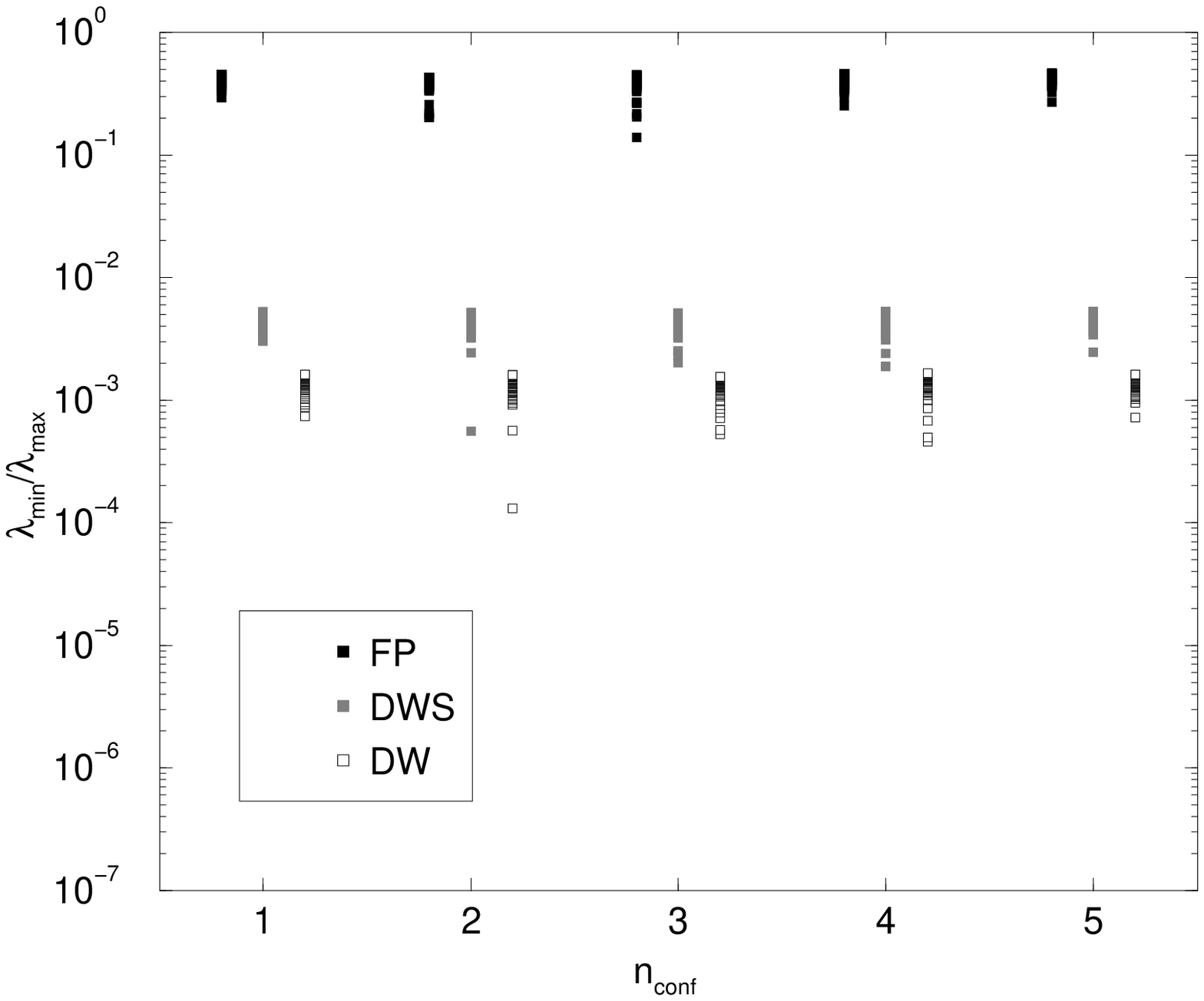}
    \caption{The ratio of the 50 smallest $A^\dagger A$ eigenvalues to the
      largest $A^\dagger A$ eigenvalue on 5 different $12^4$
      configurations at $\beta=3.0$ (top) and $\beta=3.5$ (bottom) is
      shown for $\Dtpar$, $D_W^S$ and $D_W$.  As expected this ratio
      is clearly closer to 1 for $\Dtpar$ than for $D_W$. Notice that
      the legend is valid for both figures.}
    \label{fig:AdaggerA}
  \end{center}
\end{figure}

The ratio of the smallest $\lambda_{\rm min}$ to the largest
eigenvalue $\lambda_{\rm max}$ of $A^\dagger A$ is related to the
asymptotic convergence of approximations to the overlap construction
\cite{Hernandez:2001yd}.  Hence, this ratio can be used to
indicate how expensive the overlap construction is approximately for a
certain Dirac operator and gauge action. In Figure \ref{fig:AdaggerA},
we plot the 50 smallest eigenvalues of $A^{\dagger} A$ divided by the
largest eigenvalue of $A^{\dagger} A$ for $5$ different gauge
configurations at gauge coupling $\beta = 3.0$ and $\beta = 3.5$,
respectively.  We see that the $A^{\dagger} A$ eigenvalues are much
closer to 1 for $\Dtpar$ than for $D_W$ at both gauge couplings. At
this point it should be mentioned that relatively to $D_W$ the
calculation of the $A^{\dagger} A$ eigenvalues, which is an important
ingredient in the numerical implementation of the overlap Dirac
operator (see Chapter \ref{cha:overlap}), is by far less expensive for
$\Dtpar$ than one would naively think. Because $\Dtpar$ is
substantially closer to be a normal operator and because the
$A^{\dagger} A$ eigenvalue spectrum is less dense at the lower edge
than for the Wilson operator, the Arnoldi and the Ritz functional
method find the eigenvalues with considerably less matrix vector
multiplications. The difference is particularly large between $D_W$ on
unsmeared configurations and $\Dtpar$, as shown in Table
\ref{tab:AdagA_steps}, where the number of operator times vector
products is given for the Arnoldi method\footnote{A comparison of
  $\Dtpar$ and $D_{\rm W}$ with the same conclusion for the Ritz
  functional is given in \cite{Hasenfratz:2000qb}.}. At the upper edge
of the spectrum the difference is much smaller and amounts to roughly
200 matrix vector multiplications for $\Dtpar$ and 800 for $D_W$,
independent of smearing.

Let us finally remark that one might try to use the Chebyshev
acceleration technique described in \cite{Golub:1996bk} to map the
low-lying part of the $A^\dagger A$ eigenvalue spectrum onto a larger
interval, which in certain cases speeds up the calculation of the
eigenvalues and eigenvectors \cite{Neff:2001zr}. It would be very
interesting to know how $\Dtpar$ and $D_W$, respectively, can profit
from this technique for the calculation of the low-lying $A^\dagger A$
eigenvalues.

\begin{table}
  \begin{center}
    \begin{tabular}{|c|c|c|}\hline
      $\beta$               &    3.0      &    3.5     \\ \hline\hline  
      $D_W$                 &  16130(212) & 17174(154) \\
      $D_W^S$               &   6176(212) &  8726(122) \\
      ${\cal D}_{\rm par}$  &    682(18)  &   944(50)  \\  \hline         
    \end{tabular}
  \end{center}
  \caption{The average number of Dirac operator times vector multiplications to find the $50$ lowest
    $A^\dagger A$ eigenvalues for ${\cal D}_{\rm par}$ and the Wilson Dirac
    operator on smeared ($D_W^S$) and unsmeared ($D_W$) gauge configurations. This table shows
    that calculations using ${\cal D}_{\rm par}$ do not necessarily have to be as expensive as one would
    naively think, because the large difference in the number of matrix vector
    multiplications between $\Dtpar$ and $D_W$ almost compensates for the 
    overhead of $\Dtpar$ in one matrix vector multiplication.}
  \label{tab:AdagA_steps}
\end{table}

\subsection{Index Theorem}
\label{subsec:index_theorem}

\begin{figure}[p]
  \begin{center}
    \includegraphics[width=8.5cm]{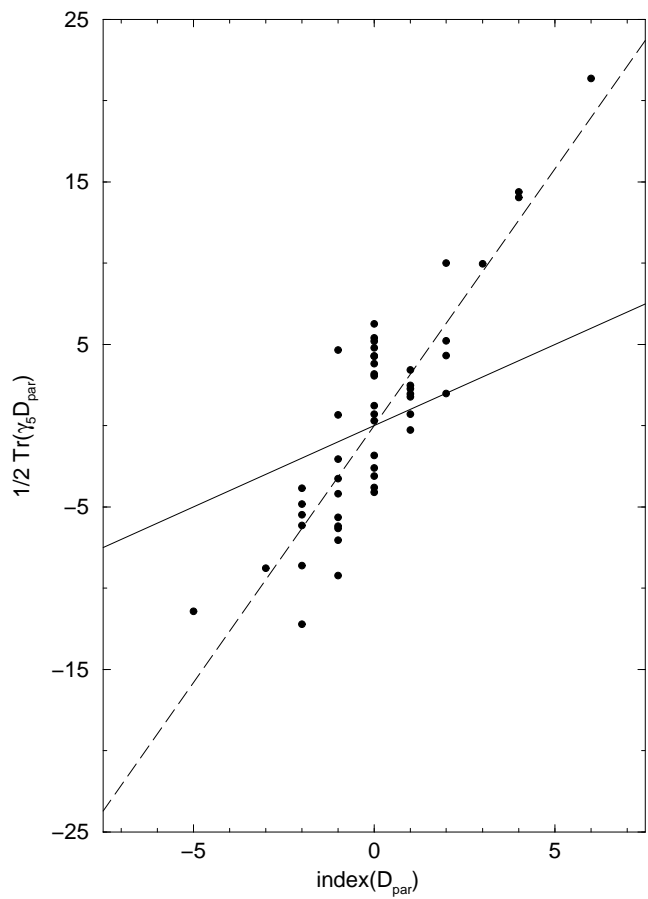}
    \caption{The correlation between the index of $\Dtpar$ and
      $1/2 \Tr ( \gamma_5 \Dtpar )$ is shown for 50 $8^4$
      configurations at $\beta = 3.0$. One can see that a certain
      correlation between the two quantities remains, but the index
      theorem clearly does not hold for the parametrized operator. The
      definition with the trace gives a result which on average is too
      large by a factor of $3.16$ as indicated by the dashed line. The
      solid line shows, where the points actually should lie if the
      index theorem would be satisfied.}
    \label{fig:index_catastrophe}
  \end{center}
\end{figure}

\begin{figure}[p]
  \begin{center}
    \includegraphics[width=8.5cm]{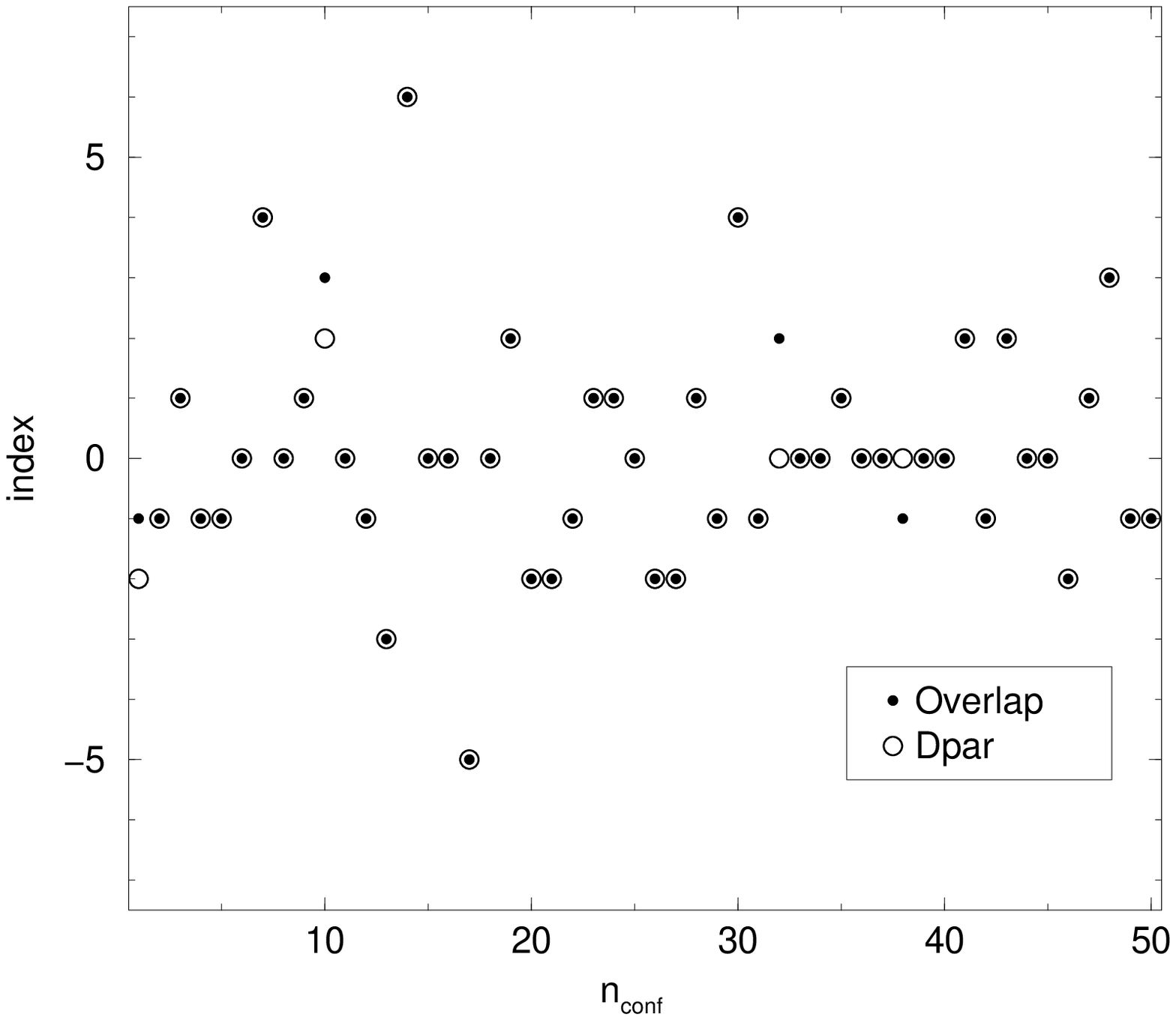}
    \caption{The correlation between the index of $\Dtpar$
      and the corresponding overlap construction with $\Dtpar$ is shown for 50 $8^4$
      configurations at $\beta = 3.0$. One can see that the value of
      the index defined by $\Dtpar$ coincides on most of the
      configurations with the overlap definition.}
      \label{fig:index_OK}
  \end{center}
\end{figure}

For GW fermions the Atiyah-Singer index theorem from
eq.~\eqref{eq:index_theorem} holds.  Recalling the previous discussion
about the distribution of the $A^\dagger A$ eigenvalues we can not
expect the index theorem to hold for $\Dtpar$ because it can be
rewritten as follows
\begin{equation}
  \label{eq:index_rewriting}
   {\rm index}(\Dt) = \frac{1}{2} \Tr ( \gamma_5 \Dt ) =
   - \frac{1}{2} \Tr (\gamma_5(1 - \Dt)) = - \frac{1}{2} \Tr (\gamma_5 A),
\end{equation}
where we use $\Tr \gamma_5 = 0$.  Hence, small eigenvalues of
$\gamma_5 A$ and, in particular, an unsymmetric distribution of
$\gamma_5 A$ eigenvalues\footnote{The distribution of the $\gamma_5 A$
  eigenvalues is symmetric for a GW Dirac operator, apart from the
  eigenmodes of $\Dt$ at 0 and 2, which lead to an asymmetry in the
  distribution that is needed for the index theorem as one can easily
  show.} tend to destroy the correspondence given by the index
theorem. We will discuss our definition of the index of $\Dtpar$, which
does not satisfy the GW relation exactly, and therefore has no exact zero modes
with definite chirality in Section \ref{sec:top_charge}.
In Figure \ref{fig:index_catastrophe} we see that the
relation is indeed distorted heavily for the production run parametrization.
The definition of the index using the trace leads to a result which on average is too large by a
factor of $3.16$.  In Figure \ref{fig:index_OK} we compare the index
of $\Dtpar$ with the index of the corresponding overlap operator
constructed from $\Dtpar$ (see Chapter \ref{cha:overlap}), which has a
very precise chiral symmetry. We see that the index of the Dirac
operator defined through the low-lying real eigenmodes and their
chirality is a much more stable quantity than the index defined through
$1/2 \Tr (\gamma_5 \Dtpar )$, because it coincides with the index of the
overlap Dirac operator in most of the cases. We will use this
observation in Chapter \ref{cha:topological_susceptibility}, where we
calculate the topological susceptibility.

In contrast to the production run parametrization the trace definition of the index works
well on the minimized configurations when the corresponding Dirac operator is used
(definition II), despite the fact that the wrong sign in the normalization condition
for the topological charge density was used in the parametrization
(see Section \ref{subsec:constraints}). This means that the linear
polynomials in the couplings really could straighten out the error in
the normalization condition and therefore we can also conclude that
the problems with the trace definition of the index, we see in the
production run parametrization, are due to other problems. Maybe an even 
richer parametrization for the pseudoscalar part is needed to give an 
accurate description of $1/2 \Tr (\gamma_5 \Dtpar )$ on configurations used in simulations.


\subsection{Residual Additive Mass Renormalization and Fluctuations of the Low-Lying Modes}
\label{sec:residual_mass_renormalization}

Another very important property of a Dirac operator satisfying the GW
relation is the absence of an additive mass renormalization and
related to this, the protection against exceptional configurations at
non-zero quark mass. This means that the pions get massless if the
bare quark mass is zero, making that no fine tuning of parameters is
needed to reach the chiral limit \cite{Hasenfratz:1998jp}. Even more
important is the fact that there are no fluctuations in the low-lying
modes of the Dirac operator that make the simulation of (quenched) QCD
in (or at least near) the region of the physical up and down quark
mass impossible.  For $\Dpar$ there is a residual additive mass
renormalization and there are fluctuations in the small eigenmodes that
do not allow to simulate at an arbitrarily small quark mass, but
the situation is definitely improved compared to the Wilson Dirac
operator, where the fluctuations in the low-lying modes make the
simulations difficult at a $\pi/\rho$ ratio of 0.5 to 0.4, depending on
the lattice spacing and the volume \cite{Bardeen:1998gv,Yoshie:2001ts}. The clover
improved Wilson Dirac operator is even worse in this respect,
i.e.~that the low-lying eigenvalues fluctuate even more. Exceptional
configurations were already found at a $\pi/\rho$ ratio of $0.54$ at a
lattice spacing of $0.1 \, {\rm fm}$ \cite{Bowler:1999ae}. Finally,
the situation does not seem to improve as much as expected in the case
of full QCD simulations \cite{Horsley:2001vm}.  In our hadron
spectroscopy study, which is described in detail in
\cite{Hauswirth_diss:2002}, we were able to simulate down to a
$\pi/\rho$ ratio of 0.28, which corresponds to a quark mass of roughly
10 MeV in the $\overline{\rm MS}$ scheme at $\mu = 2 \, \gev$; notably, without facing the
problem of exceptional configurations in an ensemble of 200 $16^3
\times 32$ configurations at a lattice spacing $a = 0.155\, {\rm fm}$
($\beta = 3.0$).  From this study we can extract the residual additive
quark mass renormalization $a \Delta_{\rm m}$ and an upper limit for
the minimal bare quark mass $a m_{\rm min}$ that is needed to prevent
from the appearance of exceptional configurations. The results are
given in Table \ref{tab:additive_renormalization} and show that the
additive mass renormalization and the fluctuations of the low-lying
modes are small. The result on the largest physical volume at
$\beta=3.0$ is the most impressive as the bare mass can be chosen to
be very small.
\begin{table}[tbhp]
  \begin{center}
    \begin{tabular}{|c|c|c|}\hline
      $\beta$ &  $a \Delta_{\rm m}$ &  $a m_{\rm min}$   \\ \hline\hline  
      3.0     &     -0.0006(4)      &      0.013         \\    
      3.4     &     -0.0180(4)      &      0.029         \\
      3.7     &     -0.0194(4)      &      0.0235        \\ \hline
    \end{tabular}
  \end{center}
  \caption{The additive mass renormalization and the smallest bare mass used in 
    the hadron spectroscopy simulations in \cite{Hauswirth_diss:2002}. The values
    at $\beta=3.0$ are from a larger physical volume and therefore it is
    possible to go to even smaller quark masses.}
  \label{tab:additive_renormalization}
\end{table}

\section{Scaling Properties of the Parametrized FP Dirac Operator}
\label{sec:scaling_properties}

The FP Dirac operator is expected to have small scaling violations.
This is a crucial property that allows one to simulate at larger
lattice spacings with only very small distortions of the continuum
physics. We do not cover the scaling properties of $\Dpar$ as detailed
as the chiral properties; the reason for this is not the significance
of scaling properties, but simply the fact that it is computationally
much more expensive to get information on scaling.  This explains why
there is not much data on scaling properties, yet.

\begin{figure}[ptph] 
  \begin{center}
    \includegraphics[width=8.5cm]{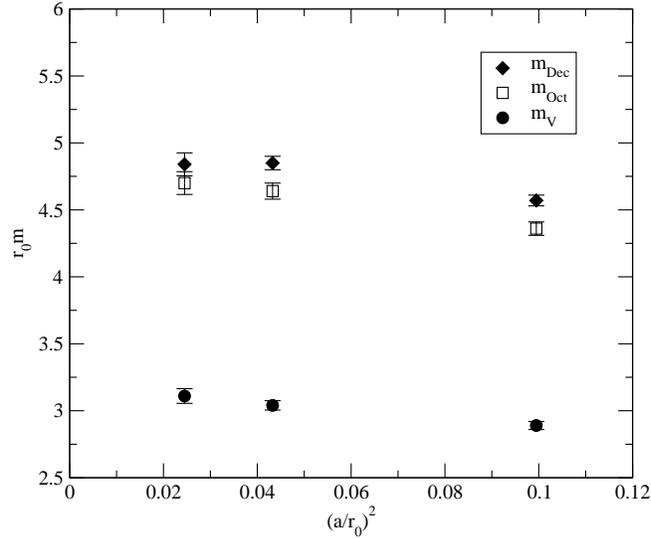}
    \caption{Scaling of the hadron masses (vector meson, octet and decuplet baryon)
      calculated with $\Dpar$ at 3 different lattice spacings. The
      data suggests that the remaining scaling violations are mostly
      ${\cal O} (a^2)$ effects and are rather small.}
    \label{fig:scaling_hadron}
  \end{center}
\end{figure}

\begin{figure}[ptph] 
  \begin{center}
    \includegraphics[width=8.5cm]{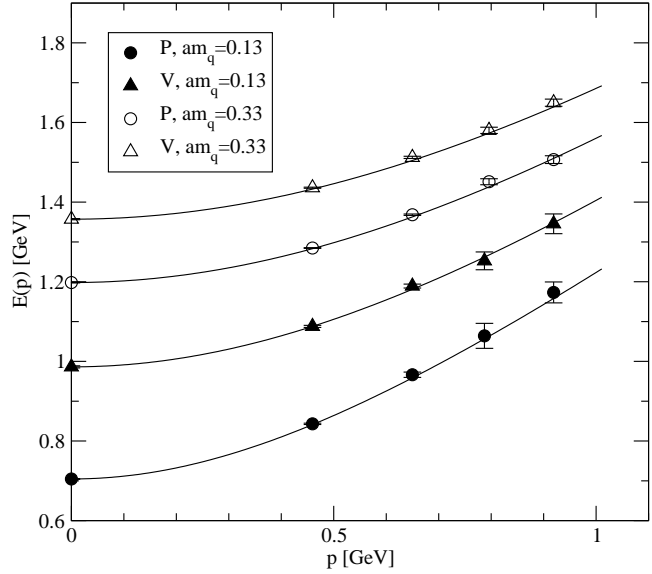}
    \caption{Energy-momentum dispersion relation $E(\vec p)$ for pseudoscalar (P) and
      vector mesons (V) on the $16^3\times 32$ lattice at $\beta=3.0$. The
      solid lines show the continuum value with $E(0)$ given by the measured mass.}
    \label{fig:EM_dispersion_pi_rho}
  \end{center}
\end{figure}

Presently, our only source of data on the scaling behaviour of $\Dpar$
is the hadron mass spectroscopy measurement described in
\cite{Hauswirth_diss:2002}. The simulations are performed at $3$
different lattice spacings. The extent of the lattice is chosen such
that all $3$ measurements are done in the same physical volume with a
spatial extent of $L \approx 1.22 \, {\rm fm}$, in order to allow for a
continuum extrapolation. This volume is rather small, which
makes that the masses of the particles (especially of the baryons) are
distorted by finite volume effects \cite{Aoki:2000kp}. But because the
finite volume distortions are physical effects, the data is
nevertheless suited for a scaling analysis, even though uncertainties
in the scale determination lead to larger uncertainties in the scaling
analysis than for larger volumes, where the dependence of the particle
masses on the volume is negligible. The data at $\beta=3.0, 3.4$ and
$3.7$ for the vector meson, the decuplet and octet baryons is shown in
Figure \ref{fig:scaling_hadron}. It indicates a remaining ${\cal O}
(a^2)$ scaling violation. However, a more careful analysis is needed, because
there are several sources of systematic errors. Most significantly is probably
the uncertainty in the scale, as the interpolating formula  
the Sommer scale $r_0$ in \cite{Wenger_diss:2000, Rufenacht:2001qy} covers
the range $\beta=2.361,\ldots,3.4$ and therefore an extrapolation for the value at $\beta=3.7$ has
been used. Furthermore, this scale determination has a systematic error which is 
not accounted for in Figure \ref{fig:scaling_hadron}. The data is not accurate 
enough to completely exclude ${\cal O}(a)$ scaling violations -- which are definitely not present in
the hadron spectrum for a GW Dirac operator \cite{Niedermayer:1998bi} -- but, if there
are remaining ${\cal O}(a)$ scaling violations, then they are very small. This is actually supported by
the fact that also the energy-momentum dispersion relation for
the pseudoscalar and the vector meson do not show any visible cut-off effect, as shown in Figure
\ref{fig:EM_dispersion_pi_rho}.

\section{Discussion}
\label{sec:discussion}

The parametrized FP Dirac operator, which we obtained from the fitting
procedure in Chapter \ref{cha:parametrization}, is clearly improved in
many respects in comparison to the Wilson Dirac operator. It has a
relatively small breaking of chiral symmetry that allows to calculate
the hadron mass spectrum down to quark masses of roughly 10 MeV in the
$\overline{\rm MS}$ scheme at $\mu=2\,\gev$. Even though the physical
point can not yet be reached with the present parametrization, it is a
remarkable progress in comparison to Wilson fermions. Furthermore, it
is possible to go to such small quark masses that there is a clear signal
for quenched chiral logarithms, even in the data of the pseudoscalar
spectroscopy with degenerate quark masses \cite{Hauswirth_diss:2002}.
The comparison of the parametrization with the blocked operator
obtained from the RGT shows the limitations of the present
parametrization, because the blocked operator has smaller breaking of
chiral symmetry.  Furthermore, the present parametrization does not
satisfy the index theorem. But one should keep in mind that the
topological charge density is a quantity which is of small order and
in the present parametrization no particular effort was undertaken to
make the fitting procedure more sensitive to the pseudoscalar
contribution and therefore some improvement might be achieved in
future parametrizations.  The data for the scaling of the hadron
masses suggests that the present parametrization leads to remaining
cut-off effects which are mainly ${\cal O}(a^2)$ and as shown in
\cite{Hauswirth_diss:2002} the cut-off effects are on the order of
various other ${\cal O}(a)$ improved Dirac operators. As discussed in Section 
\ref{sec:scaling_properties}, the sources of systematic errors are too large to make
a clear statement on the size of the ${\cal O}(a^2)$ scaling violations, apart from the fact that they
are rather small. The data, however, indicates that order ${\cal O}(a)$ cut-off effects are very small,
as one can see no sign of their presence. These statements about the small
scaling violations are supported by the fact that the energy-momentum dispersion relation for
the pseudoscalar and the vector meson do not show any visible cut-off effect.
 

\cpages
\chapter{Overlap Construction with the Parametrized FP Dirac Operator}
\label{cha:overlap}

One of the very important properties of $\Dfp$ is its exact chiral
symmetry on the lattice \cite{Hasenfratz:1998ft}. Unfortunately, this
exact symmetry gets lost through the parametrization.  As expected,
the violation of chiral symmetry is much smaller for $\Dpar$ than for
Wilson fermions. In addition, Neuberger's overlap construction
\cite{Neuberger:1998fp,Neuberger:1998wv} gives a recipe, how a Dirac
operator that respects chiral symmetry on the lattice can be
constructed from any Dirac operator that has no doublers in the
continuum limit and is local \cite{Bietenholz:1998ut}. In numerical
simulations approximations to Neuberger's construction can be used to
obtain Dirac operators whose violations of chiral symmetry are, in
principle, below an arbitrarily chosen limit. Very precise chiral
symmetry is required for some measurements, such as the chiral
condensate \cite{Hernandez:1999cu}. However, the decision to what
extent chiral symmetry is needed for a particular calculation has to
be taken from case to case, since (approximately) chiral actions are
computationally very expensive \cite{Hernandez:2000sb,Eshof:2002ms}.

In this chapter we first discuss Neuberger's overlap formula and then,
in particular, the properties of the overlap construction with $\Dpar$
and $R_{\rm par}$.  The actual construction with $R \ne 1/2$ is
numerically somewhat more involved than the standard case $R=1/2$ and
we give the technical details in Appendix \ref{cha:implementation}.
Below, we explain the details of the Legendre expansion of the
operator $(A^\dagger A)^{-1/2}$ in the overlap formula, because in our
simulations with the overlap we exclusively use this type of
approximation. We present results on the properties of the overlap
construction with $\Dpar$ and $R_{\rm par}$ and make some comparisons
to the standard overlap construction with the Wilson Dirac operator.
Finally, we conclude with a remark on scaling properties of the
overlap construction with $\Dtpar$.

\section{General Overlap Construction}
\label{sec:general_overlap}

The overlap formula in terms of a Dirac operator $D$ and the local
gauge operator $R$ from the GW relation eq.~\eqref{eq:GW} in the
massless case is given by
\begin{align}
  \label{eq:overlap_DTR_simple}
  {\cal D} &= 1 - A (A^\dagger A)^{-1/2} = 1 -\gamma_5 \epsilon(\gamma_5 A) \\
  \intertext{with}
  \label{eq:adaga_def}
  A &= 1 + s - (2 R)^{1/2} D (2 R)^{1/2} \, ,
\end{align}
where $\epsilon(\gamma_5 A)$ is the matrix sign function of the
operator $\gamma_5 A$\footnote{The matrix sign function of $\gamma_5
  A$ and the square root of $A^\dagger A$ are well defined, because
  $\gamma_5 A$ and $A^\dagger A$ are hermitian operators and therefore
  the matrix functions can be defined in terms of the eigenvalues and eigenfunctions of
  the corresponding hermitian operator.}. The real parameter $s$ is
important for the overlap construction with the Wilson Dirac operator,
where it can be used to optimize e.g.~the localization range or the
convergence rate of an approximation \cite{Hernandez:1998et}.  In
connection with the overlap with $\Dtpar$ the parameter $s$ does not
play a r\^{o}le, because there is no need to shift the $A^\dagger A$
eigenvalues, since they are already centered around 1 as shown in
Figure \ref{fig:density_AdaggerA}.  It is easy to see that $\Dt$
satisfies the GW relation. More generally, it possible to define
\begin{equation}
 \label{eq:overlap_DTR_not_simple}
  \D  =  (2 R)^{-1/2}(1 - A (A^\dagger A)^{-1/2} ) (2 R)^{-1/2}
  = (2 R)^{-1/2} (1 -\gamma_5 \epsilon(\gamma_5 A))(2 R)^{-1/2} \, ,
\end{equation}
where the operator $A$ is again as defined in eq.~\eqref{eq:adaga_def}, 
such that $\D$ satisfies the GW relation
with a general $R$. The inverse square root of the hermitian operator
$A^\dagger A$ or the matrix sign function of $\gamma_5 A$ are the
objects that cause all the trouble in the overlap formula. They have
to be approximated in a certain way, because a full decomposition of
these operators into their eigenmodes is computationally not feasible,
even for modest lattice sizes. In the last $5$ years many different
approximation schemes have been worked out and applied to this
problem. One of the main directions of approximation schemes deals
with polynomial expansions, like Legendre \cite{Hernandez:1998et},
Gegenbauer \cite{Bunk:1998wj} or Chebyshev polynomials
\cite{Hernandez:2000sb}, while the other main direction is concerned
with rational approximations, like the Remes method
\cite{Edwards:1998yw}, the polar decomposition \cite{Neuberger:1998my}
or the Zolotarev polynomials \cite{vandenEshof:2001hp,Eshof:2002ms}.
Furthermore, there are also direct Krylow subspace methods used to
approximate the matrix sign function
\cite{Borici:1998mr,Borici:1999ws} and some attempts to formulate
algorithms in a 5-dimensional space in order to avoid nested Krylow
methods, which are used in the rational approximations. Such
5-dimensional approximations have a close relation to the domain-wall
fermion approach \cite{Neuberger:1998ms,Borici:2001ua}.

It is useful to introduce the mass term of a Dirac operator satisfying
the GW relation in a particular way, namely
\begin{gather}
  \label{eq:overlap_mass}
  {\cal D}(m) = \Big(1 -\frac{m}{2}\Big){\cal D}(0) + m \\
  \intertext{and}
  \label{eq:overlap_mass_D}
  \D(m) = \Big(1 -\frac{m}{2}\Big)\D(0) + m (2 R)^{-1} \, ,
\end{gather}
respectively. With this definition the scalar density $S$ is defined
as follows
\begin{gather}
  \label{eq:scalar_density}
  S = \bar{\psi} \Big(1 - \frac{1}{2}{\cal D}(0)\Big) \psi \\
  \intertext{and} 
  S = \bar{\psi} \Big((2R)^{-1} -  \frac{1}{2} \D(0)\Big) \psi
\end{gather}
for the general case. Defined this way the scalar density transforms
under chiral transformations like the corresponding continuum density
\cite{hasenfratz:_testin_qcd}. This implies that this density, which
is a dim=3 operator, is ${\cal O}(a)$ improved, since it does not mix
with any dim=4 operator. A detailed discussion of these densities and
the construction of chiral covariant currents, which are also ${\cal
  O}(a)$ improved, is given in \cite{hasenfratz:_testin_qcd}.

Our choice of the Legendre expansion might surprise a little bit when
consulting the literature about the different approximations, because
the Legendre expansion does not figure among the most efficient
schemes \cite{Eshof:2002ms}.  However, the discussion of these
approximations is almost entirely concentrated on the overlap
construction with Wilson fermions, which is completely different due
to the fact that the operator $A^\dagger A$ deviates much more from
$1$ than in the case of $\Dtpar$ (see Figure \ref{fig:AdaggerA}), and
therefore high order polynomials have to be used in the
approximations. On the contrary, $\Dtpar$ is already quite close to
satisfy the GW relation and therefore low order polynomial
approximations can be used to approximate e.g.~the operator $(A^\dagger
A)^{-1/2}$. For low order approximations, however, the difference between the
different polynomial approximations is much less pronounced than at
high orders. In fact, our experiments with the Chebyshev expansion
didn't show a visible advantage for the Chebyshev expansion at
polynomial orders of $\leq 4$, but clearly more precise studies are
needed to clarify this issue.


\subsection{Legendre Expansion}
\label{subsec:legendre}

Following the original work \cite{Hernandez:1998et}, we give a short
description of Legendre expansion of the operator $(A^\dagger A)^{-1/2}$.

In order to ensure convergence of the Legendre expansion we assume
that the bounds
\begin{equation}
  \label{eq:bounds_legendre}
  u \le \langle \psi | A^\dagger A | \psi \rangle \le v
\end{equation}
hold for some strictly positive constants $u < v$ and arbitrary
normalized states $|\psi \rangle $.

The Legendre polynomials $P_k(z)$ may be defined through the
generating functional
\begin{equation}
  \label{eq:generating_functional_legendre}
  (1 - 2 t z + t^2)^{-1/2} = \sum_{k=0}^\infty t^k P_k(z) \, .
\end{equation}
Usually $z$ is taken to be a number, but
\eqref{eq:generating_functional_legendre} remains meaningful if we
substitute
\begin{equation}
  \label{eq:def_of_z}
  z = \frac{v + u - 2 A^\dagger A}{v - u} \, .
\end{equation}
The expansion \eqref{eq:generating_functional_legendre} is convergent
for all $t$ satisfying $|t|<1$.

Introducing the parameter $\theta$ through
\begin{equation}
  \label{eq:theta}
  \cosh \theta = \frac{v+u}{v-u}\, ,\qquad \theta > 0 \, ,  
\end{equation}
and $t=e^{-\theta}$ \eqref{eq:generating_functional_legendre} takes
the form
\begin{equation}
  \label{eq:adaga_expansion}
  (A^\dagger A)^{-1/2} = \kappa \sum_{k=0}^\infty t^k P_k(z)\, 
  ,\qquad \kappa = \sqrt{\frac{4 t}{v - u}}\,,
\end{equation}
since $1 - 2 t z +t^2$ is proportional to $A^\dagger A$ for this
choice of $t$. In practical applications the sum in
\eqref{eq:adaga_expansion} will be truncated after a finite number of
terms and this then defines the Legendre expansion of the overlap
operator to a certain order $N$, which we denote by ${\cal D}^{(N)}$
or ${\D}^{(N)}$, respectively.

\subsection{Exact Projection of $A^\dagger A$ Eigenvalues}
\label{subsec:exact_projection}

For large volumes and small values of $\beta$ the eigenvalues of
$A^\dagger A$ tend to scatter further away from $1$, even for
$\Dtpar$. This eventually leads to big problems in the numerical
expansions used for the inverse square root or the sign function,
because the parameter $c$ governing the asymptotic convergence of
these expansions is \cite{Hernandez:2001yd}
\begin{equation}
  \label{eq:condition_number}
  ||{\cal D}^{(N)} - {\cal D}|| = e^{- c N}, \quad c = 
  \sqrt{\frac{\lambda_{\rm min}}{\lambda_{\rm max}}} \, ,
\end{equation}
where $\lambda_{\rm min}$ is the smallest eigenvalue of $A^\dagger A$
and $\lambda_{\rm max}$ the largest.  Because the distribution of the
eigenmodes of $A^\dagger A$ is typically not very dense at the lower
edge of the spectrum, one can use exact projection methods to decrease
the condition number $c$ substantially.  The same is not true for the
upper edge of the spectrum of $A^\dagger A$, because the modes lie
much denser there.  Let us specify what we mean by the exact
projection of the $n$ eigenmodes $|\psi_i \rangle$ of $\gamma_5 A$
with the eigenvalues $\lambda_i$ of smallest modulus:
\begin{gather}
  \label{eq:exact_projection}
  (A^\dagger A)^{-1/2} = \sum_{i=1}^{n} \frac{1}{|\lambda_i|}
  |\psi_i \rangle \langle \psi_i| + {\cal P}^n_\perp \,
  {\rm App} \big[(A^\dagger A)^{-1/2}\big] {\cal P}^n_\perp, \\
  \intertext{for the square root and} \epsilon(\gamma_5 A) =
  \sum_{i=1}^{n} \epsilon(\lambda_i) |\psi_i \rangle \langle \psi_i| +
  {\cal P}^n_\perp \, {\rm App} [\epsilon(\gamma_5 A)] {\cal
    P}^n_\perp
   \\
  \intertext{for the matrix sign function, where the projection to the
    subspace orthogonal to the lowest $n$ eigenmodes is defined by}
  {\cal P}^n_\perp = 1 - \sum_{i=1}^{n} |\psi_i \rangle \langle
  \psi_i|
\end{gather}
and where ${\rm App}$ stands for one of the possible expansions which
may be used in practice. The calculation of the $A^\dagger A$
eigenvalues and eigenmodes can be performed e.g.~with the Ritz
functional \cite{Bunk:1994,Kalkreuter:1996mm} or the implicitly
restarted Arnoldi method \cite{Sorensen:1992,Lehoucq:1998}. In our
overlap studies we mainly use the Arnoldi method to calculate the
eigenmodes, because it showed to be efficient in the search of rather
large numbers ($\leq 200$) of small $A^\dagger A$ eigenvectors.

\section{Properties of the Overlap with FP Kernel}
\label{sec:FP_overlap}

In the following discussion of the properties of the overlap
construction with $\Dtpar$ as kernel we focus on two topics: the
locality of the resulting operator and the dependence of the remaining
breaking of chiral symmetry on the precision of the approximation.

\subsection{Locality}
\label{subsec:locality}

\begin{figure}[htbp]
  \begin{center}
    \psfrag{FP}{\footnotesize $\Dtpar$}
    \psfrag{DWS}{\footnotesize $D_W^S$} \psfrag{DW}{\footnotesize
      $D_W$} \includegraphics[width=9.5cm]{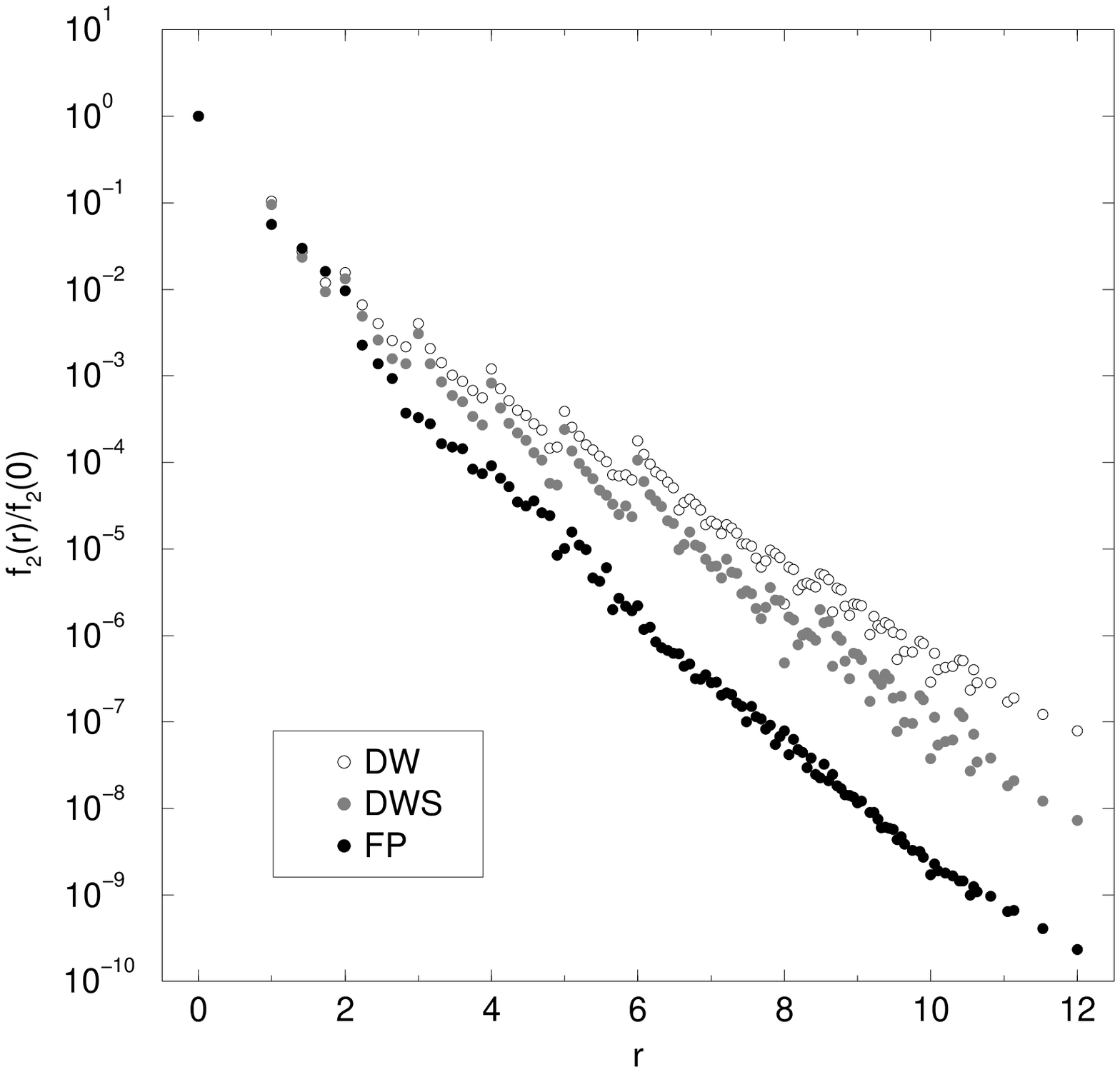}
    \includegraphics[width=9.5cm]{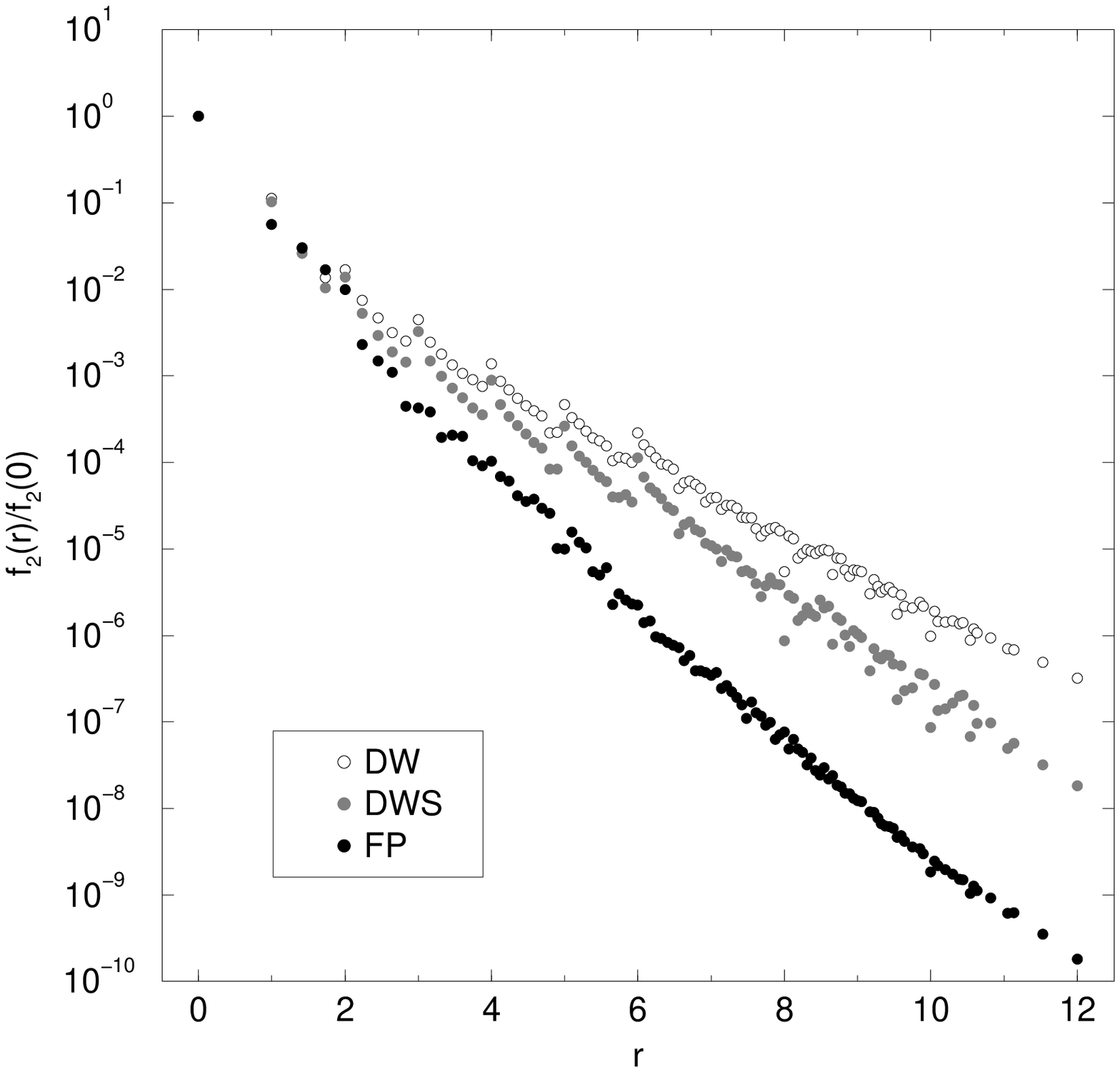}
    \caption{Locality of the overlap Dirac operator using $D_W$, $D_W^S$ and $\Dtpar$ on 
      $12^4$ configurations at $\beta = 3.5$ (top) and $\beta = 3.0$
      (bottom). The overlap construction with $\Dtpar$ is clearly more
      local than the Wilson overlap, independent of smearing. While the
      exponential fall-off of the $\Dtpar$ overlap stays the same at
      both gauge couplings, this is not the case for the Wilson
      overlap, whose locality gets worse with stronger coupling.}
    \label{fig:locality}
  \end{center} 
\end{figure}

A lattice action must be local for the continuum results to be
universal, i.e.~independent of the details of the lattice action.
Locality means that fields at large separations $r = |y - x| \gg 1$ have an exponentially
small coupling $\rho(r)$, i.e.~$\rho(r) \sim \exp(-\nu r)$ with $\nu={\cal O}(1)$. 
Optimizing the locality is essential so that e.g.~the
exponential fall-off of correlation functions can be separated from
direct couplings in the lattice action even on coarse lattices. The locality of a lattice
Dirac operator can be measured by
\begin{equation}
  f_p(r) = {\rm max}\{ ||D v||, ||y-x||_p = r \},
\end{equation}
where $v$ is a vector with point source at $x$ and $||v||_p =
(\sum_{i=1,n} |v_i|^p)^{1/p}$ is the vector norm. We use the usual
vector norm (p=2) and the taxi driver norm (p=1) to discuss the data.
These two norms are also the most common choices in the literature
\cite{Hernandez:1998et, DeGrand:2000tf}.  In Figure
\ref{fig:locality}, we present the locality measured by the
expectation value of the normalized function $f_2(r)/f_2(0)$ of the overlap construction
with $\Dtpar$ and the Wilson Dirac operator on smeared ($D_W^S$) and
unsmeared configurations ($D_W$) at $\beta = 3.0$ and $\beta = 3.5$ on
$12^4$ lattices.  We use a Legendre expansion of order $250$ for the
Wilson overlap and order $25$ for the overlap with $\Dtpar$ and
project out the 50 lowest $A^\dagger A$ eigenmodes to make sure that
the deviations from the exact overlap Dirac operator are smaller than
the couplings at $r/a=12$. The locality for the Wilson overlap is
optimized by adjusting the parameter $s$ in
eq.~\eqref{eq:overlap_DTR_simple}. The overlap construction with
$\Dtpar$ is clearly more local than the Wilson overlap, independent of
smearing. While the exponential fall-off of the $\Dtpar$ overlap stays
the same on both gauge couplings, this is not the case with the Wilson
overlap, whose locality gets slightly worse with stronger coupling. In
Table \ref{tab:locality} we list the exponents $\nu$ obtained from an
exponential fit to the expectation value of $f_p(r)$
\begin{equation}
  \label{eq:exponential_fit_locality}
  \langle f_p(r) \rangle \propto e^{- \nu r/a} \, ,
\end{equation}
where $r/a > 13$ for $p = 1$ and $r/a > 7$ for $p = 2$. The results in Table \ref{tab:locality} show
that with the overlap with $\Dtpar$ one can investigate masses up to $m a \approx 1.5$ in spectroscopy
calculations.
Finally, the
overlap construction from eq.~\eqref{eq:overlap_DTR_not_simple} with
$\Dpar$ and $R_{\rm par}$ has the same exponential fall-off as the
construction from eq.~\eqref{eq:overlap_DTR_simple}. 
\begin{table}
  \begin{center}
    \begin{tabular}{|c|c|c|c|c|}\hline
      $\beta$               & \multicolumn{2}{|c|}{3.5} & \multicolumn{2}{|c|}{3.0} \\ \hline 
      p                     &      1    &      2     &     1      &     2       \\ \hline
      $D_W$                 &  0.48(1)  &    1.14(5) &  0.42(1)   &   0.94(3)  \\
      $D_W^S$               &  0.58(1)  &    1.32(5) &  0.55(1)   &   1.22(4)  \\
      ${\cal D}_{\rm par}$  &  0.81(2)  &    1.60(2) &  0.82(1)   &   1.60(2) \\ \hline         
    \end{tabular}
  \end{center}
  \caption{Value of the exponent $\nu$ as defined in eq.~\eqref{eq:exponential_fit_locality} for 
    the overlap construction from eq.~\eqref{eq:overlap_DTR_simple} with different Dirac operators and
    for the taxi driver norm (p=1) as well as the usual vector norm (p=2), which gives the locality in
    terms of physical distances.}
  \label{tab:locality}
\end{table}

\subsection{Chiral Properties} 
\label{subsec:breaking_of_chiral}

In this section we consider the breaking of chiral symmetry defined
through various quantities. We use again the breaking of the GW
relation $\Delta_{\rm GW}(N)$ and the breaking of normality
$\Delta_{\rm n}(N)$ on normalized random vectors, as defined in
eqs.~\eqref{eq:GW_breaking} and \eqref{eq:normality_breaking};
however, with the additional dependence on the order $N$ of the
Legendre expansion. But first we try to characterize the deviations of
the eigenvalues from the GW circle.

\begin{figure}[tbfp]
  \begin{center}
    \includegraphics[width=9cm]{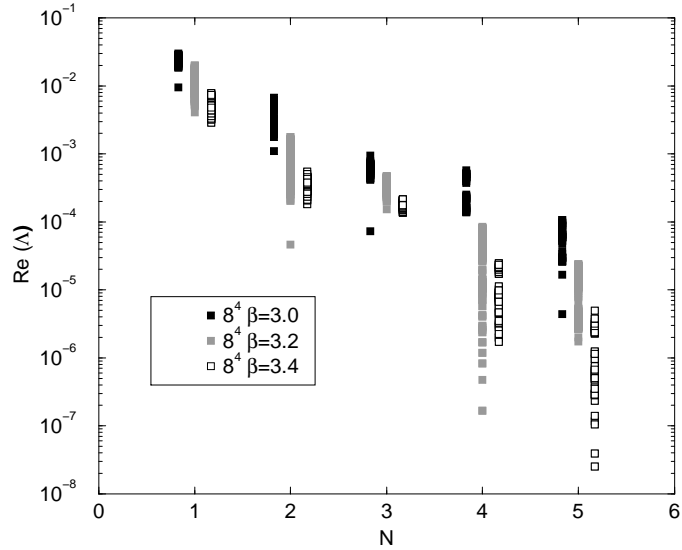}
    \caption{Deviation of the ${\cal D}^{(N)}$ eigenvalues from the GW circle 
      defined by the stereographic projection in
      eq.~\eqref{eq:stereographic_projection}.}
    \label{fig:stereo}
  \end{center}
\end{figure}

The eigenvalue spectrum is suited to show deviations from the GW
circle. However, when the GW breaking $\Delta_{\rm GW}$ drops below
$\sim 10^{-2}$, it gets difficult to judge the quality of the
approximation by eye, which makes it necessary to use a different
measure to visualize the deviations from the circle. Hence, we use the
fact that if the eigenvalues $\lambda$ of $\Dt$ lie exactly on the GW
circle, then the stereographic projection
\begin{equation}
  \label{eq:stereographic_projection}
  \Lambda = \frac{\lambda}{1-\frac{\lambda}{2}}
\end{equation}
is purely imaginary. In Figure \ref{fig:stereo}, we see that ${\rm
  Re}(\Lambda)$ falls off exponentially as we increase the polynomial
order, showing that the eigenvalues $\lambda$ get closer and closer to
the GW circle. The most rapid decrease is at the finest lattice
spacing $a \approx 0.10 \, {\rm fm}$ ($\beta = 3.4$), which is 
what one expects.

\begin{figure}[htbf]
  \begin{center}

    \includegraphics[width=9cm]{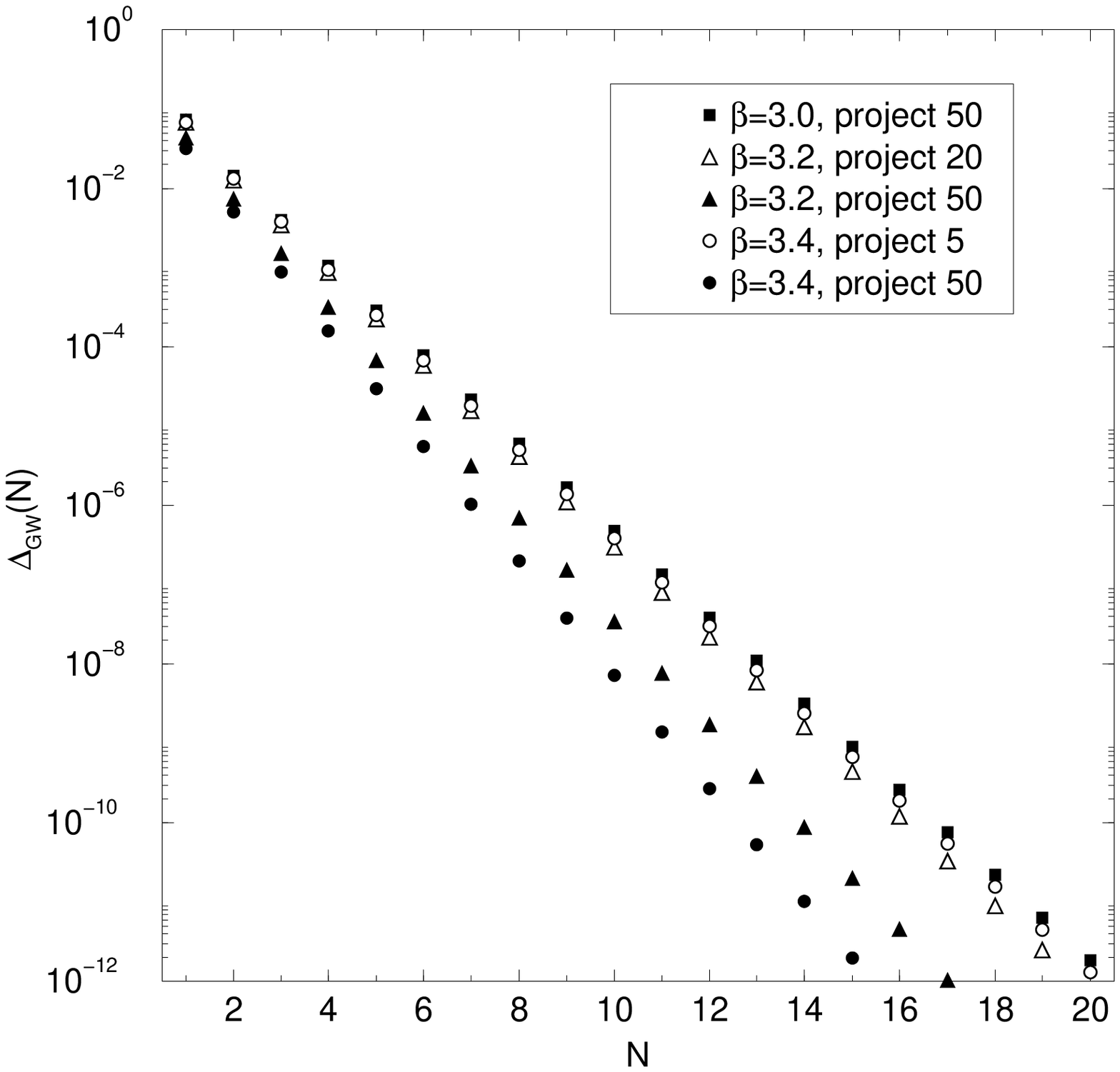}\vspace{0.5cm}
    \includegraphics[width=9cm]{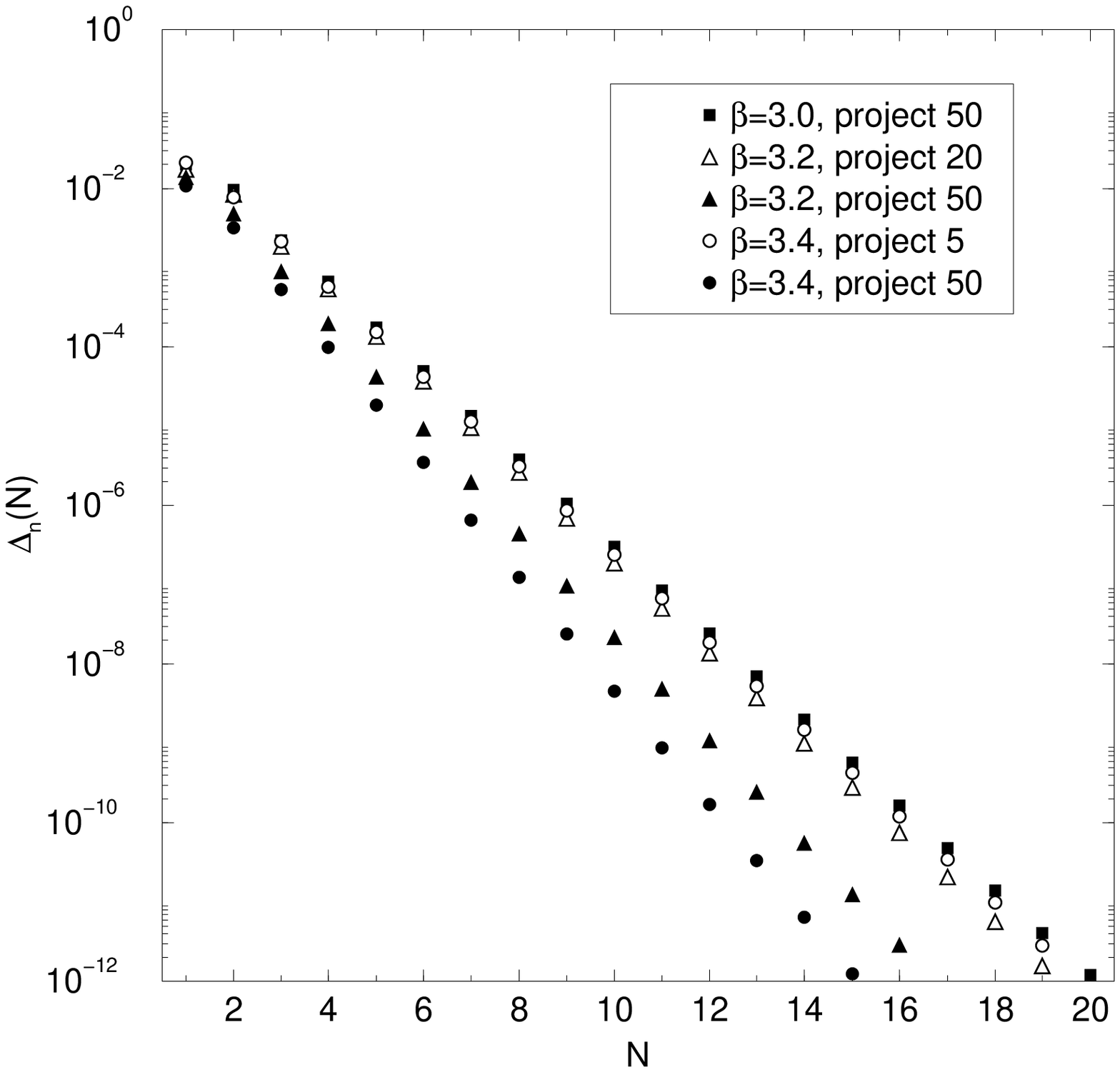}
    \caption{Breaking of Ginsparg-Wilson relation $\Delta_{\rm GW}(N)$ (top) and the breaking of
      normality $\Delta_{\rm n}(N)$(bottom) on $10^4$ lattices. Both quantities have
      almost exactly the same exponential fall-off, even though
      $\Delta_{\rm n}(N)$ remains slightly smaller for all the orders
      of the Legendre approximation shown on this figure. The figure
      also shows that the exponential fall-off is rather governed by
      the condition number $c$, defined in
      eq.~\eqref{eq:condition_number}, because the exponent is roughly
      the same for the $3$ different lattice spacings, where the lower
      boundary of the projected $A^\dagger A$ eigenvalues is chosen
      such that the condition number $c$ is approximately the same.}
    \label{fig:gwbreaking}
  \end{center}
\end{figure}

In Figure \ref{fig:gwbreaking} the breaking of the GW relation and the
breaking of the normality is shown on $10^4$ lattices at $\beta =
3.0$, $\beta = 3.2$ and $\beta = 3.4$. First of all we see that the
breaking $\Delta_{\rm GW}(N)$ and $\Delta_{\rm n}(N)$ fall off
exponentially as we increase the order $N$ of the Legendre polynomial
approximation of $(A^{\dagger} A)^{-1/2}$. Treating the the same
number of $A^{\dagger} A$ eigenvalues exactly, the fall-off is clearly
the steepest for the smallest lattice spacing ($\beta = 3.4$), which
is what one expects. When the number of projected $A^{\dagger} A$
eigenvalues, however, is decreased such that the ratio $\lambda_{\rm
  min}/ \lambda_{\rm max}$ is approximately the same on all lattice
spacings (this actually corresponds to projecting only the 20 lowest
eigenvalues of $A^{\dagger} A$ at $\beta = 3.2$ and the 5 lowest at
$\beta = 3.4$) the approximation is of the same quality to a very high
degree. In Table \ref{tab:exponent_overlap_order} we collect the
exponents $\mu$ describing the fall-off of $\Delta_{\rm GW}(N)$ with
increasing order of the Legendre expansion and compare it to the
predicted asymptotic value $c$ from eq.~\eqref{eq:condition_number}.
Notice that the exponent $\mu$ is clearly larger than the asymptotic
value in all cases, which indicates that the regime of asymptotic
convergence is by far not reached, making the convergence in the
region that is interesting for simulations much faster than in the
worst case. But, all the same the value $c$ seems to be in clear
correlation with the measured fall-off $\mu$. The results are based on
rather low statistics and in order to make the statement more
quantitative clearly more statistics is needed.
\begin{table}[tbhp]
  \begin{center}
    \begin{tabular}{|c|c|c|}\hline
      $\beta$ & $\mu$ &   c   \\ \hline\hline 
      3.4     & 1.66  &  0.67 \\ 
      3.2     & 1.51  &  0.63 \\
      3.0     & 1.26  &  0.55 \\ \hline
    \end{tabular}
  \end{center}
  \caption{Value of the exponent $\mu$ giving the exponential fall-off of the breaking of the GW relation
    in terms of $\Delta_{\rm GW}(N)$ is compared to the asymptotic fall-off given by the condition 
    number $c$, defined in eq.~\eqref{eq:condition_number}. The exponential fall-off of $\Delta_{\rm GW}(N)$
    is clearly steeper than the asymptotic fall-off, but $c$ is obviously correlated to $\mu$. The data
    is obtained from $10^4$ lattices.}
  \label{tab:exponent_overlap_order}
\end{table}

\section{A Final Remark}
\label{sec:final_remark}

\begin{figure}[tbfp]
  \begin{center}
    \includegraphics[width=4cm]{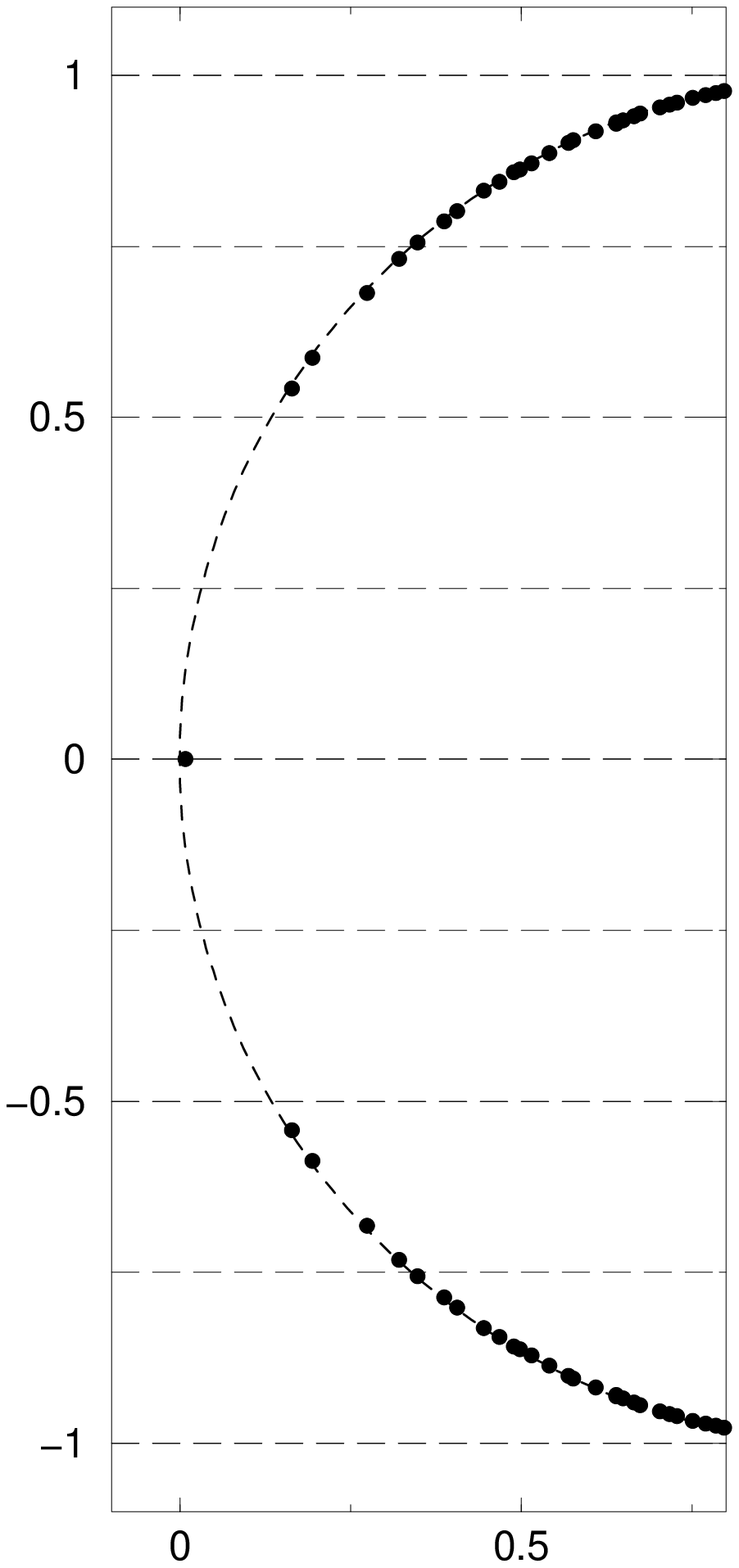}\hspace{0.5cm}
    \includegraphics[width=4cm]{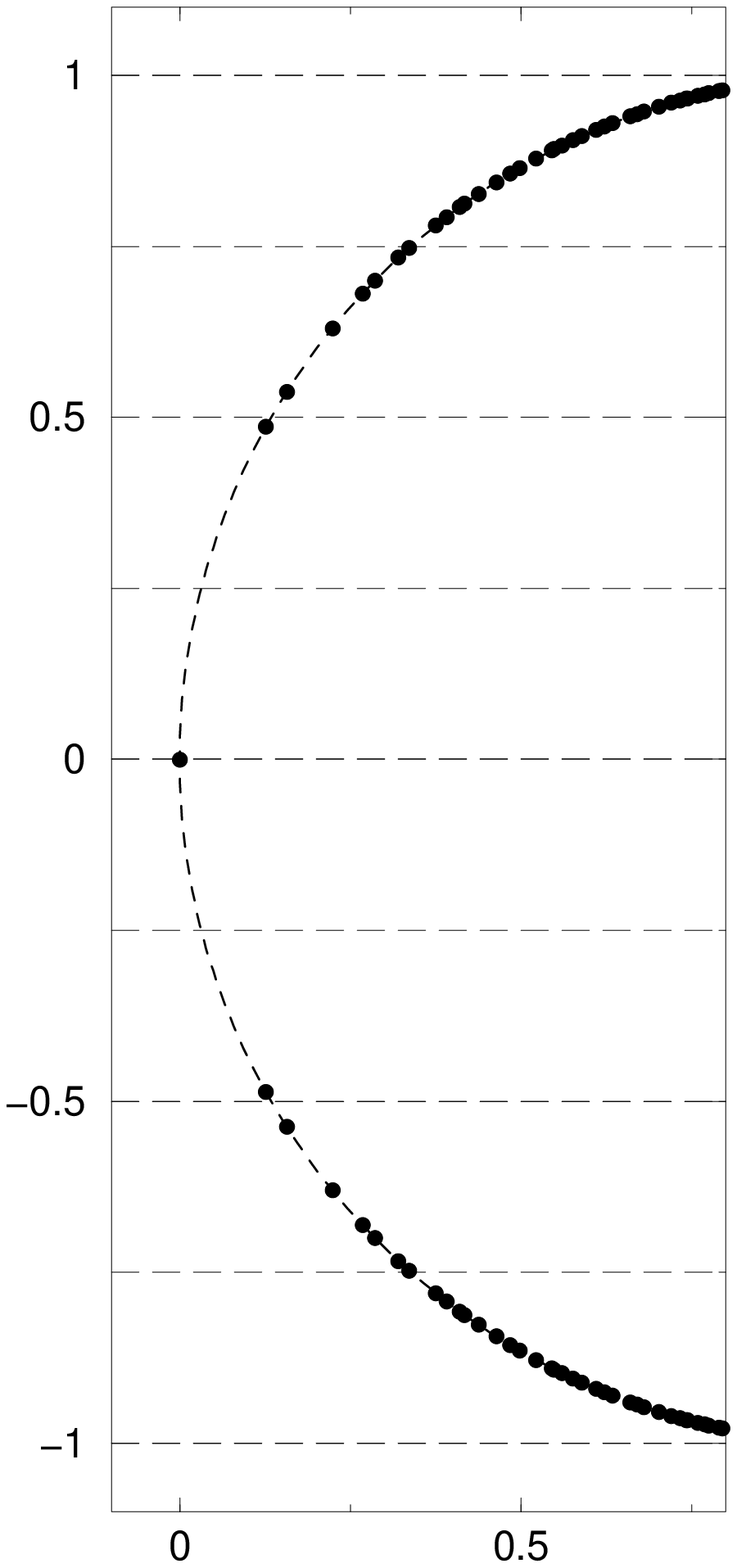}
    \caption{The physical branch of the eigenvalue spectra of the blocked Dirac operator (left) and
      the overlap Dirac operator with $\Dtpar$ as kernel (right) are
      shown. One observes that the eigenvalues of the blocked operator
      are shifted away slightly more from the origin than the
      eigenvalues of the overlap Dirac operator. This indicates that,
      even though the overlap construction cures the remaining breaking of
      chiral symmetry in $\Dtpar$, the overlap Dirac operator with
      $\Dtpar$ might show different cut-off effects than the
      (approximate) FP Dirac operator, defined through the block
      transformation in eqs.~\eqref{eq:RG_Dirac} and \eqref{eq:RGT_R}.
      The lines on the figure are shown to make the different locations of the eigenvalues 
      more evident.}
    \label{fig:eigenvalue_comparison_block_over}
  \end{center}
\end{figure}

The overlap construction with $\Dtpar$ can be looked upon as a tool to
cure the breaking of chiral symmetry that remains in the parametrized
FP Dirac operator.  Clearly, one hopes that the good scaling
properties of the FP Dirac operator, which to a certain extent are
still present after the parametrization process, are not distorted too
much by the overlap construction. In the best case the overlap cures
some small remaining ${\cal O}(a)$ cut-off effects in hadron
spectroscopy measurements, without creating larger ${\cal O}(a^2)$
scaling violations. The hadron spectroscopy measurements performed in
\cite{Hauswirth_diss:2002}, which do not only contain simulations with
$\Dpar$, but also include simulations with an order $3$ Legendre polynomial
expansion of the overlap Dirac operator from eq.~\eqref{eq:overlap_DTR_not_simple}
with $\Dpar$ and $R_{\rm par}$ at a lattice spacing of $0.155 \, {\rm fm}$ ($\beta =
3.0$), may be interpreted this way. It is, however, very risky to make such a conclusion
with so little data. The results on the dependence of the vector meson mass with respect to
the pseudoscalar mass show that there are larger discrepancies
between the measurements with $\Dpar$ and the corresponding overlap construction than one
might expect. Furthermore, the energy-momentum relation for the overlap construction indicates
cut-off effects. Even though the differences cancel in the Edinburgh plot, the data
indicates different cut-off effects for $\Dpar$ and the corresponding overlap construction.
To conclude this section we compare the physical branch of the eigenvalue spectrum of the blocked
Dirac operator obtained from the RGT equations \eqref{eq:RG_Dirac} and \eqref{eq:RGT_R} (Dirac operator II
in the nomenclature of Chapter \ref{cha:properties})
with the overlap construction with $\Dtpar$ in Figure
\ref{fig:eigenvalue_comparison_block_over}. It shows that even if the
overlap construction shifts the eigenvalues to the circle, there
remains a visible difference between the two spectra, which indicates
that the cut-off effects for the two Dirac operators might, indeed, be different.
The question how large the cut-off effects are for the different Dirac 
operators discussed in this section has surely to be answered in future studies.
\cpages
\chapter{Topological Susceptibility and Local Chirality of Near-Zero Modes}
\label{cha:topological_susceptibility}

The zero modes as well as the low-lying modes of the Dirac operator
have a particularly important influence on several important phenomena
of QCD, because their contribution to the quark propagator gets very
large in the chiral limit. Topological excitations like instantons
might be responsible for the spontaneous breaking of chiral symmetry
by building up a non-zero density of eigenvalues at zero in the
infinite volume limit; a fact which is encoded in the Banks-Casher
relation \cite{Banks:1980yr}. Instantons also provide an explanation
for the large mass of the $\eta^\prime$ particle
\cite{'tHooft:1986nc}. Related explanations for the mass of the
$\eta^\prime$ predict a connection between the quenched topological
susceptibility and the mass of the $\eta^\prime$
\cite{Witten:1979vv,Veneziano:1979ec}.

The question has recently been raised if it is possible to show that
the near-zero modes are dominated by instantons or whether instantons
are not important to explain the non-perturbative properties of the
QCD vacuum.

Answers to questions related to gauge field topology and its influence
on the low-lying modes of the fermions living on such a gauge
background can be explored within lattice QCD. Lattice Dirac operators
which satisfy the GW relation provide an excellent framework to study
topological excitations of the QCD vacuum.  They allow a definition of
the topological charge and the topological charge density on the
lattice which is unambiguous in the continuum limit (see Section
\ref{subsec:GW}).  Hence, GW fermions can be used to
study the topological susceptibility and properties of the low-lying
modes of the Dirac operator on the lattice in a very clean way.

In the following we present some facts about the topological
susceptibility.
Then we will collect the data for the topological susceptibility which
we obtained from different measurements with approximations to the
overlap construction with $\Dtpar$ as defined in Chapter
\ref{cha:overlap}. Finally, we present our results on the local
chirality of near-zero modes.

\section{Quenched Topological Susceptibility}
\label{sec:quenched_topological_susceptibility}

A mechanism that gives rise to the large mass of the $\eta^\prime$ was
given in \cite{Witten:1979vv,Veneziano:1979ec} and involves large
vacuum fluctuations associated with confinement. In the leading order
of the large $N_c$ expansion it allows to relate the mass of the
$\eta^\prime$ to the strength of the topological fluctuations through
the following relation
\begin{gather}
  \label{eq:eta_prime}
  \chi_{\rm t} \doteq \frac{\langle Q^2 \rangle}{V} \simeq \frac{m_{0}^2 f_{\pi}^2}{2 N_f}\\
  \intertext{with} m_{0}^2 = m_{\eta^\prime}^2 + m_{\eta}^2 - 2 m_{\rm
    K}^2 \, ,
\end{gather}
where $Q$ is the topological charge, $\chi_t$ the topological
susceptibility and where $\langle . \rangle$ denotes the expectation
value.  Using the physical values for $m_{\eta^\prime}$, $m_{\eta}$,
$m_{\rm K}$ and $f_\pi$ one obtains $\chi_{\rm t} \approx (180\,
\mev)^4$ and lattice measurements typically yield $\chi_{\rm t} \sim
(200 \, \mev)^4$ \cite{Teper:1999wp}.  Note that on the left hand side
of eq.~\eqref{eq:eta_prime} the quenched topological susceptibility
$\chi_{\rm t}$ enters as this equation has been derived in the large
$N_c$ limit; for practical applications, however, one assumes that the
physical value $N_c = 3$ is already close to $N_c = \infty$. This
assumption which is used quite frequently, clearly needs further
testing, even though some results point into the direction that this
approximation is reasonable, e.g.~\cite{Cundy:2002hv}.

The topological susceptibility in quenched QCD differs substantially
from the topological susceptibility in full QCD which vanishes in the
chiral limit. The reason for this can be understood quite easily.
After integrating out the fermion fields in the QCD partition function
one is left with the determinant of the Dirac operator with mass $m$.
As this mass goes to zero in the chiral limit the determinant vanishes,
if the Dirac operator has an eigenmode at $m$, i.e.~a zero mode in the
chiral limit, and therefore configurations with a non-zero topological
charge will be suppressed.  In contrast to this the determinant in
quenched QCD is set to $1$ for purely technical reasons and hence
there is no suppression of the zero modes in the quenched case.

\subsection{Lattice Determination of the Topological Charge}
\label{sec:top_charge}

The determination of the topological charge on the lattice has a long
history and several different techniques have been developed in this
time. The techniques vary from determinations from purely gluonic
quantities like lattice discretizations of the $F_{\mu\nu}
\tilde{F}_{\mu\nu}$ operator \cite{Alles:1997nm} to determinations
from fermionic observables as the method by Smit and Vink
\cite{Smit:1987fn}, which in a certain way tries to make use of
remnants of the Atiyah-Singer index theorem \cite{Atiyah:1971rm}.
While the traditional gluonic methods run into trouble because the
discretizations of $F_{\mu\nu} \tilde{F}_{\mu\nu}$ typically are very
sensitive to ultraviolet fluctuations and require either cooling or
smearing, the fermionic methods have the problem that the
Atiyah-Singer index theorem does not hold for non-chiral lattice
fermions. A solution is clearly provided by a chiral formulation of
lattice fermions as they satisfy the Atiyah-Singer index theorem also
at finite cut-off and therefore a definition of the index of a gauge
configuration can be given, even though the result may vary from Dirac
operator to Dirac operator \cite{Hasenfratz:1998ri}. In the continuum
limit, however, the results will ultimately agree. The FP Dirac
operator offers, in principle, an even better solution, because the
result for the fermionic charge agrees with the FP gauge charge at
finite cut-off \cite{Hasenfratz:1998ri}.  While the discretizations of
$F_{\mu\nu} \tilde{F}_{\mu\nu}$ have to deal with large ultraviolet
fluctuations even at small lattice spacing, the fermionic
determination using chiral fermions and the chirality of the zero
modes, i.e.~the index theorem, does not have any problems with
ultraviolet fluctuations because the zero modes are completely
unaffected by the fluctuations at the cut-off.  In the following we
explain the method we use to determine the topological charge in our
test studies.


\subsubsection*{Counting Zero Modes and their Chirality}

The simplest way to make use of the lattice index theorem to determine
the topological charge is to calculate the zero modes and to
determine their chirality. In real simulations this actually means
that one has to calculate the chirality of the small real modes,
because in numerical simulations one cannot have an exact GW fermion
and therefore the ``zero modes'' will scatter a little bit and become
real modes $|\psi\rangle$ with eigenvalues with a small modulus. Furthermore, their
chirality, i.e.~$\langle\psi|\gamma_5|\psi\rangle$, will not be exactly $\pm1$,
but e.g.~something like $\pm 0.7$ to $\pm 1$ for the 
small real modes of $\Dtpar$. In rare cases the real modes of $\Dtpar$ can have 
an eigenvalue on the order of $0.3$ and then also $\langle\psi|\gamma_5|\psi\rangle$
can be on the order $\pm 0.2$ to $\pm 0.4$. In our definition of the index of $\Dtpar$ 
we will sum up the chiralities of the small real modes with eigenvalues smaller than 
approximately $0.5$ by simply counting 
the signs, independent of the magnitude of $\langle\psi|\gamma_5|\psi\rangle$.  
As we showed in
Figure \ref{fig:index_OK} the real modes are quite stable, because
their number and chirality does change only rarely when one compares
the results from $\Dtpar$ and the corresponding overlap construction
with $\Dtpar$.  This already gives a hint that the level to which
chiral symmetry is required to determine the number of small modes and
their chirality is not very high. Using a moderate approximation of the overlap construction
the small real eigenvalues will approach the origin and their chirality gets clearly closer
to $\pm 1$, making the determination of the index by counting the modes less 
ambiguous than with $\Dtpar$. In order to calculate the low-lying
eigenmodes we use again the implicitly restarted Arnoldi solver
\cite{Sorensen:1992,Lehoucq:1998}.

\subsection{Determination of the Quenched Topological Susceptibility}
\label{sec:determination_topological_susceptibility}

In this determination of the quenched topological susceptibility, where we use
the method described above to extract the topological charge of a
gauge configuration, we collect data from various measurements that we
have done with different approximations to the overlap with $\Dtpar$
(see Chapter \ref{cha:overlap}) and therefore the accuracy in the
approximations varies substantially. In order to assure that for the
counting of small real eigenmodes and their chirality there are
essentially no differences between approximations with lower accuracy
and such with higher accuracy we performed several tests, where we
compare the outcome of the mode counting on samples of up to $100$
configurations.  The tests all lead to the same conclusion that no
difference in the counting of the modes and their chirality was found
between the different approximations we used. It shows that in the
volumes we use it is actually enough to treat the $75 - 200$ lowest
$A^\dagger A$ eigenmodes -- the eigenmodes which are responsible for
the major part of the breaking of chirality -- exactly and leave the
rest of the $\Dtpar$ as it is, i.e.~no Legendre expansion is used for
the approximation for the square root in
eq.~\eqref{eq:exact_projection}. This shows that the degree to which
chirality is needed to measure the topological charge with this
definition is actually not very high. Obviously, the number of $A^\dagger A$ eigenmodes which
have to be treated exactly in order to provide enough accurate chiral symmetry
depends on the gauge coupling and the volume.

\begin{figure}[p]
  \begin{center}
    \includegraphics[width=10cm]{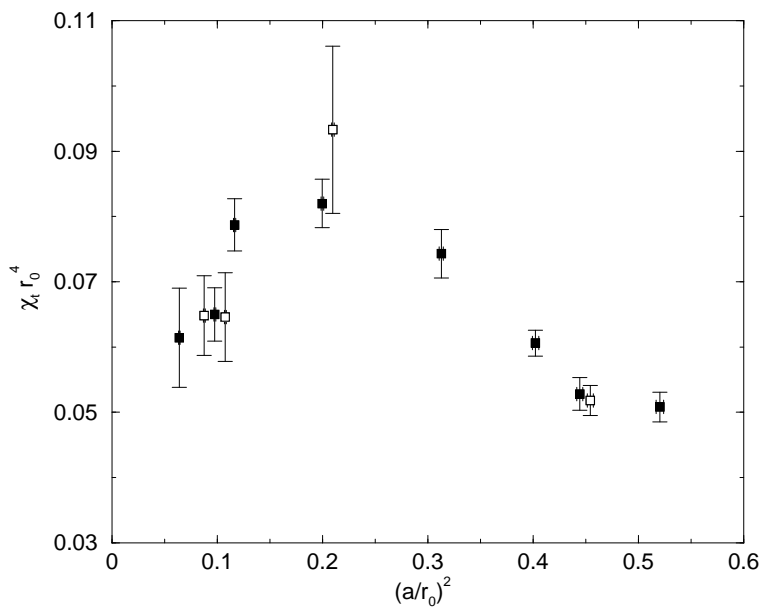}
    \caption{The quenched topological susceptibility $\chi_t$ measured with the overlap 
      construction with $\Dtpar$. The open squares are used to denote  in
      a clearer way that the corresponding symbol is shifted horizontally in order to
      display measurements done at the same scale,
      but with either different accuracy of the overlap construction or on different volumes.
      The decrease of $\chi_t$ towards coarse lattices (large $a$) is an artifact
      due to the fact that instantons fall through the lattice, i.e.~they can no longer be resolved. The 
      decrease towards small lattice spacings is much more important as it indicates clearly
      a cut-off effect.}
    \label{fig:topological_susc_fp}
  \end{center}
\end{figure}

We determine the quenched topological susceptibility at several values of the
gauge coupling $\beta$, which lie in a range between $3.2 \leq \beta
\leq 2.3 $, and on different lattice sizes. We give the details in
Table \ref{tab:topological_susceptibility}. The scale is determined
from the interpolating formula in \cite{Rufenacht:2001qy}. The results
are also shown in Figure \ref{fig:topological_susc_fp}. Notice that these
measurements are pushed to the extreme: the resolution lies in between
$a=0.13 \, {\rm fm}$ and $a=0.36 \, {\rm fm}$, i.e.~very coarse
lattices are included and some of the measurements are on $4^4$ and
$6^4$ lattices.  Further, there does not exist, not even
approximately, a unique way to fix $r_0$ (and so $a$) on such coarse
lattices. We believe, however, that Figure \ref{fig:topological_susc_fp}
contains interesting information.

\begin{table}[t]
  \begin{center}
    \begin{tabular}{|c|c|c|c|c|c|c|c|c|} \hline
      $\beta$ & $V/a^4$ & $r_0/a$ & $L a [{\rm fm}]$ & $n_{\rm conf}$ & $\langle Q^2\rangle$ & $\chi_t r_0^4$ & $N$ \\ \hline\hline
      2.300   &  $4^4$ & 1.386(5) & 1.443(5) & 1000 & 3.53(16)   &  0.0508(23) &  5 \\
      2.361   &  $4^4$ & 1.500(5) & 1.333(4) & 1000 & 2.67(13)   &  0.0528(25) &  2 \\
      2.361   &  $6^4$ & 1.500(5) & 2.000(7) &  954 &13.28(62)   &  0.0518(23) &  2 \\
      2.400   &  $4^4$ & 1.576(6) & 1.269(5) & 2000 & 2.51(8)    &  0.0606(20) &  4 \\
      2.500   &  $4^4$ & 1.787(6) & 1.119(4) & 1000 & 1.87(9)    &  0.0743(37) &  4 \\
      2.680   &  $6^4$ & 2.237(7) & 1.341(4) & 1000 & 4.24(19)   &  0.0820(37) &  0 \\
      2.680   &  $9^4$ & 2.237(7) & 2.012(6) &  100 &24.46(3.37) &  0.0933(128)&  2 \\
      2.927   &  $8^4$ & 2.969(14)& 1.347(6) &  990 & 4.14(20)   &  0.0787(40) &  0 \\
      3.000   &  $8^4$ & 3.197(16)& 1.251(6) &  600 & 2.55(16)   &  0.0650(41) &  4 \\
      3.000   &  $8^4$ & 3.197(16)& 1.251(6) &  340 & 2.54(24)   &  0.0648(61) &  0 \\
      3.000   & $10^4$ & 3.197(16)& 1.564(8) &  192 & 6.18(65)   &  0.0646(68) &  4 \\
      3.200   & $10^4$ & 3.947(26)& 1.267(8) &  200 & 2.53(27)   &  0.0614(76) &  2 \\ \hline
    \end{tabular}
  \end{center}
  \caption{In this table the different measurements of the quenched topological 
    susceptibility are collected. It shows the 
    different gauge couplings $\beta$, volumes $V$, the scale $r_0/a$, the length
    of the box in fm (the error of the matching of $r_0$ to a physical scale is not 
    included), the number of configurations $n_{\rm conf}$, the measured expectation 
    value of the square of the topological charge $\langle Q^2\rangle$, the topological 
    susceptibility measured in units of $r_0$ and finally $N$ gives the order of the Legendre
    polynomial approximation used for the overlap construction with $\Dtpar$ kernel. Where no Legendre
    expansion is used, 75 to 200 $A^\dagger A$ eigenmodes are treated exactly.}
  \label{tab:topological_susceptibility}
\end{table}

\subsection{Discussion}

\begin{figure}[htbp]
  \begin{center}
    \includegraphics[width=10cm]{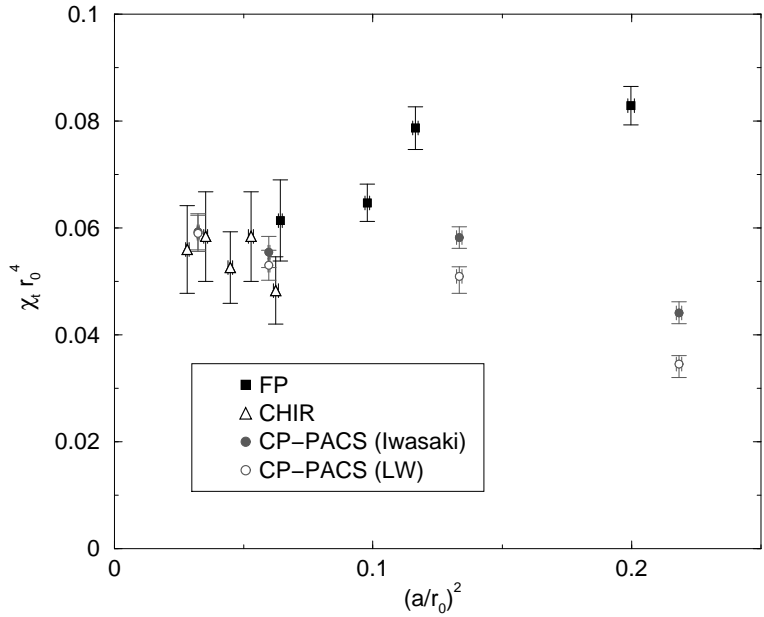}
    \caption{The quenched topological susceptibility obtained from different recent 
      determinations. In this figure only the physically interesting part towards
      smaller lattice spacings is shown. We compare our results (FP), where we have combined
      the data from different determinations at the same lattice spacing in this 
      figure, to the very recent determination in \cite{Gattringer:2002mr},
    which uses a truncated solution of the GW relation and counts the chirality of the
    small real modes (CHIR). The other two measurements are done by the CP-PACS collaboration
    and use traditional cooling techniques with the Iwasaki and the L\"uscher-Weisz action
    \cite{AliKhan:2001ym}.}
    \label{fig:topological_susc}
  \end{center}
\end{figure}

The results shown in Figure \ref{fig:topological_susc} give clearly
the impression that the topological susceptibility measured with the
overlap construction with $\Dtpar$ has substantial cut-off effects.
But one has to keep several things in mind.  The decrease towards
coarse lattice spacings is expected as the lattice gets too coarse to
resolve topological objects, i.e.~instantons fall through the lattice,
and the results for the very coarse lattice spacings are only given
for completeness. Obviously more important is the decrease towards
small lattice spacings, which really indicates a relevant cut-off
effect. It can be understood in different ways. The measurement at the
smallest lattice spacing, which with $a=0.13 \,{\rm fm}$ is still not
very fine, is not very accurate and the picture might change with more
statistics. Furthermore, the determination of the scale enters with
$(r_0/a)^4$ which makes that systematic errors become quite relevant.
Clearly, the finite volume effects are not under control in this
compilation of data and we can not exclude that the points at $\beta =
3.2$ and $\beta = 3.0$ are affected more by finite volume effects than
the points at $\beta = 2.927$ and $\beta = 2.68$, which are measured on
a larger volume.  However, the finite volume effects are not expected
to be large when the size of the box is $1.2\, {\rm fm}$ or more.
Finally, we give a comparison to other recent measurements
\cite{AliKhan:2001ym,Gattringer:2002mr} of the topological
susceptibility. Our results do not show any obvious discrepancy with
the other data as the lattice spacing gets small, however, neither our
data nor the data of the other groups are as accurate that a more
precise statement can be made at this point.

\section{Local Chirality of Near-Zero Modes}
\label{sec:local_chirality}

Exact zero modes of the Dirac operator tell us, via the index theorem,
about the topological charge $Q$ of the background gauge
configuration.  However, the exact zero modes alone cannot break
chiral symmetry spontaneously. According to the Banks-Casher relation
\cite{Banks:1980yr} $\langle \bar{\psi} \psi \rangle = - \pi \rho(0)
\ne 0$, the Dirac operator must build up a finite density of near-zero
modes, which does not vanish as $V \rightarrow \infty$.

One mechanism which explains the formation of near-zero modes involves
instantons.  Consider a gauge configuration containing one instanton
and one anti-instanton. If the instanton and anti-instanton are
separated by a large distance, the Dirac operator has a pair of
complex eigenvalues lying close to 0. The farther the instanton and
anti-instanton are separated from each other, the closer the complex
eigenvalue pair moves to the origin. If the instanton and
anti-instanton are brought closer together, the complex eigenvalue
pair moves away from the origin and disappears into the bulk of the
eigenvalue spectrum. If the gauge configurations contain many
instantons and anti-instantons, this could produce a non-zero density
of near-zero modes, giving $\rho(0) \ne 0$ in the infinite volume
limit.

The question has recently been raised if it is possible to show that
the near-zero modes are dominated by instantons. From instanton
physics, it is expected that the modes are highly localized where the
instantons and anti-instantons sit. If this is so, then in these
regions the modes should be close to chiral i.e.~mostly either left-
or right-handed, depending on whether it is sitting on an instanton or
anti-instanton. In \cite{Horvath:2001ir}, the authors defined a
measure of local chirality at lattice site $x$ by
\begin{equation}
\tan\left[\frac{\pi}{4}(1+X(x))\right] = \sqrt{\frac{{\psi_{\rm L}}^{\dagger} 
     \psi_{\rm L}(x)}{{\psi_{\rm R}}^{\dagger} \psi_{\rm R}(x)}}.
\end{equation}
An exact zero mode is purely either left- or right-handed, giving
$X(x)=\pm 1$ at all lattice sites $x$. If a near-zero mode is
localized around instanton-anti-instanton lumps, then $X(x)$ should be
close to $\pm 1$ for the sites $x$ where the probability density
$\psi^{\dagger} \psi (x)$ is largest.

In the original paper \cite{Horvath:2001ir}, many near-zero modes of
the Wilson operator $D_W$ were analyzed for many gauge configurations
and the finding was that in the regions where the modes are localized,
the distribution for $X(x)$ is peaked around 0 and the modes do not
display local chirality. This led to the conclusion that the near-zero
modes are not dominated by instantons. Since then, several other
groups have found the opposite conclusion
\cite{DeGrand:2001pj,Gattringer:2001ia,Edwards:2001nd,Blum:2001qg},
using Dirac operators with much better chiral symmetry than $D_W$ (or
even just an alternative definition of a complete basis for the
non-normal operator $D_W$). They find the distribution of $X(x)$ is
double-peaked with maxima at large positive and negative values of
$X$, indicating that the modes are locally chiral.

\subsection{Results}
\label{sec:results_instanton_dominance}

\begin{figure}[htbp]
  \begin{center}
    \includegraphics[width=10cm]{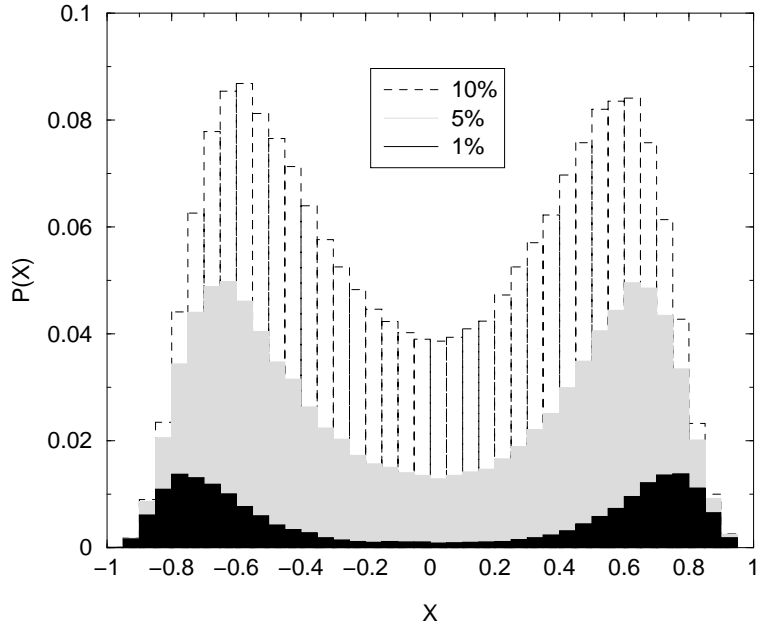}
    \caption{The distribution $P(X)$ for the top 1\%,5\%,10\% lattice sites $x$ with the
      largest $\psi^\dagger \psi(x)$.}
    \label{fig:localchirality}
  \end{center}
\end{figure}

In our study we analyze the 10 smallest near-zero modes of the overlap
operator with $\Dpar$ as kernel for 60 different $10^4$ gauge
configurations at $\beta=3.2$. We use a Legendre polynomial of order 2
to approximate the operator $(A^{\dagger}A)^{-1/2}$ in overlap
formula, with the 10 smallest $A^{\dagger}A$ being projected out and
treated exactly (see Chapter \ref{cha:overlap} for details).  The
eigenvalues $\lambda$ and eigenvectors $\psi$ are found using the
implicitly restarted Arnoldi method \cite{Sorensen:1992,Lehoucq:1998}.
In Figure~\ref{fig:localchirality}, we plot the distribution $P(X)$ of
the measure of local chirality $X$ at the lattice sites where the
density $\psi^{\dagger} \psi (x)$ of a mode is largest. The three
distributions correspond to taking, for each mode, 1\%, 5\% and 10\%
of all lattice sites that have the largest density $\psi^{\dagger}
\psi (x)$. We do not include exact zero modes, for which $X(x)=\pm 1$
at all lattice sites. We see a clear double-peaked distribution, whose
maxima are farther from zero if we only include the sites where the
modes are most localized. The maxima are not at $X=\pm 1$ as the modes
are not exactly chiral. We find the same conclusion as
\cite{DeGrand:2001pj,Gattringer:2001ia,Edwards:2001nd,Blum:2001qg}:
where the near-zero modes are most localized, they are also very
chiral. Recent work \cite{Gattringer:2002gn} has shown that, using
near-zero modes of the Dirac operator, the gauge field strength
appears to contain lumps which are mostly either self-dual or
anti-self-dual. This behavior is consistent with the picture of
instanton-dominance of the near-zero modes, however it is no
conclusive evidence that instantons are the driving mechanism for
chiral symmetry breaking \cite{Horvath:2002gk,Horvath:2002yn}.

\cpages
\chapter{Determination of the Low-Energy Constant $\Sigma$}
\label{cha:condensate}

It is the common expectation that for QCD with a number $N_f \ge 2$ of
massless quark flavors, the chiral symmetry is spontaneously broken by
a non-zero expectation value for the chiral condensate $\langle
\bar{\psi} \psi \rangle$.  Chiral perturbation theory ($\chi {\rm
  PT}$), which is based on this assumption, is an excellent
description of many low-energy QCD phenomena \cite{Gasser:1984yg}.
However, it is only possible via lattice QCD to test from first
principles if the symmetry is spontaneously broken.

The leading order effective theory of $\chi {\rm PT}$ contains the
low-energy constants $f_{\pi}$ and $\Sigma$. In full QCD the
Gell-Mann-Oakes-Renner relation (GMOR)
\begin{equation}
  \label{eq:gmor_2}
  f_{\pi}^2 m_{\pi}^2 = 4 m \Sigma,
\end{equation}
becomes exact in the $m \rightarrow 0$ chiral limit and the chiral
condensate $\langle \bar{\psi} \psi \rangle = \langle \bar{u} u
\rangle = \langle \bar{d} d \rangle = \dots$, defined at zero quark
mass, is equal to -$\Sigma$. The low energy constant $\Sigma$ depends
on the number of massless flavors $N_f$.

In quenched QCD, the relation $\langle \bar{\psi} \psi \rangle =
-\Sigma$ and eq.~\eqref{eq:gmor_2} receive corrections even in the
chiral limit \cite{Sharpe:1992ft,Bernard:1992mk,Osborn:1998qb}.
Actually, $\langle \bar{\psi} \psi \rangle$ and $m_{\pi}^2/m$ are not
even defined in the chiral limit due to diverging quenched chiral
logarithms. In the case of $m_{\pi}^2/m$ the logarithms, which are
characterized by the parameter $\delta$, are connected to the mass of
the flavor singlet particle $\eta^\prime$\footnote{The mass of the
  $\eta^\prime$ in the quenched QCD Lagrangian does not correspond to
  the mass of the physical $\eta^\prime$, because the physical
  $\eta^\prime$ is mixed with the $\eta$.}, which is an additional
Goldstone boson in Q$\chi {\rm PT}$
\cite{Bernard:1992mk,Sharpe:1992ft}.  The mass of the $\eta^\prime$ is
an additional, third leading order low-energy constant $m_0$ of Q$\chi
{\rm PT}$, which is not present in the normal $\chi {\rm PT}$
Lagrangian. The expectation that $m_{\pi}^2/m$ diverges in Q$\chi {\rm
  PT}$ in the chiral limit is confirmed by different measurements
\cite{Kanaya:1998sd,Bardeen:2000cz, Dong:2001yf}, including our hadron
spectroscopy measurements with $\Dpar$ \cite{Hauswirth_diss:2002}.
Similarly, the divergence of $\langle \bar{\psi} \psi \rangle$ in the
infinite volume limit seems to be confirmed in numerical studies of
the Banks-Casher relation \cite{Kiskis:2001zt}.  On the other hand, it
is possible to study and determine the low-energy constant $\Sigma$ in
the quenched theory.  Under the assumption that $\Sigma(N_f)$ is a
smooth function of $N_f$ and $\Sigma(N_f=0)$ is close to
$\Sigma(N_f=3)$ (which is the standard assumption when quenched
results are used to estimate full QCD quantities) we get an estimate
for the chiral condensate $-\langle \bar{\psi} \psi \rangle(N_f=3) =
\Sigma(N_f=3) \approx \Sigma(N_f=0)$, even though it is hard to give a
systematic error for this estimate.  Typically, the quenching error
are said to be of order $10-20\%$ \cite{Gupta:1997nd}.

On the lattice the calculation of the low-energy constant $\Sigma$ of
(quenched) chiral perturbation theory requires good control over the
chiral properties of the theory and therefore it has notoriously been
difficult to measure it directly. Still, early attempts with staggered
fermions have been tried already in 1985 \cite{Barkai:1985gy} and the
calculation can also be performed with other traditional
discretizations, like ${\cal O}(a)$ improved Wilson fermions using
Ward identities and the GMOR relation eq.~\eqref{eq:gmor_2}
\cite{Bochicchio:1985xa,Giusti:1998wy}.  However, the calculation is
more transparent and cleaner in the framework of GW fermions due to
many reasons. With Wilson fermions one has to deal with a power
divergent subtraction that has to be determined numerically
\cite{Bochicchio:1985xa}. Furthermore, the occurrence of exceptional
configurations makes it difficult to go to quark masses which are
small enough to provide an accurate estimate for $\Sigma$ in the
chiral limit.  In the case of GW fermions chiral symmetry is already
realized on the lattice \cite{Luscher:1998pq}. This makes it possible
to define scalar and pseudoscalar densities which transform under
chiral transformations exactly like the corresponding continuum
densities \cite{Niedermayer:1998bi}. Using such a scalar density the
power divergent subtraction, which has to be performed on the bare
condensate, is exactly known and can be done without problems in
numerical simulations \cite{Hasenfratz:1998jp}.  A further advantage
of GW fermions is that one does not encounter any exceptional
configurations in the small quark mass regime. These facts altogether
allow a much better control over the calculations.

There are different techniques to calculate $\Sigma$ with GW fermions.
One possibility makes use of the GMOR relation and data from
pseudoscalar spectroscopy \cite{Giusti:2001pk}. Another possibility is
to use the distribution of the lowest-lying eigenmodes of the massless
Dirac operator, because these distribution can be described in the
framework of random matrix theory (RMT) \cite{Damgaard:2000ah} and
provide a completely different way to extract $\Sigma$. Actually, the
distributions of the low-lying eigenmodes contain a lot more of
information than the value of $\Sigma$ as explained in
\cite{Damgaard:2001js}, but in leading order they are essentially
described by $\Sigma$. A technique which is closely related to the
determination of $\Sigma$ from the eigenvalue distributions is the so
called finite-volume scaling technique, which has been worked out in
\cite{Osborn:1998qb}, and has been used for the first time in
\cite{Hernandez:1999cu}.  In our calculation of $\Sigma$ we use a
slightly modified version of this technique.

All the 3 different methods provide a bare value of $\Sigma$ that
subsequently has to be renormalized in order to give the
phenomenologically relevant predictions $\hat{\Sigma}$ and
$\Sigma_{\overline{\rm MS}}(2 { \rm GeV})$, respectively. Again, there
are different possibilities how the scalar renormalization factor
$Z_S$ can be calculated. It is known that lattice perturbation theory
converges only slowly in many cases and is therefore very difficult to
control in the range of gauge couplings where typical simulations are
performed \cite{Jansen:1996ck}. The way to circumvent this problem is
the use a non-perturbative renormalization scheme. There are
essentially two different non-perturbative renormalization schemes
used in the present day lattice simulations. One scheme is the RI/MOM
technique \cite{Martinelli:1995ty}, which uses the fact that QCD can
be treated perturbatively at large momentum scales, in order to match
amputated quark Green functions to a perturbative scheme
(e.g.~$\overline{\rm MS}$), while the other scheme, the Schr\"odinger
functional technique, is based on the fact that QCD in a very small
volume can be treated perturbatively \cite{Jansen:1996ck}. The size
$L$ of the box acts as a reference scale and can e.g.~be connected to
the normalization mass in the $\overline{\rm MS}$ scheme using
perturbative renormalization group arguments.  We will follow the
renormalization scheme proposed in \cite{Hernandez:2001yn}, which is
based -- even though not directly -- on the Schr\"odinger functional
technique.

But, before we start to explain our determination of $\Sigma$, let us
mention that more details of the calculation, in particular about the
stochastic evaluation of the trace of the subtracted quark propagator,
are given in Appendix \ref{cha:details_of_sigma}.

\section{Determination of the Bare $\Sigma$}
\label{sec:bare_sigma}

One possibility to determine $\Sigma$ is to study the chiral
condensate in a fixed topological sector with charge $Q$ in a finite
volume $V$ at finite quark mass $M$ \cite{Osborn:1998qb}. The volume
and the quark mass are chosen such that the finite size effects are
dominated by the pions with zero momentum. In this situation the
partition function simplifies considerably and can even be evaluated
analytically in certain cases \cite{Leutwyler:1992yt, Osborn:1998qb}.
Using (Q)$\chi {\rm PT}$, or random matrix theory (RMT), the fermion
condensate at finite volume and quark mass has been calculated in the
continuum, both for full and quenched QCD.  The quenched QCD
condensate $\langle \bar{\psi} \psi \rangle_{M,V,Q}$ can be written in
terms of the derivative of the logarithm of the partition function
${\cal Z}_Q$ in fixed topology
\begin{gather}
  \label{eq:partition_function_Znu}
  - \langle \bar{\psi} \psi \rangle_{M,V,Q} = \frac{\partial}{\partial M} \ln {\cal Z_Q} \, ,\\
  \intertext{with} {\cal Z}_Q = \int_{{\rm Gl}(1|1)} dU_0 {\rm
    Sdet}(U_0^Q) \exp[V \Sigma {\rm Re \,Str [M U_0]}] \,,
\end{gather}
where ${\rm Sdet}$ and ${\rm Str}$ are the supersymmetric
generalizations of the determinant and the trace, respectively. The
integration is over the graded Lie group ${\rm Gl}(1|1)$. For details
of the supersymmetric formulation of Q$\chi {\rm PT}$ see
\cite{Morel:1987xk,Bernard:1992mk,Sharpe:1992ft,Osborn:1998qb,Sharpe:2001fh,
  Damgaard:2001js}.  The analytic expression for this partition
function is known \cite{Osborn:1998qb} and therefore the condensate in
fixed topology at mass $M$ is given by
\begin{equation}
  \label{eq:bessel_sigma}
  - \langle \bar{\psi} \psi \rangle_{M,V,Q} 
  = M V \Sigma^2 [I_{|Q|}(z) K_{|Q|}(z) + I_{|Q|+1}(z) K_{|Q|-1}(z)] + \frac{|Q|}{M V},
\end{equation}
where $I_Q$ and $K_Q$ are modified Bessel functions, $z = M \Sigma V$
and the low-energy constant $\Sigma$ is the quantity we wish to
measure. By measuring $\langle \bar{\psi} \psi \rangle_{M,V,Q}$ in
different topological sectors, at different masses and volumes, the
continuum prediction of the $M$ and $V$ dependence can be used to
extract $\Sigma$.

On the lattice the subtracted condensate in a fixed topological sector
in a volume $V$ and with bare mass $m$ can be calculated by measuring
the trace \cite{Hasenfratz:1998jp,Hernandez:1999cu}
\begin{gather}
  \label{eq:condensate_trace}
  - \langle \bar{\psi} \psi \rangle^{\rm sub}_{m, V, Q} = \frac{1}{V}
  \Big \langle {\rm Tr}^\prime
  \Big[ \D^{-1}(m) - R \Big] \Big \rangle_Q \\
  \intertext{with} \D(m)= \Big(1 - \frac{m}{2} \Big) \D(0) +
  m (2 R)^{-1}
  \label{eq:condensate_mass}
\end{gather}
and where the primed trace is without the divergent contribution of
the zero modes and $\langle . \rangle_Q$ is the expectation value on
the ensemble of gauge fields with topological charge $Q$.  Because we
want to make use of the renormalization scheme proposed by
Hern\'{a}ndez et al.~in \cite{Hernandez:2001yn}, which is based on
pseudoscalar spectroscopy, we have to modify the strategy of the
measurements performed in \cite{Hernandez:1999cu,DeGrand:2000tf}
slightly.  The reason for this is that in our hadron spectroscopy
study \cite{Hauswirth_diss:2002} a mass definition $m$ is used which
differs from the usual definition $M$ by a factor $2 R$ (see
e.g.~eq.~\eqref{eq:condensate_mass}). Instead of the trace in
eq.~\eqref{eq:condensate_trace}, we use
\begin{equation}
  \label{eq:condensate_modified}
  - \langle S \rangle_{m,V,Q} = \frac{\partial}{\partial m} \ln {\cal Z_Q} = 
  \frac{1}{V} \bigg \langle {\rm Tr}^\prime \bigg[
  (\D(m) 2 R)^{-1} - \frac{1}{2} \D(0) \D^{-1}(m)\bigg] \bigg\rangle_Q \, . 
\end{equation}
The scalar density $S$ is the same as introduced in Chapter
\ref{cha:overlap} and its properties are discussed in more details in
\cite{hasenfratz:_testin_qcd}, i.e.~
\begin{gather}
  S = \bar{\psi} \Big( (2 R)^{-1} - \frac{1}{2}\D(0) \Big) \psi \,. \\
  \intertext{The partition function ${\cal Z_Q}$ is given by} {\cal
    Z_Q} = \int DU_Q D\bar{\psi} D\psi \exp[-(\bar{\psi} \D(0) \psi +
  m S) - {\cal A}_g] \, ,
\end{gather}
with the gauge action ${\cal A}_g$. Furthermore, the path integral is
restricted to gauge fields with topological charge $Q$.  For masses
$m$ smaller than the modulus of the smallest eigenvalue $\lambda_{\rm
  min}$ of the massless Dirac operator $\Dt$(0), i.e~$m^2 <
|\lambda_{\rm min}|^2$, the trace in
eq.~\eqref{eq:condensate_modified} reduces to
\begin{equation}
  \label{eq:condensate_modified_measured}
  - \langle S \rangle_{m,V,Q} =  \frac{1}{V} \bigg \langle {\rm Tr}^\prime \bigg[
  (\D(m) 2 R)^{-1} - \frac{1}{2}\bigg] \bigg\rangle_Q
\end{equation}
up to terms of ${\cal O}(m^2)$, as one can show easily. Because the numerical
results show to be less sensitive to a residual breaking of chiral
symmetry with the definition from
eq.~\eqref{eq:condensate_modified_measured} than with
eq.~\eqref{eq:condensate_modified}, we use the definition from
eq.~\eqref{eq:condensate_modified_measured} for our measurements,
which clearly agrees with the definition in
eq.~\eqref{eq:condensate_modified} in the chiral limit. The
correlation function $\langle S \rangle_{m,V,Q}$ differs from $\langle
\bar{\psi} \psi \rangle^{\rm sub}_{M,V,Q}$ by a factor $(2 R)^{-1}$.
Together with the scalar renormalization factor $Z_S$, which for
chiral actions is given by $Z_S = 1/Z_m$, however, the additional
factor drops out in the physical prediction, because the additional
factors from the mass definition and $\langle S \rangle_{m,V,Q}$
cancel. Let us be a bit more specific about what is meant with this:
In our case the eigenvalue spectrum of the operator $2 R$ is roughly
bounded by $2/\kappa$ and $4/\kappa$ with $\kappa = 3.45$ for our
parametrization and therefore its eigenvalues are close to 1 and no
exceptionally small or large eigenvalues can occur, which in a
simulation might spoil the fact that the two additional factors of $(2
R)^{-1}$ and $2 R$ in $\langle S \rangle_{m,V,Q}$ and $m$,
respectively, cancel to a high degree.

\subsection{Details of the Simulation}

We measure the trace in eq.~\eqref{eq:condensate_modified_measured}
stochastically using random $Z(2)$ vectors \cite{Dong:1994pk}. In
order to measure the condensate at very small quark mass, the remnant
explicit chiral symmetry breaking must be very small, so we use the
overlap operator with the kernel $\Dtpar$, as discussed in Chapter
\ref{cha:overlap}.  Due to the $|Q|$ chiral zero modes of the Dirac
operator, the quenched condensate contains a term $|Q|/M V$ which
diverges as the mass becomes small. The $|Q|$ zero modes have the same
chirality and their contribution to the condensate is removed by
measuring the trace ${\rm Tr}'$ in the chiral sector opposite to the
zero modes \cite{Edwards:1998wx,Hernandez:1999cu}, i.e.~if the $|Q|$
modes have chirality $+$, the $Z(2)$ vectors used to measure the trace
are chosen to have chirality $-$. To determine $\langle S
\rangle_{m,V,Q}$, the stochastic trace is doubled to include both
chiral sectors.

\begin{figure}[p]
  \begin{center}
    \includegraphics[width=9cm]{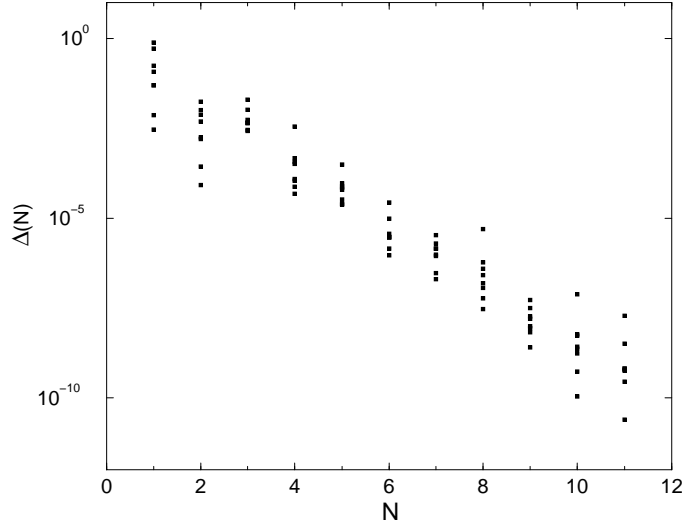}
    \caption{In this figure the dependence of the relative difference $\Delta(N)$, defined
      in eq.~\eqref{eq:delta_difference_sigma}, between the (nearly)
      exact overlap and lower order Legendre approximations with order
      $N$ is shown for several different $8^4$ configurations at
      $\beta = 3.2$.}
    \label{fig:condensate_difference}
  \end{center}
\end{figure}

\begin{figure}[p]
  \begin{center}
    \includegraphics[width=9cm]{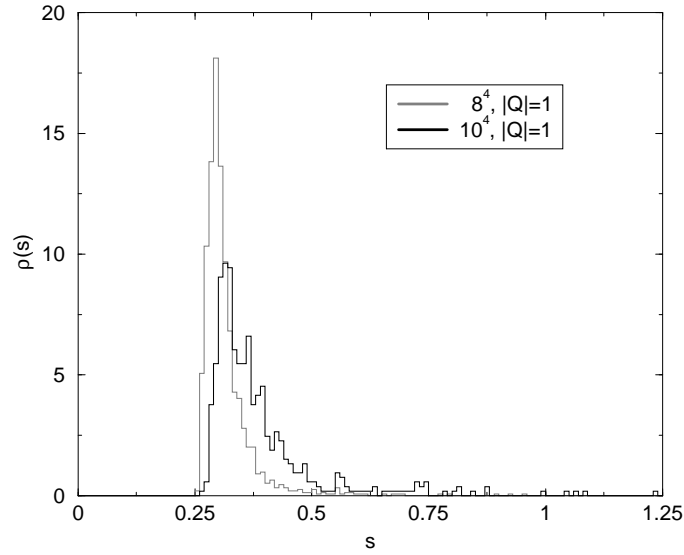}
    \caption{Normalized distribution $\rho(s)$ of the estimate of
      $s = (\langle S \rangle_{m,V,Q} a^3)/(m a)$ at $m a = 10^{-4}$
      for the different $Z(2)$ random vectors and configurations used
      in the measurements is shown for the $Q = \pm 1$ sector on $8^4$
      and $10^4$ configurations.}
    \label{fig:condensate_distribution}
    \end{center} 
\end{figure}

We measure the condensate in volumes $8^4$ and $10^4$ at $\beta=3.2$.
We use 10 random $Z(2)$ vectors to measure the trace for each
configuration and a BiCGstab multi-mass solver to invert the Dirac
operator at all 12 masses in the range $m a = 0.0001, \ldots, 0.2048$
simultaneously.  In the overlap operator, we approximate
$(A^{\dagger}A)^{-1/2}$ with Legendre polynomials of order 7 and 10
for volumes $8^4$ and $10^4$, respectively. We project out the 10
smallest $A^{\dagger} A$ eigenvalues, which are treated exactly.  This
altogether gives sufficiently precise chiral symmetry: increasing the
polynomial order further, the relative change in the estimate of
$(\langle S \rangle_{m,V,Q} a^3)/(m a)$ is $\leq {\cal O}(10^{-4})$.
This fact is shown in Figure \ref{fig:condensate_difference}, where
the relative difference between the (nearly) exact result with a
Legendre expansion of order $12$ and lower order expansions with order
$N$ at $m a = 10^{-4}$ for the quantity
\begin{equation}
  \label{eq:delta_difference_sigma}
  \Delta(N) = \frac{\langle \phi|(\D^{(N)}(m) 2 R)^{-1} - \frac{1}{2}| \phi \rangle -
    \langle \phi |(\D^{(12)}(m) 2 R)^{-1} - \frac{1}{2}| \phi \rangle}
  {\langle \phi |(\D^{(12)}(m) 2 R)^{-1} - \frac{1}{2}| \phi \rangle} \, , 
\end{equation}
where $\phi$ is a $Z(2)$ random vector, is shown for several $8^4$
configurations.  Our statistics are given in Table
\ref{tab:condensate_meas}.  In Figure
\ref{fig:condensate_distribution}, we show the distribution of the
estimates of $(\langle S \rangle_{m,V,Q} a^3)/(m a)$ obtained from the
different $Z(2)$ random vectors in $|Q|=1$ topological sector on the
$8^4$ and $10^4$ configurations, respectively. One observes that the
distributions have quite a long tail towards large values, especially
on the $10^4$ lattice.
\begin{table}[tb]
  \begin{center}
    \begin{tabular}{|c|c|c|c|} \hline 
      $L$ & order & $|Q|$ & $N_{\rm conf}$ \\ \hline \hline
      8   & 7     & 1     & 154 \\
          &       & 2     &  41 \\ \hline
      10  & 10    & 1     &  53 \\
          &       & 2     &  43 \\ \hline      
    \end{tabular}
  \end{center}
  \caption{Polynomial order of $(A^{\dagger} A)^{-1/2}$ approximation and statistics
    for the measurement of $\Sigma$.}
 \label{tab:condensate_meas} 
\end{table}

\subsection{Results}

\begin{figure}[p]
  \begin{center}
    \includegraphics[width=9cm]{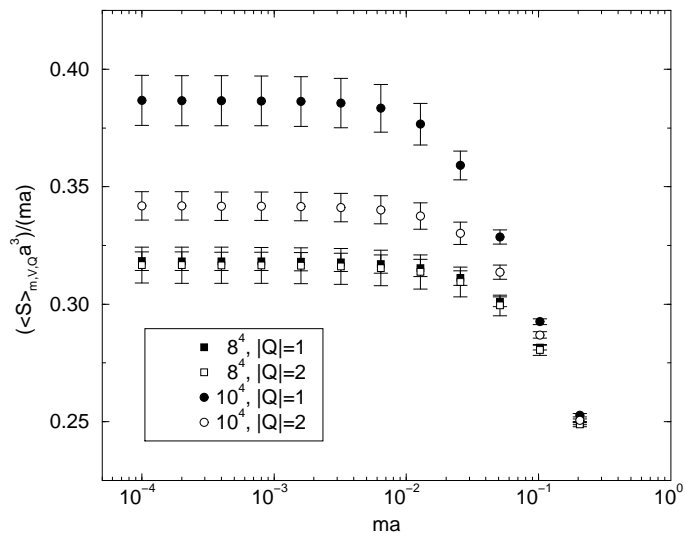}
    \caption{The estimated $(\langle S \rangle_{m,V,Q} a^3)/(m a)$ for different volumes, 
      masses and topological charges.}
    \label{fig:condensate_plateau}
    \end{center}
\end{figure}

\begin{figure}[p]
  \begin{center}
    \includegraphics[width=9cm]{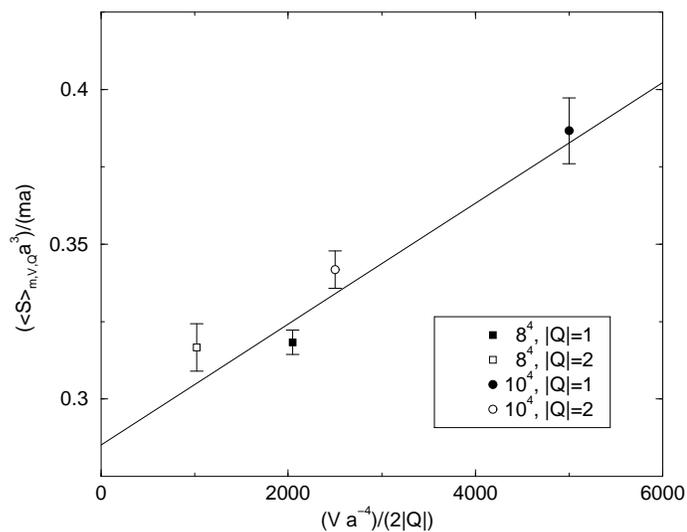}
    \caption{$(\langle S \rangle_{m,V,Q} a^3)/(m a)$ for the smallest value of the mass
      $m a = 10^{-4}$ plotted versus $(V a^{-4})/(2 |Q|)$.  The fit
      used to determine the slope is shown by the line.}
    \label{fig:condensate_fit}
    \end{center}
\end{figure}

Let us first explain how the bare value of $\Sigma$ can be extracted
from eq.~\eqref{eq:bessel_sigma} and the measurements of the trace in
eq.~\eqref{eq:condensate_modified_measured}.  The bare quark
condensate at finite quark mass contains a $\sim m/a^2$ cut-off
effect. As the quark mass $m \rightarrow 0$, one can rewrite
eq.~\eqref{eq:bessel_sigma} such that
\begin{equation}
  - \frac{\langle \bar{\psi} \psi \rangle^{\rm sub}_{M,V,Q}}{M} =
  - \alpha^2 \frac{\langle S \rangle_{m,V,Q}}{m}
  = \frac{\Sigma^2 V}{2 |Q|} + \frac{c_1}{a^2} ,
  \label{eq:condensate_fit}
\end{equation}
where $c_1$ is an unknown coefficient, which has to be fitted, and
$\alpha$ denotes the ratio between the different mass definitions:
$\alpha = m/M$. As the coefficient $c_1$ comes from ultraviolet
fluctuations, it is natural to assume that it is independent of the
topological charge $Q$.
The contribution of the $|Q|$ zero modes is removed by taking the
trace as described above, and there is no artifact $1/(m a^3)$ in
eq.~\eqref{eq:condensate_fit} due to the fact that the condensate is
defined as the expectation value of a scalar operator that transforms
covariantly under chiral transformations and has no mixing with the
unit operator \cite{Hasenfratz:1998jp}. From Figure
\ref{fig:condensate_plateau}, we see that $-(\langle S \rangle_{m,V,Q}
a^3)/(m a)$ reaches a plateau at very small quark mass. In Figure
\ref{fig:condensate_fit}, we plot the value of $-(\langle S
\rangle_{m,V,Q} a^3)/(m a)$ at $m a=10^{-4}$ versus $(V a^{-4})/2|Q|$,
which we fit to the form of eq.~\eqref{eq:condensate_fit}.  From the
slope, we extract the bare low-energy constant as $\Sigma a^3 =
4.42(36) \times 10^{-3}$ with $\chi^2/df = 3.55$.
We can convert this result using the Sommer parameter
($r_0/a=3.943(60)$ at $\beta=3.2$ from the interpolating formula in
\cite{Wenger_diss:2000}), giving $\Sigma r_0^3 = 0.271(22)(12)$, where
the first error is statistical and the second the uncertainty in the
scale.

\section{Renormalization of $\Sigma$}
\label{sec:renormalization}

In order to turn this bare result into the renormalized low-energy
constant $\hat{\Sigma}$ we need the scalar renormalization factor
$Z_S$.  In the present test study we obtained $Z_S$ combining the
continuum extrapolated renormalization group invariant (RGI) quark
mass of the ALPHA collaboration \cite{Garden:1999fg} with our
spectroscopy data.  This method has been suggested recently by
Hern\'andez et al.~\cite{Hernandez:2001yn}. Even though the method is
based on the Schr\"odinger functional, the special boundary conditions
of Schr\"odinger functional make a direct application to GW fermions
difficult, if not impossible, and therefore improved Wilson fermions
were used in an intermediate step. In the following we give a short
description of the calculations done in \cite{Hernandez:2001yn} and
apply the result to our data.

\subsection{Short Introduction to the Technique}

The idea of this technique is to connect the bare quark mass $m$ in
the given discretization to the renormalization group invariant (RGI)
quark mass $M$
\begin{equation}
  \label{eq:renorm_quark}
  M = Z_M(g_0) m(g_0)
\end{equation}
and to use the fact that for lattice regularizations that preserve
chiral symmetry the renormalization factor of the scalar density is
the inverse of that for the quark mass (see,
e.g.~\cite{Alexandrou:2000kj})
\begin{equation}
  \label{eq:renorm_factor_identity}
  Z_P(g_0) = Z_S(g_0) = \frac{1}{Z_M(g_0)}\, .
\end{equation}
In order to fix $Z_M(g_0)$ one can rewrite the ratio $M/m(g_0)$ as
\begin{align}
  \label{eq:eq:quark_mass_ratio}
  \frac{M}{m(g_0)} &= \frac{M}{m_W(g_0^\prime)} \frac{m_W(g_0^\prime)}{m(g_0)} \\
  &= Z_M^W(g_0^\prime) \frac{(r_0 m_W)(g_0^\prime)}{(r_0 m)(g_0)} \, ,
\end{align}
where $g_0^\prime \ne g_0$ in general, and the hadronic radius $r_0$
is used to set the scale.  The factor $Z_M^W$ has been computed in
\cite{Capitani:1998mq} for a large range of couplings. The ratio
$M/m(g_0)$ is then obtained by determining $(r_0 m)$ and $(r_0 m_W)$
at a reference value of $x_{\rm ref}$ of a physical observable, say
$x_{\rm ref} = (r_0 m_P)^2$, where $m_P$ is the mass of the
pseudoscalar meson. In order to get rid of discretization errors the
combination $Z_M^W(g_0^\prime)(r_0 m_W)(g_0^\prime)$ was calculated in
the continuum limit, i.e.~
\begin{equation}
  \label{eq:cont_lim_U_factor} 
  U_M = \lim_{g_0^\prime \to 0} Z_M^W(g_0^\prime)(r_0 m_W)(g_0^\prime) \, ,
\end{equation}
leading to the following results \cite{Hernandez:2001yn}
\begin{equation}
  \label{universal_Um_factor}
  U_M =
  \begin{cases}
    0.181(6),  &     x_{\rm ref} = 1.5736, \\
    0.349(9),  &     x_{\rm ref} = 3.0,    \\
    0.580(12), & x_{\rm ref} = 5.0.
  \end{cases}
\end{equation}
By combining \eqref{eq:renorm_quark}, \eqref{eq:eq:quark_mass_ratio}
and \eqref{eq:cont_lim_U_factor} one obtains
\begin{equation}
  \label{eq:ZM_factor}
  Z_M(g_0) = U_M \frac{1}{(r_0 m)}\Big|_{(r_0 m_P)^2 = x_{\rm ref}} \, ,
\end{equation}
where now all reference to the bare coupling $g^\prime_0$ and the use
of the intermediate Wilson fermions has disappeared. Apart from
cut-off effects, $Z_M(g_0)$ is expected to be independent of the
reference point $x_{\rm ref}$.

\subsection{Results}

\begin{figure}[tbp]
  \begin{center}
    \includegraphics[width=10cm]{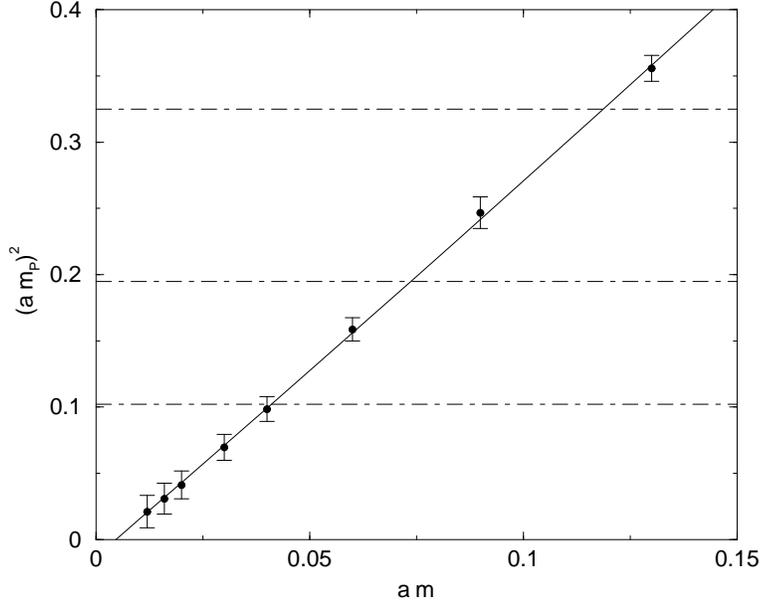}
    \caption{The $(m_P a)^2$ versus $m a$ data used to determine the scalar
      renormalization factor at $\beta=3.2$.  The solid line is a fit
      using a second order polynomial, while the dot-dashed lines
      indicate the three reference values for the pseudoscalar mass
      $x_{\rm ref} = (r_0 m_P)^2 = 1.5736, 3.0, 5.0$ used to extract
      $Z_M$.}
    \label{fig:renorm_b320}
  \end{center}
\end{figure}

Using the results of the hadron spectroscopy with the overlap with
$\Dpar$ at $\beta=3.2$ \cite{Hauswirth_diss:2002}, we are able to
calculate the scalar renormalization factor $Z_S$.  The pseudoscalar
spectroscopy data is obtained from $32$ $9^3 \times 24$ lattices.
Because of the small statistics and the rather small volume the
determination of the renormalization constants at $\beta = 3.2$ has a
rather large error. For our determination we combine data from the
pseudoscalar - scalar (PS) and the axial-vector (A) correlator
measurements of the pseudoscalar mass.  The reason for this is that
the PS correlator gets contributions from the scalar at high bare
quark masses, while the A correlator is sensitive to the topological
finite size effects caused by the zero modes at small bare quark
masses \cite{Blum:2000kn}. Using this combined data, we fit a second
order polynomial through the data points in the region $m = 0.01,
\ldots, 0.17$ as shown in Figure \ref{fig:renorm_b320}.
With these fits we determine the bare quark at which the pseudoscalar
takes on the three reference values $x_{\rm ref} = (r_0 m_P)^2 =
1.5736, 3.0, 5.0$.  We obtain
\begin{equation}
  \label{results_mr0_b320}
  (r_0 m)|_{(r_0 m_P)^2=x_{\rm ref}} = 
  \begin{cases}
    0.141(24),  &     x_{\rm ref} = 1.5736, \\
    0.271(36),  &     x_{\rm ref} = 3.0,    \\
    0.447(24), & x_{\rm ref} = 5.0,
  \end{cases}
\end{equation}
where we use the hadronic scale $r_0/a = 3.943(60)$ from the
interpolating formula in \cite{Wenger_diss:2000} to set the scale. In
the determination of the $r_0 m$ values a small additive mass
renormalization $a \Delta_m = 0.005(4)$ is taken into account.  This
residual additive mass renormalization is due to the fact that in the
determination of the pseudoscalar spectrum a Legendre expansion of
order 3 was used. This leaves place for a breaking of the GW relation
which is large enough to give rise to a residual mass on the order of
$10^{-3}$.

Using \eqref{eq:ZM_factor}, we can combine the results for the bare
quark masses in \eqref{results_mr0_b320} with the universal factor
$U_M$ from \eqref{universal_Um_factor} and obtain the mass and scalar
renormalization factors. The results are given in the Table
\ref{tab:renorm_factor_b320}.

\begin{table}[tbp]
  \begin{center}
    \begin{tabular}[h]{|c|c|c|}\hline
      $x_{\rm ref}$ & $Z_M(g_0)$   & $Z_S(g_0)$ \\ \hline \hline
      1.5736        & 1.28(25)     &  0.78(15) \\ 
      3.0           & 1.29(12)     &  0.78(7)  \\
      5.0           & 1.30(7)      &  0.77(4)  \\\hline
    \end{tabular}
    \caption{Renormalization factors at $\beta$=3.2 for the different reference values of
      the pseudoscalar mass.}
    \label{tab:renorm_factor_b320}
  \end{center}
\end{table}
The dependence on the value of $x_{\rm ref}$ is weak and well covered
by the statistical uncertainty.  Following the arguments in
\cite{Hernandez:2001yn} we choose the value at $x_{\rm ref} = 3.0$ as
our best estimate of the renormalization constant, and thus we obtain
the following final result for the renormalization factor $Z_M(g_0) =
1.29(12)$ and $Z_S(g_0) = 0.869(32)$, respectively.  These
renormalization factors can be converted into the $\overline{\rm MS}$
scheme at the reference scale $\mu = 2$ GeV using the results from
\cite{Garden:1999fg}
\begin{equation}
  \label{eq:conversion_factor_msbar}
  \frac{\overline{m}_{\overline{\rm MS}}(\mu)}{M} =  \frac{Z_m(g_0,\mu)}{Z_M(g_0)} = 0.72076, 
  \qquad \mu = 2 \, \text{GeV}\, ,
\end{equation}
with an error of $1.5\%$ due to the uncertainty in the quenched value
of $\Lambda_{\overline{\rm MS}}$. This leads to $Z_M(g_0,\mu) =
0.93(9)$ and $Z_S(g_0,\mu) = 1.08(10)$, respectively.

We can combine the scalar renormalization factor with the bare value
of $\Sigma$, which leads to the final prediction for the physical
value of $\hat{\Sigma}$ and $\Sigma_{\overline {\rm MS}}(2 {\rm
  GeV})$. We get
\begin{align}
  \label{eq:final_result_sigma_hat}
  r_0^3 \hat{\Sigma} &= 0.210(17)(9)(20) \, , \\
  r_0^3 \Sigma_{\overline {\rm MS}}(2 {\rm GeV}) &=
  0.291(24)(12)(28)\, ,
\end{align}
with statistical, scale and renormalization errors, respectively.
Combining all errors, but the error associated with the scale
ambiguity in quenched QCD in quadrature and using $r_0 = 0.5 \,{\rm
  fm}$, we find at $\beta = 3.2$
\begin{align}
  \label{eq:final_result_sigma_hat}
  \hat{\Sigma} &= (235 \pm 11 \,\mev)^3 \times \bigg( \frac{a^{-1}[\mev]}{1556\, \mev}\bigg)^3\, , \\
  \Sigma_{\overline {\rm MS}}(2 {\rm GeV}) &= (262 \pm 12 \,
  \mev)^3 \times \bigg( \frac{a^{-1}[\mev]}{1556\, \mev}\bigg)^3\, .
\end{align}
These results are still subject to discretization errors and clearly
it would be desirable to have data at smaller lattice spacings in
order to make a continuum extrapolation.

Using the GMOR relation from eq.~\eqref{eq:gmor_2}, our result for
$\Sigma$ and the slope $B$ in the $m_\pi^2$ vs.~$m$ plot, which we
determine to $a B = 2.90(12)$, we can make a consistency check of our
measurements. We get $a f_\pi = 78(4) \times 10^{-3}$, which
corresponds to
\begin{equation}
  \label{eq:fpi}
  f_\pi = (121.5 \pm 5.9\, \mev ) \times \bigg( \frac{a^{-1}[\mev]}{1556\, \mev}\bigg) \,,
\end{equation}
i.e.~we get sufficiently close to the value from other determinations of $f_\pi$ in quenched 
QCD, e.g.~\cite{Aoki:1999av}, which are quite close to the experimental value $f_\pi = 131
\, \mev$, to be confident that no big systematic errors are induced
through the low statistics or the small volumes we used in this study.

\section{Discussion}
\label{sec:discussion_sigma}

\begin{figure}[tbp]
  \begin{center}
    \includegraphics[width=10cm]{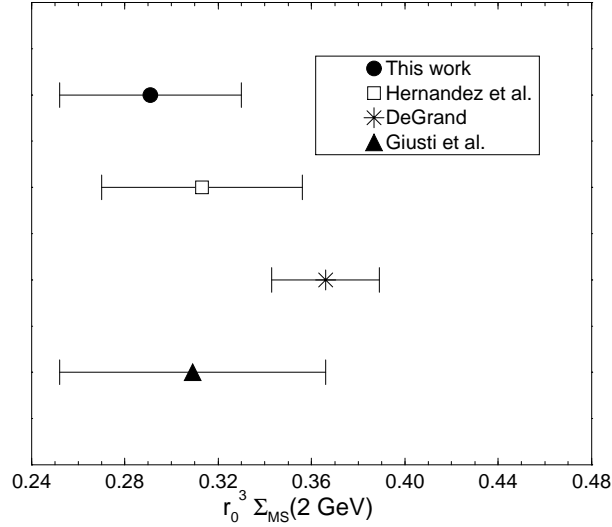}
    \caption{Comparison of different recent determinations of $r_0^3 
      \Sigma_{\overline {\rm MS}}(2 {\rm GeV})$.}
    \label{fig:sigma_comparison}
  \end{center}
\end{figure}

The result of this determination of the low-energy constant $\Sigma$
in quenched QCD is in very good agreement with other recent
determinations with chiral fermions
\cite{Hernandez:2001yn,DeGrand:2001ie,Giusti:2001pk}, as shown in
Figure \ref{fig:sigma_comparison}. One should, however, mention that
these studies -- including ours -- are based on poor statistics and
the simulations were done in small physical volumes, which means that
all these studies have a certain test character. Furthermore, none of
these studies contains a continuum extrapolation.  The small volume is
a problem because the RMT prediction of the distribution of the
low-lying eigenmodes is only valid up to the Thouless energy
\cite{Osborn:1998nm,Damgaard:2001ep}, which means that only a small
fraction of the eigenmodes are described properly. This clearly leads
to a systematic error. The problem of small statistics is enhanced
through the fact the trace in
eq.~\eqref{eq:condensate_modified_measured} gets large contributions
from the smallest eigenvalues of the Dirac operator. Hence, it is
probable that the results are distorted by the very rare but sizeable
contributions from very small eigenmodes and therefore we think that
the statement in \cite{Damgaard:2001ep} that the distributions of the
lowest-lying eigenvalues of the Dirac operator would provide an
approach to the determination of $\Sigma$, which has smaller systematic
errors, is most likely true.

\cpages
\chapter{Conclusions and Prospects}
\label{cha:conclusions}

\subsubsection*{The parametrized FP Dirac operator}

In this work we have presented the construction and parametrization of
a FP Dirac operator using a general hypercubic ansatz for the
parametrization. We have shown how general Dirac operators can be
constructed in an efficient way such that it is feasible to use them
in numerical simulations. We have discussed the details of the
parametrization procedure and have shown where the main difficulties
lie.  The properties of the resulting parametrization of the FP Dirac
operator $\Dpar$ have been investigated. We have found that the
parametrization has clearly improved chiral behaviour with respect to
the most frequently used lattice Dirac operator, the Wilson Dirac
operator $D_W$. This fact is most evidently seen in the eigenvalue
distribution of the operator $A^\dagger A$, which for a Dirac operator
satisfying the Ginsparg-Wilson relation is equal to $1$. It deviates
much less from $1$ for $\Dpar$ than for $D_W$.  Furthermore, we have
found that the Krylow space methods used to calculate the eigenvalues
and eigenvectors of the operator $A^\dagger A$, which are important
for an efficient numerical implementation of the overlap construction,
use substantially less operator times vector multiplications in the
case of $\Dpar$ than for $D_W$. We have found that the difference in
the number of operator vector products can get so large that the
determination of the low-lying eigenmodes of $A^\dagger A$ actually is only very little
more expensive for $\Dpar$ than for $D_W$, even though one matrix
vector multiplication is ${\cal O}(40)$ times more expensive than for
$D_W$.  We have shown that the additive mass renormalization and the
fluctuations of the low-lying eigenmodes of the Dirac operator are
small, such that hadron spectroscopy simulations at remarkably small
quark masses can be performed. Because of the possibility to go to very small
quark masses clear signs of quenched chiral logarithms have been 
found in our hadron spectroscopy, which is described
in detail in the thesis of Simon Hauswirth \cite{Hauswirth_diss:2002}. 
In the same study we have found that the scaling properties of $\Dpar$ are,
indeed, clearly improved to $D_W$ and ${\cal O}(a)$ scaling violations can not
not be seen, even though more precise studies are needed to give a more quantitative statement.
Finally, we have found that the
definition of the topological charge through $\Tr (\gamma_5 R_{\rm par} \Dpar)$
is problematic with the present parametrization as it deviates
substantially from the index of $\Dpar$, i.e.~it does not satisfy the
index theorem.

\subsubsection*{Overlap construction with $\Dpar$}

We have shown that the overlap construction with $\Dpar$ is clearly
more local in the sense that its couplings fall off faster than the
common construction with $D_W$. In contrast to the Wilson overlap
there is no need to fine tune a mass parameter to get optimal
convergence or optimal locality in the overlap expansion with $\Dpar$.
We have found that polynomial approximations to the overlap
construction with $\Dpar$ converge very fast, i.e.~that chiral
symmetry up to machine precision can be reached with Legendre polynomial expansions of 
${\cal O}(20)$ for the lattice sizes and gauge couplings used in this work.
It has shown that such a precision, however, is
barely needed in most simulations. We have found that low order
expansions can be used in many cases to get enough accuracy for the
chiral properties of the Dirac operator.

\subsubsection*{Physical Results}

We have used low order polynomial approximations to the overlap construction with
$\Dpar$ to calculate the quenched topological susceptibility at
various lattice spacings and volumes. The results point in the same
direction as other recent results. However, our data is not yet
accurate enough to allow for a controlled continuum extrapolation and shows
cut-off effects.  We also confirm the findings of various other groups about the local
chirality of the near-zero modes of a Dirac operator, which might
result from instanton dominance in the QCD vacuum.  Finally, we have
determined the low-energy constant $\Sigma$ of quenched chiral
perturbation theory using a finite-volume scaling technique in a test
study. Our result for the renormalized $\Sigma$ is
\begin{equation}
  r_0^3 \Sigma_{\overline {\rm MS}}(2 \, {\rm GeV}) = 0.291(24)(12)(28)
 \nonumber \,\,\,  \text{at}\,\,  \beta =3.2 \, ,
\end{equation}
with statistical, scale and renormalization errors, respectively. This
result does not include a continuum extrapolation and therefore is
still subject to discretization errors.  Converting to physical units by using
the scale $r_0 = 0.5 \, {\rm fm}$,
which is somewhat problematic due to the scale ambiguity in quenched
QCD, we get
\begin{equation}
  \Sigma_{\overline {\rm MS}}(2 \,{\rm GeV}) = 
  (262 \pm 12 \, \mev)^3  \times \bigg( \frac{a^{-1}[\mev]}{1556\, \mev}\bigg)^3 
 \nonumber \,\,\,  \text{at} \,\,\beta =3.2 \, ,
\end{equation}
where all errors are combined in quadrature. This result is in good
agreement with other recent studies of $\Sigma$ using chiral fermions.
It should be mentioned that all these studies, including ours, are
based on rather small statistics. Nevertheless, we have shown that
with $\Dpar$ we can use low order polynomial approximations to the
overlap construction to perform calculations where chiral symmetry is
needed to high precision.

\subsubsection*{Prospects}

Let us start with the most obvious weakness of the present
parametrization: The topological charge definition through $\Tr
(\gamma_5 R_{\rm par} \Dpar)$. In the present parametrization no special
attention has been given to the pseudoscalar part of the
parametrization and it is presently under study whether this weakness
can be fixed with additional care in the parametrization process. If
this is possible it would provide a numerically very cheap definition
of the topological charge and moreover of the topological charge
density. The overlap construction with $\Dpar$ is suited to study 
the structure of topological excitations of the QCD vacuum, because
there are indications that this overlap construction can resolve the underlying structure of
the gauge fields better than the Wilson overlap.  Using approximations
to the overlap operator with $\Dpar$, or simply $\Dpar$ itself,
further problems where chiral symmetry plays an essential r\^{o}le
might be investigated, because the list of such problems is long. Let
us mention just a few: Determination of further low-energy constants
of (quenched) chiral perturbation theory, determination of quark
masses of the light quarks, the resolution of the Kaplan-Manohar
ambiguity in the first non-leading order Lagrangian of chiral
perturbation theory and the calculation of the pion scattering length.
Using the available data from the present study on hadron spectroscopy
some of these problems can already be tackled. The data contains information
about the pion decay constant, the light quark masses and further
information about the parameter $\delta$ describing the quenched
chiral logarithms. In order to extract some of this data one might use
e.g.~the RI/MOM scheme to calculate renormalization factors accurately.
In connection with this renormalization scheme, which is sensitive to momentum 
cut-off effects, the good energy-momentum dispersion relation of $\Dpar$ might
be a big virtue. Furthermore, first applications of conserved chiral currents, 
which we proposed in \cite{Hasenfratz:2001qp,hasenfratz:_testin_qcd}, might be 
interesting and useful.

\cpages
\chapter*{Acknowledgements}
\label{cha:acknowledgements}

I would like to thank Peter Hasenfratz, not only because he made it
possible for me to work on this interesting topic of FP fermions,
where so many different aspects come together, but also because he
helped a lot to lay the foundations for this project.  It is also a
pleasure to thank Ferenc Niedermayer for many interesting discussions
and valuable advice. I would like to thank both of them for carefully
reading the manuscript.

Then, I am grateful to all the people I
collaborated with during my thesis, in particular, my fellow PhD
student Simon Hauswirth, Kieran Holland, Philipp R\"ufenacht and Urs
Wenger, who were always helpful and also made that the social part at
the workshops and conferences always was pleasant.  Furthermore, I
would like to thank Simon Hauswirth, Julia Schweizer and Kay Bieri for
the pleasant atmosphere in the office and, in particular, Julia for
helping me with the basics of Chiral Perturbation Theory, Simon and
Kay for taking care of the computing environment.  A special thank
goes to Michael Marti and Alexander Gall for making all those hours we
spent together in trying to keep the computers at the institute alive
much more pleasant than one might fear. I would also like to thank
Ottilia H\"anni and Ruth Bestgen for making wheels turn and for many
interesting discussions.  My thanks also go to Christoph Gattringer
and the other people from the BGR collaboration for the pleasant stay
in Regensburg and the possibility to use the Munich supercomputer.
Further thanks go to all the other members of the institute for the
good social environment.  

Finally, I will not miss the opportunity to
thank Anne Tscherter for her support and for much more than can be
written on a few lines.

\cpages

\begin{appendix}
\chapter{Representation of the Clifford Algebra}
\label{cha:clifford}

In this appendix we give the explicit representation of the $16$ elements of the basis of the 
Clifford algebra in the Weyl representation, as they are used throughout this work. In the
 current implementation
of the programs for the parametrized FP Dirac operator the same convetions are used.

\subsection*{Scalar}

\begin{equation}
  \Gamma_1 = 1 =
  \begin{pmatrix}
    1&0&0&0\\
    0&1&0&0\\
    0&0&1&0\\
    0&0&0&1\\
  \end{pmatrix} 
\end{equation} 

\subsection*{Vector}

\begin{align}
  \Gamma_2 &= \gamma_1 =
  \begin{pmatrix}
    0&0&0&-\i\\
    0&0&-\i&0\\
    0&\i&0&0\\
    \i&0&0&0\\
  \end{pmatrix}&
  \Gamma_3 &= \gamma_2 =
  \begin{pmatrix}
    0&0&0&-1\\
    0&0&1&0\\
    0&1&0&0\\
    -1&0&0&0\\
  \end{pmatrix}\notag \\ \phantom{|} \notag \\
  \Gamma_4 &= \gamma_3 =
  \begin{pmatrix}
    0&0&-\i&0\\
    0&0&0&\i\\
    \i&0&0&0\\
    0&-\i&0&0\\
  \end{pmatrix}&
  \Gamma_5 &= \gamma_4 =
  \begin{pmatrix}
    0&0&1&0\\
    0&0&0&1\\
    1&0&0&0\\
    0&1&0&0\\
  \end{pmatrix}
\end{align}

\subsection*{Tensor}

\begin{align}
  \Gamma_6 &= \i \sigma_{12} = 
  \begin{pmatrix}
    \i&0&0&0\\
    0&-\i&0&0\\
    0&0&\i&0\\
    0&0&0&-\i\\
  \end{pmatrix}&
  \Gamma_7 &= \i \sigma_{13} = 
  \begin{pmatrix}
    0&-1&0&0\\
    1&0&0&0\\
    0&0&0&-1\\
    0&0&1&0\\
  \end{pmatrix}\notag \\ \phantom{|} \notag \\
  \Gamma_8 &= \i \sigma_{14} = 
  \begin{pmatrix}
    0&-\i&0&0\\
    -\i&0&0&0\\
    0&0&0&\i\\
    0&0&\i&0\\
  \end{pmatrix}&
  \Gamma_9 & = \i \sigma_{23}  = 
  \begin{pmatrix}
    0&\i&0&0\\
    \i&0&0&0\\
    0&0&0&\i\\
    0&0&\i&0\\
  \end{pmatrix}\notag \\  \phantom{|} \notag \\
  \Gamma_{10} &=  \i \sigma_{24} = 
  \begin{pmatrix}
    0&-1&0&0\\
    1&0&0&0\\
    0&0&0&1\\
    0&0&-1&0\\
  \end{pmatrix}&
  \Gamma_{11} &= \i \sigma_{34} = 
  \begin{pmatrix}
    -\i&0&0&0\\
    0&\i&0&0\\
    0&0&\i&0\\
    0&0&0&-\i\\
  \end{pmatrix}
\end{align}

\subsection*{Pseudoscalar}

\begin{equation}
  \Gamma_{12} = \gamma_5= -\gamma_1 \gamma_2 \gamma_3 \gamma_4 = 
  \begin{pmatrix}
    -1&0&0&0\\
    0&-1&0&0\\
    0&0&1&0\\
    0&0&0&1\\
  \end{pmatrix}
\end{equation}

\subsection*{Axial-vector}

\begin{align}
  \Gamma_{13} &= \gamma_1 \gamma_5=
  \begin{pmatrix}
    0&0&0&-\i\\
    0&0&-\i&0\\
    0&-\i&0&0\\
    -\i&0&0&0\\
  \end{pmatrix}&
  \Gamma_{14} &= \gamma_2 \gamma_5=
  \begin{pmatrix}
    0&0&0&-1\\
    0&0&1&0\\
    0&-1&0&0\\
    1&0&0&0\\ 
  \end{pmatrix}\notag \\ \phantom{|} \notag \\
  \Gamma_{15} &= \gamma_3 \gamma_5=
  \begin{pmatrix}
    0&0&-\i&0\\
    0&0&0&\i\\
    -\i&0&0&0\\
    0&\i&0&0\\
  \end{pmatrix}&
  \Gamma_{16} &= \gamma_4 \gamma_5=
  \begin{pmatrix}
    0&0&1&0\\
    0&0&0&1\\
    -1&0&0&0\\
    0&-1&0&0\\
  \end{pmatrix}
\end{align}

\cpages
\chapter{The List of Factorized Contributions}
\label{cha:factorized_contribs}

In this appendix we provide a factorization of the paths used in the program code of 
the parametrized FP Dirac operator. The correspondig reference paths and their properties 
are given in the Tables~\ref{tab:refpath0}-\ref{tab:refpath4}.

\section{Definitions}
\label{sec:definitions}

We introduce the operator $\hat{U}_\mu$ of the parallel
transport for direction $\mu$
\begin{equation}
  \label{Uhat}
  \left(\hat{U}_\mu\right)_{nn'}=U_\mu(n)\delta_{n+\hat{\mu},n'} \,,
\end{equation}
and analogously for the opposite direction
\begin{equation}
  \label{Uhatd}
  \left(\hat{U}_{-\mu}\right)_{nn'}=U_\mu(n-\hat{\mu})^\dagger
  \delta_{n-\hat{\mu},n'} \,.
\end{equation}
It is also useful to introduce the operator $\hat{U}(l)$ of the
parallel transport along some path $l=[l_1,l_2,\ldots,l_k]$ where
$l_i=\pm 1,\ldots,\pm 4$, by
\begin{equation}
  \label{Ul}
  \hat{U}(l)=\hat{U}_{l_1}\hat{U}_{l_2}\ldots\hat{U}_{l_k} \,.
\end{equation}
In terms of gauge links this is
\begin{equation}
  \label{Ul1}
  \left( \hat{U}(l)\right)_{nn'}=
  \left( U_{l_1}(n)U_{l_2}(n+\hat{l}_1)\ldots \right)\delta_{n+r_l,n'}\,,
\end{equation}

We define the plaquette products (i.e.~the operators of parallel
transport along a plaquette) in the following way:
\begin{equation}
  \label{Plaq}
  P_{l_1 ,l_2}=\hat{U}([l_1,l_2,-l_1,-l_2]) \,,
\end{equation}
where $l_i=\pm 1,\ldots,\pm4$.  Their hermitian conjugate is given by
$(P_{l_1 ,l_2})^\dagger=P_{l_2 ,l_1}$.  Reflections and permutations
act on them in an obvious way.

We also define the staple products as
\begin{equation}
  \label{Stap}
  S_{l_1 ,l_2}=\hat{U}([l_1,l_2,-l_1]) \,.
\end{equation}
We have $(S_{l_1 ,l_2})^\dagger = S_{l_1 ,-l_2}$.

To describe the shortest path to an offset we introduce the notation
\begin{eqnarray}
  \label{line}
  & & V_{l_1}=\hat{U}([l_1]) \,, \nonumber \\
  & & V_{l_1 ,l_2}=\hat{U}([l_1,l_2]) \,, \nonumber \\
  & & V_{l_1 ,l_2,l_3}=\hat{U}([l_1,l_2,l_3]) \,,  \\
  & & V_{l_1 ,l_2,l_3,l_4}=\hat{U}([l_1,l_2,l_3,l_4]) \,. \nonumber 
\end{eqnarray}

Introduce the following linear combinations transforming in a simple
way under reflections:
\begin{eqnarray}
  \label{ppp}
  & &P^{(++)}_{\mu\nu} = 
  P_{\mu,\nu}+P_{\mu,-\nu}+P_{-\mu,\nu}+P_{-\mu,-\nu} \,, \nonumber \\
  & &P^{(+-)}_{\mu\nu} = 
  P_{\mu,\nu}-P_{\mu,-\nu}+P_{-\mu,\nu}-P_{-\mu,-\nu} \,,  \\
  & &P^{(-+)}_{\mu\nu} = 
  P_{\mu,\nu}+P_{\mu,-\nu}-P_{-\mu,\nu}-P_{-\mu,-\nu} \,, \nonumber \\
  & &P^{(--)}_{\mu\nu} = 
  P_{\mu,\nu}-P_{\mu,-\nu}-P_{-\mu,\nu}+P_{-\mu,-\nu} \,. \nonumber 
\end{eqnarray}
The signs in the superscript denote the parity for reflections of the
$\mu$, $\nu$ axes, respectively.  Hermitian conjugation acts as
interchanging both upper and lower indices, permutations simply by
$\mu \to p_\mu$, $\nu \to p_\nu$.

It is also useful to denote combinations which are
symmetric/antisymmetric with respect to interchanging the axes:
\begin{eqnarray}
  \label{psa}
  & &P^{\rm (sym)}_{\mu\nu} = 
  P^{(++)}_{\mu,\nu}+P^{(++)}_{\nu,\mu}
  = P^{(++)}_{\mu,\nu}+{\rm h.c.} \,, \\
  & &P^{\rm (as)}_{\mu\nu} = 
  P^{(--)}_{\mu,\nu}-P^{(--)}_{\nu,\mu}
  = P^{(--)}_{\mu,\nu}-{\rm h.c.} \,. \nonumber 
\end{eqnarray}

For the staples we write
\begin{eqnarray}
  \label{sd1}
  & &S^{(\nu, +)}_{\mu}  = S_{\nu,\mu}  + S_{-\nu,\mu} \,, \nonumber \\
  & &S^{(\nu, -)}_{\mu}  = S_{\nu,\mu}  - S_{-\nu,\mu} \,, \\
  & &S^{(\nu, +)}_{-\mu} = S_{\nu,-\mu} + S_{-\nu,-\mu}\,, \nonumber \\
  & &S^{(\nu, -)}_{-\mu} = S_{\nu,-\mu} - S_{-\nu,-\mu}\,. \nonumber 
\end{eqnarray}
The subscript $\pm\mu$ denotes the direction of the staple, the
superscript specifies the plane $\mu\nu$ and the parity in $\nu$.

For a line of length 2 we have
\begin{eqnarray}
  \label{line1}
  & &V^{(++)}_{\mu\nu} = 
  V_{\mu,\nu}+V_{\mu,-\nu}+V_{-\mu,\nu}+V_{-\mu,-\nu} \,, \nonumber \\
  & &V^{(+-)}_{\mu\nu} = 
  V_{\mu,\nu}-V_{\mu,-\nu}+V_{-\mu,\nu}-V_{-\mu,-\nu} \,,  \\
  & &V^{(-+)}_{\mu\nu} = 
  V_{\mu,\nu}+V_{\mu,-\nu}-V_{-\mu,\nu}-V_{-\mu,-\nu} \,, \nonumber \\
  & &V^{(--)}_{\mu\nu} = 
  V_{\mu,\nu}-V_{\mu,-\nu}-V_{-\mu,\nu}+V_{-\mu,-\nu} \,, \nonumber 
\end{eqnarray}
and analogously for longer lines. In analogy to eq.~(\ref{psa}) we
introduce the completely (anti)symmetric combinations $V^{\rm
  (sym)}_{\mu\nu\ldots}$, $V^{\rm (as)}_{\mu\nu\ldots}$ for the line
products.

We also introduce the ``4d plaquette''
\begin{equation}
  Q_{l_1,l_2,l_3,l_4}=\hat{U}([l_1,l_2,l_3,l_4,-l_1,-l_2,-l_3,-l_4]) \,,
\end{equation}
and the odd combination $Q_{l_1,l_2,l_3,l_4}^{(----)}$.

To simplify the expressions below, we require that all the directions
entering $P_{l_1,l_2}$, $S_{l_1,l_2}$, $V_{l_1,l_2,\ldots}$, etc.~are
different, i.e.~they are taken to be zero if e.g. $|l_1|=|l_2|$.

Below we list terms for the different choices of
$\Gamma_0$ and $l_0$ used in current implementation of the parametrized FP Dirac operator code. 
The notation $\left. \sum \right.'$ indicates
that all indices in the corresponding sum are taken to be different.

\section{The Offset $r_0=(0,0,0,0)$}

\noindent{$\Gamma_0=1$, $l_0=[\,]$}
\begin{equation}
  \label{g0a_0}
  1 \,.
\end{equation}

\noindent{$\Gamma_0=1$, $l_0=[1,2,-1,-2]$}
\begin{equation}
  \label{g0b_0}
  \frac{1}{48}  \sum_{\mu<\nu} P^{\rm (sym)}_{\mu\nu} \,.
\end{equation}

\noindent{$\Gamma_0 = \gamma_1$, $l_0=[1,2,-1,-2]$}
\begin{equation}
  \label{g1_0}
  \frac{1}{24} \left. \sum_{\mu\nu} \right.'
  \gamma_\mu \left( P^{(-+)}_{\mu\nu} - \mbox{h.c.} \right) \,.
\end{equation}

\noindent{$\Gamma_0= i\sigma_{12}$, $l_0=[1,2,-1,-2]$}
\begin{equation}
  \label{s12_0}
  \frac{1}{8} \sum_{\mu<\nu} i\sigma_{\mu\nu} P^{\rm (as)}_{\mu\nu} \,.
\end{equation}

\noindent{$\Gamma_0= \gamma_5$, $l_0=[1,2,-1,-2,3,4,-3,-4]$}
\begin{equation}
  \label{g5a_0}
  \frac{1}{384} \gamma_5 \left. \sum_{\mu\nu\rho\sigma} \right.'
  \frac{1}{4} \epsilon_{\mu\nu\rho\sigma}
  P^{\rm (as)}_{\mu\nu}P^{\rm (as)}_{\rho\sigma} \,.
\end{equation}

\noindent{$\Gamma_0= \gamma_5$, $l_0=[1,2,3,4,-1,-2,-3,-4]$}
\begin{equation}
  \label{g5b_0}
  \frac{1}{384}  \gamma_5 \left. \sum_{\mu\nu\rho\sigma} \right.'
  \epsilon_{\mu\nu\rho\sigma} Q_{\mu\nu\rho\sigma}^{(----)} \,.
\end{equation}

\noindent{$\Gamma_0= \gamma_1\gamma_5$, $l_0=[1,2,-1,-2,3,4,-3,-4]$}
\begin{equation}
  \label{g15a_0}
  \frac{1}{192} \left. \sum_{\mu\nu\rho\sigma} \right.'
  \gamma_\mu\gamma_5 \frac{1}{2}\epsilon_{\mu\nu\rho\sigma}
  \left( P^{(+-)}_{\mu\nu}P^{\rm (as)}_{\rho\sigma} + \mbox{h.c.} \right) \,.
\end{equation}

\noindent{$\Gamma_0= \gamma_1\gamma_5$, $l_0=[2,1,-2,-1,3,4,-3,-4]$}
\begin{equation}
  \label{g15b_0}
  \frac{1}{192} \left. \sum_{\mu\nu\rho\sigma} \right.'
  \gamma_\mu\gamma_5 \frac{1}{2} \epsilon_{\mu\nu\rho\sigma}
  \left( P^{(-+)}_{\nu\mu}P^{\rm (as)}_{\rho\sigma} + \mbox{h.c.} \right) \,.
\end{equation}

\section{The Offset $r_0=(1,0,0,0)$}

\noindent{$\Gamma_0 = 1$,  $l_0=[1]$}\\
\begin{equation}
  \label{g0a_1}
  \sum_\mu \left( V_\mu + V_{-\mu} \right) \,.
\end{equation}

\noindent{$\Gamma_0 = 1$, $l_0=[2,1,-2]$}\\
\begin{equation}
  \label{g0b_1}
  \frac{1}{6} \left. \sum_{\mu\nu} \right.'
  \left( S^{(\nu,+)}_\mu + S^{(\nu,+)}_{-\mu} \right) \,.
\end{equation}

\noindent{$\Gamma_0 = 1$, $l_0=[2,3,1,-3,-2]$}\\
\begin{equation}
  \label{g0c_1}
  \frac{1}{24} \left. \sum_{\mu\nu\rho} \right.'
  \left( V_\rho  S^{(\nu,+)}_\mu  V_{-\rho} +  V_{-\rho} S^{(\nu,+)}_{\mu} V_\rho + {\rm h.c.} \right) \,.
\end{equation}

\noindent{$\Gamma_0 = \gamma_1$, $l_0=[1]$}\\
\begin{equation}
  \label{g1a_1}
  \sum_\mu \gamma_\mu \left( V_\mu - V_{-\mu} \right) \,.
\end{equation}

\noindent{$\Gamma_0 = \gamma_1$, $l_0=[2,1,-2]$}\\
\begin{equation}
  \label{g1b_1}
  \frac{1}{6} \left. \sum_{\mu\nu} \right.' \gamma_\mu 
  \left( S^{(\nu,+)}_\mu - S^{(\nu,+)}_{-\mu} \right) \,.
\end{equation}

\noindent{$\Gamma_0 = \gamma_2$, $l_0=[1,2,3,-2,-3]$}\\
\begin{equation}
  \label{g2_1}
  \frac{1}{16} 
  \left. \sum_{\mu\nu\rho}\right.' \, \gamma_\nu
  \left(  V_\mu P^{(-+)}_{\nu\rho} - P^{(+-)}_{\rho\nu} V_\mu - {\rm h.c.}
  \right) \,.
\end{equation}

\noindent{$\Gamma_0 = i\sigma_{12}$, $l_0=[2,1,-2]$}\\
\begin{equation}
  \label{s12_1}
  \frac{1}{2} \left. \sum_{\mu\nu} \right.'  i\sigma_{\mu\nu} \frac{1}{2}
  \left(  S^{(\nu,-)}_\mu - S^{(\mu,-)}_\nu - \mbox{h.c.} \right) \,.
\end{equation}

\noindent{$\Gamma_0 = i\sigma_{23}$, $l_0=[1,2,3,-2,-3]$}\\
\begin{equation}
  \label{s23a_1}
  \frac{1}{16} \left. \sum_{\mu\nu\rho} \right.' 
  i\sigma_{\mu\nu} \frac{1}{2} 
  \left(  V_\rho P^{\rm (as)}_{\mu\nu} + 
    P^{\rm (as)}_{\mu\nu} V_\rho-\mbox{h.c.} \right) \,.
\end{equation}


\noindent{$\Gamma_0 = \gamma_5$, $l_0=[2,1,-2,3,4,-3,-4]$}\\
\begin{equation}
  \label{g5_1}
  \frac{1}{96} \gamma_5 \left. \sum_{\mu\nu\rho\sigma} \right.'
  \frac{1}{2} \epsilon_{\mu\nu\rho\sigma}
  \left( S^{(\nu,-)}_\mu P^{\rm (as)}_{\rho\sigma}
    + P^{\rm (as)}_{\rho\sigma} S^{(\nu,-)}_\mu + \mbox{h.c.} \right) \,.
\end{equation}

\noindent{$\Gamma_0 = \gamma_1\gamma_5$, $l_0=[2,1,-2,3,4,-3,-4]$}\\
\begin{equation}
  \label{g15_1}
  \frac{1}{96} \left. \sum_{\mu\nu\rho\sigma} \right.' 
  \gamma_\mu\gamma_5 \frac{1}{2} \epsilon_{\mu\nu\rho\sigma}
  \left( S^{(\nu,-)}_\mu P^{\rm (as)}_{\rho\sigma}
    - P^{\rm (as)}_{\rho\sigma} S^{(\nu,-)}_\mu + \mbox{h.c.} \right) \,.
\end{equation}

\noindent{$\Gamma_0 = \gamma_2 \gamma_5$, $l_0=[1,3,4,-3,-4]$}\\
\begin{equation}
  \label{g25a_1}
  \frac{1}{16} \left. \sum_{\mu\nu\rho\sigma} \right.' 
  \gamma_\nu\gamma_5 \frac{1}{2} 
  \epsilon_{\mu\nu\rho\sigma}
  \left( V_\mu P^{\rm (as)}_{\rho\sigma}
    + P^{\rm (as)}_{\rho\sigma} V_\mu + \mbox{h.c.} \right) \,.
\end{equation}


\section{The Offset $r_0=(1,1,0,0)$}

\noindent{$\Gamma_0=1$, $l_0=[1,2]$}\\
\begin{equation}
  \label{g0a_2}
  \frac{1}{2} \sum_{\mu<\nu} V_{\mu\nu}^{\rm (sym)} \,.
\end{equation}


\noindent{$\Gamma_0=\gamma_1$, $l_0=[1,2]$}\\
\begin{equation}
  \label{g1a_2}
  \frac{1}{2} \left. \sum_{\mu\nu} \right.' 
  \gamma_\mu \left( V_{\mu\nu}^{(-+)} - \mbox{h.c.} \right) \,.
\end{equation}


\noindent{$\Gamma_0=\gamma_3$, $l_0=[1,3,2,-3]$}\\
\begin{equation}
  \label{g3_2}
  \frac{1}{8} \left. \sum_{\mu\nu\rho} \right.'
  \gamma_\rho \left( 
    V_{\mu} S_{\nu}^{(\rho,-)} + V_{-\mu} S_{\nu}^{(\rho,-)}
    - S_{\mu}^{(\rho,-)} V_{\nu} - S_{\mu}^{(\rho,-)} V_{-\nu}
    - \mbox{h.c.} \right) \,.
\end{equation}

\noindent{$\Gamma_0=i\sigma_{12}$, $l_0=[1,2]$}\\
\begin{equation}
  \label{s12a_2}
  \frac{1}{2} \sum_{\mu < \nu}  i\sigma_{\mu\nu} V_{\mu\nu}^{\rm (as)} \,.
\end{equation}


\noindent{$\Gamma_0=i\sigma_{13}$, $l_0=[1,3,2,-3]$}\\
\begin{equation}
  \label{s13_2}
  \frac{1}{4} \left. \sum_{\mu\nu\rho} \right.'
  i\sigma_{\mu\rho} \left(
    V_{\mu} S_{\nu}^{(\rho,-)} - V_{-\mu} S_{\nu}^{(\rho,-)}
    +V_{\mu} S_{-\nu}^{(\rho,-)} - V_{-\mu} S_{-\nu}^{(\rho,-)}
    - \mbox{h.c.} \right) \,.
\end{equation}

\noindent{$\Gamma_0=i\sigma_{34}$, $l_0=[1,2,3,4,-3,-4]$}\\
\begin{equation}
  \label{s34a_2}
  \frac{1}{32} \left. \sum_{\mu\nu\rho\sigma} \right.'
  i\sigma_{\rho\sigma} \frac{1}{2} \left(
    V_{\mu\nu}^{(++)} P_{\rho\sigma}^{\rm (as)} - \mbox{h.c.} \right) \,.
\end{equation}


\noindent{$\Gamma_0=\gamma_5$, $l_0=[1,2,3,4,-3,-4]$}\\
\begin{equation}
  \label{g5a_2}
  \frac{1}{32} \gamma_5 \left. \sum_{\mu\nu\rho\sigma} \right.'
  \frac{1}{2} \epsilon_{\mu\nu\rho\sigma} \left(
    V_{\mu\nu}^{(--)} P_{\rho\sigma}^{\rm (as)} + \mbox{h.c.} \right) \,.
\end{equation}


\noindent{$\Gamma_0=\gamma_1\gamma_5$, $l_0=[1,2,3,4,-3,-4]$}\\
\begin{equation}
  \label{g15a_2}
  \frac{1}{16} \left. \sum_{\mu\nu\rho\sigma} \right.'
  \gamma_\mu \gamma_5 \frac{1}{2} \epsilon_{\mu\nu\rho\sigma} \left(
    V_{\mu\nu}^{(+-)} P_{\rho\sigma}^{\rm (as)} + \mbox{h.c.} \right) \,.
\end{equation}


\noindent{$\Gamma_0=\gamma_3\gamma_5$, $l_0=[1,4,2,-4]$}\\
\begin{multline}
  \label{g35_2}
  \frac{1}{8} \left. \sum_{\mu\nu\rho\sigma} \right.'  \gamma_\rho
  \gamma_5 \epsilon_{\mu\nu\rho\sigma} \left(
    V_{\mu} S_{\nu}^{(\sigma,-)}  - V_{-\mu} S_{\nu}^{(\sigma,-)} \right. \\
  \left. -V_{\mu} S_{-\nu}^{(\sigma,-)} + V_{-\mu}
    S_{-\nu}^{(\sigma,-)} + \mbox{h.c.} \right) \,.
\end{multline}

\section{The Offset $r_0=(1,1,1,0)$}

\noindent{$\Gamma_0 = 1$,  $l_0=[1,2,3]$}\\
\begin{equation}
  \label{g0_3}
  \frac{1}{6} \sum_{\mu<\nu<\rho} V_{\mu\nu\rho}^{\rm (sym)} \,.
\end{equation}

\noindent{$\Gamma_0 = \gamma_1$,  $l_0=[1,2,3]$}\\
\begin{equation}
  \label{g1_3}
  \frac{1}{4} \left. \sum_{\mu\nu\rho} \right.'
  \gamma_\mu \left(  V_{\mu\nu\rho}^{(-++)}+V_{\nu\rho\mu}^{(++-)} \right) \,.
\end{equation}

\noindent{$\Gamma_0 = \gamma_4$,  $l_0=[1,2,4,3,-4]$}\\
\begin{equation}
  \label{g4_3}
  \frac{1}{24} \left. \sum_{\mu\nu\rho\sigma} \right.'
  \gamma_\nu \frac{1}{2}
  \left( 
    V^{\rm (sym)}_{\rho\sigma}S_\mu^{(\nu,-)}
    + V^{\rm (sym)}_{\rho\sigma}S_{-\mu}^{(\nu,-)} - \mbox{h.c.}
  \right) \,.
\end{equation}

\noindent{$\Gamma_0 = i\sigma_{12}$,  $l_0=[1,2,3]$}\\
\begin{equation}
  \label{s12_3}
  \frac{1}{4} \left. \sum_{\mu\nu\rho} \right.'
  i\sigma_{\mu\nu} \frac{1}{2} \left(
    V_{\mu\nu\rho}^{(--+)}+V_{\rho\mu\nu}^{(+--)} -  \mbox{h.c.}
  \right) \,.
\end{equation}

\noindent{$\Gamma_0 = i\sigma_{14}$,  $l_0=[1,4,2,-4,3]$}\\
\begin{multline}
  \label{s14_3}
  \frac{1}{8} \left. \sum_{\mu\nu\rho\sigma} \right.'
  i\sigma_{\mu\nu} \left( V_\mu S_\rho^{(\nu,-)}V_\sigma + V_\sigma
    S_\rho^{(\nu,-)}V_\mu -V_{-\mu} S_\rho^{(\nu,-)}V_\sigma -
    V_\sigma S_\rho^{(\nu,-)}V_{-\mu}
  \right. \\
  \left.  +V_\mu S_\rho^{(\nu,-)}V_{-\sigma} + V_{-\sigma}
    S_\rho^{(\nu,-)}V_\mu -V_{-\mu} S_\rho^{(\nu,-)}V_{-\sigma} -
    V_{-\sigma} S_\rho^{(\nu,-)}V_{-\mu} - \mbox{h.c.} \right) \,.
\end{multline}

\noindent{$\Gamma_0 = \gamma_5$,  $l_0=[1,4,2,-4,3]$}\\
\begin{multline}
  \label{g5_3}
  \frac{1}{12} \gamma_5 \left. \sum_{\mu\nu\rho\sigma} \right.'
  \frac{1}{2} \epsilon_{\mu\nu\rho\sigma} \left( V_\mu
    S_\rho^{(\nu,-)}V_\sigma - V_\sigma S_\rho^{(\nu,-)}V_\mu
    -V_{-\mu} S_\rho^{(\nu,-)}V_\sigma + V_\sigma
    S_\rho^{(\nu,-)}V_{-\mu}
  \right. \\
  \left.  -V_\mu S_\rho^{(\nu,-)}V_{-\sigma} + V_{-\sigma}
    S_\rho^{(\nu,-)}V_\mu +V_{-\mu} S_\rho^{(\nu,-)}V_{-\sigma} -
    V_{-\sigma} S_\rho^{(\nu,-)}V_{-\mu} + \mbox{h.c.} \right) \,.
\end{multline}

\noindent{$\Gamma_0 = \gamma_1\gamma_5$,  $l_0=[1,4,2,-4,3]$}\\
\begin{multline}
  \label{g15_3}
  \frac{1}{8} \left. \sum_{\mu\nu\rho\sigma} \right.'  \gamma_\mu
  \gamma_5 \epsilon_{\mu\nu\rho\sigma} \left( V_\mu
    S_\rho^{(\nu,-)}V_\sigma + V_\sigma S_\rho^{(\nu,-)}V_\mu
    +V_{-\mu} S_\rho^{(\nu,-)}V_\sigma + V_\sigma
    S_\rho^{(\nu,-)}V_{-\mu}
  \right. \\
  \left.  -V_\mu S_\rho^{(\nu,-)}V_{-\sigma} - V_{-\sigma}
    S_\rho^{(\nu,-)}V_\mu -V_{-\mu} S_\rho^{(\nu,-)}V_{-\sigma} -
    V_{-\sigma} S_\rho^{(\nu,-)}V_{-\mu} + \mbox{h.c.} \right) \,.
\end{multline}

\noindent{$\Gamma_0 = \gamma_4\gamma_5$,  $l_0=[1,2,3]$}\\
\begin{equation}
  \label{g45_3}
  \frac{1}{6} \left. \sum_{\mu\nu\rho\sigma} \right.'
  \gamma_\sigma \gamma_5 \frac{1}{2} \epsilon_{\mu\nu\rho\sigma}
  \left( V_{\mu\nu\rho}^{(---)} + \mbox{h.c.} \right) \,.
\end{equation}

\section{The Offset $r_0=(1,1,1,1)$}

\noindent{$\Gamma_0 = 1$,  $l_0=[1,2,3,4]$}\\
\begin{equation}
  \label{g0_4}
  \frac{1}{24} V_{1234}^{\rm (sym)} \,.
\end{equation}

\noindent{$\Gamma_0 = \gamma_1$,  $l_0=[1,2,3,4]$}\\
\begin{equation}
  \frac{1}{12} \left. \sum_{\mu\nu\rho\sigma} \right.'
  \gamma_\mu \left( V_{\mu\nu\rho\sigma}^{(-+++)} - \mbox{h.c.} \right) \,.
  \label{g1_4}
\end{equation}

\noindent{$\Gamma_0 = i\sigma_{12}$,  $l_0=[1,2,3,4]$}\\
\begin{equation}
  \label{s12_4}
  \frac{1}{8} \left. \sum_{\mu\nu\rho\sigma} \right.'
  i\sigma_{\mu\nu} \left( V_{\mu\nu\rho\sigma}^{(--++)}- \mbox{h.c.} \right) \,.
\end{equation}

\noindent{$\Gamma_0 = \gamma_5$,  $l_0=[1,2,3,4]$}\\
\begin{equation}
  \label{g5_4}
  \frac{1}{24} \gamma_5 \left. \sum_{\mu\nu\rho\sigma} \right.'
  \frac{1}{2} \epsilon_{\mu\nu\rho\sigma} 
  \left( V_{\mu\nu\rho\sigma}^{(----)} + \mbox{h.c.} \right)\,.
\end{equation}

\noindent{$\Gamma_0 = \gamma_1\gamma_5$,  $l_0=[1,2,3,4]$}\\
\begin{equation}
  \label{g15_4}
  \frac{1}{12} \left. \sum_{\mu\nu\rho\sigma} \right.'
  \gamma_\mu \gamma_5  \epsilon_{\mu\nu\rho\sigma} 
  \left( V_{\mu\nu\rho\sigma}^{(+---)} + \mbox{h.c.} \right) \,.
\end{equation}


\cpages
\chapter{Parametrizations}
\label{cha:params}

In this appendix we give the parameters of $\Dpar$ and $R_{\rm par}$ at different stages of the
parametrization procedure\footnote{Note that the parameters of 
the $R$ operator do not exactly correspond to the parameters in the parameter files for the production
run code, because the definitions of the operators in this code differs slightly from the ones 
given in this work. The parameters given in this appendix, however, are rewritten such that they
conform to the conventions used in this work.}. We give only the parameters of the operators used in the most important steps
of the parametrization. These are the parametrizations of the free operators, which are
the starting point of the parametrization procedures described in Chapter \ref{cha:parametrization}, 
the parametrizations on the minimized $\beta \approx 3.0$ configurations and finally we specify
the parametrizations for $\Dfp$ and $R_{\rm FP}$ obtained from the last step of the parametrization 
procedure. These are the operators which used in all the production runs. 

\newpage

\section{Parametrizations for $R_{\rm FP}$}
\label{sec:r_parametrizations}

\begin{table}[H]
  \begin{center}
    \begin{tabular}{|c|c|c|}
      \hline
      offset & reference path $l_0$  &  $c_0$      \\
      \hline\hline
      (0000) & $[]$                  & \phm0.36647189 \\      
      \hline                                                                   
      (1000) & $[1]$                 & \phm0.01491600 \\
      \hline
      (1100) & $[1,2]$               & \phm0.00299651 \\
      \hline                                                     
      (1110) & $[1,2,3]$             & \phm0.00062104 \\
      \hline                                                     
      (1111) & $[1,2,3,4]$           & \phm0.00013253 \\
      \hline                                                       
    \end{tabular}
    \caption{{}The reference paths and the corresponding couplings for all offsets of
      the operator $R_{\rm par}=R_{\rm FP}$ in the free case. Note
      that the determination of the parameters of the free $R_{\rm FP}$ can be done analytically. 
      Together with the fact that $R_{\rm FP}$ is strictly a hypercubic operator this makes that this 
      parametrization is exact, except for the truncation of the coefficients.} 
    \label{tab:free_R_refpath}
  \end{center}
\end{table}

\begin{table}[H]
  \begin{center}
    \begin{tabular}{|c|c|c|c|}
      \hline
      offset & reference path $l_0$  &  $c_0$          & $c_1$   \\
      \hline\hline
      (0000) & $[]$                  &  0.363762     &  0.000240 \\
             & $[1,2,-1,-2]$         &  0.002710     &  - \\
      \hline                                                         
      (1000) & $[1]$                 &  0.011249     &  0.001769 \\
             & $[2,1,-2]$            &  0.003667     & - \\
      \hline                               
      (1100) & $[1,2]$               &  0.002996     &  0.000710 \\
      \hline                                           
      (1110) & $[1,2,3]$             &  0.000621     &  0.000247 \\
      \hline                                           
      (1111) & $[1,2,3,4]$           &  0.000132     &  0.000062 \\
      \hline                                                       
    \end{tabular}
    \caption{{}The reference paths and the corresponding couplings for all offsets of
      the operator $R_{\rm par}$ used on minimized $\beta \approx 3.0$ configurations. It is used 
      in connection with a 1-level APE smearing 
      with coefficients $c_1 = 0.356$ and $c_2 = -0.079$ (see Section \ref{sec:smearing}). } 
    \label{tab:free_R_refpath}
  \end{center}
\end{table}

\vspace{5cm}
\newpage

\label{sec:final}
\begin{table}[h]
  \begin{center}
    \begin{tabular}{|c|c|c|c|}
      \hline
      offset & reference path $l_0$  &  $c_0$          & $c_1$  \\
      \hline\hline
      (0000) & $[]$                  & \phm0.35793322 & \phm0.00105816 \\
             & $[1,2,-1,-2]$         & \phm0.00853867 & \phm- \\
      \hline                                                         
      (1000) & $[1]$                 & \phm0.01948398 &    -0.00971484 \\
             & $[2,1,-2]$            &    -0.00456798 & \phm- \\
      \hline                               
      (1100) & $[1,2]$               & \phm0.00299651 &    -0.00069294 \\
      \hline                                           
      (1110) & $[1,2,3]$             & \phm0.00062104 & \phm0.00009383 \\
      \hline                                           
      (1111) & $[1,2,3,4]$           & \phm0.00013253 & \phm0.00004382 \\
      \hline                                                       
    \end{tabular}
    \caption{{}The reference paths and the corresponding couplings for all offsets of
      the operator $R_{\rm par}$ used in production runs.} 
    \label{tab:free_R_refpath}
  \end{center}
\end{table}

\section{Parametrizations for $\Dfp$}
\label{sec:Dparametrizations}

\begin{table}[H]
  \begin{center}
    \begin{tabular}{|c|c|c|c|}
      \hline
      offset & $\Gamma_0  $        &  reference path $l_0$  &  $c_0$     \\
      \hline\hline
      (0000) & $1$                 & $[]$                    & \phantom{-}2.68037621 \\      
      \hline\hline                                                                 
      (1000) & $1$                 & $[1]$                   & -0.15397946 \\
      \hline
             & $\gamma_1$          & $[1]$                   &  \phantom{-}0.10830529 \\
      \hline\hline
      (1100) & $1$                 & $[1,2]$                 & -0.03798073  \\
      \hline                                                     
             & $\gamma_1$          & $[1,2]$                 &  \phantom{-}0.03545779  \\
      \hline\hline                                                     
      (1110) & $1$                 & $[1,2,3]$               & -0.01355349  \\
      \hline                                                      
             & $\gamma_1$          & $[1,2,3]$               & \phantom{-}0.01206385  \\
      \hline\hline                                                      
      (1111) & $1$                 & $[1,2,3,4]$             & -0.00645569  \\
      \hline                                                       
             & $\gamma_1$          & $[1,2,3,4]$             & \phantom{-}0.00427270 \\
      \hline                                                       
    \end{tabular}
    \caption{{}The reference paths and the corresponding couplings for the
      different $\Gamma_0$'s and offsets of the hypercubic
      parametrization of the free massless $\Dfp$.}
    \label{tab:free_Dfp}
  \end{center}
\end{table}

\vspace{5cm}
\newpage

\begin{table}[H]
  \begin{center}
    \begin{tabular}{|c|c|c|c|c|}
      \hline
      offset & $\Gamma_0  $        &  reference path $l_0$  &  $c_0$     &  $c_1$  \\
      \hline\hline
      (0000) & $1$                 & $[]$            &  \phm1.94363679 & \phm11.19824151    \\
             &             & $[1,2,-1,-2]$           &  \phm0.73666167 &    -11.29811978    \\           
      \hline                                                                   
      &$\gamma_1$          & $[1,2,-1,-2]$           &     -0.11480487 & \phm \, 0.10127720 \\
      \hline                                                                   
      &$i\sigma_{12}$      & $[1,2,-1,-2]$           &     -0.04800436 & \phm \, 0.06833909 \\
      \hline                                                                   
      &$\gamma_5$          & $[1,2,-1,-2,3,4,-3,-4]$ &     -0.04851795 & \phm \, 4.32121617 \\
      \hline                                                                   
      &$\gamma_1 \gamma_5$ & $[1,2,-1,-2,3,4,-3,-4]$ &  \phm0.15675378 & \phm \, 3.07442815 \\
              &            & $[2,1,-2,-1,3,4,-3,-4]$ &  \phm0.44943367 & \phm \, 0.51046076 \\ 
      \hline\hline
      (1000) & $1$         & $[1]$                   &     -0.09051958 & \phone \,-0.25661340 \\
             &             & $[2,1,-2]$              &     -0.06354147 & \phm \, 0.23665490 \\
      \hline
      &$\gamma_1$          & $[1]$                   &  \phm0.01888183 & \phone    \,-0.01941104 \\
              &            & $[2,1,-2]$              &  \phm0.08936156 & \phm \, 0.05080964 \\
      \hline
      &$\gamma_2$          & $[1,2,3,-2,-3]$         &     -0.08485598 & \phone \,-0.17419947 \\
      \hline                                                                   
      &$i\sigma_{12}$      & $[2,1,-2]$              &  \phm0.02286974 & \phone \,-0.01987646 \\
      \hline                                                                   
      &$i\sigma_{23}$      & $[1,2,3,-2,-3]$         &     -0.03402500 & \phone \,-0.02913557 \\
      \hline                                                                   
      &$\gamma_5$          & $[2,1,-2,3,4,-3,-4]$    &  \phm0.01152283 & \phone \,-0.92662206 \\
      \hline                                                                   
      &$\gamma_1 \gamma_5$ & $[2,1,-2,3,4,-3,-4]$    &     -0.06931361 & \phm \, 0.88455911 \\
      \hline                                                                               
      &$\gamma_2 \gamma_5$ & $[1,3,4,-3,-4]$         &  \phm0.01892036 & \phm \, 0.05246687 \\
      \hline\hline                                                                         
      (1100) & $1$         & $[1,2]$                 &     -0.03810518 & \phone \,-0.01157131 \\
      \hline                                                                               
      &$\gamma_1$          & $[1,2]$                 &  \phm0.03481370 & \phone \,-0.01010100 \\
      \hline                                                                               
      &$\gamma_3$          & $[1,3,2,-3]$            &     -0.00216685 & \phone \,-0.10400256 \\
      \hline                                                                               
      &$i\sigma_{12}$      & $[1,2]$                 &     -0.01946887 & \phm \, 0.15803834 \\
      \hline                                                                               
      &$i\sigma_{13}$      & $[1,3,2,-3]$            &  \phm0.00035914 & \phm \, 0.05117233 \\
      \hline                                                                               
      &$i\sigma_{34}$      & $[1,2,3,4,-3,-4]$       &     -0.02584512 & \phone \,-0.05780439 \\
      \hline                                                                               
      &$\gamma_5$          & $[1,2,3,4,-3,-4]$       &  \phm0.00833421 & \phone \,-0.23649769 \\
      \hline                                                                               
      &$\gamma_1 \gamma_5$ & $[1,2,3,4,-3,-4]$       &     -0.01619093 & \phone \,-0.01034868 \\
      \hline                                                                               
      &$\gamma_3 \gamma_5$ & $[1,4,2,-4]$            &  \phm0.02107499 & \phm \, 0.07475660 \\ 
      \hline\hline                                                                         
      (1110) & $1$         & $[1,2,3]$               &     -0.01355233 & \phone \,-0.00847175 \\
      \hline                                                                               
      &$\gamma_1$          & $[1,2,3]$               &  \phm0.01221388 & \phm \, 0.01717281 \\
      \hline                                                                               
      &$\gamma_4$          & $[1,2,4,3,-4]$          &     -0.00696166 & \phm \, 0.08095624 \\
      \hline                                                                               
      &$i\sigma_{12}$      & $[1,2,3]$               &     -0.01690089 & \phm \, 0.04723674 \\
      \hline                                                                               
      &$i\sigma_{14}$      & $[1,4,2,-4,3]$          &  \phm0.00098380 & \phm \, 0.04858656 \\
      \hline                                                                               
      &$\gamma_5$          & $[1,4,2,-4,3]$          &     -0.00265353 & \phm \, 0.03418916 \\
      \hline                                                                               
      &$\gamma_1 \gamma_5$ & $[1,4,2,-4,3]$          &     -0.00878439 & \phone \,-0.01941618 \\
      \hline                                                                               
      &$\gamma_4 \gamma_5$ & $[1,2,3]$               &  \phm0.00614657 & \phm \, 0.19247807 \\
      \hline\hline                                                                         
      (1111) & $1$         & $[1,2,3,4]$             &     -0.00622566 & \phone \,-0.00847456 \\
      \hline                                                                                
      &$\gamma_1$          & $[1,2,3,4]$             &  \phm0.00453820 & \phm \, 0.01531472 \\
      \hline                                                                               
      &$i\sigma_{12}$      & $[1,2,3,4]$             &     -0.00525155 & \phone \,-0.15768691 \\
      \hline                                                                               
      &$\gamma_5$          & $[1,2,3,4]$             &     -0.20323184 & \phm \, 3.11220088 \\
      \hline                                                                               
      &$\gamma_1 \gamma_5$ & $[1,2,3,4]$             &  \phm0.00347359 & \phone \,-0.23403519 \\
      \hline
    \end{tabular}
    \caption{{}The reference paths and the corresponding couplings for the different $\Gamma_0$'s
      and offsets of the hypercubic parametrization $\Dpar$ used on minimized $\beta \approx 3.0$ 
      configurations. It is obtained from the overlap reparametrization step discussed in Section 
      \ref{subsec:parametrization_details}. It is used in connection with a 2-level APE smearing 
      with coefficients $c_1 = -0.10$ and $c_2 = -0.70$ (see Section \ref{sec:smearing}).}
    \label{tab:final_parametrization_Dfp}
  \end{center}
\end{table}

\begin{table}[h]
  \begin{center}
    \begin{tabular}{|c|c|c|c|c|}
      \hline
      offset & $\Gamma_0  $        &  reference path $l_0$  &  $c_0$     &  $c_1$  \\
      \hline\hline
      (0000) & $1$                 & $[]$                    &  \phm2.83651159 & -1.40064002 \\
             &             & $[1,2,-1,-2]$           & -0.22445666 &  \phm1.98366915 \\           
      \hline                                                                   
      &$\gamma_1$          & $[1,2,-1,-2]$           & -0.12076995 & -0.42014243 \\
      \hline                                                                   
      &$i\sigma_{12}$      & $[1,2,-1,-2]$           & -0.03056282 & -0.07057469 \\
      \hline                                                                   
      &$\gamma_5$          & $[1,2,-1,-2,3,4,-3,-4]$ & -0.03056051 &  \phm0.31628438 \\
             &             & $[1,2,3,4,-1,-2,-3,-4]$ &  \phm0.64742088 &  \phm0.40015556 \\           
      \hline                                                                   
      &$\gamma_1 \gamma_5$ & $[1,2,-1,-2,3,4,-3,-4]$ &  \phm0.15555202 & -0.60960766 \\
              &            & $[2,1,-2,-1,3,4,-3,-4]$ &  \phm0.03179317 &  \phm1.99545374 \\ 
      \hline\hline
      (1000) & $1$                 & $[1]$                   & -0.17409523 & -0.02912516 \\
             &             & $[2,1,-2]$              &  \phm0.02792864 & -0.04200253 \\
             &             & $[2,3,1,-3,-2]$         & -0.00981706 &  \phm0.13878744 \\
      \hline
      &$\gamma_1$          & $[1]$                   &  \phm0.06080162 & -0.10246365 \\
              &            & $[2,1,-2]$              &  \phm0.04196345 &  \phm0.16430626 \\
      \hline
      &$\gamma_2$          & $[1,2,3,-2,-3]$         & -0.07123397 & -0.09855541 \\
      \hline                                                                   
      &$i\sigma_{12}$      & $[2,1,-2]$              &  \phm0.00687342 &  \phm0.05725091 \\
      \hline                                                                   
      &$i\sigma_{23}$      & $[1,2,3,-2,-3]$         & -0.04266141 & -0.06541919 \\
      \hline                                                                   
      &$\gamma_5$          & $[2,1,-2,3,4,-3,-4]$    &  \phm0.04175240 & -0.66992860 \\
      \hline                                                                   
      &$\gamma_1 \gamma_5$ & $[2,1,-2,3,4,-3,-4]$    & -0.04050605 &  \phm0.02898812 \\
      \hline                                                                   
      &$\gamma_2 \gamma_5$ & $[1,3,4,-3,-4]$         &  \phm0.01942004 &  \phm0.06487760 \\
      \hline\hline
      (1100) & $1$                 & $[1,2]$                 & -0.04134473 & -0.01726090 \\
      \hline                                                                   
      &$\gamma_1$          & $[1,2]$                 &  \phm0.03393463 &  \phm0.04718603 \\
      \hline                                                                   
      &$\gamma_3$          & $[1,3,2,-3]$            &  \phm0.00194291 & -0.00255123 \\
      \hline                                                                   
      &$i\sigma_{12}$      & $[1,2]$                 & -0.01086183 & -0.00693550 \\
      \hline                                                                   
      &$i\sigma_{13}$      & $[1,3,2,-3]$            & -0.00116166 &  \phm0.05587536 \\
      \hline                                                                   
      &$i\sigma_{34}$      & $[1,2,3,4,-3,-4]$       & -0.04012812 & -0.00457675 \\
      \hline                                                                   
      &$\gamma_5$          & $[1,2,3,4,-3,-4]$       & -0.01627782 &  \phm0.34101473 \\
      \hline                                                                   
      &$\gamma_1 \gamma_5$ & $[1,2,3,4,-3,-4]$       & -0.01478707 & -0.11230535 \\
      \hline                                                                   
      &$\gamma_3 \gamma_5$ & $[1,4,2,-4]$            &  \phm0.01667895 &  \phm0.14258977 \\ 
      \hline\hline
      (1110) & $1$                 & $ [1,2,3]$              & -0.01219653 & -0.03212035 \\
      \hline                                                                   
      &$\gamma_1$          & $ [1,2,3]$              &  \phm0.01204799 &  \phm0.02121544 \\
      \hline                                                                   
      &$\gamma_4$          & $[1,2,4,3,-4]$          &  \phm0.00081979 &  \phm0.04339593 \\
      \hline                                                                   
      &$i\sigma_{12}$      & $[1,2,3]$               & -0.00948029 & -0.06624570 \\
      \hline                                                                   
      &$i\sigma_{14}$      & $[1,4,2,-4,3]$          & -0.00300787 &  \phm0.05319220 \\
      \hline                                                                   
      &$\gamma_5$          & $[1,4,2,-4,3]$          & -0.00900855 &  \phm0.14650243 \\
      \hline                                                                   
      &$\gamma_1 \gamma_5$ & $[1,4,2,-4,3]$          & -0.00253705 & -0.08197206 \\
      \hline                                                                   
      &$\gamma_4 \gamma_5$ & $[1,2,3]$               &  \phm0.00112681 &  \phm0.11754621 \\
      \hline\hline
      (1111) & $1$                 & $[1,2,3,4]$              &  \phm0.00119452 & -0.04279686 \\
      \hline
      &$\gamma_1$          & $[1,2,3,4]$              &  \phm0.00613421 &  \phm0.00906434 \\
      \hline                                                                    
      &$i\sigma_{12}$      & $[1,2,3,4]$              &  \phm0.00474526 & -0.07095205 \\
      \hline                                                                    
      &$\gamma_5$          & $[1,2,3,4]$              & -0.23018075 &  \phm0.71623195 \\
      \hline                                                                    
      &$\gamma_1 \gamma_5$ & $[1,2,3,4]$              &  \phm0.00053879 & -0.07524030 \\
      \hline
    \end{tabular}
    \caption{{}The reference paths and the corresponding couplings for the different $\Gamma_0$'s
      and offsets of the hypercubic parametrization $\Dpar$ used in the production runs.}
    \label{tab:final_parametrization_Dfp}
  \end{center}
\end{table}

\cpages
\chapter{Implementation of the Overlap Dirac Operator with $R \ne 1/2$}
\label{cha:implementation}

In this appendix we give an overview of the details of the
implementation of Neuberger's construction with a general $R$, as
defined in eqs.~\eqref{eq:overlap_DTR_simple} and
\eqref{eq:overlap_DTR_not_simple}. We assume that the operator $R$ from the GW relation eq.~\eqref{eq:GW}
commutes with $\gamma_5$, is trivial in
Dirac space, local, invertible and hermitian.  The main problem in
numerical simulations is the fact that the general overlap
construction, starting from a Dirac operator $D$ and the 
operator $R$, involves the square
root of the operator $R$ or the inverse of its square root. This
becomes numerically an expensive part of the whole construction and
therefore one would definitely like to circumvent the calculation of
this square root as often as possible.  In the following we give the
general ideas, how one can perform calculations with only very few
applications of the square root of $R$, if any at all. These ideas
that are based on a simple change of basis have been successfully been
implemented in all our codes where the chiral properties of $\Dpar$ or
$\Dtpar$ had to be improved.

\section{Change of Basis}
\label{sec:change_fo_basis}

Let us first of all recall the facts about the change of basis in
linear algebra, before we make use of it in the overlap construction.
We denote quantities in the changed basis by primes.  In matrix form a
linear map of the space $C^n$ onto itself in a certain basis can be
written as
\begin{equation}
  \label{eq:linear_map}
  y =  A x \, , 
\end{equation}
where $A$ is a $n\times n$-matrix. A change of basis can be expressed
also in terms of a matrix $S\in GL(C^n)$. This matrix describes the
old basis ${\textfrak B}$ in terms of the new basis ${\textfrak
  B}^\prime$. This makes clear that a vector $x$ under a change of
basis becomes
\begin{equation}
  \label{eq:linear_map}
  x^\prime =  S x \,. 
\end{equation}
The linear map itself does not depend on the choice of basis and
therefore we have
\begin{align}
  \label{eq:change_of_linear_map}
  y^\prime &= A^\prime x^\prime\, \notag \\
  A^\prime &= S A S^{-1}\, .
\end{align}
Next we have a look at the effect a change of basis has on hermitian
bilinear forms. In matrix form a hermitian form in a given basis can
be written as:
\begin{equation}
  \label{eq:bilinear_form}
  f(x,y) \dot{=} \langle x,y\rangle_B = x^\dagger B y \, . 
\end{equation}
Using again the fact that the choice of basis does not have an
influence on the hermitian form, i.e.~
\begin{equation}
  \label{eq:invariance_of_hermitian_form}
  \langle x,y \rangle_B = \langle x^\prime,y^\prime \rangle_{B^\prime} \, ,
\end{equation}
one sees that the transformation behaviour of $B$ is:
\begin{gather}
  \label{eq:change_of_linear_map}
  x^\dagger B y = x^{\prime \dagger} (S^{-1})^\dagger B^{\prime} S^{-1} y^{\prime} \notag \\
  B^\prime = (S^{-1})^\dagger B S^{-1}\, .
\end{gather}
This means that the standard inner product, where $B = 1$, under a
change of basis goes over into
\begin{equation}
  \label{eq:standard_inner}
  \langle x^\prime,y^\prime \rangle_{B^\prime} = x^{\prime \dagger} (S S^\dagger)^{-1} y^{\prime}\,,
\end{equation}
which means that the form of the standard inner product is only
invariant under unitary transformations, i.e.~under a change to a
different orthonormal basis.

With the form of the inner product also the form of the norm changes
under a change of basis. The norm of a vector is defined through the
standard inner product in an orthonormal basis:
\begin{equation}
  \label{eq:norm}
  ||x|| \dot{=} \sqrt{ x^\dagger x} \, .
\end{equation}
Under a change of basis this amounts to
\begin{equation}
  \label{eq:norm_change_of_basis}
  ||x^\prime|| =  \sqrt{ x^{\prime \dagger} (S S^\dagger)^{-1} x^{\prime}} \, ,
\end{equation}
which leaves the norm of the vector invariant under the
transformation.

\section{Application to the Overlap Expansion}
\label{application_to_overlap}
 
In this section we discuss the use of a specific change of the basis,
which will be used to get rid of the square roots of $R$ in the
overlap construction. First we explain the benefits of this change of
basis for the operator ${\cal D}$ and later for $\D$.

\subsection{Tricks for ${\cal D}$}
\label{subsec:DT}

The overlap formula in terms of $D$ and $R$ is given by:
\begin{align}
  \label{eq:overlap_DTR}
  {\cal D} &= 1 - A (A^\dagger A)^{-1/2} \notag\\
  A &= 1 + s - (2 R)^{1/2} D (2 R)^{1/2} \, ,
\end{align}
where the square root of the operator $A^\dagger A$\footnote{Even though the discussion 
is done with the square root of $A^\dagger A$, the statements are also valid for approximations
to the matrix sign function $\epsilon$.} has to be
approximated in a certain way, as discussed in Chapter \ref{cha:overlap}.
The approximation schemes used in the literature \cite{Hernandez:1998et,
Bunk:1998wj,Hernandez:2000sb,Edwards:1998yw,Neuberger:1998my,vandenEshof:2001hp
,Eshof:2002ms} have in common that the
square root term is rewritten in a more or less complicated way as a
function $f(A^\dagger A)$ that under a change of basis transforms like
$S f(A^\dagger A) S^{-1} = f( S A^\dagger S^{-1} S A S^{-1})$. This
fact can now be used to rewrite the overlap Dirac operator in a new
basis, with the change of basis given by\footnote{$S = (2 R)^{1/2}$
  is also possible as change of basis, but it has the drawback
  that the inner product in this basis is given by $x^\dagger (2
  R)^{-1} y$, while for the choice in the text it is $x^\dagger 2 R
  y$, which is clearly simpler to handle numerically.}
\begin{equation}
  \label{eq:explicit_change_of_basis}
  S = (2 R)^{-1/2} \, .
\end{equation}
In this basis the operator $A^\prime$ takes the form:
\begin{equation}
  \label{eq:A_prime}
  A^\prime = 1 + s -  D 2 R
\end{equation}
and the overlap Dirac operator
\begin{equation}
  \label{eq:DT_prime}
  {\cal D}^\prime = 1 -  (1 + s - D 2 R) f(( 1 + s - D^\dagger 2 R)(1 + s - D 2 R)) \, .
\end{equation}
This means that the multiplication of ${\cal D}$ times a vector $x$
can be done as a sequence of the following three steps:
\begin{enumerate}
\item Change the vector $x$ to the new basis: $x^\prime = S x$.
\item Multiply $x^\prime$ with ${\cal D}^\prime$ from
  eq.~\eqref{eq:DT_prime}: $y^\prime = {\cal D}^\prime x^\prime$.
\item Change the resulting vector $y$ back to the old basis: $y =
  S^{-1} y^\prime$.
\end{enumerate}
At the first sight, it doesn't seem as we would have gained much, because we
are still left with a multiplication of $(2 R)^{1/2}$ and an other one
with $(2 R)^{-1/2}$. The crucial point is now that one typically
wants to calculate quantities like propagators, eigenvectors or
eigenvalues. The calculation of eigenvalues anyway becomes very simple
as the eigenvalues are invariant under a change of basis and therefore
we can simply calculate the eigenvalues of ${\cal D}^\prime$ without
ever having to change the basis. For the calculation of eigenvectors
and propagators the story is a little more involved, but here the
crucial point is that Krylov space methods involved in the inversion
of a matrix or the calculation of eigenvectors do not depend on the
choice of the basis and therefore the algorithms can simply be
rewritten in a arbitrarily chosen basis by adjusting the dot products, the norms and the matrix
vector multiplications. This finally makes that the
sequence of steps for the calculation of e.g.~a propagator is still
similar to the one shown before, however, step 2 is now replaced by
typically hundreds of multiplications with ${\cal D}^\prime$. That one
still has to perform steps 1 and 3 becomes negligible in terms of
computer time.

\subsection{Tricks for $\D$}
\label{subsec:D}

In the overlap construction $\D$ is defined as
\begin{align}
  \label{eq:overlap_DR}
  \D &= (2 R)^{-1/2} (1 -
  A (A^\dagger A)^{-1/2}) (2 R)^{-1/2} \notag \\
  A &= 1 +s - (2 R)^{1/2} D (2 R)^{1/2} \, .
\end{align}
In terms of the basis change matrix $S = (2 R)^{-1/2}$ the
relation between $\D$ and ${\cal D}$ is:
\begin{equation}
  \label{eq:D_DT_relation}
  \D = S {\cal D} S 
\end{equation}
and using the representation of ${\cal D}$ in the changed basis we
have
\begin{equation}
  \label{eq:D_DT_relation}
  \D = {\cal D}^\prime S^2 =  {\cal D}^\prime (2 R)^{-1} \, .
\end{equation}
In this form the multiplication of $\D$ with a vector involves no calculation of the
square root of $R$.  The only inconvenience is the inverse
of the operator $R$. However, one of the most important and also most expensive numerical tasks is the
calculation of propagators; and for this particular calculation the calculation of the inverse of $R$
is not even needed, because we have
\begin{equation}
  \label{eq:D_inverse} 
   \D^{-1} = 2 R ({\cal D}^\prime)^{-1} = 2 R \big[1 -
  (1 - D 2 R) f((1 - D^\dagger 2 R)(1 - D 2 R))\big]^{-1} \, ,
\end{equation}
where the function $f$ is again an approximation to the square root of
$A^\dagger A$ defined in the previous section and therefore we simply
have to invert the same operator as for the calculation of the inverse 
of ${\cal D}^\prime$. It is very
important to note that here one does actually not perform a change of
basis, one just makes use of the simpler structure of ${\cal
  D}^\prime$. Hence, contrarily to the case of $\Dt$, the algorithms must 
not be adapted to another choice of basis.

One of the biggest advantages of writing $\D$ in the form
\eqref{eq:D_DT_relation} is that its massive version (see Section \ref{sec:general_overlap} for more details)
\begin{equation}
  \label{eq:D_massive_prime} 
  \D(m) = \Big(1 -\frac{m}{2}\Big)\D(0) + m (2R)^{-1}
\end{equation}
can be written as
\begin{equation}
  \label{eq:D_massive_trick}
  \D(m) = \Big[\Big(1 -\frac{m}{2}\Big){\cal D}^\prime + m \Big](2 R)^{-1} = {\cal D}^\prime(m) (2 R)^{-1}
\end{equation}
and in this form it can be used in the multi-mass Krylov space
inverters \cite{Frommer:1995ik, Jegerlehner:1996pm}.  The reason for
this is clearly that ${\cal D}^\prime(m)$ and ${\cal D}(m)$, respectively, have the shift structure which
is needed in multi-mass Krylov space inverters \cite{Edwards:1998wx}, because
\begin{equation}
  \label{eq:shift_structure_multimass}
  {\cal D}^\prime(m) = \Big(1 -\frac{m}{2}\Big){\cal D}^\prime(0) + m
\end{equation}
can be rewritten in the form
\begin{gather}
  \label{eq:shift_multimass_structure_explicit}
  {\cal D}^\prime(m) = \alpha(m)( {\cal D}^\prime(0) + \mu(m))\, , \\
  \intertext{with}
  \alpha(m) = 1 -\frac{m}{2} \quad \text{and} \quad \mu(m)= \frac{m}{1-\frac{m}{2}}
\end{gather}
and therefore the inverse is
\begin{equation}
  \label{eq:inverse_linear_shift}
  {\cal D}^\prime(m)^{-1} = \alpha(m)^{-1}( {\cal D}^\prime(0) + \mu(m))^{-1}\, ,
\end{equation}
which makes the linear shift structure obvious. The corresponding expression for $\D(m)^{-1}$ is
\begin{equation}
  \label{eq:inverse_linear_shift}
  \D(m)^{-1} = \alpha(m)^{-1} 2 R ( {\cal D}^\prime(0) + \mu(m))^{-1}\, ,
\end{equation}
which allows again the use of a multi-mass solver.      
\cpages
\chapter{Details of the Determination of the Low-Energy Constant $\Sigma$}
\label{cha:details_of_sigma}

In this appendix we give a detailed description of our determination
of the low-energy constant $\Sigma$ in quenched QCD from the technical
point of view. We focus especially on the calculation of the trace of
the subtracted quark propagator, which is used to extract $\Sigma$.

\section{Calculating the Trace}
\label{sec:trace_calculation}

The finite-volume and finite-mass behaviour of $\langle S
\rangle_{m,V,Q}$ (see Chapter \ref{cha:condensate} for definitions)
can be extracted from the trace
\begin{equation}
  \label{eq:sigma_trace}
 \langle S \rangle_{m,V,Q} = \frac{1}{V}{\rm Tr}^\prime \bigg[(\D(m) 2 R)^{-1} -
 \frac{1}{2} \bigg] \, ,
\end{equation}
where the prime denotes the trace without the contribution of the zero
modes\footnote{Actually, with zero modes we mean the modes that in the
  limit $m \to 0$ become zero modes.}, because these cause a divergence
in the quantity $\langle S \rangle_{m,V,Q}/m$ which is a topological finite volume 
effect. The contribution of the zero modes is exactly known in the analytical formula
for $\langle S \rangle_{m,V,Q}$ (see eq.~\eqref{eq:bessel_sigma}, where the corresponding divergence
is given for $\langle \bar \psi \psi \rangle_{M,V,Q}$) and therefore we can compare the 
subtracted values for $\langle S \rangle_{m,V,Q}/m$. The calculation of the
full trace in eq.~\eqref{eq:sigma_trace} is computationally too expensive to be done for lattice
sizes used in real simulations.  Hence, one has to resort to other
techniques. Stochastic estimators with random sources are the method
of choice in cases, where the operator of which one wants to calculate
the trace is essentially dominated by its diagonal elements. This is
actually the case for the operator $(\D(m) 2 R)^{-1} - 1/2$.

Let us be a little bit more specific about this method: We take a set
of $N$ random vectors $ |\eta^{(\alpha)} \rangle, \alpha = 1,\ldots,N$
such that their elements $\eta^{(\alpha)}_i, i=1,\ldots,n$ have the
properties
\begin{align}
  \label{random_vectors}
  \lim_{N\to\infty} \frac{1}{N} \sum_{\alpha=1}^{N} \eta^{(\alpha)}_i \dot{=} \langle \eta_i \rangle &= 0 \notag\\
  \lim_{N\to\infty} \frac{1}{N} \sum_{\alpha=1}^{N}
  {\eta^{(\alpha)}_i}^* \eta^{(\alpha)}_j \dot{=} \langle \eta_i^*
  \eta_j \rangle &= \delta_{ij} \, ,
\end{align}
where the brackets denote the expectation value.  These properties are
fulfilled by, for example, Gaussian or $Z(2)$ (see \cite{Dong:1994pk}
for details) random number distributions. Using vectors whose elements
are taken out of such a random number distribution one can show that
trace of a $n\times n$ matrix $M$ can be written as
\begin{multline}
  \label{trace_estimate}
  {\rm Tr}\, M = \lim_{N\to\infty} \frac{1}{N} \sum_{\alpha=1}^{N}
  \langle \eta^{(\alpha)}| M | \eta^{(\alpha)} \rangle =
  \lim_{N\to\infty} \frac{1}{N} \sum_{\alpha=1}^{N} \sum_{i,j=1}^{n} M_{ij} (\eta^{(\alpha)}_i)^* \eta^{(\alpha)}_j =\\
  \lim_{N\to\infty} \frac{1}{N} \sum_{\alpha=1}^{N}\big(\sum_{i=1}^{n}
  M_{ii} (\eta^{(\alpha)}_i)^* \eta^{(\alpha)}_i
  + \sum_{i \ne j}^{n} M_{ij} (\eta^{(\alpha)}_i)^* \eta^{(\alpha)}_j\big) = \\
  \sum_{i=1}^{n} M_{ii} \lim_{N\to\infty} \frac{1}{N}
  \sum_{\alpha=1}^{N} (\eta^{(\alpha)}_i)^* \eta^{(\alpha)}_i +
  \sum_{i \ne j}^{n} M_{ij} \lim_{N\to\infty} \frac{1}{N}
  \sum_{\alpha=1}^{N} (\eta^{(\alpha)}_i)^* \eta^{(\alpha)}_j =
  \sum_{i}^{n} M_{ii}\, ,
\end{multline}
where we have used the properties from eq.~\eqref{random_vectors} to make
the reduction to the sum over the diagonal elements of $M$.

In order to calculate the trace without the zero mode contributions
there are two different ways one can proceed. The most obvious way is
to calculate the zero modes and to project them out of the random
vectors, using a projection operator that projects orthogonal to the subspace of the zero modes.
This direct method, however, shows
to be rather expensive because the precision to which the Dirac
operator has satisfy to the GW relation has to be very high in order that
this method works properly \cite{Hernandez:2000sb} and therefore, we
apply the second method, which makes use of the definite chirality of
the zero modes of a GW Dirac operator (see e.g.~Section
\ref{subsec:GW}). The crucial point is that one can calculate
the trace in the chiral sector that has opposite chirality with
respect to the zero modes. This is achieved by using only random
vectors whose chirality are opposite to the zero modes. The chiral
random vectors from one single sector obviously no longer form a complete basis. But the contributions
of the two chiral sectors are equal, apart from the contributions of
the real modes of ${\cal D}$\footnote{For reasons of simplicity we use
  the fact that ${\rm Tr}^\prime \big[(\D(m) 2R)^{-1} - 1/2 \big]$ is
  equal to ${\rm Tr}^\prime \big[{\cal D}(m)^{-1} - 1/2 \big]$ to do
  the derivation in terms of ${\cal D}$. The equality of the two
  traces can be shown easily using the change of basis used in
  eq.~\eqref{eq:explicit_change_of_basis} and the invariance of the
  trace under a change of basis.}. In order to show this we use a
complete set of eigenstates $| \psi_i \rangle, i=1,\ldots,N$ of ${\cal
  D}$\footnote{Note that the operator ${\cal D}$ is normal and
  therefore a complete set of eigenstates exists.}  and define the
eigenstates projected to the two chiral sectors
\begin{align}
  | \psi_i^{+} \rangle &= \frac{1}{2} (1+\gamma_5) | \psi_i \rangle\,, \\
  | \psi_i^{-} \rangle &= \frac{1}{2} (1-\gamma_5) | \psi_i \rangle
  \,.
\end{align}
Having defined these projected eigenstates we can evaluate the
contributions of the two chiral sectors to the trace of the operator
$M = {\cal D}(m)^{-1} - 1/2$
\begin{align}
  \label{eq:chiral_pm}
  \sum_{i=1}^N \langle \psi^{+}_i| M | \psi^{+}_i \rangle &=
  \frac{1}{4}
  \sum_{i=1}^N \langle \psi_i| M + \gamma_5 M + M \gamma_5 + M^\dagger| \psi_i \rangle =\\
  \frac{1}{2} ({\rm Tr}\, M  + {\rm Tr}\, \gamma_5 M)\,, \\
  \sum_{i=1}^N \langle \psi^{-}_i| M | \psi^{-}_i \rangle &=
  \frac{1}{4}
  \sum_{i=1}^N \langle \psi_i| M - \gamma_5 M - M \gamma_5 + M^\dagger| \psi_i \rangle = \\
  \frac{1}{2} ({\rm Tr}\, M - {\rm Tr}\, \gamma_5 M)\,.
\end{align} 
Hence the difference between the trace in the two chiral sectors is
given by ${\rm Tr}\, \gamma_5 M$ which for $M={\cal D}(m)^{-1} - 1/2$
is exactly the contribution of the zero modes, since the contributions
of all the non-real eigenvalues cancel as they come in pairs of
opposite chirality and the contribution of the modes at $\lambda = 2$
are cancelled by the subtraction of $1/2$. Thus for GW fermions we can
indeed take the trace in the chiral sector opposite to the zero modes
and then double the result at the end to get the full trace in
eq.~\eqref{eq:sigma_trace}.

Technically, there remains still a problem, which is that the zero
modes are only exactly chiral when the GW relation is fulfilled
exactly and this is not possible for an ultralocal Dirac operator
\cite{Bietenholz:1999dg, Horvath:1999bk,Horvath:2000az}. In order to
make use of the method for the calculation of the trace discussed
above, we rewrite the trace from eq.~\eqref{eq:sigma_trace} as follows
\begin{multline}
  \label{eq:trace_rewriting}
  {\rm Tr}^\prime \bigg[(\D(m) 2 R)^{-1} - \frac{1}{2} \bigg] =
  {\rm Tr}^\prime \bigg[\D(m)^\dagger \big(\D(m) 2 R \D(m)^\dagger\big)^{-1} - \frac{1}{2} \bigg] =\\
  {\rm Tr}^\prime \bigg[\D(m)^\dagger \big(\D(m) +\D(m)^\dagger\big)^{-1} -
  \frac{1}{2} \bigg] =\\ {\rm Tr}^\prime \Bigg[\D(m)^\dagger 2 R \big( \big(\D(m)
  +\D(m)^\dagger \big) 2 R \big)^{-1} - \frac{1}{2} \Bigg] \, ,
\end{multline}
where the last line of the rewriting is only exact, when $\D(m)$
satisfies the GW relation.  We use the form
of the trace from eq.~\eqref{eq:trace_rewriting} in our calculation,
because the zero modes of $\D(m) +\D(m)^\dagger$ have definite
chirality and therefore the divergent contribution of the zero
modes\footnote{In numerical simulations one always has to use
  approximations to GW fermions and therefore the zero modes are not
  exactly zero modes, but real modes with an eigenvalue very close to
  $0$. The statement that the divergent contribution to the trace
  eq.~\eqref{eq:trace_rewriting} is eliminated from the calculation of
  the trace, however, also holds for these approximate zero modes.} is
completely eliminated from the calculation. Furthermore, the insertion
of a $1$ in form of $2 R (2 R)^{-1}$ in eq.~\eqref{eq:trace_rewriting}
has a purely technical reason. It makes that the operator, which has
to be inverted, can be brought in such a form that a multi-mass Krylow
space solver can be used for the inversion (see Appendix
\ref{cha:implementation}).

Forcing the GW relation to hold for a Dirac operator which does not
fulfill it exactly, however, introduces a systematic error in the
calculation of the trace. But it is possible to control this
systematic error by increasing the accuracy of the approximation of
the overlap operator to machine precision and then compare it to the
result of a less accurate approximation.  In our calculations we
choose the accuracy of the appoximation to the overlap operator such
that the systematic error is roughly $2$ orders of magnitude smaller
than the statistical error of our calculation and therefore can safely
be neglected. Figure \ref{fig:condensate_difference} in Chapter
\ref{cha:condensate} shows the dependence of the relative difference
between a high order Legendre expansion and lower order approximations
and clearly shows that the systematic error introduced by the
rewriting of the trace in eq.~\eqref{eq:trace_rewriting} is under
control.  Let us finally remark that after rewriting
the trace it is possible to reduce the precision to which the Dirac operator
used in the calculations has to satisfy the GW relation.  
This is surely supports the hope that for the practical implementation of chiral 
fermions on the lattice the GW relation has not always to be satisfied to
machine precision in order perform chiral measurements with small systematic errors.

\cpages
\end{appendix}

\backmatter
\addcontentsline{toc}{chapter}{Bibliography}
\bibliography{main}
\bibliographystyle{h-physrev4}
\newpage
\cleardoublepage
\newpage

\end{document}